\documentclass[a4paper,11pt]{article}
\usepackage{jheppub} 
\usepackage{comment}
\usepackage{lineno}
\usepackage[dvipsnames]{xcolor}
\usepackage[utf8]{inputenc}  
\usepackage[inkscapelatex=false]{svg}
\usepackage{fix-cm}  
\usepackage{tikz}    
\usepackage{float}
\usepackage{longtable}
\usepackage{array}
\usetikzlibrary{arrows}
\usepackage{tikz-cd} 
\usepackage{tkz-euclide}
\usepackage{circuitikz}
\usepgfmodule{shapes}
\usetikzlibrary{decorations.pathmorphing}
\usetikzlibrary{backgrounds, arrows,calc,shapes,decorations.pathreplacing, automata,positioning}
\usepackage{cancel} 
\usepackage{graphicx}
\usepackage{mathdots}
\usepackage{slashed}
\usetikzlibrary{decorations.markings,backgrounds}
\usetikzlibrary{matrix}
\usetikzlibrary{shapes.misc}
\usepackage[most]{tcolorbox}
\usepackage{hyperref}
\usepackage{amsmath} 
\usepackage{amssymb}
\usepackage{graphicx}
\usepackage{caption}
\usepackage{subcaption}
\usepackage{mathrsfs}
\usepackage{mathtools}

\graphicspath{ {images/} }

\definecolor{CScolor}{rgb}{1.0, 0.0, 0.0} 
\definecolor{BFcolor} {RGB} {0, 158, 96} 
\definecolor{FIcolor}{RGB}{250, 95, 85} 

\definecolor{bcolor}{RGB}{4, 55, 242} 
\definecolor{fcolor}{RGB}{255, 36, 0} 
\definecolor{labelcolor}{HTML}{808080} 

\def\Twall[#1]{%
  \ifx\\#1\\%
    \mathcal{T}_{Wall}%
  \else%
    \mathcal{T}_{Wall}^{(#1)}%
  \fi%
}

\newcommand{\zz}{\mathbb{Z}}

\newcommand{\NN}{\mathcal{N}}

\newcommand{\G}{{\Gamma}}
\newcommand{\Ge}{{\Gamma_{ele}}}

\def\identityN[#1,#2]{
	_{#1}\mathbb{I}_{#2}^{[N]}
}

\def\identityOne[#1,#2]{
	_{#1}\mathbb{I}_{#2}^{[1^N]}
}

\def\DeltaN[#1]{
	\Delta^{[N]}\left( #1 \right)
}

\def\DeltaOne[#1]{
	\Delta^{[1^N]}\left( #1 \right)
}

\def\Dirac[#1,#2]{\left\langle #1,#2 \right\rangle}
\def\Arg[#1]{\text{Arg}\left(#1 \right)}

\DeclareMathOperator{\Tr}{Tr}

\tikzstyle{BPSquivernode}=[circle,draw=black,thick,minimum size=3mm]
\tikzstyle{BPSquiverarrow}=[draw,black,->]

\newcommand{\tikznode}[2]{\relax
\ifmmode%
  \tikz[remember picture,baseline=(#1.base),inner sep=0pt]\node(#1){$#2$};
\else
  \tikz[remember picture,baseline=(#1.base),inner sep=0pt]\node(#1){#2};%
\fi}

\tikzstyle{BFline}=[dashed,black,draw]
\tikzstyle{farrowstyle}=[fcolor,thick,draw]
\tikzstyle{barrowstyle}=[bcolor,thick,draw]

\tikzstyle{flavor}=[rectangle,draw=black,thick,inner sep = 0pt, minimum size = 6mm]
\tikzstyle{manifest}=[rectangle,draw=blue!50,thick,inner sep = 0pt, minimum size = 6mm]
\tikzstyle{gauge}=[circle,draw=black,thick,inner sep = 0pt, minimum size = 6mm] 
\tikzstyle{gauge2}=[circle,draw=black!50,thick,inner sep = 0pt, minimum size = 4.5mm] 
\tikzstyle{gauge3}=[rounded rectangle, draw=black!100, thick, minimum size=5mm] 
\tikzset{->-/.style={decoration={
  markings,
  mark=at position .5 with {\arrow{>}}},postaction={decorate}}}
\tikzset{-<-/.style={decoration={
  markings,
  mark=at position .5 with {\arrow{<}}},postaction={decorate}}}
\tikzstyle{BFline}=[dashed]
  
\def\nodeCS(#1,#2)(#3,#4,#5){ 	
	\node at (#1,#2) (#3) [gauge,black] {#4};
	\draw[CScolor] (#3) ++(6pt,-7pt) node[anchor=west] {\tiny#5};
}

\def\flavorCS(#1,#2)(#3,#4,#5){ 	
	\node at (#1,#2) (#3) [flavor,black] {#4};
	\draw[CScolor] (#3) ++(6pt,-8pt) node[anchor=west] {\tiny#5};
}

\def\arrowBF(#1,#2)(#3){ 
	\begin{scope}[every node/.style={auto, outer sep=-1pt}]
	\path (#1) edge[->-] node[BFcolor,midway] {\tiny#3} (#2);
	\end{scope}
}

\def\arrowBFlr(#1,#2)(#3,#4){ 
	\begin{scope}[every node/.style={auto=#4, outer sep=-1pt}]
	\path (#1) edge[->-] node[BFcolor,midway] {\tiny#3} (#2);
	\end{scope}
}

\def\dottedBF(#1,#2)(#3){ 
	\begin{scope}[every node/.style={auto, outer sep=-1pt}]
	\path (#1) edge[BFline] node[BFcolor,midway] {\tiny#3} (#2);
	\end{scope}
}

\def\farrow(#1,#2,#3){
	\begin{scope}[every node/.style={auto=#3, outer sep=-1pt}]
	\path (#1) edge[->-,fcolor] node[fcolor,midway] {\tiny$\psi$} (#2);
	\end{scope}
}

\def\barrow(#1,#2,#3){
	\begin{scope}[every node/.style={auto=#3, outer sep=-1pt}]
	\path (#1) edge[->-,bcolor] node[bcolor,midway] {\tiny$\phi$} (#2);
	\end{scope}
}

\def\barrowBF(#1,#2,#3,#4){
	\begin{scope}[every node/.style={auto=#3, outer sep=-1pt}]
	\path (#1) edge[->-,bcolor] node[bcolor,midway] {\tiny$\phi$}  node[BFcolor,midway,swap] {\tiny#4} (#2);
	\end{scope}
}

\def\farrowBF(#1,#2,#3,#4){
	\begin{scope}[every node/.style={auto=#3, outer sep=-1pt}]
	\path (#1) edge[->-,fcolor] node[fcolor,midway] {\tiny$\psi$}  node[BFcolor,midway,swap] {\tiny#4} (#2);
	\end{scope}
}

\def\fQED(#1,#2){ 
\begin{tikzpicture}{baseline=(current bounding box).center}
	\node at (0,0) (g) [gauge,black] {$1$};
	\draw[CScolor] (g.south east) ++(-3pt,3pt) node[anchor=north west] {\tiny$#2$};
	\node at (1.5,0) (f) [flavor,black] {$#1$};
	\farrow(g,f,left)
\end{tikzpicture}
}

\def\sQED(#1,#2){ 
\begin{tikzpicture}{baseline=(current bounding box).center}
	\node at (0,0) (g) [gauge,black] {$1$};
	\draw[CScolor] (g.south east) ++(-3pt,3pt) node[anchor=north west] {\tiny$#2$};
	\node at (1.5,0) (f) [flavor,black] {$#1$};
	\barrow(g,f,left)
\end{tikzpicture}
}
\def\bQED(#1,#2){ \sQED(#1,#2) }

\tikzset{cross/.style={cross out, draw=black, minimum size=5*(#1-\pgflinewidth), inner sep=0pt, outer sep=0pt},
cross/.default={2pt}}

\tikzset{snake it/.style={decorate, decoration=snake}}

\tikzset{mid arrow/.style={postaction={decorate,decoration={
        markings,
        mark = at position .55 with {\arrow[#1]{Straight Barb[width=5pt]}}
      }}}}
      
\tikzset{mid arrowsm/.style={postaction={decorate,decoration={
        markings,
        mark = at position .55 with {\arrow[#1]{Straight Barb[width=3pt]}}
      }}}}
      
\tikzset{middx arrowsm/.style={postaction={decorate,decoration={
        markings,
        mark = at position .7 with {\arrow[#1]{Straight Barb[width=3pt]}}
      }}}}
      
\tikzset{midsx arrowsm/.style={postaction={decorate,decoration={
        markings,
        mark = at position .4 with {\arrow[#1]{Straight Barb[width=3pt]}}
      }}}}

\author[a]{Riccardo Comi,}
\author[b,c]{Simone Rota}

\affiliation[a]{Abdus Salam Centre for Theoretical Physics, Imperial College London, London SW7 2AZ, UK}
\affiliation[b]{INFN, Sezione di Trieste, Via Valerio 2, I-34127 Trieste, Italy}
\affiliation[c]{SISSA, Via Bonomea 265, I-34136 Trieste, Italy}

\emailAdd{rcomi@ic.ac.uk} 
\emailAdd{srota@sissa.it}

\title{\boldmath Half-BPS Boundaries and the RG-Wall of $\mathcal{N}=2$ $SU(N)$ SYM
\\
\centering 
\vspace{0.4cm}
\begin{tikzpicture}[scale=0.3,line width=2pt, line cap=round, line join=round]
\draw[xshift=-1.5cm] (60:1) -- (180:1) -- (240:1);
\draw[xshift=0cm]  (-60:1) -- (120:1) -- (60:1) -- (-120:1);
\draw[xshift=2cm]  (-60:1) -- (0:1) -- (60:1) --  (-120:1) -- (180:1) -- (1200:1) ;
\end{tikzpicture}
}

\abstract{We identify the 3d theory that realizes the RG-wall interface of 4d $\mathcal{N}=2$ $SU(N)$ super-Yang-Mills, interpolating between the UV Lagrangian and the IR Seiberg-Witten effective description. The same theory also describes the low-energy boundary condition that corresponds to giving half-BPS Dirichlet boundary condition in the Lagrangian description of 4d $\mathcal{N}=2$ $SU(N)$ super-Yang-Mills.
The theory is a 3d $\mathcal{N}=2$ SCFT that can be obtained as a massive deformation of the $T[SU(N)]$ theory, which is the S-duality interface of 4d $\mathcal{N}=4$ $SU(N)$ SYM. 
As the main validating tests, we match half-indices and discuss non-trivial consistency conditions when colliding such interfaces.

}

\begin{document}
\maketitle
\flushbottom

\section{Introduction and Summary}

Quantum field theories admit natural constructions involving coupled $D$-dimensional bulk theories and $(D-1)$-dimensional theories supported on codimension-one loci. 
These setups
include interfaces between different theories, codimension-1 defects and boundaries, and
provide a framework in which the dynamics of the $D$-dimensional and $(D-1)$-dimensional degrees of freedom are tightly interrelated.

Interesting interfaces involving strongly coupled $(D-1)$-dimensional theories can arise naturally when considering dualities or deformations of the $D$-dimensional theory.
A prominent example is provided by S-duality walls in 4d $\mathcal{N}=4$ super-Yang-Mills \cite{Gaiotto:2008sa,Gaiotto:2008ak,Hosomichi:2010vh}, realized as interacting 3d theories, denoted as $T[G]$, that implement the action of S-duality on supersymmetric boundary conditions for the theory in the bulk. In this sense, S-duality walls provide the realization of S-duality as an interface between two equivalent descriptions of the 4d theory.

A different, naturally occurring, type of interface can be obtained by turning on an RG-flow on half of spacetime. 
The resulting interface, which now bridges between a UV and an IR theory, is denoted as “RG-wall" or “RG-interface" and is the main topic of this paper.

The concept of RG-walls was first formalized for 2d CFTs \cite{Gaiotto:2012np} and the first example for a 4d SQFT was provided by Dimofte, Gaiotto and van der Veen in \cite{Dimofte:2013lba} for $\mathcal{N}=2$ $SU(2)$ gauge theories. 
In the case of pure $SU(2)$ SYM, the RG-wall is described by a simple three-dimensional model with four supercharges, consisting of two free chirals and a background Chern-Simons interaction.
This simple theory naturally appears as the RG-interface between the UV Lagrangian description of $\mathcal{N}=2$ $SU(2)$ SYM and the low energy effective description provided by the Seiberg-Witten (SW) solution in the strong-coupling region of the Coulomb branch \cite{Seiberg:1994rs,Seiberg:1994aj}.

In this work we present the 3d $\mathcal{N}=2$ theory that describes the RG-wall for pure $\mathcal{N}=2$ $SU(N)$ super-Yang-Mills, thus generalizing to higher rank the result for $SU(2)$ of \cite{Dimofte:2013lba}.
The RG-wall theory, that we denote as $\Twall[]$, can be obtained as a mass deformation of the S-duality wall of $\NN=4$ SYM, namely $T[SU(N)]$, followed by gauging some Abelian symmetries.
The particular mass deformation of $T[SU(N)]$ was first introduced in \cite{Benvenuti:2024seb,Benvenuti:2025huk} and breaks 3d SUSY from $\NN=4$ to $\NN=2$, while simultaneously breaking the global symmetry to $SU(N)\times U(1)^{N-1}$.
The resulting 3d theory, denoted as $G[SU(N)]$ in \cite{Benvenuti:2024seb,Benvenuti:2025huk}, has a Lagrangian UV completion in terms of a quiver with unitary gauge groups.
We find that $G[SU(N)]$ implements the RG-wall of $\NN=2$ SYM when coupled to the IR in an electromagnetic (EM) duality frame where the monopoles are electric.
We can then go to the canonical “electric" EM duality frame by an IR duality which induces a Witten's $SL(2,\zz)$ action on the 3d theory \cite{Witten:2003ya}.
In our case, this action effectively ungauges the diagonal $U(1)$s of all the gauge nodes, resulting in the quiver theory in Figure \ref{fig: twall_lagrangian_intro} which is our proposal for the RG-wall of pure $\NN=2$ SYM with gauge group $SU(N)$.

\begin{figure}[t]
    \centering
    \begin{equation*}
    \Twall[] : \quad
    \begin{tikzpicture}[baseline=(current bounding box).center]
        \node at (0,0) (n1) [flavor,black] {$1$};
        \node at (1.5,0) (n2) [gauge,double,black] {$2$};
        \node at (3,0) (n3) [gauge,double,black] {$3$};
        \node at (4.5,0) (n4) {$\cdots$};
        \node at (6,0) (nnm1) [gauge,double,black] {$\scriptstyle N\text{-}1$};
        \node at (7.5,0) (nn) [flavor,black] {$N$};

        \draw[->-] (n1) --  (n2);
        \draw[->-] (n2) --  (n3);
        \draw[-<-] (n3)++(1,0) --  (n3);
        \draw[->-] (nnm1)++(-1,0) --  (nnm1);
        \draw[->-] (nnm1) --  (nn);

    \end{tikzpicture}
    \end{equation*}
    \caption{RG-wall of $\NN=2$ $SU(N)$ pure SYM. 
    Each double circle indicates a $SU(n)$ gauge group, and arrows are chiral fields in the bifundamental representation.
    There are background CS interactions for the global symmetries discussed in the main body of the paper.
    }
    \label{fig: twall_lagrangian_intro}
\end{figure}

This theory can couple to the UV $SU(N)$ gauge field through its $SU(N)$ global symmetry, and can couple to the IR effective theory through the $U(1)^{N-1}$ baryonic global symmetry, together with boundary superpotential interactions.
\\

The same theory implements the IR boundary conditions that are obtained by giving half-BPS Dirichlet b.c.~to the UV gauge field.
Putting $\NN=2$ SYM on a spacetime with boundary, with Dirichlet b.c.~for the gauge field, provides a setup with an $SU(N)$ boundary global symmetry.
Flowing to the IR, in the bulk one obtains the Seiberg-Witten effective description, coupled to the $\Twall[]$ theory at the boundary:
\begin{equation}
    \begin{tikzpicture}
    
    \begin{scope}[xshift=-.5cm]
        \node at (0,0) (UV) {$\begin{array}{c} \text{UV} \\ \\ SU(N) \\ \NN=2 \; \text{SYM}\end{array}$};
        \path[draw] (1.5,-1.5) -- node[midway,right,anchor=west] {Dirichlet} (1.5,1.5) node[above] {$x_3=0$};
    \end{scope}
    
    \begin{scope}[xshift=\linewidth/2]
        \node at (0,0) (IR) {$\begin{array}{c} \text{IR} \\ \\ \text{Seiberg-Witten} \\ \text{effective theory} \end{array}$};
        \path[draw] (2,-1.5) -- node[midway,right,anchor=west] {Neumann$ + \Twall[]$} (2,1.5) node[above] {$x_3=0$};
    \end{scope}

    \begin{scope} [xshift=\linewidth/4]
        \path[draw,->] (0,0) -- node[midway,above] {RG flow} (1.5,0);
    \end{scope}
    
    \end{tikzpicture}
\end{equation}

Here the
$U(1)^{N-1}$ symmetry of $\Twall[]$ couples to the bulk photons, which have Neumann b.c., and the
$SU(N)$ symmetry is mapped to the UV boundary symmetry.

This picture is the main result presented in this work. We provide high-precision checks of this proposal through the matching of half-indices \cite{Gang:2012ff,Dimofte:2011py,Cordova:2016uwk,Gaiotto:2019jvo}, that are a type of conformal blocks \cite{Pasquetti:2011fj,Benini:2012ui,Nieri:2015yia,Doroud:2012xw,Beem:2012mb}.
In the UV the index can be computed via standard localization techniques.
In the IR Seiberg-Witten theory the index can be computed with the techniques developed in \cite{Cordova:2015nma,Cecotti:2015lab,Cordova:2016uwk}, within the framework of the quantum torus algebra of Kontsevic and Soibelman \cite{Kontsevich:2008fj,Gaiotto:2010okc,Dimofte:2009bv,Dimofte:2009tm}.
We show that these two (very different looking) computations match perturbatively up to high order in $q$ for various values of the rank.
\\

There are various interesting directions that one can follow from here. 
A first natural extension of the proposal, as well as an additional check of it, would be to study the analogous problem from the point of view of the half-four-sphere partition function \cite{Pestun:2007rz,Gava:2016oep}. Possibly, studying the mass deformation of the S-dual setup of $\mathcal{N}=4$ SYM \cite{Hosomichi:2010vh}. 
On a more physical note, it would also be interesting to study implications of the RG-wall on 4d $\mathcal{N}=2$ theories. For example, it was shown that RG-walls between 2d minimal models \cite{Gaiotto:2012np} encode the mapping of operators between the two theories. It would be interesting to extend this analysis to the case of RG-walls of 4d SQFTs. Similarly, we expect the RG-wall to play a role in the study of consequences on different possible choices of the global form of the gauge group of 4d $\mathcal{N}=2$ SYM \cite{Gaiotto:2010be,Gaiotto:2014kfa,Aharony:2013hda}, in the Seiberg-Witten theory, see e.g. \cite{DelZotto:2022ras,Argyres:2022kon,Closset:2023pmc,Garding:2023unh,Arias-Tamargo:2023duo}.

Another natural direction is to study analogous boundary problems for different $\mathcal{N}=2$ gauge theories, such as the SQCD and even quiver theories. 
We also expect this broader class of RG-walls to be related to other types of walls, such as the braid duality wall of $\mathcal{N}=2$ SQCD of \cite{LeFloch:2015bto,Garozzo:2019xzi}, duality walls for quiver theories \cite{Bason:2026qbc}, or to the R-flow theories of \cite{Cecotti:2011iy}. It is also quite intriguing to consider the possibility to study theories with lower SUSY such as $\mathcal{N}=1$ SYM and their relation with the domain walls of \cite{Acharya:2001dz}.

A second direction regards the possible geometric construction of the $\Twall[]$ theory. 
In their first appearance, the RG-walls for 4d $\mathcal{N}=2$ $SU(2)$ gauge theories were obtained from the compactification of the 6d $\mathfrak{a}_1$ $\mathcal{N}=(2,0)$ theory on particular framed three-manifolds with boundary \cite{Dimofte:2011ju,Dimofte:2013lba,Dimofte:2011py}. 
It is then quite natural to expect that the $\Twall[]$, our proposal for the RG-wall of $SU(N)$ SYM, can be obtained by a suitable compactification of the 6d $\mathfrak{a}_{N-1}$ $\NN=(2,0)$ theory on the same three-manifolds. 
On more general grounds, it would be interesting to further explore the landscape of 3d $\mathcal{N}=2$ theories resulting from generic compactifications of the 6d $\mathfrak{a}_{N-1}$ theory for $N>1$, for example by relating the $\Twall[]$ theory to the construction of \cite{Dimofte:2013iv}. In this context, the 3d $\mathcal{N}=2$ mirror-like dualities of \cite{Benvenuti:2024seb,Benvenuti:2025huk,Benvenuti:2026usm,Benvenuti:2026xcv}, as well as the alternative IR-dual Lagrangian description of the $\Twall[]$ theory that we provide in this work, could facilitate this task and reveal connections with geometry.
\\

The paper is organized as follows. In Section \ref{sec: bc_proposal} we describe our proposal for the RG boundary problem in $\NN=2$ SYM, providing a heuristic derivation and discussing the details of the proposal itself.
In Section \ref{sec: interfaces} we tackle interface setups and provide consistency checks of our proposal for the RG-wall of $\NN=2$ SYM by considering collision of interfaces.
We discuss a possible connection with the Acharya-Vafa domain walls of $\NN=1$ SYM.
In Section \ref{sec: half-index} we compute the half-indices related to the boundary setups. The perturbative match of such half-indices between the UV and the IR constitutes the most stringent test of our results.
In Section \ref{sec: half_index_interfaces} we provide similar checks in the context of the interfaces.
The appendix \ref{app: 3dwalls} contains a review of the 3d theories of main interest in this paper. We also discuss novel 3d dualities that have not yet appeared in the literature.

\newpage

\section{Bridging between the UV and IR of 4d $\mathcal{N}=2$ SYM} \label{sec: bc_proposal}

In this section we present the main result of this paper, which is an explicit Lagrangian for the 3d $\mathcal{N}=2$ theory describing the RG-wall of 4d $\mathcal{N}=2$ $SU(N)$ super-Yang-Mills. This result generalizes to higher rank the $SU(2)$ case proposed by Dimofte-Gaiotto-van der Veen in \cite{Dimofte:2013lba}. We start by describing half-BPS boundary conditions for $\mathcal{N}=2$ SYM. 

We consider $\NN=2$ SYM on half of flat space $\mathbb{R}^3\times \mathbb{R}_{+}$ where $\mathbb{R}^3$ is parametrized by $x_{0,1,2}$ and $\mathbb{R}_{+}$ is parametrized by $x_3\geq0$, with the boundary sitting at $x_3=0$.
One can define half-BPS boundary conditions for 4d $\NN=2$ gauge theories which preserve 3d $\NN=2$ supersymmetry at the boundary, in the spirit of \cite{Gaiotto:2008ak} (see also \cite{Gaiotto:2010okc,Dimofte:2013lba,Hashimoto:2014nwa,Bason:2025qsw,Bason:2025zpy,Bason:2025sxb,Bason:2026qbc,Erdmenger:2002ex}). 
On the boundary the $SU(2)_R$ R-symmetry breaks to a Cartan $U(1)_R$ and the 4d $\mathcal{N}=2$ vector multiplet reorganize into 3d $\mathcal{N}=2$ vector and chiral multiplets.
A 4d $\NN=2$ vector multiplet has on-shell bosonic field content $(A_\mu, \phi)$, where we parameterize the complex scalar in terms of two real fields as: $\phi = X + iY$.
With a suitable choice of embedding of the 3d $\mathcal{N}=2$ superconformal group into the 4d $\mathcal{N}=2$ one, the 3d $\NN=2$ vector has bosonic field content $(A_{i},X)$, with $i=0,1,2$, and the chiral has bosonic field content $(A_3, Y)$ which combine into a complex scalar\footnote{This 3d chiral multiplet has non-canonical R-charge $0$. It can be thought of as the product of a chiral and an antichiral multiplet with canonical R-charges $\pm 1$, in analogy with the $\NN=4$ case \cite{Gaiotto:2008sa}.}. 
The gaugini similarly split between the two 3d multiplets according to supersymmetry transformations (see \cite{Erdmenger:2002ex,Cordova:2016emh} for more details). 

To preserve half of supersymmetry on the boundary one has to choose consistent b.c.~for the fields sitting in the same 3d supermultiplet.
We consider half-BPS boundary conditions for $\NN=2$ $SU(N)$ SYM which involve Dirichlet boundary conditions for the boundary 3d $\NN=2$ vector multiplet and Neumann for the chiral multiplet. We denote this b.c.~as Dirichlet-like, or simply as $\mathcal{D}$:
\begin{equation}    \label{eq:D_bc_N2}
    \mathcal{D}: 
    \quad
    \left.F_{i, j}\right|_{x_3=0} = 0,
    \quad
    \left.X\right|_{x_3=0} = 0,
    \quad
    \left.\partial_3 Y\right|_{x_3=0} = 0,
    \quad
    \partial_3 A_3|_{x_3=0}=0.
\end{equation}
suitably extended to the fermions as in \cite{Erdmenger:2002ex}.
This setup enjoys a boundary $SU(N)$ global symmetry because of the Dirichlet b.c.~on the gauge field. The corresponding transformations are bulk gauge transformations whose parameter is constant at the boundary, quotient by bulk gauge transformations that are trivial on the boundary.
The corresponding conserved current is $\left.F_{3,i}\right|_{x_3=0}$.

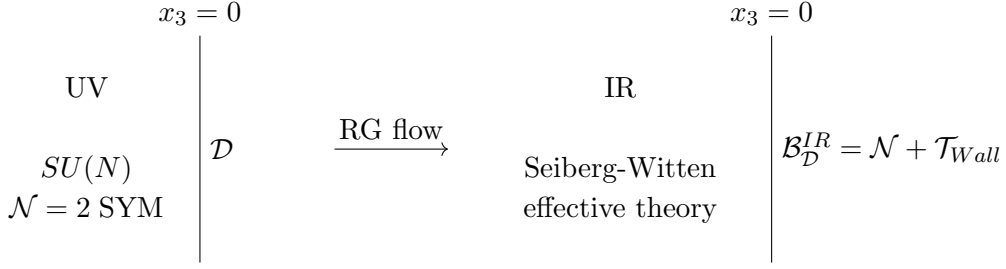
\begin{figure}
    \centering
    \begin{tikzpicture}
    
    \begin{scope}[xshift=.5cm]
        \node at (0,0) (UV) {$\begin{array}{c} \text{UV} \\ \\ SU(N) \\ \NN=2 \; \text{SYM}\end{array}$};
        \path[draw] (1.5,-1.5) -- node[midway,right,anchor=west] {$\mathcal{D}$} (1.5,1.5) node[above] {$x_3=0$};
    \end{scope}
    
    \begin{scope}[xshift=\linewidth/2]
        \node at (0,0) (IR) {$\begin{array}{c} \text{IR} \\ \\ \text{Seiberg-Witten} \\ \text{effective theory} \end{array}$};
        \path[draw] (2,-1.5) -- node[midway,right,anchor=west] {$\mathcal{B}_{\mathcal{D}}^{IR} = \mathcal{N} + \Twall[]$} (2,1.5) node[above] {$x_3=0$};
    \end{scope}

    \begin{scope} [xshift=\linewidth/4]
        \path[draw,->] (0,0) -- node[midway,above] {RG flow} (1.5,0);
    \end{scope}
    
    \end{tikzpicture}
    \caption{A schematic representation of the RG flow for $SU(N)$ SYM with SUSY completion of Dirichlet b.c.~for gauge fields, denoted as $\mathcal{D}$. The IR b.c.~can be described by giving Neumann b.c.~to the $U(1)^{N-1}$ gauge fields in the Seiberg-Witten low energy solution, suitably SUSY completed and denoted as $\mathcal{N}$, and coupling them to a 3d SCFT $\Twall[]$ living at the boundary. The Lagrangian for $\Twall[]$ is presented in Figure \ref{fig:Twall}.}
    \label{fig:schematic_RG}
\end{figure}

Under RG flow, the theory in the bulk flows to strong coupling and, correspondingly, we also expect that the $\mathcal{D}$ b.c.~flows to some b.c.~$\mathcal{B}_{\mathcal{D}}^{IR}$ in the IR, as schematically depicted in Figure \ref{fig:schematic_RG}. 
The goal of this paper is to provide a description of $\mathcal{B}_{\mathcal{D}}^{IR}$.

In the bulk, the effective low energy description of $\mathcal{N}=2$ SYM is described by the Seiberg-Witten solution. At a generic point of the Coulomb branch, as well as at singularities where mutually local BPS hypermultiplets become massless, the low energy theory is an IR free Abelian $\NN=2$ gauge theory\footnote{It would be interesting to study the analogous boundary setup in other special points of the Coulomb branch, such as Argyres-Douglas points \cite{Argyres:1995jj}. We leave this analysis to future work. }. 
One may hope to provide a description of $\mathcal{B}_{\mathcal{D}}^{IR}$ from the perspective of the low energy effective theory. Immediately, we observe that $\mathcal{B}_{\mathcal{D}}^{IR}$ can not be regular (SUSY completion of) Dirichlet or Neumann boundary conditions for the gauge fields, as they would not realize the expected $SU(N)$ global symmetry on the boundary.

In the spirit of \cite{Gaiotto:2008ak}, we propose that $\mathcal{B}_{\mathcal{D}}^{IR}$ can be described by Neumann b.c.~for the gauge fields, which we denote as Neumann-like or simply $\mathcal{N}$, in the low energy $U(1)^{N-1}$ effective theory, with extra 3d degrees of freedom on the boundary coupled to the bulk trough a diagonal gauging of the $U(1)^{N-1}$ symmetry.
Indeed, the non-trivial problem to solve is to identify correctly this 3d $\mathcal{N}=2$ SCFT, which we denote as the $\Twall[]$ theory. Before solving this problem, simply from the consistency of the setup, we can already observe that the $\Twall[]$ theory must possess a $U(1)^{N-1} \times SU(N)$ global symmetry. This is because we expect a $U(1)^{N-1}$ symmetry to be able to couple the theory to the bulk via the diagonal gauging on the boundary, while the $SU(N)$ symmetry provides the expected global symmetry on the boundary to match the expectation given by the UV setup.

The rest of this section is devoted to the identification of the 3d $\mathcal{N}=2$ Lagrangian quiver theory that is expected to flow to the $\Twall[]$ SCFT at the IR fixed point\footnote{We actually identify many possible IR-dual Lagrangian UV descriptions. One is the $\mathcal{N}=2$ mirror-like dual found in \cite{Benvenuti:2024seb,Benvenuti:2025huk,Benvenuti:2026usm}. More duality frames are also discussed in Appendix \ref{app: gsun}.}.
A schematic picture of our proposal is shown in Figure \ref{fig:schematic_RG}, while the identification of the 3d $\mathcal{N}=2$ $\Twall[]$ theory is provided in Subsection \ref{subsec: N=2_bc}\footnote{Further details of the $\Twall[]$ theory can be found in Appendix \ref{app: gsun}.}. While the description of the solution provided in this section remains somehow qualitative, we provide an extensive analysis of consistencies in Section \ref{sec: interfaces}, as well as solid checks in the form of the matching between half-index identities in Section \ref{sec: half-index} and Section \ref{sec: half_index_interfaces}.

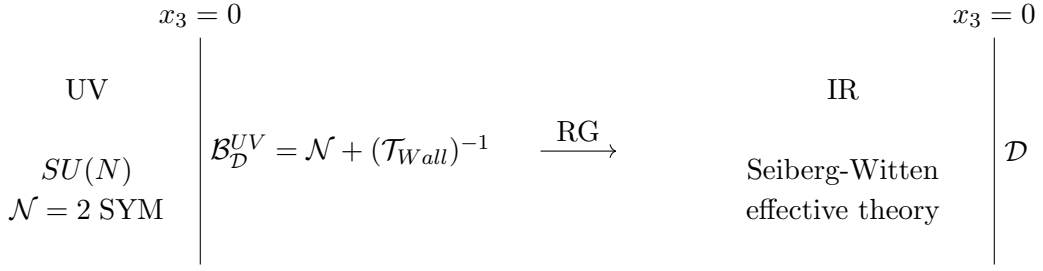
\begin{figure}
    \centering
    \begin{tikzpicture}
    
    \begin{scope}[xshift=0cm]
        \node at (0,0) (UV) {$\begin{array}{c} \text{UV} \\ \\ SU(N) \\ \NN=2 \; \text{SYM}\end{array}$};
        \path[draw] (1.5,-1.5) -- node[midway,right,anchor=west] {$\mathcal{B}_{\mathcal{D}}^{UV} = \mathcal{N} + (\Twall[])^{-1}$} (1.5,1.5) node[above] {$x_3=0$};
    \end{scope}
    
    \begin{scope}[xshift=10cm]
        \node at (0,0) (IR) {$\begin{array}{c} \text{IR} \\ \\ \text{Seiberg-Witten} \\ \text{effective theory}\end{array}$};
        \path[draw] (2,-1.5) -- node[midway,right,anchor=west] {$\mathcal{D}$} (2,1.5) node[above] {$x_3=0$};
    \end{scope}

    \begin{scope} [xshift=6cm]
        \path[draw,->] (0,0) -- node[midway,above] {RG} (1,0);
    \end{scope}
    
    \end{tikzpicture}
    \caption{A schematic representation of the RG flow for $SU(N)$ SYM coupled to the 3d “inverse" RG-wall theory $(\Twall[])^{-1}$ at the boundary. 
    At the end of the RG flow this b.c.~flows to Dirichlet boundary condition for the $U(1)^{N-1}$ gauge fields in the low energy Seiberg-Witten theory.}
    \label{fig:schematic_RG_inverse}
\end{figure}

One can also study the complementary problem, where we impose Dirichlet b.c.~for the gauge fields of the $U(1)^{N-1}$ low-energy effective description and look for a UV completion in terms of b.c.~$\mathcal{B}_{\mathcal{D}}^{UV}$ for $\mathcal{N}=2$ SYM, as schematically represented in Figure \ref{fig:schematic_RG_inverse}. We find that that $\mathcal{B}_{\mathcal{D}}^{UV}$ consists of $\mathcal{N}$ b.c.~for $\NN=2$ SYM plus 3d degrees of freedom coupled to the bulk trough  the gauging of the $SU(N)$ symmetry. 
We find that this theory is the $CP$ transformed of $\Twall[]$ and we denote it as the “inverse" RG-wall $(\Twall[])^{-1}$. 
As the name suggests, this theory is further related to the $\Twall[]$ by a property that schematically reads: $\Twall[] \cdot (\Twall[])^{-1} \sim 1$. Here the $\cdot$ product is a short notation that indicate the diagonal gauging of a global symmetry, while the ``$1$" element indicates an identity-wall: a special theory that implements a defect that is IR-equivalent to the trivial defect.
Indeed, there are two such relations depending on whether we glue $\Twall[]$ and $(\Twall[])^{-1}$ through their $SU(N)$ or $U(1)^{N-1}$ symmetries.
These relations will be made more precise later in Section \ref{sec: interfaces}.

The proposed solutions for $\mathcal{B}_{\mathcal{D}}^{IR}$ and $\mathcal{B}_{\mathcal{D}}^{UV}$ are reminiscent and tightly linked to that proposed by Gaiotto and Witten for S-duality of Dirichlet b.c.~in $\mathcal{N}=4$ SYM \cite{Gaiotto:2008ak}. 
We propose that one can think of the RG-wall as the result of a massive deformation of the S-wall, which breaks half of SUSY.
We will then start with a brief review of the Gaiotto-Witten description of S-duality for Dirichlet b.c.~in $\mathcal{N}=4$ SYM and then proceed to describe the proposal for $\mathcal{N}=2$ SYM.

\subsection{S-duality of Dirichlet b.c.~in $\mathcal{N}=4$ SYM} \label{subsec: N4bc}

Let us consider the maximally supersymmetric setup given by $\mathcal{N}=4$ $SU(N)$ SYM on $\mathbb{R}^3 \times \mathbb{R}_+$ with half-BPS Dirichlet boundary conditions for the gauge fields.
In $\mathcal{N}=4$ SYM the coupling $\tau$ is exactly marginal and we can relate the theory at weak coupling $\text{Im}(\tau) >> 1$ to strong coupling $\text{Im}(\tau') << 1$  using S-duality which maps $\tau \to \tau'=-1/\tau$.
A problem similar to that posed for $\mathcal{N}=2$ SYM previously, consists of starting from $\mathcal{N}=4$ SYM at fixed coupling $\tau$ with a suitable SUSY extension of Dirichlet b.c.~for the gauge field and asking what is the S-dual setup.

The solution was proposed in \cite{Gaiotto:2008ak} using the Type IIB construction of $\mathcal{N}=4$ SYM with boundary conditions and Hanany-Witten brane setups \cite{Hanany:1996ie}. In particular, it was shown that half-BPS Dirichlet b.c.~at fixed coupling $\tau$ are S-dual to half-BPS Neumann b.c.~for $SU(N)$ SYM at coupling $\tau' =-1/\tau$\footnote{Under S-duality the gauge group typically change global form, for example starting from $SU(N)$ we obtain $PSU(N)$ after S-duality. However, we will not discuss subtleties arising from the global structure of the gauge group. Thus we will avoid giving precise details about it.}, coupled to the $T[SU(N)]$ 3d $\mathcal{N}=4$ SCFT, which admits a UV Lagrangian description depicted in Figure \ref{fig:TSUN}. This S-dual boundary setup is schematically reported in Figure \ref{fig:N4_Sduality}. 
\begin{figure}
    \centering
    \begin{tikzpicture}
    
    \begin{scope}[xshift=0cm]
        \node at (0,0) (UV) {$\begin{array}{c} SU(N) \\ \NN=4 \; \text{SYM} \\ \text{w/ $\tau$} \end{array}$};
        \path[draw] (1.5,-1.5) -- node[midway,right,anchor=west] {$\mathcal{D}$} (1.5,1.5) node[above] {$x_3=0$};
    \end{scope}
    
    \begin{scope}[xshift=8cm]
        \node at (0,0) (IR) {$\begin{array}{c} SU(N) \\ \NN=4 \; \text{SYM} \\ \text{w/ $\tau' = -\frac{1}{\tau}$} \end{array}$};
        \path[draw] (1.5,-1.5) -- node[midway,right,anchor=west] {$\mathcal{N} + T[SU(N)]$} (1.5,1.5) node[above] {$x_3=0$};
    \end{scope}

    \begin{scope} [xshift=4cm]
        \path[draw,<->] (0,0) -- node[midway,above] {S-duality} (1,0);
    \end{scope}
    
    \end{tikzpicture}
    \caption{
    S-duality for $\NN=4$ SYM with half-BPS Dirichlet boundary conditions \cite{Gaiotto:2008ak}.
    The boundary global symmetry is $SU(N)$, which arises from Dirichlet b.c.~for the gauge fields on the l.h.s.~and as a global symmetry of the 3d $T[SU(N)]$ theory on the r.h.s.
    }
    \label{fig:N4_Sduality}
\end{figure}

\begin{figure}
    \centering
    \begin{equation} 
    T[SU(N)] \quad : \qquad\qquad
    \begin{tikzpicture}[baseline=(current bounding box).center]
        
        \node at (0,0) (n1) [gauge,black] {$1$};
        \node at (1.5,0) (n2) [gauge,black] {$2$};
        \node at (3,0) (n3) {$\cdots$};
        \node at (4.5,0) (nnm1) [gauge,black] {$\scriptstyle N-1$};
        \node at (6,0) (nn) [flavor,black] {$N$};
        
        \draw[-] (n1.north east)      arc[start angle=-45, end angle=225, radius=3mm];
        \draw[-] (n2.north east)      arc[start angle=-45, end angle=225, radius=3mm];
        \draw[-] (nnm1.north east)    arc[start angle=-45, end angle=225, radius=3.1mm];
        
        \draw[->-] (n1.20) --   (n2.160);
        \draw[-<-] (n1.-20) --  (n2.-160);
        \draw[->-] (n2.20) --  ++(.75,0);
        \draw[-<-] (n2.-20) -- ++(.75,0);
        \draw[->-] (nnm1.160) --  ++(-.75,0);
        \draw[-<-] (nnm1.-160) -- ++(-.75,0);
        \draw[->-] (nnm1.20) --   (nn.160);
        \draw[-<-] (nnm1.-20) --  (nn.-160);
        
    \end{tikzpicture}
    \end{equation}
    \caption{Lagrangian UV completion of the S-duality wall theory $T[SU(N)]$. We use $\NN=2$ quiver notation: each node is an $\NN=2$ vector multiplet and each arrow is a chiral field. Arcs are adjoint chiral fields completing the $\NN=2$ vector multiplets to $\NN=4$, two chirals in conjugate representation form an htpermultiplet. In $\NN=2$ language, the theory has a cubic superpotential coupling hypermultiplets to adjoint chirals. }
    \label{fig:TSUN}
\end{figure}

Let us review the S-dual boundary setups in Figure \ref{fig:N4_Sduality} in more detail, which also allows us to set up our notation.
The bosonic content of $\NN=4$ SYM consists of a gauge field $A_\mu$ and six real scalars $\phi_i$ transforming in the six-dimensional representation of the $SO(6)$ R-symmetry. 
Half-BPS boundary conditions break the R-symmetry to $SO(6) \to SO(3)\times SO(3)$, therefore it is useful to reparameterize these scalar fields into two triplets, each rotated by one of the two $SO(3)$ factors:
\begin{equation}
    \vec{X} = \{\phi_1,\phi_2,\phi_3\} \quad,\quad \vec{Y} = \{\phi_4,\phi_5,\phi_6\} \,.
\end{equation}

There are two natural choices of boundary conditions:
\begin{itemize}
    \item \textbf{NS5-like:} The parallel components of the gauge field $A_{012}$ and the $\vec{X}$ fields satisfy Neumann b.c.~while $A_3$ and $\vec{Y}$ satisfy Dirichlet b.c. From the 3d point of view $A_{012}$ and $\vec{X}$ form a $\mathcal{N}=4$ vector multiplet, while $A_3$ and $\vec{Y}$ form an hypermultiplet. \\
    With a slight abuse of notation we will refer to these b.c.~as Neumann-like or $\mathcal{N}$ in short, in analogy with the $\NN=2$ b.c.~in \eqref{eq:D_bc_N2}. 
    The distinction between the two should always be clear from the context.

    \item \textbf{D5-like:} The parallel components of the gauge field $A_{012}$ and the $\vec{Y}$ scalars satisfy Dirichlet b.c.~while $A_3$ and $\vec{X}$ satisfy Neumann b.c. \\
    With the same slight abuse of no will refer to these b.c.~as Dichelet-like or $\mathcal{D}$ in short.
\end{itemize}
In both cases the 4d $\NN=4$ vector multiplet splits at the boundary into a 3d $\NN=4$ vector multiplet $V^{\NN=4}$ and a 3d adjoint hyper $H$:
\begin{equation}    \label{eq:N4_multiplet_bdy_decomposition}
    \text{4d $\NN=4$ vector }
    \xrightarrow[]{\text{boundary}}
    \begin{cases}
        \text{3d $\NN=4$ vector } V^{\NN=4} = (V,\chi)
        \\
        \text{3d $\NN=4$ hyper } H = (H^+, H^-)
    \end{cases}
\end{equation}
where in parenthesis we wrote the multiplets also in 3d $\NN=2$ components for a choice a $\NN=2$ subalgebra: $V$ is a 3d $\NN=2$ vector and $H^{\pm}$ and $\chi$ are 3d $\NN=2$ chirals. The two half-BPS boundary conditions described above correspond to giving Neumann b.c.~to one of the 3d $\NN=4$ multiplet and Dirichlet to the other. 
The bosonic components of the 3d multiplets are\footnote{In this paper for any hypermultiplet $H$ we denote its chiral and antichiral components as $H^+$ and $H^-$, respectively, under our choice of $\NN=2$ subalgebra and write:
\begin{equation}
    H = (H^+;H^-)
\end{equation}}:
\begin{equation}
    V^{\NN=4} \supset (\overbracket{A_{0,1,2},X_1}^{V} , \overbracket{X_2, X_3}^{\chi}),
    \qquad
    H \supset (\overbracket{A_3, Y_1}^{H^+}; \;\overbracket{Y_2, Y_3}^{H^-})
\end{equation}
While, under the corresponding choice of 4d $\NN=2$ subalgebra, the $\NN=4$ multiplet splits into an $\NN=2$ vector multiplet and an $\NN=2$ adjoint hyper $\Phi$, with the bosonic flieds splitting as:
\begin{equation}
\text{4d $\NN=4$ vector} \supset
    ( \overbracket{A_{0,1,2}, A_3, X_1, Y_1}^{\text{4d $\NN=2$ vector}},
    \overbracket{X_2, X_3; \; Y_2, Y_3}^{\text{4d $\NN=2$ hyper } \Phi} )
\end{equation}
Notice in particular that in terms of bosonic fields we have $\Phi^+ \simeq \chi$ and $\Phi^- \simeq H^-$. 
In this paper we consider two half-BPS b.c.~for 4d a $\NN=2$ hyper $\Phi$:
\begin{itemize}
    \item Boundary condition $\mathcal{B}_{\chi}$: the chiral $\Phi^+$ has Neumann b.c. and the antichiral $\Phi^-$ has Dirichlet b.c.
    \item Boundary condition $\mathcal{B}_{\overline{\chi}}$: the chiral $\Phi^+$ has Dirichlet b.c. and the antichiral $\Phi^-$ has Neumann b.c.
\end{itemize}
Therefore the Dirichlet-like boundary condition for $\NN=4$ SYM corresponds to $(\mathcal{D},\mathcal{B}_{\chi})$ in 4d $\NN=2$ language. Similarly, $\NN=4$ Neumann-like b.c.~corresponds to $(\mathcal{N},\mathcal{B}_{\overline{\chi}})$.
\\

Let us now go back to the boundary duality for $\NN=4$ SYM in Figure \ref{fig:N4_Sduality}.
On the l.h.s.~of Figure \ref{fig:N4_Sduality} we give $\mathcal{D}$ b.c., therefore the setup has a $SU(N)$ boundary global symmetry.
Instead, the boundary setup on the r.h.s.~of Figure \ref{fig:N4_Sduality} is more involved. Indeed 
in \cite{Gaiotto:2008ak} it was shown that the set of D5-like and NS5-line boundary conditions summarized above is not closed under S-duality and should be extended.
Neumann boundary conditions can be extended e.g.~by including additional 3d degrees of freedom on the boundary and Dirichlet b.c.~can be extended by considering singular scalar field configurations at the boundary.
Considering such generalized b.c.~it is possible to determine sets of half-BPS b.c.~that are closed under the action of S-duality.

In particular the b.c. S-dual to $\mathcal{D}$ is given by $\mathcal{N}$ plus the $T[SU(N)]$ 3d SCFT coupled to the boundary. The $T[SU(N)]$ theory is characterized by a $SU(N)_H \times SU(N)_C$ global symmetry\footnote{The faithful global symmetry group is $PSU(N)_H \times PSU(N)_C$ with a mixed anomaly \cite{Aharony:2013hda}. However, we will leave out details regarding the global form and anomalies for the moment.}. $SU(N)_H$ and $SU(N)_C$ are respectively the Higgs and Coulomb branch symmetries, realized as the flavor symmetry and topological symmetry in the UV Lagrangian completion in Figure \ref{fig:TSUN}, where the latter arises from the enhancement of the $U(1)^{N-1}$ topological symmetries.
The $T[SU(N)]$ theory is coupled to the bulk through one of its $SU(N)$ symmetry, for concreteness let us consider the Higgs Branch symmetry $SU(N)_H$. The coupling at the boundary can be written as an interaction between the corresponding conserved current $J_{SU(N)_H}^{3d}$ and the 4d gauge field $A$:
\begin{equation}    \label{eq:boundary_gauging_N4}
    \int d^3 x J_{SU(N)_H}^{3d} \left. A \right|_{x_3=0}
    \; + \; \text{3d $\NN=4$ SUSY completion}
\end{equation}
The other $SU(N)$ symmetry of $T[SU(N)]$, the Coulomb branch symmetry $SU(N)_C$, it is mapped to the $SU(N)$ symmetry of the dual boundary setup, arising from Dirichlet boundary conditions for the dual gauge field.
For later convenience we note here that the SUSY completion of this interaction contains a term that can be written as a 3d $\NN=2$ superpotential:
\begin{equation}    \label{eq:bdy_W_N4}
    \mathcal{W}_{\text{boundary}} = \chi \mu_{H}
\end{equation}
where $\chi$ is the chiral component of the bulk vectormultiplet at the boundary that takes $\mathcal{N}$ b.c.~(see \eqref{eq:N4_multiplet_bdy_decomposition}), and $\mu_H$ is the moment map of the $SU(N)_H$ symmetry of $T[SU(N)]$.

\subsection{The RG-wall as a massive deformation of the S-wall} \label{subsec: N=2_bc}

One may hope that a suitable SUSY-breaking deformation of the 4d $\NN=4$ setup in Figure \ref{fig:N4_Sduality} can help understand the behavior of $\NN=2$ SYM. 
While this is a heuristic line of thought, in this paper we take it seriously and follow it to develop a proposal for the RG-wall of $\NN=2$ $SU(N)$ SYM. 
Later, in Section \ref{sec: half-index}, we perform a precision check of this proposal by matching half-indices of the holomorphic block computed both in the UV and the IR setups, providing a strong check of our result.

Our goal is to try to reproduce the boundary setups for $\NN=2$ SYM described before, represented in figures \ref{fig:schematic_RG} and \ref{fig:schematic_RG_inverse}, starting from that of $\NN=4$ SYM in Figure \ref{fig:N4_Sduality} and performing two different SUSY breaking deformations. 
We find it more pedagogical to start with the latter, where we look for a 3d theory $\Twall[]^{-1}$ that, when coupled to $\NN=2$ SYM with $\mathcal{N}$ b.c., it flows to $\mathcal{D}$ b.c.~for the effective IR description.

We start from $\mathcal{N}=4$ $SU(N)$ SYM at weak coupling with fixed $\text{Im}(\tau) >> 1$ on $\mathbb{R}^3 \times \mathbb{R}_+$ with $\mathcal{N}$ b.c.~and coupled to $T[SU(N)]$, the r.h.s.~of Figure \ref{fig:N4_Sduality}.
To study SUSY breaking deformations, it is convenient to rewrite the $\NN=4$ setup in $\NN=2$ language.
The $\NN=4$ vector splits in a $\NN=2$ vector plus an hypermultiplet in the adjoint representation. 
We aim to turn on a mass for the adjoint hyper, which would leave just an $\NN=2$ vector in the bulk, which is $\NN=2$ SYM at weak coupling $1/g^2_{YM} = \text{Im}(\tau) >> 1$.

To perform the deformation, we first notice that in the $\mathcal{N}=2$ language there is a $U(1)_A$ global symmetry, which is the commutant of the $\NN=2$ R-symmetry inside the $\NN=4$ R-symmetry. $U(1)_A$ acts on the adjoint hyper as an axial  symmetry, assigning $+1$ charge to both the chiral and anti-chiral.
The complex mass $m$ of the adjoint hyper can be though as the VEV of the complex scalar field sitting inside a background $\mathcal{N}=2$ vector multiplet associated to the $U(1)_A$ symmetry. We can thus give a mass to the adjoint hyper by weakly gauging $U(1)_A$ and turning on a suitable VEV for its complex scalar component.
What is left to understand is how this deformation is projected on the boundary and what is its effect. To solve this problem we note the following points:
\begin{itemize}
    \item The 4d $U(1)_A$ global symmetry of the bulk remains as a global symmetry also on the boundary and it is identified with the commutant of the $U(1)_R$ 3d $\mathcal{N}=2$ R-symmetry subgroup inside the $SU(2)\times SU(2)$ $\mathcal{N}=4$ R-symmetry.

    \item The $U(1)_A$ background $\NN=2$ vector multiplet, involved in the weakly gauging procedure, decomposes on the boundary into a 3d $\NN=2$ vector and a 3d $\NN=2$ chiral, i.e.~a $\mathcal{N}=4$ vector.
    In particular, as described at the beginning of Section \ref{sec: bc_proposal}, 
    the complex scalar $m$ splits on the boundary so that its real part becomes the real scalar sitting inside the 3d $\mathcal{N}=2$ vector, while the imaginary part, together with the orthogonal component of the gauge field, parameterize the complex scalar of the 3d chiral multiplet. 
    
    \item By giving  $\mathcal{N}$ boundary conditions to the 4d $\mathcal{N}=2$ vector, the 3d $\mathcal{N}=2$ vector also receives Neumann-like b.c.~and thus we are allowed to turn on a VEV for its real scalar component. 
    This implies that we can turn on a \textit{real} non-zero value for $m$ on all $\mathbb{R}^3 \times \mathbb{R}^+$, including the boundary, in an half-BPS way.

    \item Similarly, suitable real VEVs for the scalar in the dynamical 4d $\mathcal{N}=2$ vector multiplet correspond to real masses for the $SU(N)_H$ symmetry of $T[SU(N)]$, the one that is coupled to the bulk trough the boundary gauging. 
    As we do not want to Higgs the 4d gauge group, we do not turn on these VEVs.
\end{itemize}
All in all, we can conclude that turning on a real value of $m$ in the bulk, it corresponds to an axial real mass deformation associated to the $U(1)_A$ subgroup of the 3d $\mathcal{N}=4$ R-symmetry, which breaks SUSY to $\mathcal{N}=2$.
The $T[SU(N)]$ theory is then deformed by this real mass and,
in principle, we expect it to flow to a 3d $\mathcal{N}=2$ interacting SCFT as it should possess, for example, conserved currents and operators to be able to properly couple it to the bulk.

\begin{figure}
    \centering
    \begin{tikzpicture}[baseline=(current bounding box).center]
        \begin{scope}
        \node at (-2,0) {$T[SU(N)] \quad : \qquad \qquad $};
        \node at (0,0) (n1) [gauge,black] {$1$};
        \node at (2,0) (n2) [gauge,black] {$2$};
        \node at (4,0) (n3) {$\cdots$};
        \node at (6,0) (nnm1) [gauge,black] {$\scriptstyle N-1$};
        \node at (8,0) (nn) [flavor,black] {$N$};
        
        \draw[-] (n1.north east)      arc[start angle=-45, end angle=225, radius=3mm];
        \draw[-] (n2.north east)      arc[start angle=-45, end angle=225, radius=3mm];
        \draw[-] (nnm1.north east)    arc[start angle=-45, end angle=225, radius=3.1mm];
        
        \draw[->-] (n1.20) --   (n2.160);
        \draw[-<-] (n1.-20) --  (n2.-160);
        \draw[->-] (n2.20) --  ++(1.25,0);
        \draw[-<-] (n2.-20) -- ++(1.25,0);
        
        \draw[->-] (nnm1.160) --  ++(-1.25,0);
        \draw[-<-] (nnm1.-160) -- ++(-1.25,0);
        \draw[->-] (nnm1.20) --   (nn.160);
        \draw[-<-] (nnm1.-20) --  (nn.-160);
        \end{scope}

        \draw[->] (3,-1) -- node[midway,left]{axial} node[midway,right]{mass} (3,-2);

        \begin{scope}[yshift=-3cm]
        \node at (-2,0) {$G[SU(N)] \quad : \qquad \qquad $};
        \node at (0,0) (n1) [gauge,black] {$1$};
        \node at (2,0) (n2) [gauge,black] {$2$};
        \node at (4,0) (n3) {$\cdots$};
        \node at (6,0) (nnm1) [gauge,black] {$\scriptstyle N-1$};
        \node at (8,0) (nn) [flavor,black] {$N$};

        \draw[CScolor] (n1.south east) node[anchor=west] {$\scriptstyle 1$};
        \draw[CScolor] (n2.south east) node[anchor=west] {$\scriptstyle (0,2)$};
        \draw[CScolor] (nnm1.south east) node[anchor=west] {$\scriptstyle (0,N-1)$};
        \draw[CScolor] (nn.south east) node[anchor=west] {$\scriptstyle \tfrac{N-1}{2}$};
        
        \draw[-<-] (n1) -- (n2);
        \draw[->-] (n2)++(1.25,0) -- (n2);
        \draw[-<-] (nnm1)++(-1.25,0) -- (nnm1);
        \draw[-<-] (nnm1) --  (nn);

        \path (n1) -- node[midway,above,BFcolor] {$\scriptstyle -1$} (n2);
        \path (n2)++(1.25,0) -- node[midway,above,BFcolor] {$\scriptstyle -1$} (n2);
        \path (nnm1)++(-1.25,0) -- node[midway,above,BFcolor] {$\scriptstyle -1$} (nnm1);
        \path (nnm1) --  (nn);
        \end{scope}
        
    \end{tikzpicture}
    \caption{Axial mass deformation of the 3d $\NN=4$ theory $T[SU(N)]$. The resulting 3d $\NN=2$ theory is the (inverse) RG-wall for 4d $\NN=2$ SYM.
    The deformations breaks the global symmetry from $SU(N)\times SU(N)$ to $SU(N) \times U(1)^{N-1}$.
    All bifundamental fields have $R$-charge $0$.
    There are CS level $+1$ for the $U(1) \in U(K)$ diagonal subgroups of the gauge symmetries
    and mixed CS level $-1$ between adjacent $U(K)$ gauge groups, as labeled in red and green respectively.
    More details regarding the theory $G[SU(N)]$ are given later in subsection \ref{subsec: twall_gsun_generalities}.
    }
    \label{fig:axial_mass_tsun_gsun}
\end{figure}
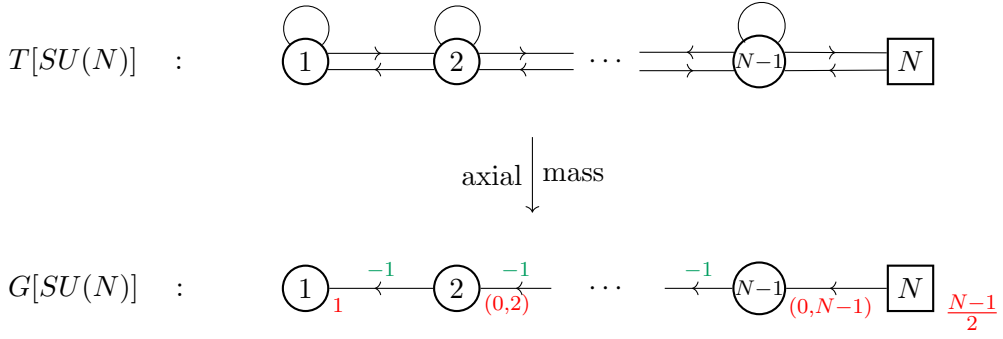

Such a deformation of a 3d $\NN=4$ SCFT is precisely of the type studied in \cite{Benvenuti:2024seb,Benvenuti:2025huk,Benvenuti:2026usm}. 
In particular, it was observed that this axial real mass deformation for the $T[SU(N)]$ theory leads to an interacting theory only if it is 
paired with non-homogeneous and finely tuned real mass for the Coulomb branch symmetry $SU(N)_C$, the global symmetry on the boundary. More precisely, we can turn on $N-1$ real mass parameters, associated to the $U(1)^{N-1}$ maximal subtorus of $SU(N)_C$. These masses can be parameterized in terms of $N$ parameters $t_i$ satisfying $\sum_{i=1}^N t_i = 0$. To find an interacting fixed point for non-zero value $m$ of the axial real mass, we have to turn on also the $t_i$ parameters as:
\begin{equation} \label{eq: cbmasses}
    t_{i+1} - t_i = m
\end{equation}
This deformation breaks $\NN=4$ to $\NN=2$ as well as $SU(N)_C$ to its Cartan $U(1)^{N-1}$, while the Higgs branch symmetry $SU(N)_H$ is preserved. Starting from the $T[SU(N)]$, the 3d $\mathcal{N}=2$ theory resulting from this flow was named $G[SU(N)]$, which is a non-Abelian gauge theory depicted in Figure \ref{fig:axial_mass_tsun_gsun}\footnote{More precisely, the theory considered here and shown in Figure \ref{fig:axial_mass_tsun_gsun} corresponds to a redefinition of the originally defined $G[SU(N)]$ in \cite{Benvenuti:2025huk}. The $G[SU(N)]$ theory, as defined in this work, was observed to coincide with a theory generating $\mathcal{N}=2$ mirror-like dualities for an extremely wide variety of 3d $\mathcal{N}=2$ and, recently, raised some interest as the Higgs branch of this theory is a complete flag manifold \cite{Closset:2025akk,Closset:2026bnk}.}
that carries a $SU(N)$ flavor symmetry and a $U(1)^{N-1}$ topological symmetry\footnote{There are more UV completions realizing a manifest $SU(N)$ flavor symmetry. These are obtained by combining the deformation with the $S_N$ Weyl group of the emergent $SU(N)_C$ symmetry. We present them in Appendix \ref{app: gsun}. Another UV Lagrangian description also exists, which is simply obtained by combining mirror duality with the mass deformation. The result is a purely Abelian planar quiver, that has a $U(1)^{N-1}$ flavor symmetry and an emergent $SU(N)$ topological symmetry. The two descriptions are related by an $\NN=2$ mirror-like duality \cite{Benvenuti:2025huk} which we describe in Appendix \ref{app: gsun}.}.

All in all, the setup obtained after the deformation is given, in the bulk, by 4d $\mathcal{N}=2$ SYM at weak coupling $1/g^2_{YM} >> 1$ with $\mathcal{N}$ b.c. coupled to the $G[SU(N)]$ theory on the boundary via the diagonal gauging of the $SU(N)$ symmetry, so that we are left with a $U(1)^{N-1}$ global symmetry on the boundary. This flow is schematically summarized as:
\begin{equation} \label{eq: N4toN2_schematic}
    \centering
    \begin{tikzpicture}
    
    \begin{scope}[xshift=0cm]
        \node at (0,0) (UV) {$\begin{array}{c} SU(N) \\ \NN=4 \; \text{SYM} \end{array}$};
        \path[draw] (1.5,-1.5) -- node[midway,right,anchor=west] {$\mathcal{N} + T[SU(N)]$} (1.5,1.5) node[above] {$x_3=0$};
    \end{scope}
    
    \begin{scope}[xshift=8.5cm]
        \node at (0,0) (UV) {$\begin{array}{c} SU(N) \\ \NN=2 \; \text{SYM} \end{array}$};
        \path[draw] (1.5,-1.5) -- node[midway,right,anchor=west] {$\mathcal{N} + G[SU(N)]$} (1.5,1.5) node[above] {$x_3=0$};
    \end{scope}

    \begin{scope} [xshift=5.5cm]
        \path[draw,->] (0,0) -- node[midway,above] {axial} node[midway, below] {mass} (1,0);
    \end{scope}
    
    \end{tikzpicture}
\end{equation}

We now want to study the S-dual side of this setup, that is the effect of the mass deformation of the $\mathcal{N}=4$ setup on the l.h.s.~of Figure \ref{fig:N4_Sduality}, which is $\NN=4$ SYM at strong coupling $\tau' <<1$ with Dirichlet b.c.
As a guideline we note the following properties of the setup in \eqref{eq: N4toN2_schematic}.
\begin{itemize}
    \item The $SU(N)$ boundary symmetry is broken to its maximal torus $U(1)^{N-1}$ due to the real mass deformation associated to the $SU(N)_C$ Coulomb branch symmetry reported in \eqref{eq: cbmasses}. In the S-dual setup, the parameters $t_i$ correspond to the VEVs of the complex scalar in the 4d $\mathcal{N}=2$ $SU(N)$ vector multiplet.
    
    \item There are $N-1$ chiral operators localized on the boundary, each carrying charge under a single $U(1)$ factor of the $U(1)^{N-1}$ symmetry. They arise as monopole operators of the 3d theory $G[SU(N)]$ in Figure \ref{fig:axial_mass_tsun_gsun}.
\end{itemize}
We are lead to the conclusion that in the S-dual setup, the l.h.s.~of Figure \ref{fig:N4_Sduality}, we are giving a mass $m$ to the adjoint hyper and, at the same time, 
breaking explicitly the gauge symmetry as $SU(N) \to U(1)^{N-1}$ due to the VEVs for the dynamical scalar fields. The result is qualitatively an effective theory with $U(1)^{N-1}$ gauge group, where the gauge fields $\mathcal{A}^{\NN=2}$ receive Dirichlet b.c.~and provide a $U(1)^{N-1}$ global symmetry on the boundary. 

Notice that, each component of the adjoint hyper $\Phi_{i,j}$, sitting inside the $\mathcal{N}=4$ vector, have an effective mass $t_i - t_j + m$. Therefore, when we turn on the deformation dictated by the axial mass $m$ and the Coulomb branch deformation as in \eqref{eq: cbmasses}, only the $N-1$ components $\Phi_{i,i+1}$ on the upper-diagonal of the $N \times N$ adjoint matrix remains massless.

All in all, the 4d bosonic fields that remain massless are:
\begin{equation}
    (\mathcal{A}^{\NN=2})_{i,i} \supset 
    (\underbrace{A_{0,1,2},X_1}_{\mathcal{D}},
    \underbrace{A_3,Y_1}_{\mathcal{N}})_{i,i},
    \qquad
    \Phi_{i,i+1} \supset
    (\underbrace{X_2,X_3}_{\mathcal{D}}, 
    \underbrace{Y_2, Y_3}_{\mathcal{N}})_{i,i+1}
    ,
    \qquad i=1,\dots,N-1
\end{equation}
where we also reported their boundary conditions.
In terms of the half-BPS boundary conditions described at the beginning of this Section, $\mathcal{A}^{\NN=2}$ has $\mathcal{D}$ boundary conditions and $\Phi_{i,i+1}$ have $\mathcal{B}_{\overline{\chi}}$ boundary conditions.

The antichiral component $\Phi_{i,i+1}^{-}$ of each of these $N-1$ hypermultiplets has Neumann boundary condition, providing a chiral boundary operator. Moreover, in a suitable basis for the gauge fields, each  of these chiral operators carries charge $+1$ under only one of the $U(1)$ factors in the $U(1)^{N-1}$ global symmetry.
These operators correspond to the $N-1$ chiral monopole of the $G[SU(N)]$ theory discussed above.
Putting it all together, on the S-dual side we find $N-1$ copies of QED with one charged hyper. 
All of the $U(1)$ vector multiplets have $\mathcal{D}$ b.c.~while the hypermultiples have $\mathcal{B}_{\chi}$ b.c. This flow can be schematically represented as follows.
\begin{equation}
    \begin{tikzpicture}
         \begin{scope}[xshift=0cm]
        \node at (0,0) (UV) {$\begin{array}{c} SU(N) \\ \NN=4 \; \text{SYM} \end{array}$};
        \path[draw] (1.5,-1.5) -- node[midway,right,anchor=west] {$\mathcal{D}$} (1.5,1.5) node[above] {$x_3=0$};
    \end{scope}

    \begin{scope} [xshift=4cm]
        \path[draw,->] (0,0) -- node[midway,above] {axial} node[midway,below] {mass} (1,0);
    \end{scope}

    \begin{scope}[xshift=10cm]
        \node at (-1.5,0) (IR) {$\begin{array}{c} N \text{ copies of} \\ \NN=2 \; \text{SQED w/ 1 hyper}  \end{array}$};
        \path[draw] (1.5,-1.5) -- node[midway,right,anchor=west] {$\mathcal{B}_{\chi},\;\mathcal{D}$} (1.5,1.5) node[above] {$x_3=0$};
    \end{scope}

    \end{tikzpicture}
    \label{eq:schematic_RG_RHS}
\end{equation}

The analysis so far has been purely semiclassical, but surprisingly we note that the bulk theory on the magnetic side looks exactly like the Seiberg-Witten theory of $\NN=2$ SYM at the multi-monopole point!\footnote{The multimonopole point for higher rank was first discussed in \cite{Argyres:1994xh} for $SU(3)$, later generalized to $SU(N)$ in \cite{Douglas:1995nw}. In this special point of the CB mutually local hypermultiplets become massive, in contrast to the AD SCFT points \cite{Argyres:1995jj} where mutually \textit{non}-local particles become massless.} 
Indeed, close to the multimonopole point, the IR SW theory describes $N-1$ copies of QED, where the charged hypermultiplets are monopoles of the UV theory. The fact that the operators $\Phi_{i,i+1}$ on the magnetic side, r.h.s.~of \eqref{eq:schematic_RG_RHS}, are mapped to monopole operators of the electric side, l.h.s.~of \eqref{eq: N4toN2_schematic}, is not surprising, as our starting point was a pair of $\NN=4$ boundary setups related by S-duality.
This leads us to conjecture that $G[SU(N)]$ is the RG-wall between the UV of $\NN=2$ SYM and the IR SW effective theory close to the multimonopole point:
\begin{equation} \label{eq: schematic_invSTwall_final}
    \boxed{
    \begin{tikzpicture}
    
    \begin{scope}[xshift=0cm]
        \node at (0,0) (UV) {$\begin{array}{c} \text{UV} \\ \\ SU(N) \\ \NN=2 \; \text{SYM}\end{array}$};
        \path[draw] (1.5,-1.5) -- node[midway,right,anchor=west] {$\mathcal{N} + G[SU(N)]$} (1.5,1.5) node[above] {$x_3=0$};
    \end{scope}
    
    \begin{scope}[xshift=9cm]
        \node at (0,0) (IR) {$\begin{array}{c} \text{IR} \\ \\\text{Seiberg-Witten} \\ \text{at multimonopole point}\end{array}$};
        \path[draw] (2.5,-1.5) -- node[midway,right,anchor=west] {$\mathcal{B}_{\chi},\;\mathcal{D}$} (2.5,1.5) node[above] {$x_3=0$};
    \end{scope}

    \begin{scope} [xshift=5cm]
        \path[draw,->] (0,0) -- node[midway,above] {RG flow} (1.5,0);
    \end{scope}
    
    \end{tikzpicture}
    }
\end{equation}
The picture above can be made more precise by recalling that pure $\NN=2$ SYM has a strongly coupled chamber \cite{Alim:2011kw} where the spectrum is finite and includes the monopoles.
In this chamber the $N-1$ monopoles and the corresponding $N-1$ dyons provide a basis for all BPS particles.
Morally, we can then define a b.c.~in all of this chamber by giving $\mathcal{B}_{\chi}$ b.c.~to all the BPS particles and extend our proposal to the full strongly coupled chamber. This requires some care, as already discussed in \cite{Dimofte:2013lba}, because the BPS particles are not mutually local, we will come back to this issue in Section \ref{sec: half-index}.

Furthermore, as discussed in \cite{Dimofte:2013lba,Cordova:2016uwk}, the IR b.c.~transform in a controlled way under wall-crossing, therefore this proposal is not limited to the multimonopole point or the strongly coupled chamber, but can be extended to the full Coulomb branch as long as one avoids Argyres-Douglas strongly coupled points \cite{Argyres:1995jj}.
In the rest of this paper we always consider b.c. defined in the strongly coupled chamber of \cite{Alim:2011kw}, with the understanding that they can then be extended to the rest of the CB by tracking the wall-crossing happening across walls of marginal stability.
Indeed, we provide a detailed check of this proposal from localization computations in Section \ref{sec: half-index}, where we also discuss such b.c.~in more detail. 

The discussion above further leads to the fact that the $G[SU(N)]$ is not precisely the (inverse) RG wall theory $(\Twall[])^{-1}$ as in Figure \ref{fig:schematic_RG_inverse}, but the ``$S$-transformed" of it\footnote{We by $S$-transformed we mean that we perform an $S \in SL(2,\mathbb{Z})$ transformation on all the $U(1)^{N-1}$ gauge groups in the bulk and its consequent transformation on the 3d theory on the boundary, as in \cite{Witten:2003ya}.}. 
Indeed, the semiclassical analysis above is compatible with a local description of the SW solution at the multimonopole point \textit{and} in an EM duality frame where these monopoles are electric.
We will discuss this point more in depth in the next Section where we study the complementary boundary problem in Figure \ref{fig:schematic_RG}.

\subsubsection*{The RG-wall in the IR}

We can now move to the complementary picture to derive the boundary setup depicted in Figure \ref{fig:schematic_RG}: we give $\mathcal{D}$ b.c.~to the gauge field of $\NN=2$ SYM and look for the corresponding b.c.~in the IR Seiberg-Witten theory.
To do so, we start from the S-dual boundary conditions for $\NN=4$ SYM in Figure \ref{fig:N4_Sduality} at strong coupling and turn on the axial mass of the adjoint hypermultiplet. 

On the l.h.s., where the theory is weakly coupled and there are $\mathcal{D}$ b.c., the adjoint hyper takes a mass and we obtain $\NN=2$ SYM at weak coupling with $\mathcal{D}$ b.c.:
\begin{equation}
    \begin{tikzpicture}
        \begin{scope}[xshift=0cm]
        \node at (0,0) (UV) {$\begin{array}{c} SU(N) \\ \NN=4 \; \text{SYM} \end{array}$};
        \path[draw] (1.5,-1.5) -- node[midway,right,anchor=west] {$\mathcal{D}$} (1.5,1.5) node[above] {$x_3=0$};
    \end{scope}

    \begin{scope} [xshift=4cm]
        \path[draw,->] (0,0) -- node[midway,above] {axial} node[midway,below] {mass} (1,0);
    \end{scope}

     \begin{scope}[xshift=8cm]
        \node at (0,0) (UV) {$\begin{array}{c} SU(N) \\ \NN=2 \; \text{SYM} \end{array}$};
        \path[draw] (1.5,-1.5) -- node[midway,right,anchor=west] {$\mathcal{D}$} (1.5,1.5) node[above] {$x_3=0$};
    \end{scope}

    \end{tikzpicture}
    \label{fig:schematic_RG_RHS_2}
\end{equation}

On the r.h.s.~of Figure \ref{fig:N4_Sduality}, borrowing the analysis in the previous section, we propose that after the axial mass deformation we flow in the bulk to the Seiberg-Witten theory of $\mathcal{N}=2$ SYM at the multimonopole point, which is coupled on the boundary to the $G[SU(N)]$ theory.
More precisely, due to the change in the deformation data and the precise coupling of the symmetries between the 3d and 4d theories, we obtain the $CP$-transformed of $G[SU(N)]$. 
The action of $P$ inverts the sign of all CS levels, includind mixed CS and therefore FI terms. 
The action of $C$ changes the sign of all conserved currents, and at the level of the quiver corresponds to changing the direction of all the arrows and the sign of all the FIs.
The symmetry that couples the 3d theory to the bulk is broken to $SU(N) \to U(1)^{N-1}$, therefore the bulk gauge group should be Higgsed to its maximal torus.
We expect some components of the adjoint hyper to remain massless, resulting in $N-1$ copies of 4d $\NN=2$ QED that we identify with the multimonopole point of the Seiberg-Witten theory. 
We can then identify the $CP$ transformed of $G[SU(N)]$ as the $S$-transformed of the $\Twall[]$  or, in short, the $S \cdot \Twall[]$ theory.
Schematically we propose that the r.h.s.~of Figure \ref{fig:N4_Sduality} is deformed to:
\begin{equation}
    \centering
    \begin{tikzpicture}
    
     \begin{scope}[xshift=0cm]
        \node at (0,0) (UV) {$\begin{array}{c} SU(N) \\ \NN=4 \; \text{SYM} \end{array}$};
        \path[draw] (1.5,-1.5) -- node[midway,right,anchor=west] {$\mathcal{N} + T[SU(N)]$} (1.5,1.5) node[above] {$x_3=0$};
    \end{scope}
    
    \begin{scope}[xshift=9cm]
        \node at (-1,0) (UV) {$\begin{array}{c} \text{Seiberg-Witten} \\ \text{at multimonopole point} \end{array}$};
        \path[draw] (1.5,-1.5) -- node[midway,right,anchor=west] {$\mathcal{B}_{\chi},\;\mathcal{N} + S\cdot  \Twall[]$} (1.5,1.5) node[above] {$x_3=0$};
    \end{scope}

    \begin{scope} [xshift=4.5cm]
        \path[draw,->] (0,0) -- node[midway,above] {axial} node[midway,below] {mass} (1,0);
    \end{scope}
    
    \end{tikzpicture}
\end{equation}
After the axial mass deformation the boundary global symmetry is the $SU(N)$ global symmetry of the $S\cdot \Twall[]$.

The coupling between the 4d and 3d theories is partially broken by this deformation. The $S\cdot \Twall[]$ and the low energy SW theory are coupled trough the diagonal gauging of the $U(1)^{N-1}$ symmetry:
\begin{equation}
    \sum_{i=1}^{N-1} \int d^3x J^{3d}_{i} \left.\widetilde{A}_i \right|_{x_3=0}
    + \text{ 3d $\NN=2$ SUSY competion}
\end{equation}
Furthermore the $N-1$ 4d hypermultiplets with electric charges, let us name them $\Phi_i$, split into a pair of 3d $\NN=2$ chiral/anti-chiral multiplets $(\Phi_i^+,\Phi^-_i)$. The bosonic component of $\Phi_i^+$ receive Neumann b.c.~so that they can be coupled to monopole operators $M_i$ of the $S \cdot \Twall[]$. The coupling is encoded in a boundary superpotential term:
\begin{equation}    \label{eq:bdy_W_N2}
    \mathcal{W}_{\text{bdry}} \supset \sum_{i=1}^{N-1} \Phi^+_i M_i
\end{equation}
which is a remnant of the coupling \eqref{eq:bdy_W_N4} between $T[SU(N)]$ and $\NN=4$ SYM.
Consequently, the $\Phi_i^-$ receive Dirichlet b.c.~and are thus identified with the $M_i$ operators on the boundary.
As a non-trivial check of our proposal one can compute the R-charge of the monopoles $M_i$ to be 1.
We denote this boundary condition on the hypers as “modified" $\mathcal{B}_{\chi}$ because the antichiral component of the hyper if fixed at the boundary\footnote{Similarly one can consider “modified"   $\mathcal{B}_{\overline{\chi}}$ boundary conditions where the chiral component is identified with some operator of the boundary theory.}.

On top of the boundary superpotential \eqref{eq:bdy_W_N2}, which is a direct consequence of the semiclassical analysis, one can consider couplings involving BPS particles other then the monopoles. As an example the strong coupling spectrum also involve (massive) BPS dyons. These particles are not electric in this duality frame, thus their coupling can not be written as a superpotential. Nevertheless, we find that the RG-wall theory includes non-genuine operators that only exists when turning on a non-zero background flux for the topological symmetries. These operators have the correct global and R-charges to couple to the dyons. We will come back to this issue in Section \ref{sec: half-index} when we consider examples at fixed rank. 
\\

Putting it all together we propose that the $CP$-transformed of the $G[SU(N)]$ theory is the RG-wall of $\NN=2$ $SU(N)$ SYM in the low-energy EM duality frame where the monopoles are electric, or simply the $S \cdot \Twall[]$:
\begin{equation} \label{eq: schematic_STwall_final}
    \boxed{
    \begin{tikzpicture}
    
    \begin{scope}[xshift=0cm]
        \node at (0,0) (UV) {$\begin{array}{c} \text{UV} \\ \\ SU(N) \\ \NN=2 \; \text{SYM}\end{array}$};
        \path[draw] (1.5,-1.5) -- node[midway,right,anchor=west] {$\mathcal{D}$} (1.5,1.5) node[above] {$x_3=0$};
    \end{scope}
    
    \begin{scope}[xshift=7.25cm]
        \node at (0,0) (IR) {$\begin{array}{c} \text{IR} \\ \\ \text{Seiberg-Witten} \\ \text{at multimonopole point}\end{array}$};
        \path[draw] (2.2,-1.5) -- node[midway,right,anchor=west] {$\mathcal{B}_{\chi},\;\mathcal{N} + S \cdot \Twall[]$} (2.2,1.5) node[above] {$x_3=0$};
    \end{scope}

    \begin{scope} [xshift=3.5cm]
        \path[draw,->] (0,0) -- node[midway,above] {RG flow} (1,0);
    \end{scope}
    
    \end{tikzpicture}
    }
\end{equation}

\begin{figure}
    \centering
    \begin{equation*}
    \Twall[] \quad : \qquad\qquad
    \begin{tikzpicture}[baseline=(current bounding box).center]
        \node at (0,0) (n1) [flavor,black] {$1$};
        \node at (2,0) (n2) [gauge,double,black] {$2$};
        \node at (5,0) {$\cdots$};
        \node at (8,0) (nnm1) [gauge,double,black] {$\scriptstyle N-1$};
        \node at (10,0) (nn) [flavor,black] {$N$};

        \draw[->-] (n1) -- node[midway,above,gray] {$b_1 b_2^{-\tfrac{1}{2}}$} (n2);
        \draw[-<-] (n2)++(2,0) -- node[midway,above,gray] {$b_2^{\tfrac{1}{2}} b_3^{-\tfrac{1}{3}}$} (n2);
        \draw[->-] (nnm1)++(-2,0) -- node[midway,above,gray] {$b_{N-2}^{\tfrac{1}{N-2}} b_{N-1}^{-\tfrac{1}{N-1}}$} (nnm1);
        \draw[->-] (nnm1) -- node[midway,above,gray] {$b_{N-1}^{\tfrac{1}{N-1}}$} (nn);
    \end{tikzpicture}
    \end{equation*}
    \caption{The UV Lagrangian for the theory $\Twall[]$.
    This is the theory on the RG-wall in the canonical electromagnetic duality frame of the low energy SW theory.
    Each double circle corresponds to an $SU(K)$ gauge group while arrows are bifundamental chirals. The gray labels are the fugacities assigned to the corresponding chiral, the power of the fugacity $b_i$ is the charge of the chiral under the $U(1)_{b_i}$ symmetry. 
    The CS levels of all the gauge groups vanish while there are background CS levels for the $U(1)^{N-1}\times SU(N)$ global symmetry discussed in the main text. All the chiral fields have R-charge 0.
    }
    \label{fig:Twall}
\end{figure}

While this picture is already a full proposal for the RG-wall, one may wish to have a description of this IR boundary condition in terms of the ``electric" gauge fields.
Indeed, as already pointed out, the hypermultiplets in the low-energy effective description are identified with monopoles of the UV Lagrangian.
Such a change in EM duality frame can be achieved by performing an $S^{-1}$-duality transformation on every Abelian gauge field. 
As observed in \cite{Witten:2003ya}, this operation on the boundary consists of the gauging of the $N-1$ $U(1)$ global symmetries of the 3d theory at the boundary, namely the $CP \cdot G[SU(N)]$ theory.
These symmetries are the topological symmetries of the quiver and the effect of gauging them is to ``ungauge" the diagonal $U(1) \subset U(K)$ of the corresponding gauge group, 
leading to the Lagrangian theory in Figure \ref{fig:Twall}\footnote{Indeed, the actual $\Twall[]$ theory lives at the IR fixed point of this theory. But with a small abuse of notation we denote the quiver in Figure \ref{fig:Twall} as $\Twall[]$ as well. }.

Schematically, we propose the following boundary RG flow:
\begin{equation}
    \boxed{
    \begin{tikzpicture}
    
    \begin{scope}[xshift=0cm]
        \node at (0,0) (UV) {$\begin{array}{c} \text{UV} \\ \\ SU(N) \\ \NN=2 \; \text{SYM}\end{array}$};
        \path[draw] (1.5,-1.5) -- node[midway,right,anchor=west] {$\mathcal{D}$} (1.5,1.5) node[above] {$x_3=0$};
    \end{scope}
    
    \begin{scope}[xshift=7.25cm]
        \node at (0,0) (IR) {$\begin{array}{c} \text{IR} \\ \\ \text{Seiberg-Witten} \\ \text{in electric EM frame} \end{array}$};
        \path[draw] (2,-1.5) -- node[midway,right,anchor=west] {$\mathcal{B}_{\chi},\;\mathcal{N} + \Twall[]$} (2,1.5) node[above] {$x_3=0$};
    \end{scope}

    \begin{scope} [xshift=3.5cm]
        \path[draw,->] (0,0) -- node[midway,above] {RG flow} (1,0);
    \end{scope}
    
    \end{tikzpicture}
    }
\end{equation}
The theory $\Twall[]$ is our proposal for the RG-wall of $\NN=2$ $SU(N)$ SYM and, as expected, it has all the desired properties.
It has an $SU(N)$ global symmetry, which is the symmetry rotating the $N$ flavors in Figure \ref{fig:Twall}, as well as $N-1$ $U(1)$ global symmetries given by the baryonic symmetries associated to each $SU(K)$ gauge group\footnote{Here we consider the first flavor node of the quiver in Figure \ref{fig:Twall} as an $SU(1)$ gauge node for notational convenience.}.

\subsection{Generalities of the $\Twall[]$ theory} \label{subsec: twall_gsun_generalities}

We now provide a more detailed description of the $\Twall[]$ theory.
Following the logic order of the previous discussion, we find it convenient to start from the $G[SU(N)]$ theory.

\subsubsection{The S-dual frame: the $G[SU(N)]$ theory}
\label{subsec:GsuN}

As already discussed, the $G[SU(N)]$ theory is the result of a SUSY breaking deformation of the  3d $\mathcal{N}=4$ $T[SU(N)]$ theory, which is the S-duality wall of 4d $\mathcal{N}=4$ SYM. We provide more details on this deformation Appendix \ref{app: 3dwalls}.

We now describe in more detail the $G[SU(N)]$ theory, starting actually from its $CP$-transformed that we have previously identified with the $S$-transformed of the $\Twall[]$ theory:
\begin{equation}    \label{eq:quiv_ptransf_Gsun}
\begin{array}{l}
    S\cdot \Twall[]=
    CP \cdot G[SU(N)] = \quad 
    \begin{tikzpicture}[baseline=(current bounding box).center]
        
        \begin{scope}[yshift=-3cm]
        \node at (0,0) (n1) [gauge,black] {$1$};
        \node at (2,0) (n2) [gauge,black] {$2$};
        \node at (4,0) (n3) {$\cdots$};
        \node at (6,0) (nnm1) [gauge,black] {$\scriptstyle N-1$};
        \node at (8,0) (nn) [flavor,black] {$N$};

        \draw[CScolor] (n1.south east) node[anchor=west] {$\scriptstyle -1$};
        \draw[CScolor] (n2.south east) node[anchor=west] {$\scriptstyle (0,-2)$};
        \draw[CScolor] (nnm1.south east) node[anchor=west] {$\scriptstyle (0,1-N)$};
        \draw[CScolor] (nn.south east) node[anchor=west] {$\scriptstyle \tfrac{1-N}{2}$};

        \draw[FIcolor] (n1.north) node[anchor=south] {$\scriptstyle \xi_1^{-1}$};
        \draw[FIcolor] (n2.north) node[anchor=south] {$\scriptstyle \xi_2^{-1}$};
        \draw[FIcolor] (nnm1.north) node[anchor=south] {$\scriptstyle \xi_{N-1}^{-1}$};

        \draw[gray] (nn.north) node[anchor=south] {$\scriptstyle \vec{x}$};
        
        \draw[->-] (n1) -- node[midway,above,BFcolor] {$\scriptstyle +1$} (n2);
        \draw[-<-] (n2)++(1.5,0) -- node[midway,above,BFcolor] {$\scriptstyle +1$} (n2);
        \draw[->-] (nnm1)++(-1.5,0) -- node[midway,above,BFcolor] {$\scriptstyle +1$} (nnm1);
        \draw[->-] (nnm1) --  (nn);
        \end{scope}
    \end{tikzpicture}
\end{array}
\end{equation}
We assign R-charge 0 to all bifundamental chiral field. Notice that this is not necessarly the superconformal R-symmetry of the $G[SU(N)]$ theory in isolation. It is however the parameterization dictated by the embedding of the 3d $\mathcal{N}=2$ group in the 4d $\mathcal{N}=2$ group on the boundary. 
Also, we reported in gray the set $\vec{x}$ of fugacities associated to the $SU(N)$ global symmetry, satisfying $\prod_{i=1}^N x_i =1$. Only the rightmost bifundamental transforms non trivially under the $SU(N)$ global symmetry: in the antifundamental representation.
The orange labels are associated to the topological symmetries so that, for later convenience, we parameterize the theory such that the topological symmetry of the $U(i)$ gauge node is identified with minus the $U(1)_{\xi_i}$ global symmetry, thus the label $\xi_i^{-1}$. 
Also, red labels are CS levels and green labels are mixed CS levels, for the two gauge groups connected by the arrow. 
The notation $(k_1,k_2)$ indicate the CS level of, respectively, $SU(n) \subset U(n)$ and of $U(1) \subset U(n)$, where it must be true the $k_2 = k_1 + ln$ with $l \in \mathbb{Z}$. We use the following normalization of CS interactions:
\begin{equation}
    U(n)_{k,k+ln} \quad : \quad -i\frac{k}{4\pi}\int \text{tr}(A \wedge dA) - i\frac{l}{4\pi} \int \text{tr}(A) \wedge \text{tr}(dA) \, + \text{SUSY completion} \,.
\end{equation}
The theory also possesses background CS terms with levels:
\begin{equation}
    k_{\xi_i} = k_{\xi_i \xi_j} = 0 \,, \quad k_{SU(N)}=\frac{1-N}{2} \,,
\end{equation}
notice in particular that although they are admitted to be non-zero, we have no background CS interactions for the topological symmetries.

The theory possesses $N-1$ gauge invariant chiral monopole operators $M_i$, each carrying $-1$ unit of magnetic flux with respect to the $i$-th gauge group of the quiver. These monopoles thus carry $-1$ charge under $U(1)_{\xi_i}$, recalling that in this convention an FI parameter is a level $-2$ interaction between the topological and the gauge symmetry, thus a monopole with $-1$ magnetic flux carries $+1$ topological symmetry, and also the topological symmetry is identified with minus the $U(1)_{\xi_i}$ symmetry.
Each of these monopoles carries R-charge 1, which can be computed knowing that given a $U(n)$ gauge theory with $F$ fundamental/antifundamental chirals with R-charge $r$, the monopole carrying magnetic flux $\vec{m} = \{ m_1,m_2,\ldots,m_n \}$ has R-charge:
\begin{equation}
    R[M^{(\vec{m})}] = -\sum_{i < j}^n |m_i-m_j| + \frac{F}{2}(1-r) \sum_{i=1}^n |m_i|   
\end{equation}
where the first contribution arises from gaugino zero-modes, while the second from fermionic zero-modes of the fundamental/antifundamental chiral fields.

As a reference, we also report the quiver describing the $(S \cdot \Twall[])^{-1}$, which is described by the following quiver theory.
\begin{equation}    \label{eq:quiv_Gsun}
\begin{array}{l}
    (S\cdot \Twall[])^{-1} =
    G[SU(N)] =
    \quad
    \begin{tikzpicture}[baseline=(current bounding box).center]
        \begin{scope}[yshift=-3cm]
        \node at (0,0) (n1) [gauge,black] {$1$};
        \node at (2,0) (n2) [gauge,black] {$2$};
        \node at (4,0) {$\cdots$};
        \node at (6,0) (nnm1) [gauge,black] {$\scriptstyle N-1$};
        \node at (8,0) (nn) [flavor,black] {$N$};

        \draw[CScolor] (n1.south east) node[anchor=west] {$\scriptstyle 1$};
        \draw[CScolor] (n2.south east) node[anchor=west] {$\scriptstyle (0,2)$};
        \draw[CScolor] (nnm1.south east) node[anchor=west] {$\scriptstyle (0,N-1)$};
        \draw[CScolor] (nn.south east) node[anchor=west] {$\scriptstyle \tfrac{N-1}{2}$};

        \draw[FIcolor] (n1.north) node[anchor=south] {$\scriptstyle \xi_{1}$};
        \draw[FIcolor] (n2.north) node[anchor=south] {$\scriptstyle \xi_{2}$};
        \draw[FIcolor] (nnm1.north) node[anchor=south] {$\scriptstyle \xi_{N-1}$};

        \draw[black] (nn.north) node[anchor=south] {$\scriptstyle \vec{X}$};
        
        \draw[-<-] (n1) -- node[midway,above,BFcolor] {$\scriptstyle -1$} (n2);
        \draw[->-] (n2)++(1.5,0) -- node[midway,above,BFcolor] {$\scriptstyle -1$} (n2);
        \draw[-<-] (nnm1)++(-1.5,0) -- node[midway,above,BFcolor] {$\scriptstyle -1$} (nnm1);
        \draw[-<-] (nnm1) --  (nn);
        \end{scope}
    \end{tikzpicture}
\end{array}
\end{equation}
That coincides, up to reparameterizations, with the $G[SU(N)]$ theory first described in \cite{Benvenuti:2025huk}. 
Notice that the Abelian global symmetry $U(1)_{\xi_i}$ coincide with the topological symmetries of the $U(i)$ gauge node.

\paragraph{A planar-Abelian mirror dual} As observed first in \cite{Benvenuti:2025huk}, the $G[SU(N)]$ theory enjoys an $\mathcal{N}=2$ mirror-like dual description which is an Abelian planar-shaped quiver theory. This duality is inherited from the $\mathcal{N}=4$ self-mirror duality of the $T[SU(N)]$ theory. We describe in more detail this duality in Appendix \ref{app: gsun}, while here we limit the discussion to presenting the dual theory.

The $\mathcal{N}=2$ planar-Abelian quiver that is IR dual to the $G[SU(N)]$ theory given in \eqref{eq:quiv_Gsun} is given as follows:
\begin{equation} \label{eq: planarabelian_gsun}
    \includegraphics[]{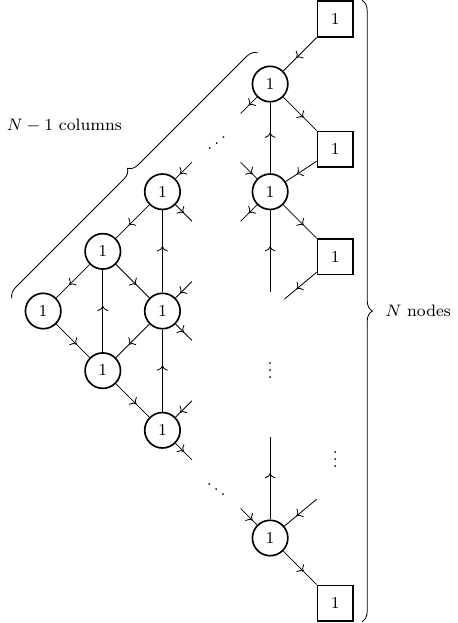}
\end{equation}
where the theory possesses $N-1$ ``columns" of gauge nodes, the first of height 1, up to height $N-1$.
This is a fully Abelian Lagrangian with $\tfrac{N(N-1)}{2}$ Abelian gauge groups.

This Lagrangian involves CS interactions, “planar" superpotential interactions and monopole superpotential terms. We do not report the here in detail to avoid clutter, and refer the reader to Appendix \ref{app: gsun} or the the original literature \cite{Benvenuti:2025huk}.
\\

We report here some of the salient features of the duality between the planar \eqref{eq: planarabelian_gsun} and chiral \eqref{eq:quiv_Gsun} Lagrangians of the RG-wall theory. The topological and flavor symmetries are exchanged, as is to be expected in a mirror-like duality. In particular, the flavor symmetry of the planar quiver is $U(1)^{N-1}$, which is mapped to the topological symmetry of $G[SU(N)]$. The $SU(N)$ symmetry is more involved. Firstly, there are monopole superpotential terms that break the $U(1)^{N(N-1)/2}$ topological symmetry of the planar quiver down to $U(1)^{N-1}$. Each $U(1)$ factor is associated to a column of the quiver. Secondly, this is believed to enhance to $SU(N)$ in the IR \cite{Benvenuti:2025huk}.
Therefore, if we were to use this Lagrangian to describe the RG-wall of $\NN=2$ SYM, the topological symmetry would couple to the UV $SU(N)$ symmetry and the flavor symmetry would couple to the IR $U(1)^{N-1}$ symmetry.
\\

We further point out that this Lagrangian is built out of chiral fields, $U(1)$ gauge groups, CS interactions and (monopole) superpotential terms, which are the basic ingredients of class $\mathcal{R}$ theories \cite{Dimofte:2011py}. 
The Lagrangian \eqref{eq: planarabelian_gsun} is therefore similar to class $\mathcal{R}$ theories and, although we do not provide any analysis in this direction, it is quite natural to expect that $G[SU(N)]$ can be engineered as the compactification of the 6d $\mathcal{N}=(2,0)$ $\mathfrak{a}_{N-1}$ theory on the RG three-manifolds of \cite{Dimofte:2013lba}, following the strategies established in \cite{Dimofte:2011ju,Dimofte:2011py,Dimofte:2013iv}. It would be interesting to study further this connection, for example attempting for the construction of the RG-wall solely from geometric data, an compare the result to \eqref{eq: planarabelian_gsun}. We leave this problem for future investigations.

\subsubsection{$Sp(2N,\mathbb{Z})$ transformations} \label{subsec: sl2z_transf}

The duality group of a 4d $U(1)^r$ effective theory is given by an $Sp(2r,\mathbb{Z})$ group, whose generators are conventionally taken to be of three ``types":
\begin{itemize}
    \item ``S-type": $\begin{pmatrix}
        I-J & -J \\ J & I-J
    \end{pmatrix}$ for any $J = \text{diag}(j_1,\ldots,j_r)$ with $j_i \in \{0,1\}$.

    \item ``T-type": $\begin{pmatrix}
        I & 0 \\ -T & I
    \end{pmatrix}$ for any $n \times n$ symmetric matrix $T$.

    \item ``U-type": $\begin{pmatrix}
        U & 0 \\ 0 & (U^{-1})^{T}
    \end{pmatrix}$ for any $U \in GL(r,\mathbb{Z})$.
\end{itemize}
where $I$ is the $n \times n$ identity matrix.

An $Sp(2N,\mathbb{Z})$ transformation on the bulk consequently transforms the $\Twall[]$ theory. Roughly speaking the actions are the following: an ``S-type" transformation has the effect of gauging the $i$-th $U(1)$ global symmetry if $j_i = 1$; a ``T-type" transformation introduces a background CS-level specified by the matrix $T$; a ``U-type" transformation is a change of basis for the $U(1)$ global symmetries dictated by the matrix $U$.

For what concerns this work, we will we be interested only in $SL(2,\mathbb{Z})_i$ transformations acting on a single $U(1)_i$ gauge symmetry. We define the generators of each $SL(2,\mathbb{Z})_i$ to be\footnote{Our choice of generator $T$ is slightly unusual and is the $S$ conjugate of the “standard" one. }:
\begin{equation}
    S_i = \begin{pmatrix}
        0 & 1 \\ -1 & 0
    \end{pmatrix} 
    \qquad,\qquad
    T_i = \begin{pmatrix}
        1 & 0 \\ -1 & 1
    \end{pmatrix} 
\end{equation}
So that the action on the 3d index of a 3d theory with $U(1)$ global symmetry is:
\begin{equation} \label{eq: sl2z_indexaction}
\begin{split}
    & \mathcal{I}^{S \cdot \mathcal{T}} (\xi',m') = \sum_m \oint \frac{d\xi}{2\pi i \xi} \xi^{m'} {\xi'}^{m} \mathcal{I}^{\mathcal{T}}(\xi,m) \\
    & \mathcal{I}^{T \cdot \mathcal{T}} (\xi,m) = \xi^m \mathcal{I}^{\mathcal{T}} (\xi,m)
\end{split}
\end{equation}
We also denote in short the following $Sp(2r,\mathbb{Z})$ operators:
\begin{equation}
    S = \prod_{i=1}^{r} S_i,
    \qquad
    T = \prod_{i=1}^{r} T_i, \qquad \text{in general:} \quad O = \prod_{i=1}^r O_i , \quad \text{for identical $O_i \in SL(2,\mathbb{Z})_i$}
\end{equation}
which consists of performing the same action on all the $U(1)$ global symmetries.

The $\Twall[]^{-1}$ theory instead transforms ``covariantly" with the inverse operator. In the sense that when we consider the setup in Figure \ref{fig:schematic_RG_inverse}, a transformation in the bulk on the r.h.s.~by an element $O \in Sp(2(N-1),\mathbb{Z})$ consequently transforms the inverse RG-wall $\Twall[]^{-1}$ with the action of $O^{-1}$ on the right. Where the meaning of acting ``on the right" is just important to remember the order of the action of the generators. Let us take an example to clarify this point. Suppose that we consider the $\Twall[2]$ theory and the following operator $S^{-1} T^2 \in SL(2, \mathbb{Z})$, which we will consider often in this work. Then the corresponding action on the $\Twall[]^{-1}$ theory is:
\begin{equation}
    (S^{-1} T^2 \cdot \Twall[2])^{-1} = (\Twall[2])^{-1} \cdot T^{-2} S =  S T^{-2} \cdot (\Twall[2])^{-1}
\end{equation}
we thus act first with $T^{-2}$ and then with $S$, with an action defined as in \eqref{eq: sl2z_indexaction}. 

Starting from the $CP\cdot G[SU(N)]$ theory we can recover the $\Twall[]$ theory by performing an $S^{-1}$ transformation in $Sp(2(N-1),\mathbb{Z})$ which is the product of the $S^{-1}$ generators of all the $SL(2,\mathbb{Z})$ acting on a single $U(1)$ symmetry.
This implies that we gauge all the topological symmetries in \eqref{eq:quiv_ptransf_Gsun} with the effect of ungauging all the $U(1) \subset U(K)$ subgroups of all the gauge groups, which become the baryonic symmetries of the $\Twall[]$ theory.

\subsubsection{The $\Twall[]$ theory}

Starting from the $S$-transformed of the $\Twall[]$ and $\Twall[]^{-1}$ theories described previously, we can perform an $S^{-1}$ transformation to obtain the quiver theories describing the $\Twall[]$ and $\Twall[]^{-1}$ theories, that are the RG-wall in the electric duality frame.

The $\Twall[]$ theory is described for generic $N$ by the following Lagrangian quiver:
\begin{equation} \label{eq: twall_quiverpic}
\Twall[] \quad : \qquad\qquad
\begin{tikzpicture}[baseline=(current bounding box).center]
    \node at (0,0) (n1) [flavor,black] {$1$};
    \node at (2,0) (n2) [gauge,double,black] {$2$};
    \node at (5,0) {$\cdots$};
    \node at (8,0) (nnm1) [gauge,double,black] {$\scriptstyle N-1$};
    \node at (10,0) (nn) [flavor,black] {$N$};

    \draw[->-] (n1) -- node[midway,above,gray] {$b_1 b_2^{-\tfrac{1}{2}}$} (n2);
    \draw[-<-] (n2)++(2,0) -- node[midway,above,gray] {$b_2^{\tfrac{1}{2}} b_3^{-\tfrac{1}{3}}$} (n2);
    \draw[->-] (nnm1)++(-2,0) -- node[midway,above,gray] {$b_{N-2}^{\tfrac{1}{N-2}} b_{N-1}^{-\tfrac{1}{N-1}}$} (nnm1);
    \draw[->-] (nnm1) -- node[midway,above,gray] {$b_{N-1}^{\tfrac{1}{N-1}}$} (nn);
\end{tikzpicture}
\end{equation}
Each bifundamental has R-charge 0 and is charged under the baryonic symmetries as expressed by the fugacities in gray\footnote{Notice that there is a rather simple way to find this parameterization. If one formally starts from a quiver where all gauge nodes are $U(n)$ symmetries instead of $SU(n)$, we simply freeze the subgroup $U(1) \in SU(n)$ as $\tfrac{1}{n}U(1)_{b_i}$.}. The global symmetry is $SU(N) \times U(1)^{N-1}$, where $SU(N)$ rotates only the last bifundamentals while $U(1)^{N-1}$ is the total baryonic symmetry. There are background CS levels for the global symmetries as:
\begin{equation}    \label{eq:bkg_CS_Twall}
    k_{b_i} = -1 , \quad k_{b_ib_j} = 1, \quad k_{SU(N)} = -\frac{N-1}{2}
\end{equation}

We report for reference also the inverse RG-wall theory $\Twall[]^{-1}$, which is the $CP$-transformed of $\Twall[]$:
\begin{equation} \label{eq: twallinv_quiverpic}
\Twall[]^{-1} \quad : \qquad\qquad
\begin{tikzpicture}[baseline=(current bounding box).center]
    \node at (0,0) (n1) [flavor,black] {$1$};
    \node at (2,0) (n2) [gauge,double,black] {$2$};
    \node at (5,0) (n3) {$\cdots$};
    \node at (8,0) (nnm1) [gauge,double,black] {$\scriptstyle N-1$};
    \node at (10,0) (nn) [flavor,black] {$N$};

    \draw[-<-] (n1) -- (n2);
    \draw[gray] ($(n1)!0.5!(n2)+(0,.7)$) node {$b_{1}^{-1} b_{2}^{\tfrac{1}{2}}$};
    \draw[->-] (n2)++(2,0) -- (n2);
    \draw[gray] ($(n2)!0.5!(n3)+(-.3,.7)$) node {$b_{2}^{-\tfrac{1}{2}} b_{3}^{\tfrac{1}{3}}$};
    \draw[-<-] (nnm1)++(-2,0) -- (nnm1);
    \draw[gray] ($(n3)!0.5!(nnm1)+(.3,.7)$) node {$b_{N-2}^{-\tfrac{1}{N-2}} b_{N-1}^{\tfrac{1}{N-1}}$};
    \draw[-<-] (nnm1) -- (nn);
    \draw[gray] ($(nnm1)!0.5!(nn)+(0,.7)$) node {$b_{N-1}^{-\tfrac{1}{N-1}}$};
\end{tikzpicture}
\end{equation}
where again all bifundamentals have R-charge 0 and there are background CS levels for the baryonic $U(1)_{b_i}$ and $SU(N)$ global symmetries as:
\begin{equation}
    k_{b_i} = 1 , \quad k_{b_ib_j} = -1, \quad k_{SU(N)} = \frac{N-1}{2} \,.
\end{equation}
The $\Twall[]$ and $\Twall[]^{-1}$ are related by $CP$, which corresponds in the inversion of all the arrows, charges under the baryonic symmetries and background CS levels\footnote{Notice that inverting the arrows has the effect of conjugating the representations under the $SU(N)$ flavor node, which is included in the action of $C$.}. 

We now describe the $\Twall[]$ theories for low fixed values of $N$, noticing that they enjoy simplier IR-dual description in certain cases. We discuss IR-dualities for the $\Twall[]$ theory more generally in Appendix \ref{app: gsun}.

\paragraph{RG-wall for $SU(2)$:} For $N=2$ the theory $\Twall[2]$ is given by the following quiver:
\begin{equation}
    \Twall[2] \qquad = \qquad 
    \begin{tikzpicture}[baseline=(current bounding box).center]
        \node[flavor,black] at (0,0) (f1) {$1$};
        \node[flavor,black] at (2,0) (f2) {$2$};
        \path[draw,->-] (f1) -- (f2);

        \draw[->-] (n1) -- node[midway,above,gray] {$b_1$} (n2);
    \end{tikzpicture}
    \qquad = \qquad
    \text{2 free chirals}
\end{equation}
This quiver corresponds to two free chirals with charge $1$ under the $U(1)_b$ symmetry and transforming in the fundamental of the $SU(2)$ symmetry. The theory has also a background CS level $-1$ for $U(1)_b$ and $-1/2$ for $SU(2)$.
The $\Twall[2]$ theory coincides with the proposal in \cite{Dimofte:2013lba}, where the RG-wall for $SU(2)$ $\NN=2$ SYM was obtained as a compactification of the 6d $(2,0)$ theory on a three-manifold with boundary. 

\paragraph{RG-wall for $SU(3)$:}
For $N=3$ the theory $\Twall[3]$ is a non-trivial interacting CFT. The Lagrangian is given by:
\begin{equation} 
\Twall[3] \qquad = \qquad
\begin{tikzpicture}[baseline=(current bounding box).center]
        \node at (0,0) (n1) [flavor,black] {$1$};
        \node at (2,0) (n2) [gauge,double,black] {$2$};
        \node at (4,0) (n3) [flavor,black] {$3$};

        \draw[->-] (n1) -- node[midway,above,gray] {$b_1 b_2^{-\tfrac{1}{2}}$} (n2);
        \draw[->-] (n2) -- node[midway,above,gray] {$b_2^{\tfrac{1}{2}}$} (n3);
    \end{tikzpicture}
\end{equation}
which is 3d $\NN=2$ SQCD with gauge group $SU(2)$, 4 fundamentals and vanishing CS level. 
The representations of the four fundamentals under the gauge $SU(2)$ symmetry and the global $ U(1)_R \times U(1)_{b_1} \times U(1)_{b_2} \times SU(3)$ symmetry are:
\begin{equation}
    \begin{array}{c|c|cccc}
         & SU(2) & U(1)_R & U(1)_{b_1} & U(1)_{b_2} & SU(3) \\ 
         \hline
         \tilde{q} & \square & 0 & 1 & -\frac{1}{2} &-    \\
         q & \square & 0 & 0 & - \frac{1}{2} & \overline{\square}
    \end{array}
\end{equation}
where we formally separated the four fundamentals into one antifunamental $\tilde{q}$ and three fundamentals $q$ in analogy with the higher rank cases. Indeed the theory appears to enjoy a larger global symmetry as the four chirals could be rotated by a $U(4)$ global symmetry. However, this symmetry is broken explicitly by the background CS interactions:
\begin{equation}
    k_{b_1}=k_{b_2}=-1, \quad k_{b_1 b_2}=1, \quad  k_{SU(3)}=-1 
\end{equation}
This theory, when regarded as a $SU(2)$ SQCD with four fundamental chirals is IR-dual to a WZ model of 7 chirals with a cubic superpotential\cite{Aharony:1997gp,Aharony:2014uya}:
\begin{equation}
    SU(4) \, \text{w/ 4 fundamentals} \qquad \leftrightarrow \qquad  
    \begin{gathered}
        \text{7 chirals } Y, B_{ij} \qquad 1\leq i <j \leq 4
        \\
        \text{with} \quad \mathcal{W}=Y \, \text{Pf}(B)
    \end{gathered}
\end{equation}
In terms of the $\Twall[3]$ theory, the chirals $B_{ij}$ maps to the baryons and $Y$ to the monopole\footnote{Here we regard the original gauge theory as $SU(2)$ with 4 fundamentals which has $\frac{4 \times 3}{2}=6$ baryons.}. The charges of the fields under the global $ U(1)_R \times U(1)_{b_1} \times U(1)_{b_2} \times SU(3)$ symmetry are:
\begin{equation}
    \begin{array}{c|c|cccc}
         & U(1)_R & U(1)_{b_1} & U(1)_{b_2} & SU(3) \\ 
         \hline
         B_{4i}, \quad i=1,2,3& 0 & 1 & 0 & \overline{\square}    \\
         B_{ij} \quad 1 \leq i < j \leq 3& 0 & 0 & 1 & \square   \\
         Y & 2 & -1 & -1 & -
    \end{array}
\end{equation}
The WZ model can be compactly represented as the following quiver:
\begin{equation}
\Twall[3] \qquad = \qquad
    \begin{tikzpicture}[baseline=(current bounding box).center]
        \node at (0,1) (n1) [flavor,black] {$1$};
        \node at (0,-1) (n2) [flavor,black] {$1$};
        \node at (2,0) (n3) [flavor,black] {$3$};

        \path[draw,->-] (n1) -- node[midway,above,gray] {$b_1$} (n3);
        \path[draw,->-] (n3) -- node[midway,below,gray] {$b_2$} (n2);
        \path[draw,->-] (n2) -- node[midway,left,gray] {$b_1^{-1} b_2^{-1}$} (n1);
    \end{tikzpicture}
\end{equation}
where the superpotential is associated to the closed loop in the quiver.
The background CS levels are not affected by this dualization and are thus given in \eqref{eq:bkg_CS_Twall}.

Interestingly, this simple description is reminiscent of the type of theories obtained by compactifying the 6d $(2,0)$ $\mathfrak{a}_2$ theory on a three-manifold that is a bipyramid, as proposed in \cite{Dimofte:2013iv}. This manifold should be closely related to the RG manifold.
It would be interesting to explore this further, but we leave this to future work.

\paragraph{RG-wall for $SU(4)$:}
For $N=4$ the theory $\Twall[4]$ is given by:
\begin{equation} 
\Twall[4] \qquad = \qquad
\begin{tikzpicture}[baseline=(current bounding box).center]
        \node at (0,0) (n1) [flavor,black] {$1$};
        \node at (2,0) (n2) [gauge,double,black] {$2$};
        \node at (4,0) (n3) [gauge,double,black] {$3$};
        \node at (6,0) (n4) [flavor,black] {$4$};

        \draw[->-] (n1) -- node[midway,above,gray] {$b_1 b_2^{-\tfrac{1}{2}}$} (n2);
        \draw[->-] (n2) -- node[midway,above,gray] {$b_2^{\tfrac{1}{2}}b_3^{-\tfrac{1}{3}}$} (n3);
        \draw[->-] (n3) -- node[midway,above,gray] {$b_3^{\tfrac{1}{3}}$} (n4);
    \end{tikzpicture}
\end{equation}
similarly to the case of $\Twall[3]$ this Lagrangian description can be simplified via local dualization. The $SU(2)$ gauge node S-confines as in the case of $\Twall[3]$. The resulting gauge theory is an $SU(3)$ gauge theory with one antifundamental and 5 fundamentals and vanishing CS. This theory is Aharony-like dual to an $SU(2)$ gauge theory with 6 chirals that can be formally tought as 1 fundamental and 5 ``antifundamentals". \cite{Benini:2011mf,Closset:2023vos}. Keeping track of gauge singlet fields, we find the following quiver for $\Twall[4]$:
\begin{equation}
\Twall[4] \qquad = \qquad
    \begin{tikzpicture}[baseline=(current bounding box).center]
        \node at (2,-2) (n1) [flavor,black] {$1$};
        \node at (0,1) (n2) [flavor,black] {$1$};
        \node at (2,0) (n3) [gauge,double,black] {$2$};
        \node at (4,-1) (n4) [flavor,black] {$4$};

        \path[draw,-<-] (n1) -- node[midway,left,gray] {$b_1^{-1} b_2^{-1/2}$} (n3);
        \path[draw,->-] (n3) -- node[midway,above right,gray] {$b_3 b_2^{-1/2}$} (n2);
        \path[draw,->-] (n1) -- node[midway,below,gray,yshift=0cm] {$b_1$} (n4);
        \path[draw,-<-] (n3) -- node[midway,above,gray] {$b_2^{1/2}$} (n4);
    \end{tikzpicture}
\end{equation}
where some of the arrows are just a formality to make contact with more generic pictures as in this $SU(2)$ gauge theory, there is indeed no distinction between fundamentals and anti-fundamentals for the $SU(2)$ gauge group.
All the diagonal chiral fields have R-charge $0$ and the vertical field has R-charge $2$ and there is a cubic superpotential term associated to the triangle.
The background CS levels are not affected by the dualization sequence and are thus given in \eqref{eq:bkg_CS_Twall}.

\paragraph{RG-wall for $SU(5)$:}
For $N=5$ the theory $\Twall[5]$ is given by:
\begin{equation}    \label{eq:quiv_Twall4}
\Twall[5] \qquad = \qquad
\begin{tikzpicture}[baseline=(current bounding box).center]
        \node at (0,0) (n1) [flavor,black] {$1$};
        \node at (2,0) (n2) [gauge,double,black] {$2$};
        \node at (4,0) (n3) [gauge,double,black] {$3$};
        \node at (6,0) (n4) [gauge,double,black] {$4$};
        \node at (8,0) (n5) [flavor,black] {$5$};

        \draw[->-] (n1) -- node[midway,above,gray] {$b_1 b_2^{-\tfrac{1}{2}}$} (n2);
        \draw[->-] (n2) -- node[midway,above,gray] {$b_2^{\tfrac{1}{2}}b_3^{-\tfrac{1}{3}}$} (n3);
        \draw[->-] (n3) -- node[midway,above,gray] {$b_3^{\tfrac{1}{3}} b_4^{-\tfrac{1}{4}}$} (n4);
        \draw[->-] (n4) -- node[midway,above,gray] {$b_4^{\tfrac{1}{4}}$} (n5);
    \end{tikzpicture}
\end{equation}
which can be simplified by performing local dualities. We confine the $\Twall[4]$ tail as before, resulting in a quiver with an $SU(4)$ and an $SU(2)$ gauge node. The $SU(4)$ gauge group has 7 fundamentals and 1 andifundamental and is Aharony-like dual to $SU(3)$ with 7 antifundamentals and 1 fundamental. After this dualization the $SU(2)$ gauge node has 4 fundamentals and confines as in the case of $\Twall[3]$. Keeping track of the various details we find the following quiver for $\Twall[5]$:
\begin{equation}    \label{eq:quiv_Twall4_dualized}
    \Twall[5] \qquad = \qquad 
    \begin{tikzpicture}[baseline=(current bounding box).center]
        \node at (0,0) (n1) [flavor,black] {$1$};
        \node at (0,-2) (n2) [flavor,black] {$1$};
        \node at (2,-1) (n3) [gauge,double,black] {$3$};
        \node at (2,-3) (n4) [flavor,black] {$1$};
        \node at (4,-2) (n5) [flavor,black] {$5$};

        \draw[->-] (n1) -- (n2);
        \draw[->-] (n2) -- (n3);
        \draw[->-] (n3) -- (n1);
        \draw[->-] (n3) -- (n4);
        \draw[->-] (n4) -- (n5);
        \draw[->-] (n5) -- (n3);

        \path (n1) -- node[midway,gray,left] {$ \scriptstyle {b_2 b_3^{-1} b_4^{-1} }$} (n2);
        \path (n2) -- node[midway,gray,below] {$\scriptstyle { b_3 b_2^{-\frac{2}{3}} }$} (n3);
        \path (n3) -- node[midway,gray,above] {$\scriptstyle { b_4 b_2^{-\frac{1}{3}}} $} (n1);
        \path (n3) -- node[midway,gray,right] {$\scriptstyle { b_1^{-1} b_2^{-\frac{1}{3}}} $} (n4);
        \path (n4) -- node[midway,gray,below] {$\scriptstyle { b_1} $} (n5);
        \path (n5) -- node[midway,gray,above] {$\scriptstyle { b_2^{\frac{1}{3}}} $} (n3);
    \end{tikzpicture}
\end{equation}
All the diagonal chiral fields have R-charge $0$ and all the vertical fields have R-charge $2$ and there are two cubic superpotential terms associated to the two triangular loops.
The background CS levels are not affected by the dualization sequence.
The $SU(3)$ gauge group is self-dual and the dualization sequence stops here.
\\

For $N>5$ the RG-wall theory $\Twall[N]$ does not admit, to our knowledge, dual Lagrangians that involve a single gauge group.
However, by using the IR dualities presented in Appendix \ref{app: gsun}, the ``simplest"\footnote{By ``simplest" we mean with lowest total rank of the gauge groups and, in case two or more descriptions have the same total rank, the simplest description is the one with less gauge groups.} description of the $\Twall[N]$ theory is in terms of a theory with $(\lfloor N/2 \rfloor-1)$ gauge groups with ranks jumping by 2: $SU(N-2)\times SU(N-4) \times SU(N-6) \times \ldots$

\section{The RG-wall as an interface} \label{sec: interfaces}

In this section we consider setup consisting of interfaces in $\mathcal{N}=2$ SYM. Roughly speaking there are two situations at hand. We can consider an interface between the UV description of $\mathcal{N}=2$ SYM and an IR effective description, which is naturally realized by the $\Twall[]$ theory. The second possibility instead consists of considering an interface between two different IR descriptions which are related by low-energy electromagnetic dualities. In this section we construct and describe all these setups and discuss non-trivial consistency conditions arising from the collision of multiple interfaces.
In Section \ref{sec: half_index_interfaces} we provide clear computations of these setups from localization and present high-precision checks of these consistency conditions.

Indeed, one can consider setups where the collision of two RG-walls produces an interface between two identical descriptions of the theory, either between two UV descriptions or between the same IR effective theory. In these situations, we expect that the interface becomes ``transparent", which simply means that it must be IR-equivalent to having no interface at all. Even in these simpler cases, the generic proposal of the $\Twall[]$ theory given in Section \ref{sec: bc_proposal}, proves to satisfy a number of non-trivial properties as a 3d $\mathcal{N}=2$ SCFT encoding all the desired properties.

We will start the discussion with the interface between the UV and IR description, followed by the case of two IR effective descriptions and, finally, we comment the transparent interface between two UV descriptions.

\subsection{The $\Twall[]$ as an UV/IR interface}

We start considering 4d $\mathcal{N}=2$ SYM in a space without a boundary. We first consider the theory at weak coupling and let it flow to strong coupling only in half of the full space. The result is a setup given by $\mathcal{N}=2$ SYM on one half-space and the effective IR description on the other half. 
In between there is a domain wall separating the two description which, following the proposal in Section \ref{sec: bc_proposal}, it is naturally described by the $\Twall[]$ theory. 
\begin{equation}    \label{eq:RG_wall_flow_electric}
    \begin{tikzpicture}
        \begin{scope}[xshift=0cm]
        \node at (0,0) (UV) {$\begin{array}{c} \text{UV} \\ \\ SU(N) \\ \NN=2 \; \text{SYM} \end{array}$};
    \end{scope}

    \begin{scope}[xshift=2cm]
    \path[draw,->] (0,0) -- node[midway,above] {Half-space} node[midway,below] {flow} (2.5,0);
    \end{scope}
        
        \begin{scope}[xshift=2.5cm]
        \node[left] at (5.5,0) (UV2) {$\begin{array}{c} \text{UV} \\ \\ SU(N) \\ \NN=2 \; \text{SYM} \end{array}$};
        \path[draw] (6,-1.5) node[below,anchor=north] {$\Twall[]$}  -- (6,1.5) node[anchor=north east] {$\mathcal{N}$} node[anchor= north west] {$\mathcal{N},\mathcal{B}_{\chi}$};
        \node[right] at (6.5,0) (IR) {$\begin{array}{c} \text{IR} \\  \\\text{Seiberg-Witten}\\ \text{in electric EM frame }\end{array}$};
    \end{scope}
    \end{tikzpicture}
\end{equation}
where the $\Twall[]$ theory is coupled to the UV via the diagonal gauging of the $SU(N)$ symmetry, and is coupled to the IR via the diagonal gauging of the $U(1)^{N-1}$ symmetry with the boundary superpotential \eqref{eq:bdy_W_N2} discussed in the previous Section.

Indeed, one can consider the IR theory in any $Sp(2(N-1),\mathbb{Z})$ duality frame labeled by an element $\sigma$. 
This is achieved by performing the corresponding electromagnetic duality for the Abelian low energy theory on the right half of spacetime. Correspondingly, we also perform the Witten's $Sp(2(N-1),\zz)$ action on the Abelian symmetry of the $\Twall[]$ theory with the same element $\sigma$, we denote the resulting theory as $\sigma \cdot \Twall[]$.
This setup can be graphically represented as:
\begin{equation}    \label{eq:RG_wall_flow}
    \begin{tikzpicture}
        \begin{scope}[xshift=0cm]
        \node at (0,0) (UV) {$\begin{array}{c} \text{UV} \\ \\ SU(N) \\ \NN=2 \; \text{SYM} \end{array}$};
    \end{scope}

    \begin{scope}[xshift=2cm]
    \path[draw,->] (0,0) -- node[midway,above] {Half-space} node[midway,below] {flow} (2.5,0);
    \end{scope}
        
        \begin{scope}[xshift=2.5cm]
        \node[left] at (5.5,0) (UV2) {$\begin{array}{c} \text{UV} \\ \\ SU(N) \\ \NN=2 \; \text{SYM} \end{array}$};
        \path[draw] (6,-1.5) node[below,anchor=north] {$\sigma \cdot \Twall[]$}  -- (6,1.5) node[anchor=north east] {$\mathcal{N}$} node[anchor= north west] {$\mathcal{N},\mathcal{B}_{\chi}$};
        \node[right] at (6.5,0) (IR) {$\begin{array}{c} \text{IR} \\  \\ \text{Seiberg-Witten}\\ \text{in $\sigma$ EM frame} \end{array}$};
    \end{scope}
    \end{tikzpicture}
\end{equation}
The validity of this setup is expected to hold naturally as a consequence of the discussion in Section \ref{sec: bc_proposal}. In Section \ref{sec: half_index_interfaces} we discuss how this property translates as a mathematical identity between indices.

\subsection{The IR/IR interface}

Starting from the setup in \eqref{eq:RG_wall_flow}, we can construct interfaces between IR descriptions.

The idea is that starting from the setup on the r.h.s.~of \eqref{eq:RG_wall_flow} we can let the theory flow to strong coupling also in the left half-space region. 
Possibly, one can consider the two effective IR descriptions in different duality frames, which we will label by their associated $Sp(2(N-1),\mathbb{Z})$ elements $\sigma_L,\sigma_R$, thus creating an interface between different IR descriptions.
This setup can be schematically depicted as:
\begin{equation}    
\begin{tikzpicture}
    \node[left] at (5.5,0) (UV2) {$\begin{array}{c} \text{IR} \\  \\ \text{Seiberg-Witten}\\ \text{in $\sigma_L$ EM frame} \end{array}$};
    \path[draw] (6,-1.5) node[below,anchor=north] {$_{\sigma_L} \mathbb{I}_{\sigma_R}$}  -- (6,1.5) node[anchor=north east] {$\mathcal{N},\mathcal{B}_{\chi}$} node[anchor= north west] {$\mathcal{N},\mathcal{B}_{\chi}$};
    \node[right] at (7,0) (IR) {$\begin{array}{c} \text{IR} \\  \\ \text{Seiberg-Witten}\\ \text{in $\sigma_R$ EM frame} \end{array}$};
\end{tikzpicture}
\end{equation}
The 3d $\mathcal{N}=2$ theory, that we name $_{\sigma_L} \mathbb{I}_{\sigma_R}$, living on the domain wall between the two IR phases is constructed as the ``fusion" of two $\Twall[]$ theories trough the anti-diagonal gauging of their $SU(N)$ global symmetries. We now aim to describe in more detail this theory by looking at choices for $\sigma_L$ and $\sigma_R$.

\paragraph{Multimonopole/multimonopole interface}
Similarly as for the discussion in Section \ref{sec: bc_proposal}, it turns out to be conceptually convenient to start considering both the IR theories in the multimonopole frame, which is labelled by the element $S \in Sp(2(N-1),\mathbb{Z})$ that is obtained as the product of the $N-1$ $S_i$ generators of the $SL(2,\mathbb{Z})_i$ subgroups. In this way we are able to leverage results already discussed in the literature, in particular in \cite{Benvenuti:2025huk}, and construct any other generic setup building upon this example.

The $S \cdot \Twall[]$ theory is given by the $CP$-transformed of the $G[SU(N)]$ theory given in \eqref{eq:quiv_ptransf_Gsun}.
The theory describing the interface between to multimonopole frames, denoted as $_S \mathbb{I}_S$, is given by the following quiver\footnote{To be precise, one should regard the quiver in the last line has having gauge group:
\begin{equation}
    \frac{U(1)^2 \times U(2)^2 \times \ldots \times U(N-1)^2 \times U(N)}{U(1)} = \frac{U(1)^2 \times U(2)^2 \times \ldots \times U(N-1)^2 \times SU(N)}{\mathbb{Z}_N}
\end{equation}
Therefore there is an extra $\mathbb{Z}_N$ gauged that is not represented in the quiver description. Also, any non-trivial $\mathbb{Z}_N$ background is suppressed, thus we can not actually detect its presence.
}.
\begin{equation}    \label{eq: magn_ir_interf}
    \includegraphics[]{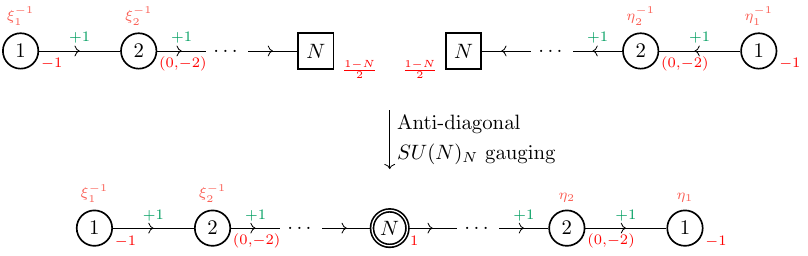}
\end{equation}
This theory is obtained from two $S\cdot \Twall[]$ theories by  the anti-diagonal gauging of the $SU(N)$ symmetry with CS level $N$. 
Notice that in the $S\cdot \Twall[]$ theory corresponding to the right tail of the quiver \eqref{eq: magn_ir_interf}, the $SU(N)$ current has opposite sign due to the anti-diagonal gauging, thus we performed a gauge transformation $A \to -A$ for all of its gauge groups to achieve a standard quiver presentation. The direction of the arrows and the sign of the FIs of the r.h.s.~tail are reversed due to this manipulation.

The quiver in \eqref{eq: magn_ir_interf} can be obtained as a defect in the low energy theory as follows. We can consider $\mathcal{N}=2$ SYM and let the theory flow to the IR effective description on two semi-infinite regions of the space, separated by a slice. We then send the width of the slice to 0, colliding the two interfaces and producing a defect in the low energy effective theory. 
\begin{equation}    \label{eq:collision_IR_def}
\resizebox{\hsize}{!}{ 
\begin{tikzpicture}[]
    \begin{scope}[xshift=0cm]
        \node[left] at (1.5,0) (IR) {$\begin{array}{c} \text{IR} \\  \\ \text{Seiberg-Witten}\\ \text{in $S$ EM frame} \end{array}$};
        \path[draw] (2,-1.5) node[below,anchor=north] {$S \cdot \Twall[]$}  -- (2,1.5) node[anchor=north east] {$\mathcal{N},\mathcal{B}_{\chi}$} node[anchor=north west] {$\mathcal{N}$};
        \node[right] at (2.5,0) (IR) {$\begin{array}{c} \text{UV} \\  \\ \text{$SU(N)$}\\ \text{$\NN=2$ SYM} \end{array}$};
        \path[draw] (5.5,-1.5) node[below,anchor=north] {$S \cdot \Twall[]$}  -- (5.5,1.5) node[anchor=north east] {$\mathcal{N}$} node[anchor=north west] {$\mathcal{N},\mathcal{B}_{\chi}$};
        \node[right] at (6,0) (IR) {$\begin{array}{c} \text{IR} \\  \\ \text{Seiberg-Witten}\\ \text{in $S$ EM frame} \end{array}$};
        \path[draw,<->] (2,1.8) -- node[midway,above] {$\Delta x_3$} (5.5,1.8);
    \end{scope}
    \path[draw,->] (9.5,0) -- node[midway,above] {$\Delta x_3 \to 0$} (11.5,0);
    \begin{scope}[xshift=13.5cm]
        \node[left] at (1.5,0) (IR) {$\begin{array}{c} \text{IR} \\  \\ \text{Seiberg-Witten}\\ \text{in $S$ EM frame} \end{array}$};
        \path[draw] (2,-1.5) node[below,anchor=north] {$_S \mathbb{I}_S$}  -- (2,1.5) node[anchor=north east] {$\mathcal{N},\mathcal{B}_{\chi}$} node[anchor=north west] {$\mathcal{N},\mathcal{B}_{\chi}$};
        \node at (4,0) (IR) {$\begin{array}{c} \text{IR} \\  \\ \text{Seiberg-Witten}\\ \text{in $S$ EM frame} \end{array}$};
    \end{scope}
\end{tikzpicture}
} 
\end{equation}
As schematically represented in the picture above, the interface between the two IR phases arises from the collision of the two IR/UV domain walls. In this picture, the theory reported in \eqref{eq: magn_ir_interf} can be read off as follows. The two tails are the IR/UV interfaces, which are two 3d $\mathcal{N}=2$ $S \cdot \Twall[]$ theories.
The two theories are glued together via the diagonal gauging of the $SU(N)$ global symmetries with an additional CS level $N$, which can be seen as the contribution of the 4d slice as $\Delta_{x_3} \to 0$. Indeed, this picture can be easily generalized for interfaces between different IR descriptions connected by Abelian electromagnetic duality.

Let us now turn the attention back to the theory describing the interface between two IR multimonopole frames, reported in \eqref{eq: magn_ir_interf}.
This theory was already studied in \cite{Benvenuti:2025huk}, where it was observed that it exibit the peculiar property of having a deformed moduli space so that the naively expected $U(1)_\xi^{N-1} \times U(1)_\eta^{N-1}$ topological symmetry is spontaneously broken down to the $U(1)^{N-1}$ diagonal subgroup. In the IR the theory is described by a set of $N-1$ free chirals, each carrying R-charge $1$ and $-1$ charge under a single unbroken $U(1)$ factor and with background CS-level $+1/2$. This relation schematically reads:
\begin{equation} \label{eq: fusion_to_id}
    \begin{tikzpicture}[baseline,font=\footnotesize,
    node distance=1.25cm
]

 \draw (-.2,0) node[right] {Interface $_S \mathbb{I}_S$};

 \draw[<->] (2.5,0)--(3.5,0);

\begin{scope}[shift={(5,-.5)}]

 \node[flavor] (g1) {$1$};
 \node[flavor, right of=g1] (g2) {$1$};
 \node[right of=g2] (g3) {$\cdots$};
 \node[flavor, right of=g3] (g4) {$1$};

 \node[flavor,above of=g1] (f1) {$1$};
 \node[flavor, above of=g2] (f2) {$1$};
 \node[flavor, above of=g4] (f4) {$1$};

 \draw[->-] (f1)--(g1);
 \draw[->-] (f2)--(g2);
 \draw[->-] (f4)--(g4);

 \draw[gray] (g1)+(0,-.5) node {$\xi_1$};
 \draw[gray] (g2)+(0,-.5) node {$\xi_2$};
 \draw[gray] (g4)+(0,-.5) node {$\xi_{N\text{-}1}$};

 \draw[CScolor] (g1)+(.2,-.4) node[right] {$+\tfrac{1}{2}$};
 \draw[CScolor] (g2)+(.2,-.4) node[right] {$+\tfrac{1}{2}$};
 \draw[CScolor] (g4)+(.2,-.4) node[right] {$+\tfrac{1}{2}$};

 \draw (4.5,.5) node[right] {$\times \quad \prod_{i=1}^{N-1} \delta(\xi_i - \eta_i)$};

\end{scope}
 
\end{tikzpicture}
\end{equation}
where the delta function on the r.h.s.~express the constraint imposed by the spontaneous breaking of the global symmetry. On an intuitive level, a gauging can be thought as an integration, which is exactly the picture in typical partition functions, therefore the delta functions represents that the gauge symmetry is spontaneously broken\footnote{This interpretation is similar to that in \cite{Spiridonov:2014cxa}, where a singular functional partition function is expected for 4d theories exibiting chiral symmetry breaking. A similar result was also found in \cite{Giacomelli:2023zkk,Comi:2025zwu}, where $\delta$-functions in 3d partition functions are interpreted, in certain circumstances, as monopole operators acquiring a VEV.}.
This property was shown to be a direct consequence of the fusion-to-identity satisfied by the 3d $\NN=4$ theory $T[SU(N)]$, obtained by turning on a suitable real mass deformation.
The fusion-to-identity of $T[SU(N)]$ can be understood from the fact that $T[SU(N)]$ is the S-duality wall of 4d $\NN=4$ SYM and it satisfies $SS^{-1}=1$. Consequently, it was observed that the partition function of the diagonal gauging of two $T[SU(N)]$ theories behaves as a delta-function \cite{Kapustin:1999ha,Benini:2010uu,Bottini:2021vms,Hwang:2021ulb,Comi:2022aqo} (see also Appendix \ref{app: tsun}).
We further notice that the fusion-to-identity, both in the $\NN=4$ and in the $\NN=2$ case, can be argued for by field theoretical arguments via a sequence of local dualization. The $\NN=4$ case was worked out in detail in \cite{Bottini:2021vms}. 
The $\NN=2$ case can be addressed in a similar fashion by performing local Giveon-Kutasov dualities \cite{Giveon:2008zn} in the quiver \eqref{eq: magn_ir_interf}.

It is important to observe that in \eqref{eq: fusion_to_id} the topological symmetries of the two $S \cdot \Twall[]$ tails are identified with a specific pattern. Namely, the topological symmetries associated to the two $U(r)$ gauge nodes with same rank $r$ are identified.
In the quiver descriptions \eqref{eq: magn_ir_interf} and \eqref{eq: fusion_to_id} this is encoded in the precise labeling of the fugacities for the topological symmetries $\xi_i$ and $\eta_i$. Notice also that the theory on the r.h.s.~of \eqref{eq: fusion_to_id} factorizes as $N-1$ copies of the same theory given by one free chiral. The implication of this observation is that the interface in the generic $SU(N)$ case breaks down to $N-1$ copies of the $SU(2)$ interface.

We expect that this simple theory of free chirals, coupled to the IR Seiberg-Witten effective theory with b.c.~$(\mathcal{N}, \mathcal{B}_{\chi})$ on both sides, flows to the trivial defect. 
Indeed, the 3d $U(1)^{N-1}$ gauging reconstructs the 4d vector multiplet at the interface while the $N-1$ 3d chiral fields introduce the d.o.f.~for the antichiral components of the 4d BPS monopoles that was set to zero by the boundary conditions $\mathcal{B}_{\chi}$. 
The CS terms at level $\tfrac{1}{2}$ compensate the natural CS on the defect that arise due to the chiral/antichiral splitting of the BPS particles with b.c. $\mathcal{B}_{\chi}$.
Overall, we expect that such defect is IR equivalent to the trivial defect $\mathbb{I}$:
\begin{equation}    
\resizebox{\hsize}{!}{ 
\begin{tikzpicture}[]
    \begin{scope}[xshift=0cm]
        \node[left] at (1.5,0) (IR) {$\begin{array}{c} \text{IR} \\  \\ \text{Seiberg-Witten}\\ \text{in $S$ EM frame} \end{array}$};
        \path[draw] (2,-1.5) node[below,anchor=north] {$S \cdot \Twall[]$}  -- (2,1.5) node[anchor=north east] {$\mathcal{N},\mathcal{B}_{\chi}$} node[anchor=north west] {$\mathcal{N}$};
        \node[right] at (2.5,0) (IR) {$\begin{array}{c} \text{UV} \\  \\ \text{$SU(N)$}\\ \text{$\NN=2$ SYM} \end{array}$};
        \path[draw] (5.5,-1.5) node[below,anchor=north] {$S \cdot \Twall[]$}  -- (5.5,1.5) node[anchor=north east] {$\mathcal{N}$} node[anchor=north west] {$\mathcal{N},\mathcal{B}_{\chi}$};
        \node[right] at (6,0) (IR) {$\begin{array}{c} \text{IR} \\  \\ \text{Seiberg-Witten}\\ \text{in $S$ EM frame} \end{array}$};
        \path[draw,<->] (2,1.8) -- node[midway,above] {$\Delta x_3$} (5.5,1.8);
    \end{scope}
    \path[draw,->] (9.5,0) -- node[midway,above] {$\Delta x_3 \to 0$} (11.5,0);
    \begin{scope}[xshift=13.5cm]
        \node[left] at (1.5,0) (IR) {$\begin{array}{c} \text{IR} \\  \\ \text{Seiberg-Witten}\\ \text{in $S$ EM frame} \end{array}$};
        \path[draw,dashed] (2,-1.5) node[below,anchor=north] {$\mathbb{I}$}  -- (2,1.5);
        \node at (4,0) (IR) {$\begin{array}{c} \text{IR} \\  \\ \text{Seiberg-Witten}\\ \text{in $S$ EM frame} \end{array}$};
    \end{scope}
\end{tikzpicture}
} 
\end{equation}

We note a subtlety in the discussion above, namely that there are more BPS particles then the $N-1$ BPS monopoles which are electric in this electromagnetic frame. 
The other BPS particles should also have half-BPS b.c.~which morally should set to zero either their chiral or antichiral component on the boundary. These are not compensated by pure 3d d.o.f.~in the defect theory \eqref{eq: fusion_to_id}.
Nevertheless,
in Section \ref{sec: half_index_interfaces} we provide a strong check that the theory of $N-1$ free chirals in \eqref{eq: fusion_to_id} flows to the trivial defect by computing the index in the presence of such interface. In particular, we compute the supersymmetric partition function on euclidean spacetime $S^3 \times S^1$ with the defect in \eqref{eq: fusion_to_id} placed along the equator of the $S^3$. We show that this reproduces the Schur index of pure $SU(N)$ SYM, therefore the defect defined by the 3d theory $_S \mathbb{I}_S$ is IR equivalent to the trivial defect. 

\paragraph{Electric/Multimonopole interface}

To construct interfaces between more generic frames it is now sufficient to perform the associated $Sp(2(N-1),\mathbb{Z})$ action on the interface theory, as defined in subsection \ref{subsec: sl2z_transf}. We want to think at interfaces as ``matrix" like objects carrying two indices that are associated to two set of $U(1)^{N-1}$ global symmetries and thus with two distinct $Sp(2(N-1),\mathbb{Z})$ groups acting on them. Indeed, the theory might exibit spontaneous symmetry breaking, such as the one already described, and in these cases we still want to think as having two set of $U(1)^{N-1}$ global symmetries that can be identified by quantum effects. These interfaces are then glued to the two semi-infinite bulk by the diagonal gauging of the two set of $U(1)^{N-1}$ symmetries.

We can then construct the interface between the electric and multimonopole frames by performing an $S^{-1}$ transformation on a single set of $U(1)^{N-1}$ symmetries in the interface in \eqref{eq: fusion_to_id}. This consists of gauging the $U(1)^{N-1}$ global symmetries with negative BF coupling. Due to the deformed moduli space, this gauging is spontaneously broken and we simply obtain a theory of $N-1$ chiral multiplets. The set of $U(1)^{N-1}$ global symmetries coupling to the multimonopole frame is the one acting on the $N-1$ chiral multiplets while their $U(1)^{N-1}$ associated magnetic symmetries are instead coupled to the electric frame. This setup can be represented as:
\begin{equation} \label{eq: elec_magn_interface}
    \begin{tikzpicture}[baseline,font=\footnotesize,
    node distance=1.5cm
]

 \draw (0,0) node[right] {\large{$_1 \mathbb{I}_S$}};

 \draw[<->] (2,0)--(3,0);

\begin{scope}[shift={(5,-.5)}]

 \node[flavor] (g1) {$1$};
 \node[flavor, right of=g1] (g2) {$1$};
 \node[right of=g2] (g3) {$\cdots$};
 \node[flavor, right of=g3] (g4) {$1$};

 \node[flavor,above of=g1] (f1) {$1$};
 \node[flavor, above of=g2] (f2) {$1$};
 \node[flavor, above of=g4] (f4) {$1$};

 \node[flavor,below of=g1] (b1) {$1$};
 \node[flavor, below of=g2] (b2) {$1$};
 \node[flavor, below of=g4] (b4) {$1$};

 \draw[->-] (f1)--(g1);
 \draw[->-] (f2)--(g2);
 \draw[->-] (f4)--(g4);
 \draw[dashed] (b1)--(g1);
 \draw[dashed] (b2)--(g2);
 \draw[dashed] (b4)--(g4);

 \draw[gray] (g1)+(-.5,0) node {$\xi_1$};
 \draw[gray] (g2)+(-.5,0) node {$\xi_2$};
 \draw[gray] (g4)+(-.6,0) node {$\xi_{N\text{-}1}$};

 \draw[gray] (b1)+(-.5,0) node {$\chi_1$};
 \draw[gray] (b2)+(-.5,0) node {$\chi_2$};
 \draw[gray] (b4)+(-.6,0) node {$\chi_{N\text{-}1}$};

 \draw[CScolor] (g1)+(.2,-.4) node[right] {$+\tfrac{1}{2}$};
 \draw[CScolor] (g2)+(.2,-.4) node[right] {$+\tfrac{1}{2}$};
 \draw[CScolor] (g4)+(.2,-.4) node[right] {$+\tfrac{1}{2}$};

 \draw[CScolor] (b1)+(.2,-.4) node[right] {$0$};
 \draw[CScolor] (b2)+(.2,-.4) node[right] {$0$};
 \draw[CScolor] (b4)+(.2,-.4) node[right] {$0$};

 \draw[BFcolor] ($(g1)!0.5!(b1)+(-.05,-.1)$) node[left] {$+2$};
 \draw[BFcolor] ($(g2)!0.5!(b2)+(-.05,-.1)$) node[left] {$+2$};
 \draw[BFcolor] ($(g4)!0.5!(b4)+(-.05,-.1)$) node[left] {$+2$};

\end{scope}
 
\end{tikzpicture}
\end{equation}
where we recall that all chirals have R-charge 1 and a dashed line represent the presence of just a BF coupling for the two connected nodes.

\paragraph{Generic interfaces}

Having understood how to generalize the results presented above, we are ready to provide a proposal of the interface between two generic frames. This will be simply constructed by picking a basic object, which is a conventional choice of pair of frames for which we can give a simple 3d $\mathcal{N}=2$ theory associated to the interface. Then, starting from this basic object we act with the appropriate pair of $Sp(2(N-1),\mathbb{Z})$ operators to generate the 3d theory associated to the interface theory.

We pick as a fundamental interface the one between two electric frames, which we can obtain by acting with a $S^{-1}$ transformation on the interface presented before in \eqref{eq: elec_magn_interface}. As the electric/multimonopole interface consists of $N-1$ free chirals, by acting with an $S^{-1}$ transformation we land on a collection of $N-1$ copies of the SQED with a single chiral with charge $-1$ and CS level $+1/2$. To simplify the picture we can use the basic $\mathcal{N}=2$ mirror duality that relates a QED with a single fundamental chiral to a free chiral. By carefully keeping track of all the details, one recovers an interface that consists of $N-1$ free chirals each with R-charge 0 and charge $+1$ under two $U(1)$ global symmetries as:
\begin{equation} 
    \begin{tikzpicture}[baseline,font=\footnotesize,
    node distance=1.5cm
]

 \draw (0,0) node[right] {\large{$_1 \mathbb{I}_1$}};

 \draw[<->] (2,0)--(3,0);

\begin{scope}[shift={(5,-.5)}]

 \node[flavor] (g1) {$1$};
 \node[flavor, right of=g1] (g2) {$1$};
 \node[right of=g2] (g3) {$\cdots$};
 \node[flavor, right of=g3] (g4) {$1$};

 \node[flavor,above of=g1] (f1) {$1$};
 \node[flavor, above of=g2] (f2) {$1$};
 \node[flavor, above of=g4] (f4) {$1$};

 \draw[-] (f1)--(g1);
 \draw[-] (f2)--(g2);
 \draw[-] (f4)--(g4);

 \draw (g1)+(-.5,0) node {$\xi_1$};
 \draw (g2)+(-.5,0) node {$\xi_2$};
 \draw (g4)+(-.6,0) node {$\xi_{N\text{-}1}$};

 \draw (f1)+(-.5,0) node {$\chi_1$};
 \draw (f2)+(-.5,0) node {$\chi_2$};
 \draw (f4)+(-.6,0) node {$\chi_{N\text{-}1}$};

 \draw[CScolor] (g1)+(.2,-.4) node[right] {$\text{-}\tfrac{1}{2}$};
 \draw[CScolor] (g2)+(.2,-.4) node[right] {$\text{-}\tfrac{1}{2}$};
 \draw[CScolor] (g4)+(.2,-.4) node[right] {$\text{-}\tfrac{1}{2}$};

 \draw[CScolor] (f1)+(.2,-.4) node[right] {$\text{-}\tfrac{1}{2}$};
 \draw[CScolor] (f2)+(.2,-.4) node[right] {$\text{-}\tfrac{1}{2}$};
 \draw[CScolor] (f4)+(.2,-.4) node[right] {$\text{-}\tfrac{1}{2}$};

 \draw[BFcolor] ($(g1)!0.5!(f1)+(-.05,-.1)$) node[left] {$-1$};
 \draw[BFcolor] ($(g2)!0.5!(f2)+(-.05,-.1)$) node[left] {$-1$};
 \draw[BFcolor] ($(g4)!0.5!(f4)+(-.05,-.1)$) node[left] {$-1$};

\end{scope}
 
\end{tikzpicture}
\end{equation}
There are also background CS levels as:
\begin{equation} \label{eq: BFcouplings_IRIRinterf}
    k_{\xi_i} = k_{\eta_i} = -\frac{1}{2} \,, \quad k_{\xi_i \eta_j} = - \delta_{i,j} \,, \quad k_{\xi_i R} = k_{\eta_i R} = -1 \,.
\end{equation}
Note in particular that there are BF couplings between $U(1)_{\xi_i}$ and $U(1)_{\eta_i}$ with the $U(1)_R$ R-symmetry. These are not typically indicated in quiver representations, but are provided in the text.
This theory still factorizes as $N-1$ identical copies of the same theory.

As anticipated, this theory can be taken to be the starting point for $Sp(2(N-1),\mathbb{Z})$ transformations that yield the interface theory between the IR effective theory at two generically different electro-magnetic duality frames. As an example that will be considered later on, we can compute the interface between the so-called magnetic and dyonic frames, that are labeled respectively with $S \in Sp(2(N-1),\mathbb{Z})$ and $S^{-1}T^2 \in Sp(2(N-1),\mathbb{Z})$. By acting with $S$ on the $U(1)^{N-1}$ symmetry corresponding to the $\xi_i$ fugacities and, at the same time, with the $S^{-1}T^2$ operator on the $U(1)^{N-1}$ symmmetry associated to $\eta_i$, we obtain:
\begin{equation} 
    \begin{tikzpicture}[baseline,font=\footnotesize,
    node distance=1.5cm
]

 \draw (0,0) node[right] {\large{$_S \mathbb{I}_{S^{-1}T^2}$}};

 \draw[<->] (2,0)--(3,0);

\begin{scope}[shift={(5,-.5)}]

 \node[gauge] (g1) {$1$};
 \node[gauge, right of=g1] (g2) {$1$};
 \node[right of=g2] (g3) {$\cdots$};
 \node[gauge, right of=g3] (g4) {$1$};

 \node[gauge,above of=g1] (f1) {$1$};
 \node[gauge, above of=g2] (f2) {$1$};
 \node[gauge, above of=g4] (f4) {$1$};

 \node[flavor, below of=g1] (a1) {$1$}; 
 \node[flavor, below of=g2] (a2) {$1$}; 
 \node[flavor, below of=g4] (a4) {$1$};

 \node[flavor, above of=f1] (b1) {$1$}; 
 \node[flavor, above of=f2] (b2) {$1$}; 
 \node[flavor, above of=f4] (b4) {$1$};

 \draw[-] (f1)--(g1);
 \draw[-] (f2)--(g2);
 \draw[-] (f4)--(g4);

 \draw[dashed] (b1)--(f1);
 \draw[dashed] (a1)--(g1);
 \draw[dashed] (b2)--(f2);
 \draw[dashed] (a2)--(g2);
 \draw[dashed] (b4)--(f4);
 \draw[dashed] (a4)--(g4);

 \draw (a1)+(-.5,0) node {$\xi_1$};
 \draw (a2)+(-.5,0) node {$\xi_2$};
 \draw (a4)+(-.6,0) node {$\xi_{N\text{-}1}$};

 \draw (b1)+(-.5,0) node {$\chi_1$};
 \draw (b2)+(-.5,0) node {$\chi_2$};
 \draw (b4)+(-.6,0) node {$\chi_{N\text{-}1}$};

 \draw[CScolor] (g1)+(.2,-.4) node[right] {$\text{-}\tfrac{1}{2}$};
 \draw[CScolor] (g2)+(.2,-.4) node[right] {$\text{-}\tfrac{1}{2}$};
 \draw[CScolor] (g4)+(.2,-.4) node[right] {$\text{-}\tfrac{1}{2}$};

 \draw[CScolor] (f1)+(.2,-.4) node[right] {$+\tfrac{3}{2}$};
 \draw[CScolor] (f2)+(.2,-.4) node[right] {$+\tfrac{3}{2}$};
 \draw[CScolor] (f4)+(.2,-.4) node[right] {$+\tfrac{3}{2}$};

 \draw[CScolor] (a1)+(.2,-.4) node[right] {$0$};
 \draw[CScolor] (a2)+(.2,-.4) node[right] {$0$};
 \draw[CScolor] (a4)+(.2,-.4) node[right] {$0$};

 \draw[CScolor] (b1)+(.2,-.4) node[right] {$0$};
 \draw[CScolor] (b2)+(.2,-.4) node[right] {$0$};
 \draw[CScolor] (b4)+(.2,-.4) node[right] {$0$};

 \draw[BFcolor] ($(g1)!0.5!(f1)+(-.05,-.1)$) node[left] {$-1$};
 \draw[BFcolor] ($(g2)!0.5!(f2)+(-.05,-.1)$) node[left] {$-1$};
 \draw[BFcolor] ($(g4)!0.5!(f4)+(-.05,-.1)$) node[left] {$-1$};

 \draw[BFcolor] ($(g1)!0.5!(a1)+(-.05,-.1)$) node[left] {$+2$};
 \draw[BFcolor] ($(g2)!0.5!(a2)+(-.05,-.1)$) node[left] {$+2$};
 \draw[BFcolor] ($(g4)!0.5!(a4)+(-.05,-.1)$) node[left] {$+2$};

 \draw[BFcolor] ($(f1)!0.5!(b1)+(-.05,-.1)$) node[left] {$-2$};
 \draw[BFcolor] ($(f2)!0.5!(b2)+(-.05,-.1)$) node[left] {$-2$};
 \draw[BFcolor] ($(f4)!0.5!(b4)+(-.05,-.1)$) node[left] {$-2$};

\end{scope}
 
\end{tikzpicture}
\end{equation}
where now the BF coupling in \eqref{eq: BFcouplings_IRIRinterf}, in addition to extra BF couplings introduces by the $Sp(2(N-1),\mathbb{Z})$ transformations, have become dynamical as we have gauged the $U(1)$ Abelian symmetries. In the final theory all the background CS levels are zero.
\\

It is particularly interesting to consider the magnetic/dyonic interface for $SU(2)$:
\begin{equation} \label{eq:interface_mon_dyon_SU2}
    \begin{tikzpicture}[baseline,font=\footnotesize,
    node distance=1.5cm
]

 \draw (0,0) node[right] {\large{$_S \mathbb{I}_{S^{-1}T^2}$}};

 \draw[<->] (2,0)--(3,0);

\begin{scope}[shift={(5,-.5)}]

 \node[gauge] (g1) {$1$};
 \node[gauge,above of=g1] (f1) {$1$};
 \node[flavor, below of=g1] (a1) {$1$}; 
 \node[flavor, above of=f1] (b1) {$1$};

 \draw[-] (f1)--(g1);

 \draw[dashed] (b1)--(f1);
 \draw[dashed] (a1)--(g1);

 \draw (a1)+(-.5,0) node {$\xi$};
 \draw (b1)+(-.5,0) node {$\chi$};

 \draw[CScolor] (g1)+(.2,-.4) node[right] {$\text{-}\tfrac{1}{2}$};

 \draw[CScolor] (f1)+(.2,-.4) node[right] {$+\tfrac{3}{2}$};

 \draw[CScolor] (a1)+(.2,-.4) node[right] {$0$};

 \draw[CScolor] (b1)+(.2,-.4) node[right] {$0$};

 \draw[BFcolor] ($(g1)!0.5!(f1)+(-.05,-.1)$) node[left] {$-1$};

 \draw[BFcolor] ($(g1)!0.5!(a1)+(-.05,-.1)$) node[left] {$+2$};

 \draw[BFcolor] ($(f1)!0.5!(b1)+(-.05,-.1)$) node[left] {$-2$};

\end{scope}
 
\end{tikzpicture}
\end{equation}
The $U(1)$ gauge node with CS level $-\tfrac{1}{2}$ confines into a free chiral field with R-charge 1, charge $-1$ under $U(1)_\xi$ and neutral otherwise. The other gauge node $U(1)_G$ has CS at level $2$ after the confinement and there are mixed CS terms:
\begin{equation}
    k_{\xi} = \frac{1}{2},
    \qquad
    k_{G} = 2,
    \qquad
    k_{\chi} = 0,
    \qquad
    k_{G\xi} = k_{G\chi} = -2,
    \qquad 
    k_{R \star } = 0
\end{equation}
In quiver notation the interface, after the confinement of the $U(1)_{-\tfrac{1}{2}}$ gauge node, reads:
\begin{equation}
    \begin{tikzpicture}[node distance=1.5cm]
     \draw (-5,0) node[right] {\large{$_S \mathbb{I}_{S^{-1}T^2}$}};

 \draw[<->] (-3,0)--(-2,0);
 \node[gauge] (g1) {$1$};
 \node[flavor, below of=g1] (a1) {$1$}; 
 \node[flavor, above of=g1] (b1) {$1$}; 
 \node[flavor,right of=a1] (f1) {$1$};

  \draw[CScolor] (g1)+(.2,-.4) node[right] {${2}$};
  \draw[CScolor] (a1)+(.2,-.4) node[right] {$\tfrac{1}{2}$};
  \draw[CScolor] (b1)+(.2,-.4) node[right] {$0$};

  \draw[->-] (a1) -- (f1);
   \draw[dashed] (b1)--(g1);
 \draw[dashed] (a1)--(g1);
 
     \draw[BFcolor] ($(g1)!0.5!(a1)+(-.05,-.1)$) node[left] {$-2$};
     \draw[BFcolor] ($(g1)!0.5!(b1)+(-.05,-.1)$) node[left] {$-2$};

     \draw (a1)+(-.5,0) node {$\xi$};
 \draw (b1)+(-.5,0) node {$\chi$};
    \end{tikzpicture}
\end{equation}

This is a good interface in the strong coupling chamber of pure $SU(2)$, and the monopole and dyon points sit at the boundary of this chamber. As there are no walls of marginal stability between the two points, excluding the wall that the two points sit on top of, we find it reasonable that this interface provides an (unstable) domain wall between the monopole and dyon points of pure $\NN=2$ $SU(2)$ SYM.
This is particularly interesting when considering the deformation to $\NN=1$ obtained by adding a microscopic superpotential $m \Tr (\Phi^2)$.
As discussed in the seminal paper by Seiberg and Witten, the effect of this deformation can be described in the IR effective theory where it induces a non-zero VEV for the monopole or the dyon depending on which neighbourhood of the CB one considers. As a result, the low energy $U(1)$ gauge field is Higgsed, the CB is lifted except for the two special points which become the two vacua of pure $\NN=1$ SYM. 
Our (unstable) domain wall is then expected to reduce to the (stable) domain wall of $SU(2)$ $\NN=1$ SYM.

Under the assumptions discussed above, our setup can be analyzed in a similar way, as the theory on the left of the interface is close to the monopole point and the theory on the right is close to the dyon point. In both cases, turning on the $\NN=1$ deformation should result in the monopole (or dyon) taking a VEV and Higgsing the Abelian gauge field in the corresponding EM duality frame $A_{\text{monop}}$ (or $A_{\text{dyon}}$). These two gauge symmetries produce the 3d symmetries $U(1)_\xi$ and $U(1)_\chi$ of the interface, which are thus expected to be spontaneously broken. 
After the spontaneous breaking the free chiral is neutral and has R-charge 1, we expect it to take a mass and be integrated out. We are thus left with a $U(1)_2$ TQFT, given by the residual 3d gauge group. Surprisingly, this is exactly the domain wall of pure $\NN=1$ SYM proposed by Acharya and Vafa \cite{Acharya:2001dz}.

Although this analysis is quite schematic, we find it very intriguing that we recover the correct 3d theory implementing the domain wall of pure $\NN=1$ SYM. It would be extremely interesting to study such a deformation for generic $SU(N)$ gauge group and to make this analysis more precise, we leave this to future work.

\subsection{The UV/UV interface}
As a last example we discuss a non-trivial property of the $\Twall[]$ theory which implies that the natural interface between two UV descriptions of $\mathcal{N}=2$ SYM, constructed using the same technique as before, flows to a transparent defect. This provides a further non-trivial check that the $\Twall[]$ has all the desired properties to the RG-wall.

We consider the following setup, where we sandwich the IR SW theory between two RG-walls:
\begin{equation}    \label{eq:collision_IR_def_UV}
\resizebox{\hsize}{!}{ 
\begin{tikzpicture}[]
    \begin{scope}[xshift=0cm]
        \node[left] at (1.5,0) (IR) {$\begin{array}{c} \text{UV} \\  \\ \text{$SU(N)$}\\ \text{$\NN=2$ SYM} \end{array}$};
        \path[draw] (2,-1.5) node[below,anchor=north] {$S \cdot \Twall[]$}  -- (2,1.5) node[anchor=north east] {$\mathcal{N}$} node[anchor=north west] {$\mathcal{N},\mathcal{B}_{\chi}$};
        \node[right] at (2.5,0) (IR) {$\begin{array}{c} \text{IR} \\  \\ \text{Seiberg-Witten}\\ \text{in $S$ EM frame} \end{array}$};
        \path[draw] (6,-1.5) node[below,anchor=north] {$S \cdot \Twall[]$}  -- (6,1.5) node[anchor=north east] {$\mathcal{N},\mathcal{B}_{\chi}$} node[anchor=north west] {$\mathcal{N}$};
        \node[right] at (6.5,0) (IR) {$\begin{array}{c} \text{UV} \\  \\ \text{$SU(N)$}\\ \text{$\NN=2$ SYM} \end{array}$};
        \path[draw,<->] (2,1.8) -- node[midway,above] {$\Delta x_3$} (6,1.8);
    \end{scope}
    \path[draw,->] (9.5,0) -- node[midway,above] {$\Delta x_3 \to 0$} (11.5,0);
    \begin{scope}[xshift=13.5cm]
        \node[left] at (1.5,0) (IR) {$\begin{array}{c} \text{UV} \\  \\ \text{$SU(N)$}\\ \text{$\NN=2$ SYM} \end{array}$};
        \path[draw] (2,-1.5) node[below,anchor=north] {$\mathbb{I}^{\text{UV}}$}  -- (2,1.5) node[anchor=north east] {$\mathcal{N}$} node[anchor=north west] {$\mathcal{N}$};
        \node[right] at (2.5,0) (IR) {$\begin{array}{c} \text{UV} \\  \\ \text{$SU(N)$}\\ \text{$\NN=2$ SYM} \end{array}$};
    \end{scope}
\end{tikzpicture}
} 
\end{equation}
The 3d defect theory $\mathbb{I}^{\text{UV}}$ is obtained by gluing two $S \cdot \Twall[]$ theories through their $U(1)^{N-1}$ symmetry.
Here the gluing induced by the 4d SW theory on the slice is more involved because of the presence of BPS hypermultiplets. 
The 4d $U(1)^{N-1}$ gauge groups effectively identify and gauge the two $U(1)^{N-1}$ global symmetries of the two $S \cdot \Twall[]$ theories. 
Furthermore the $N-1$ electric BPS particles with $\mathcal{B}_{\chi}$ boundary conditions on both walls reduce to $N-1$ 3d chiral multiplets and CS level $-\tfrac{1}{2}$ for each factor in the $U(1)^{N-1}$ symmetry.

From symmetry considerations we expect additional interactions at the defect. Indeed gauging the $U(1)^{N-1}$ symmetry at the 3d slice would produce an additional $U(1)^{N-1}$ global topological symmetry which should not be present in the setup. We thus expect that (at least) $N-1$ monopole superpotential terms are generated, breaking this additional topological symmetry. It would be interesting to further investigate the physical origin of these interactions, that is possibly the $\Delta x_3$ slice, but we leave this to future work.
Putting it all together we expect that the 4d SW theory on the thin slice induces a gluing procedure the involves identifying and gauge the $U(1)^{N-1}$ symmetries and introducing $N-1$ additional chiral fields, CS terms for the $U(1)^{N-1}$ symmetry and $N-1$ additional monopole superpotential terms. 

The resulting theory was studied in \cite{Benvenuti:2025huk}, where it was shown that it has a quantum deformed moduli space so that the theory undergoes SSB breaking of the two $SU(N)$ symmetries to the diagonal $SU(N)$\footnote{This property was denoted as "planar-planar" fusion-to-identity in \cite{Benvenuti:2025huk}.
The gluing of two $\Twall[]$ theories across their Abelian symmetry is more easily studied using an alternative Lagrangian UV completion of $\Twall[]$, which is the mirror-like dual of the quiver in \eqref{eq:quiv_ptransf_Gsun}.}. In the deep IR the theory behaves as a free $\mathcal{N}=2$ $SU(N)$ vector multiplet with BF interactions (see Appendix \ref{app: gsun} for more details).

Then, following the same reasoning as before, we expect that colliding the two walls produces the identity defect in the UV theory:
\begin{equation}   
\resizebox{\hsize}{!}{ 
\begin{tikzpicture}[]
    \begin{scope}[xshift=0cm]
        \node[left] at (1.5,0) (IR) {$\begin{array}{c} \text{UV} \\  \\ \text{$SU(N)$}\\ \text{$\NN=2$ SYM} \end{array}$};
        \path[draw] (2,-1.5) node[below,anchor=north] {$S \cdot \Twall[]$}  -- (2,1.5) node[anchor=north east] {$\mathcal{N}$} node[anchor=north west] {$\mathcal{N},\mathcal{B}_{\chi}$};
        \node[right] at (2.5,0) (IR) {$\begin{array}{c} \text{IR} \\  \\ \text{Seiberg-Witten}\\ \text{in $S$ EM frame} \end{array}$};
        \path[draw] (6,-1.5) node[below,anchor=north] {$S \cdot \Twall[]$}  -- (6,1.5) node[anchor=north east] {$\mathcal{N},\mathcal{B}_{\chi}$} node[anchor=north west] {$\mathcal{N}$};
        \node[right] at (6.5,0) (IR) {$\begin{array}{c} \text{UV} \\  \\ \text{$SU(N)$}\\ \text{$\NN=2$ SYM} \end{array}$};
        \path[draw,<->] (2,1.8) -- node[midway,above] {$\Delta x_3$} (6,1.8);
    \end{scope}
    \path[draw,->] (9.5,0) -- node[midway,above] {$\Delta x_3 \to 0$} (11.5,0);
    \begin{scope}[xshift=13.5cm]
        \node[left] at (1.5,0) (IR) {$\begin{array}{c} \text{UV} \\  \\ \text{$SU(N)$}\\ \text{$\NN=2$ SYM} \end{array}$};
        \path[draw, dashed] (2,-1.5) node[below,anchor=north] {$\mathbb{I}$}  -- (2,1.5);
        \node[right] at (2.5,0) (IR) {$\begin{array}{c} \text{UV} \\  \\ \text{$SU(N)$}\\ \text{$\NN=2$ SYM} \end{array}$};
    \end{scope}
\end{tikzpicture}
} 
\end{equation}
This property will be made more precise in Section \ref{sec: half_index_interfaces} as a partition function identity.

Similarly to the IR defects studied previously, this provides a non-trivial consistency check of our proposal. Indeed, if we let the theory flow on all of spacetime we should recover the low energy SW theory without any defect. Colliding two RG-walls to produce a defect in the UV should then produce a trivial defect, or at least a defect that flows to the trivial defect in the IR.
The discussion above shows that our proposal for the RG-wall of $\NN=2$ SYM is consistent with this expectation.

\subsection{Dirichlet boundary conditions in the IR}

Now we apply the fusion-to-indentity property of the RG-wall to the boundary setups in Figure \ref{fig:schematic_RG} and \ref{fig:schematic_RG_inverse}.
We show that such identities allow one to obtain one setup from the other, namely one can “move" the RG-wall from the IR to the UV and viceversa. 
This implies that the two proposals for the RG-walls in the previous section are not independent, and one is a consequence of the other.

In order to discuss this problem, we need to derive a second, alternative, fusion-to-identity property that we now describe.
To derive this property, we start from the relation between the theory obtained by gluing two $\Twall[]$ reported in \eqref{eq: magn_ir_interf} and $N-1$ copies of the free-hypers on the r.h.s.~of \eqref{eq: fusion_to_id}.
We first ``flip" the $N-1$ chiral fields $\phi_i$ on the r.h.s.~of \eqref{eq: fusion_to_id}, that is we introduce $N-1$ new chiral fields $t_i$ interacting as:
\begin{equation}
    \mathcal{W} = \sum_{i=1}^{N-1} t_i \phi_i \,.
\end{equation}
due to this interaction, both the $t_i$ and $\phi_i$ acquire a mass and thus decouple.
As the fields $\phi_i$ are mapped to monopoles in the quiver in \eqref{eq: magn_ir_interf}, on this side of the relation the fields $t_i$ couple to these monopoles.
We thus obtain the following relation.
\begin{equation}
    \begin{tikzpicture}[baseline=(current bounding box).center, scale = 1]
        \node at (0,0) (n1) [gauge,black] {$1$};
        \node at (2,0) (n2) [gauge,black] {$2$};
        \draw (n2)++(1.5,0) node (dots1) {$\dots$};
        \draw (dots1)++(1.5,0) node (nn) [gauge,black,double] {$N$};
        \draw (nn)++(1.5,0) node (dots2) {$\dots$};
        \draw (dots2)++(1.5,0) node (nn2) [gauge,black] {$2$};
        \draw (nn2)++(2,0) node (nn1) [gauge,black] {$1$};

        \draw[CScolor] (n1.south east) node[anchor=west] {$\scriptstyle -1$};
        \draw[CScolor] (n2.south east) node[anchor=west] {$\scriptstyle (0,-2)$};
        \draw[CScolor] (nn.south east) node[anchor=west] {$\scriptstyle 1$};
        \draw[CScolor] (nn1.south east) node[anchor=west] {$\scriptstyle -1$};
        \draw[CScolor] (nn2.south east) node[anchor=west] {$\scriptstyle (0,-2)$};

        \draw[FIcolor] (n1.north) node[anchor=south] {$\scriptstyle \xi_1^{-1}$};
        \draw[FIcolor] (n2.north) node[anchor=south] {$\scriptstyle \xi_2^{-1}$};
        \draw[FIcolor] (nn1.north) node[anchor=south] {$\scriptstyle \eta_1$};
        \draw[FIcolor] (nn2.north) node[anchor=south] {$\scriptstyle \eta_2$};
        
        \draw[->-] (n1) -- node[midway,above,BFcolor] {$\scriptstyle +1$} (n2);
        \draw[->-] (n2) -- node[midway,above,BFcolor] {$\scriptstyle +1$} (dots1);
        \draw[->-] (dots1) -- (nn);
        \draw[->-] (nn) -- (dots2);
        \draw[->-] (dots2) -- node[midway,above,BFcolor] {$\scriptstyle +1$} (nn2);
        \draw[->-] (nn2) -- node[midway,above,BFcolor] {$\scriptstyle +1$} (nn1);

        \path (n1) -- node[midway,below,yshift=-1cm] {
        $\mathcal{W} = t_1 M^{(0,\dots,0,+) }+ t_2 M^{(0,\dots,0,+,0) }+ \dots
        $
        } (nn1);
    \end{tikzpicture}
    =
    \prod_{i=1}^{N-1} \delta(\xi_i - \eta_i)
    e^{-\frac{\pi}{2} i \xi_i^2}
\end{equation}
Where the exponential factor on the r.h.s.~is a schematic way to keep track of the presence of background $1/2$ CS levels for the unbroken $U(1)^{N-1}$ symmetry, in analogy with its characteristic contribution to partition functions.

The quiver can be massaged by performing an IR duality that replaces one of the two tails, say the right one. 
The action of this duality is to flip the sign of CS levels, change orientation of the arrows and FI parameters. This duality also changes the CS-level of the central $SU(N)$ gauge node from $1$ to $0$ and it introduces $N-1$ chiral fields that couple to the $t_i$ fields, thus giving a mass to them and background $1/2$ CS-levels (more details in Appendix \ref{app: gsun}).
This duality was denoted as ``chiral-chiral flip-flip" in \cite{Benvenuti:2025huk}, where it was originally derived by iterative applications of Given-Kutasov dualities. 
The relation after performing this dualization reads:
\begin{equation}    \label{eq:FTI_chiral_modified}
    \begin{tikzpicture}[baseline=(current bounding box).center, xscale = 1]
        \node at (0,0) (n1) [gauge,black] {$1$};
        \node at (2,0) (n2) [gauge,black] {$2$};
        \draw (n2)++(1.5,0) node (dots1) {$\dots$};
        \draw (dots1)++(1.5,0) node (nn) [gauge,double,black] {$N$};
        \draw (nn)++(1.5,0) node (dots2) {$\dots$};
        \draw (dots2)++(1.5,0) node (nn2) [gauge,black] {$2$};
        \draw (nn2)++(2,0) node (nn1) [gauge,black] {$1$};

        \draw[CScolor] (n1.south east) node[anchor=west] {$\scriptstyle -1$};
        \draw[CScolor] (n2.south east) node[anchor=west] {$\scriptstyle (0,-2)$};
        \draw[CScolor] (nn.south east) node[anchor=west] {$\scriptstyle 0$};
        \draw[CScolor] (nn1.south east) node[anchor=west] {$\scriptstyle 1$};
        \draw[CScolor] (nn2.south east) node[anchor=west] {$\scriptstyle (0,2)$};

        \draw[FIcolor] (n1.north) node[anchor=south] {$\scriptstyle \xi_1^{-1}$};
        \draw[FIcolor] (n2.north) node[anchor=south] {$\scriptstyle \xi_2^{-1}$};
        \draw[FIcolor] (nn1.north) node[anchor=south] {$\scriptstyle \eta_{N-1}^{-1}$};
        \draw[FIcolor] (nn2.north) node[anchor=south] {$\scriptstyle \eta_{N-2}^{-1}$};
        
        \draw[->-] (n1) -- node[midway,above,BFcolor] {$\scriptstyle +1$} (n2);
        \draw[->-] (n2) -- node[midway,above,BFcolor] {$\scriptstyle +1$} (dots1);
        \draw[->-] (dots1) -- node[midway,above,BFcolor] {$\scriptstyle $} (nn);
        \draw[-<-] (nn) -- node[midway,above,BFcolor] {$\scriptstyle $} (dots2);
        \draw[-<-] (dots2) -- node[midway,above,BFcolor] {$\scriptstyle -1$} (nn2);
        \draw[-<-] (nn2) -- node[midway,above,BFcolor] {$\scriptstyle -1$} (nn1);


    \end{tikzpicture}
    =
    \sum_{i=1}^{N-1} \delta(\xi_i - \chi_i)
\end{equation}
The resulting quiver in \eqref{eq:FTI_chiral_modified} can be thought of as the RG-wall glued to the inverse RG wall via the diagonal gauging of their $SU(N)$ symmetries:
\begin{equation}    \label{eq:FTI_chiral_modified_split}
    \begin{tikzpicture}[baseline=(current bounding box).center, scale = .95]
        \node at (0,0) (n1) [gauge,black] {$1$};
        \node at (2,0) (n2) [gauge,black] {$2$};
        \draw (n2)++(2,0) node (dots1) {$\dots$};
        \draw (dots1)++(2,0) node (nn) [flavor,black] {$N$};
        \draw (nn)++(3,0) node (nnb) [flavor,black] {$N$};
        \draw (nnb)++(2,0) node (dots2) {$\dots$};
        \draw (dots2)++(2,0) node (nn2) [gauge,black] {$2$};
        \draw (nn2)++(2,0) node (nn1) [gauge,black] {$1$};

        \draw[CScolor] (n1.south east) node[anchor=west] {$\scriptstyle -1$};
        \draw[CScolor] (n2.south east) node[anchor=west] {$\scriptstyle (0,-2)$};
        \draw[CScolor] (nn.south west) node[anchor=east] {$\scriptstyle \frac{1-N}{2}$};
        \draw[CScolor] (nnb.south east) node[anchor=west] {$\scriptstyle \frac{N-1}{2}$};
        \draw[CScolor] (nn1.south west) node[anchor=east] {$\scriptstyle 1$};
        \draw[CScolor] (nn2.south west) node[anchor=east] {$\scriptstyle (0,2)$};

        \draw[FIcolor] (n1.north) node[anchor=south] {$\scriptstyle \xi_1^{-1}$};
        \draw[FIcolor] (n2.north) node[anchor=south] {$\scriptstyle \xi_2^{-1}$};
        \draw[FIcolor] (nn1.north) node[anchor=south] {$\scriptstyle \eta_{N-1}^{-1}$};
        \draw[FIcolor] (nn2.north) node[anchor=south] {$\scriptstyle \eta_{N-2}^{-1}$};
        
        \draw[->-] (n1) -- node[midway,above,BFcolor] {$\scriptstyle +1$} (n2);
        \draw[->-] (n2) -- node[midway,above,BFcolor] {$\scriptstyle +1$} (dots1);
        \draw[->-] (dots1) -- node[midway,above,BFcolor] {$\scriptstyle $} (nn);
        \draw[-<-] (nnb) -- node[midway,above,BFcolor] {$\scriptstyle $} (dots2);
        \draw[-<-] (dots2) -- node[midway,above,BFcolor] {$\scriptstyle -1$} (nn2);
        \draw[-<-] (nn2) -- node[midway,above,BFcolor] {$\scriptstyle -1$} (nn1);

        \draw [decorate,decoration={brace,amplitude=5pt,raise=4ex}] (nn)--(n1) node[midway,xshift=0em,yshift=-3em]{$S\cdot \Twall[]$}; 
        \draw [decorate,decoration={brace,amplitude=5pt,raise=4ex}] (nn1)--(nnb) node[midway,xshift=0em,yshift=-3em]{$(S\cdot \Twall[])^{-1}$}; 

        \draw[<->] (7,0) -- node[midway,above] {Gauge} node[midway,below] {$SU(N)_0$} (8,0);

    \end{tikzpicture}
\end{equation}
The left tail is manifestly the $S\cdot \Twall[]$ theory while the right tail is IR dual to the inverse RG-wall $(S\cdot \Twall[])^{-1}$ theory \eqref{eq:quiv_Gsun} trough an electric duality described in Appendix \ref{app: gsun}. We finally reach the proposal that the inverse $S \cdot \Twall[]$ theory is given by the following quiver:
\begin{equation}
(S \cdot \Twall[])^{-1} =
     \begin{tikzpicture}[baseline=(current bounding box).center]
        \begin{scope}[yshift=-3cm]
        \node at (0,0) (n1) [gauge,black] {$1$};
        \node at (2,0) (n2) [gauge,black] {$2$};
        \node at (4,0) (n3) {$\cdots$};
        \node at (6,0) (nnm1) [gauge,black] {$\scriptstyle N-1$};
        \node at (8,0) (nn) [flavor,black] {$N$};

        \draw[CScolor] (n1.south east) node[anchor=west] {$\scriptstyle 1$};
        \draw[CScolor] (n2.south east) node[anchor=west] {$\scriptstyle (0,2)$};
        \draw[CScolor] (nnm1.south east) node[anchor=west] {$\scriptstyle (0,N-1)$};
        \draw[CScolor] (nn.south east) node[anchor=west] {$\scriptstyle \tfrac{N-1}{2}$};

        \draw[FIcolor] (n1.north) node[anchor=south] {$\scriptstyle \eta_{1}$};
        \draw[FIcolor] (n2.north) node[anchor=south] {$\scriptstyle \eta_{2}$};
        \draw[FIcolor] (nnm1.north) node[anchor=south] {$\scriptstyle \eta_{N-1}$};

        \draw[black] (nn.north) node[anchor=south] {$\scriptstyle \vec{X}$};
        
        \draw[-<-] (n1) -- node[midway,above,BFcolor] {$\scriptstyle -1$} (n2);
        \draw[->-] (n2)++(1.5,0) -- node[midway,above,BFcolor] {$\scriptstyle -1$} (n2);
        \draw[-<-] (nnm1)++(-1.5,0) -- node[midway,above,BFcolor] {$\scriptstyle -1$} (nnm1);
        \draw[-<-] (nnm1) --  (nn);
        \end{scope}
    \end{tikzpicture}
\end{equation}
as discussed in detail in Appendix \ref{app: 3dwalls}.

\begin{figure}
    \centering
    \begin{tikzpicture}[scale=1]
    \begin{scope}
            
    \begin{scope}[xshift=0cm]
        \node[left] at (1,0) (UV) {$\begin{array}{c} \text{UV} \\  \\ \text{$SU(N)$}\\ \text{$\NN=2$ SYM} \end{array}$};
        \path[draw] (1.5,-1.5) -- node[midway,right,anchor=west] {$\mathcal{D}$} (1.5,1.5) node[above] {$x_3=0$};
    \end{scope}
    
    \begin{scope}[xshift=8.75cm]
        \node[left] at (1.5,0) (IR) {$\begin{array}{c} \text{IR} \\  \\ \text{Seiberg-Witten}\\ \text{in $S$ frame} \end{array}$};
        \path[draw] (2,-1.5) -- node[midway,right,anchor=west] {$\mathcal{B}_{\chi},\;\mathcal{N} + S \cdot \Twall[]$} (2,1.5) node[above] {$x_3=0$};
    \end{scope}

    \begin{scope} [xshift=5.5cm]
        \path[draw,->] (0,0) -- node[midway,above] {RG flow} (1,0);
    \end{scope}
    \end{scope}
    \begin{scope}[xshift=5.5cm,yshift=-3cm]
        \path[draw,->] (0,1) -- node[midway,right]
        {$\begin{array}{c} 
        \text{Couple $(S\cdot \Twall[])^{-1}$} \\ \text{via its $SU(N)$ gauging} \end{array}$}(0,-1);
    \end{scope}
    \begin{scope}[yshift=-6.5cm]
        
    \begin{scope}[xshift=0cm]
        \node at (0,0) (UV) {$\begin{array}{c} \text{UV} \\  \\ \text{$SU(N)$}\\ \text{$\NN=2$ SYM} \end{array}$};
        \path[draw] (1.5,-1.5) -- node[midway,right,anchor=west] {$\mathcal{N} + (S\cdot \Twall[])^{-1}$} (1.5,1.5) node[above] {$x_3=0$};
    \end{scope}
    
    \begin{scope}[xshift=8.75cm]
        \node[left] at (1.5,0) (IR) {$\begin{array}{c} \text{Seiberg-Witten} \\ \text{in $S$ frame}\end{array}$};
        \path[draw] (2,-1.5) -- node[midway,right,anchor=west] {$\begin{array}{c} \mathcal{B}_{\chi},\;\mathcal{N} + S \cdot \Twall[] + \\ + (S\cdot \Twall[])^{-1}\end{array}$} (2,1.5) node[above] {$x_3=0$};
    \end{scope}

    \begin{scope} [xshift=5.5cm]
        \path[draw,->] (0,0) -- node[midway,above] {RG flow} (1,0);
    \end{scope}
    \end{scope}
    \begin{scope}[xshift=5.5cm,yshift=-10cm]
        \path[draw,->] (0,1) -- node[midway,right]
        {Use \eqref{eq:FTI_chiral_modified} on the r.h.s.}(0,-1);
    \end{scope}
    \begin{scope}[yshift=-13cm]
        
    \begin{scope}[xshift=0cm]
        \node at (0,0) (UV) {$\begin{array}{c} SU(N) \\ \NN=2 \; \text{SYM}\end{array}$};
        \path[draw] (1.5,-1.5) -- node[midway,right,anchor=west] {$\mathcal{N} + (S\cdot \Twall[])^{-1}$} (1.5,1.5) node[above] {$x_3=0$};
    \end{scope}
    
    \begin{scope}[xshift=8.75cm]
        \node[left] at (1.5,0) (IR) {$\begin{array}{c} \text{Seiberg-Witten} \\ \text{in $S$ frame}\end{array}$};
        \path[draw] (2,-1.5) -- node[midway,right,anchor=west] {$\mathcal{B}_{\chi},\;\mathcal{D}$} (2,1.5) node[above] {$x_3=0$};
    \end{scope}

    \begin{scope} [xshift=5.5cm]
        \path[draw,->] (0,0) -- node[midway,above] {RG flow} (1,0);
    \end{scope}
    \end{scope}
    \end{tikzpicture}
    \caption{Schematic depiction of how the fusion-to-identity \eqref{eq:FTI_chiral_modified} connects the RG-wall and the inverse RG-wall. The first line corresponds to our proposal for the RG-wall in the multimonopole frame \eqref{eq: schematic_STwall_final}.
    In the second line, the r.h.s.~represents $\mathcal{N}$ b.c.~coupled to the quiver theory in \eqref{eq:FTI_chiral_modified}. 
    In the last line we exploited the delta function coming from \eqref{eq:FTI_chiral_modified} to obtain $\mathcal{D}$ b.c.~for the low energy theory, and obtain our proposal for the inverse RG-wall \eqref{eq: schematic_invSTwall_final}.
    }
    \label{fig:from_TWall_toTWallInverse}
\end{figure}
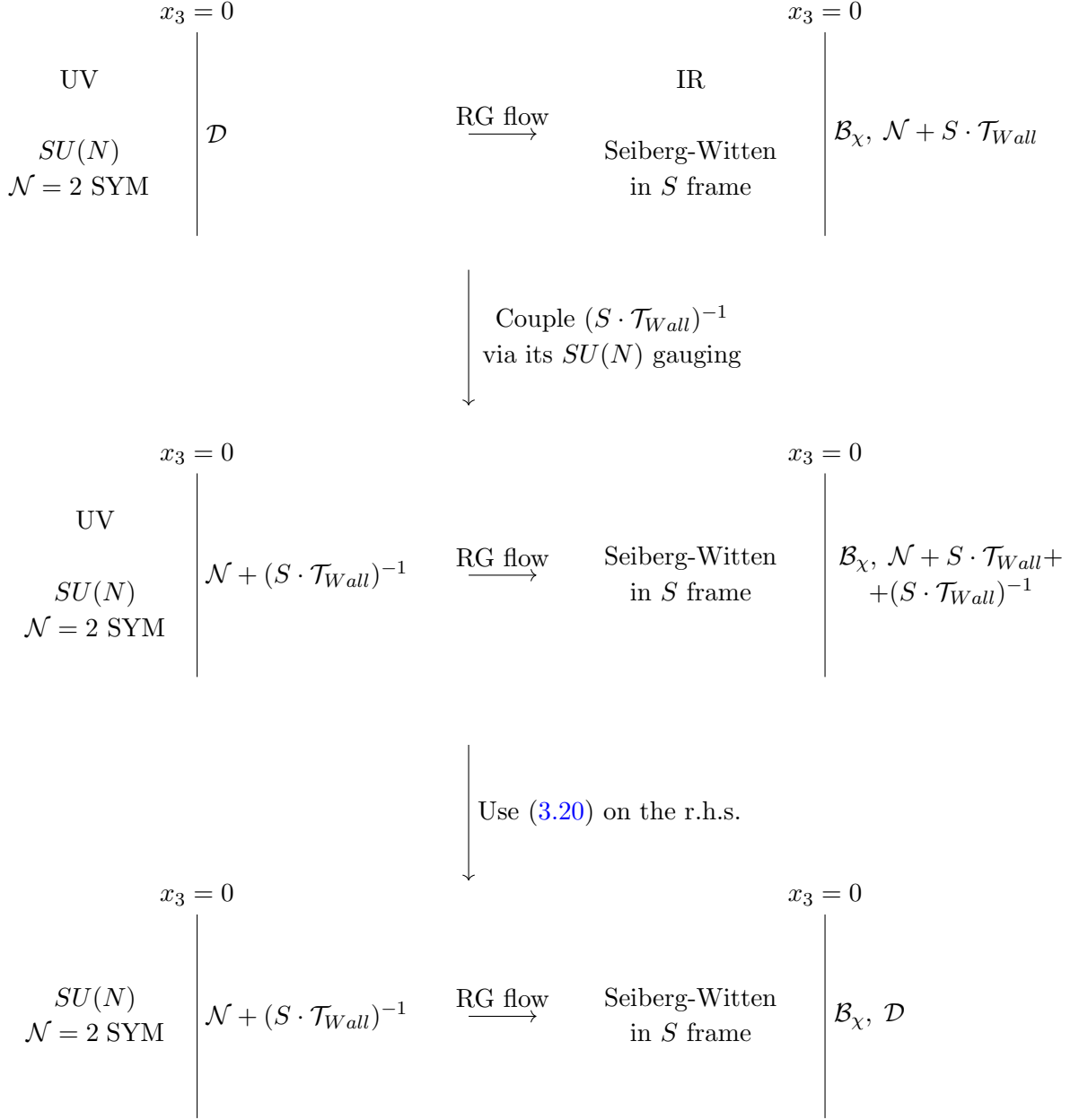

By applying the result that $(S \cdot \Twall[]) \cdot (S \cdot \Twall[])^{-1} \sim 1 $, in the sense described above, we can start from the boundary setup in Figure \ref{eq: schematic_STwall_final}, where the RG-wall is coupled to the IR theory and the UV theory has $\mathcal{D}$ boundary condition. 
We then stack the inverse RG-wall on both sides, coupling it to the boundary $SU(N)$ symmetry trough a diagonal gauging. 
In the IR we find the SW solution coupled to the quiver in \eqref{eq:FTI_chiral_modified}. Thanks to the fusion to identity property \eqref{eq:FTI_chiral_modified}, this boundary condition is equivalent to $\mathcal{D}$ boundary conditions for the low energy Abelian gauge fields. In the UV we find $\NN=2$ SYM coupled to the inverse RG-wall.
Therefore we recover the second boundary setup in Figure \ref{eq: schematic_invSTwall_final}.
This procedure is summarized in Figure \ref{fig:from_TWall_toTWallInverse}.

\clearpage

\section{Pure $SU(N)$ SYM on $HS^3 \times S^1$} \label{sec: half-index}

In this section we describe the supersymmetric partition function identities corresponding to the proposal in Section \ref{sec: bc_proposal}. Indeed, we need to compute the partition function on a manifold with a boundary in order to be sensible to the data of the setups described in Section \ref{sec: bc_proposal}.
We follow the strategies established in \cite{Gang:2012ff,Cordova:2016uwk,Gaiotto:2019jvo,Okazaki:2019ony,Hatsuda:2025yzp,Hatsuda:2025zvi,Hatsuda:2026ysv} and study the partition function on $HS^3 \times S^1$, where $HS^3$ is the half-three-sphere that has an $S^2$ boundary. 

The generic proposal is translated into the following identity:
\begin{equation} \label{eq: rgwall_index_statement}
    \mathcal{II}^{SU(N)} (\vec{u};q) = \sum_{\vec{m} \in \mathbb{Z}^{N-1}} \oint \prod_{j=1}^{N-1} \frac{d\xi_j}{2\pi i \xi_j} 
    \; \mathcal{I}^{\Twall[]}(\vec{u},\{ \vec{\xi}, \vec{m}\};q) 
    \; \mathcal{II}^{\text{SW}} (\{\vec{\xi},\vec{m}\};q)
\end{equation}
Where the various quantities appearing in the identity are:
\begin{itemize}
    \item $\mathcal{II}^{SU(N)} (\vec{u};q)$ is the Schur half-index of the 4d $\mathcal{N}=2$ pure $SU(N)$ UV theory with Dirichlet-like boundary conditions on $S^2$ \cite{Gang:2012ff}, where $\vec{u}$ is the set of fugacities associated to the $SU(N)$ gauge symmetry in the bulk, which becomes global on the boundary.

    \item $\mathcal{I}^{\Twall[]}(\vec{u},\{\vec{\xi},\vec{m}\};q)$ is the superconformal index \cite{Imamura:2011su,Kapustin:2011jm} of the 3d $\Twall[]$ theory. $\vec{u}$ is the set of fugacities associated to the $SU(N)$ global symmetry while $\{\vec{\xi},\vec{m}\}$ are the set of fugacities and the background magnetic fluxes associated to the $U(1)^{N-1}$ global symmetry.
    
    \item $\mathcal{II}^{\text{SW}} (\{\vec{\xi},\vec{m}\};q)$ is the Schur index of the low-energy Seiberg-Witten description of the 4d $SU(N)$ SYM theory.
    This can be computed from genuine IR data, namely the spectrum of BPS particles, in the framework of the Kontsevich-Soibelman operator using the strategies of \cite{Cordova:2015nma}. $\{ \vec{\xi}, \vec{m} \}$ is the fugacities/magnetic fluxes pair associated to the $U(1)^{N-1}$ symmetry.

    \item On the r.h.s.~we perform an integration over the $N-1$ fugacities associated to the $U(1)^{N-1}$ symmetry which is gauged on the boundary. Notice that this procedure is accompanied by a sum over all magnetic fluxes.
\end{itemize}
All the quantities also depend on a parameter $q$ which for all of them is the fugacity grading the R-symmetry of the operators, in a suitable definition that we will provide in detail later.

In principle, one could also turn on magnetic fluxes associated to the $SU(N)$ symmetry. However, as we will describe in a moment, the Schur (half-)index of $\NN=2$ SYM is expected to be zero for any non-zero flux. Notice that this is a prediction which is non-trivially realized on r.h.s.~of \eqref{eq: rgwall_index_statement} that we can explicitly check.

One can also consider the complementary setup where, on one side, we consider 4d $\mathcal{N}=2$ $SU(N)$ SYM with $\mathcal{N}$ boundary conditions coupled to $\Twall[]^{-1}$ theory, against the index of the low-energy effective $U(1)^{N-1}$ description with $\mathcal{D}$ b.c.:
\begin{equation} \label{eq: rgwall_index_statement_inverse}
    \mathcal{II}^{\text{SW}} (\{\vec{\xi},\vec{m}\};q) = \oint \prod_{j=1}^{N-1} \frac{du_j}{2\pi i u_j} \Delta_N(\vec{u}) 
    \mathcal{II}^{SU(N)} (\vec{u};q)
    \; \mathcal{I}^{\Twall[]^{-1}}(\vec{u};\{ \vec{\xi}, \vec{m}\};q) 
    \; 
\end{equation}
where $\Delta_N(\vec{u})$ is the $SU(N)$ Haar measure:
\begin{equation} \label{eq: sun_haar}
    \Delta_N(\vec{u}) = \frac{1}{N!} \prod_{j<k=1}^N \prod_{s=\pm1}\left[ 1-\left(\frac{u_j}{u_k}\right)^s \right] \Bigg|_{\prod_{j=1}^N u_j = 1}
\end{equation}
Notice that in \eqref{eq: rgwall_index_statement_inverse} we do not sum over magnetic fluxes because, as anticipated, the Schur (half-)index of 4d $\mathcal{N}=2$ SYM is zero for any non-zero magnetic flux, thus suppressing all these contributions. On the other hand, the matching is meant to hold for any generic value of the magnetic fluxes $\vec{m}$ related to the $U(1)^{N-1}$ symmetry.

It will be useful to introduce an hybrid 3d-4d quiver notation to represent the identities in \eqref{eq: rgwall_index_statement} and \eqref{eq: rgwall_index_statement_inverse}. Indeed, this is frame-dependent as we have to pick a specific electromagnetic duality frame for the IR to write explicitly the quiver of the RG-wall. For example in the electric frame where the RG-wall is the $\Twall[]^{-1}$ theory, corresponding to the identity \eqref{eq: rgwall_index_statement_inverse} we have: 
\begin{equation}
    \includegraphics[]{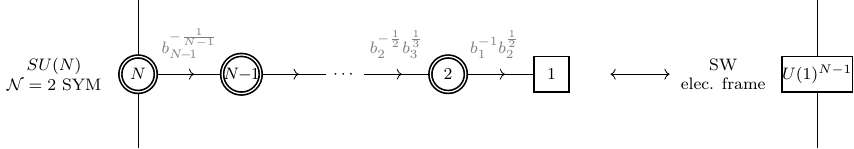}
\end{equation}
Recalling that all the bifundamentals of the $\Twall[]^{-1}$ theory carry R-charge 0 and in gray are reported the parameterization for the baryonic symmetries as the combination of fugacities associated to each bifundamental.

In the following subsections we describe how to compute the various indices appearing in \eqref{eq: rgwall_index_statement} and \eqref{eq: rgwall_index_statement_inverse}, that are in order: the Schur index on $HS^3 \times S^1$; the 3d index on $S^2 \times S^1$ and the computation of the IR index using the Kontsevich-Soibelman operator.
Later we discuss the matching as a perturbative expansion in the $q$ parameter for low value of the rank, namely for $SU(2),SU(3),SU(4)$ gauge groups. Notice that the matching for the proposal of the RG-wall in the $SU(2)$ case has been done already in \cite{Cordova:2016uwk}. We will give a brief rievew of this construction and provide further checks in different duality frames.

\subsection{The Schur (Half-)Index}
Let us start from the definition of the Schur index on the ``full" space $S^3 \times S^1$ \cite{Gadde:2011uv}. It is defined as a limit of the ``complete" 4d superconformal index \cite{Kinney:2005ej,Romelsberger:2007ec}, and it reads:
\begin{equation}
    \mathcal{I} (\mu_i;q) = \Tr (-1)^F \mu_i^{T_i} q^{E-R}
\end{equation}
where the trace is taken over the states of the theory in radial quantization. For each state: $E$ is the conformal dimension; $R$ is the spin under the $SU(2)_R$ subgroup of the R-symmetry; $T_i$ is a set of eigenvalue under the Cartan subgroup of the global symmetries, for which we turn on fugacities $\mu_i$. Notice that this index does not depend on the $U(1)_r$ subgroup of the total $U(2)$ R-symmetry and, therefore, the Schur index can be computed also for non-conformal theories.
Sometimes, to simplify the notation, we will avoid to give explicitly the $q$ dependence and simply write $\mathcal{I}(\mu_i)$.

The index can be always written as the projection over gauge invariant operators of a collection of free multiplets at zero-coupling as the index is invariant under continuous deformations and, thus, it is independent of the coupling, as long as we turn on fugacities only for non-anomalous symmetries.
Therefore, we need to know the contribution of only vector and hyper multiplets, from which we can construct Lagrangian theories.
\begin{itemize}
    \item A free $SU(N)$ vector multiplet contributes as:
    \begin{equation} \label{eq: schur_vec}
        PE\left[\frac{-2q}{1-q} \chi_{adj}^{SU(N)}(\vec{u})\right] = \prod_{i,j=1}^N (u_i/u_j  q;q)^2 
    \end{equation}
    where $\vec{u}$ is the set of fugacities associated to the $SU(N)$ gauge symmetry satisfying $\prod_{j=1}^N u_j = 1$, and $\chi_{adj}^{SU(N)}(\vec{u})$ is the character of the $SU(N)$ adjoint representation.
    
    \item A free hypermultiplet with charge $+1$ under a $U(1)$ symmetry with fugacity $\xi$ contributes as:
    \begin{equation} \label{eq: schur_hyper}
        PE\left[\frac{q^{1/2}}{1-q} \left( \xi + \xi^{-1} \right) \right] = \frac{1}{\left(q^{1/2} \xi;q\right) \left(q^{1/2} \xi^{-1};q\right)} \,.
    \end{equation}
\end{itemize}
We used the Plethystic Exponential function:
\begin{equation}\label{eq: PE_def}
    PE[x] = \exp\left[ \sum_{n=1}^{+ \infty} \frac{x^n}{n} \right] \,,
\end{equation}
and the q-Pochhammer symbol:
\begin{equation}\label{eq: qpoch_def}
    (a;b)_n = \prod_{i=0}^{n} (1-a^i b),
    \qquad
    (a;b) \equiv (a;b)_\infty = \prod_{i=0}^{+\infty} (1-a^i b) \,.
\end{equation}
As an example, the Schur index of a $SU(N)$ gauge theory with $F$ hypermultiplets in the fundamental representation reads:
\begin{equation}
    \mathcal{I}(\vec{f}) = \oint \prod_{i=1}^{N-1} \frac{du_i}{2\pi i u_i} \Delta_N(\vec{u}) \frac{\prod_{i,j=1}^N (u_i/u_j q;q)^2}{\prod_{i=1}^N \prod_{j=1}^F (q^{1/2} u_i/f_j;q) (q^{1/2} f_j/u_i;q)} \Bigg|_{\prod_{i=1}^N u_i=1} 
\end{equation}
where the integral over the gauge fugacities $\vec{u}$ with the $SU(N)$ Haar-measure $\Delta_N(\vec{u})$, defined in \eqref{eq: sun_haar}, implements the projection over gauge invariant states. Also, $\vec{f}$ is the set of fugacities associated to the $U(F)$ flavor symmetry.

In this paper we will be interested in computing indices on manifolds with boundaries. 
In general, partitions functions on manifolds with boundaries, that can be glued together to form more generic manifolds, are typically referred to as \textit{holomorphic-blocks} \cite{Pasquetti:2011fj,Benini:2012ui,Nieri:2015yia,Doroud:2012xw,Beem:2012mb} or, in the case of indices as \textit{half-indices} \cite{Gang:2012ff,Dimofte:2011py,Cordova:2016uwk,Gaiotto:2019jvo}. The Schur index on the half-space $HS^3 \times S^1$ is particular type of \textit{half-index} that is expected to coincide with the Schur limit of the superconformal index holomorphic block of \cite{Nieri:2015yia}, which is given in \cite{Gang:2012ff,Cordova:2016uwk,Dimofte:2011py}.  Although no localization computation has been done, one can still conjecture the expression of the Schur half-index from reasonable arguments. First, to define the contribution of a certain multiplet we need to decide the boundary conditions. For each supermultiplet we consider two possibilities that are $\mathcal{D}$ or $\mathcal{N}$ b.c.~(defined as in Section \ref{sec: bc_proposal}) and, in principle, each of the contribution could be different. However, we observe that regardless of the choice, the contributions remain qualitatively the same. 
\begin{itemize}
    \item A $SU(N)$ vector multiplet contributes as ``half" of the full-space contribution:
    \begin{equation}
        PE\left[\frac{-q}{1-q} \chi_{adj}^{SU(N)}(\vec{u})\right] = (q;q)^{N-1} \prod_{i<j}^N (u_i/u_j  q;q)(u_j/u_i  q;q) \Bigg|_{\prod_{i=1}^N u_i = 1}
    \end{equation}
    which can be motivated by observing that the double negative contribution on the l.h.s.~of \eqref{eq: schur_vec} represents gaugini, that are complex fermions. Regardless of the choice of boundary conditions, the gaugini are split into two real fermions on the boundary, one receiving Dirichlet and the other Neumann boundary conditions. Therefore, always only half of the gaugini is dynamical on the boundary and thus contribute to the index.

    \item An hypermultiplet with charge $+1$ under a $U(1)$ symmetry, with associated fugacity $\xi$, contributes as:
    \begin{equation}
        \begin{cases}
            PE\left[\frac{q^{1/2}}{1-q} \xi \right] = \frac{1}{ \left(\xi q^{1/2};q\right)} \quad &\text{for $\mathcal{B}_{\chi}$ b.c.} \\
            PE\left[\frac{q^{1/2}}{1-q} \xi^{-1} \right] = \frac{1}{ \left(\xi^{-1}q^{1/2};q\right)} \quad &\text{for $\mathcal{B}_{\bar{\chi}}$ b.c.}
        \end{cases}
    \end{equation}
    This formula is motivated by the fact that in \eqref{eq: schur_hyper} the two contributions on the l.h.s.~correspond to the two real components of the complex scalar field. Regardless of the choice of b.c.~we always impose Dirichlet b.c.~for one of the two and Neumann b.c.~for the other. Such that it survives, in the first case, the contribution of the chiral, carrying the same $U(1)$ charge as the hypermultiplet, and in the second case that of the anti-chiral, with opposite charge under $U(1)$. Anyway, the contributions are identical, with minor differences only in the charge assignation.
\end{itemize}
As an example, the half-Schur index of a $SU(N)$ gauge theory with $F$ fundamental hypermultiplets reads, in the two cases:
\begin{equation}
\begin{split}
    & \mathcal{II}^{(\mathcal{D})}(\vec{u},\vec{f}) = \frac{(q;q)^{N-1} \prod_{i<j}^N (u_i/u_j  q;q)(u_j/u_i  q;q)}{\prod_{i=1}^N \prod_{j=1}^F (u_i/f_j q^{1/2};q)} \Bigg|_{\prod_{i=1}^N u_i = 1} \\
    & \mathcal{II}^{(\mathcal{N})}(\vec{f}) = \oint \prod_{j=1}^N \frac{du_j}{2\pi i u_j} \Delta_N(\vec{u})\frac{(q;q)^{N-1} \prod_{i<j}^N (u_i/u_j  q;q)(u_j/u_i  q;q)}{\prod_{i=1}^N \prod_{j=1}^F (f_j/u_i q^{1/2};q)} \times \mathcal{I}(\text{3d d.o.f.}) \Bigg|_{\prod_{i=1}^N u_i = 1}
\end{split}
\end{equation}
Where in the first line we have the case of $\mathcal{D}$ b.c.~for which, we recall, the gauge symmetry becomes a global symmetry on the boundary and thus we do not take any trace over gauge invariant states. In the second line, instead, we have the case of $\mathcal{N}$ b.c.~for which the trace over gauge invariant operators is performed and we also allow for 3d d.o.f.s to live on the boundary, which enter the expression with the multiplication by the 3d superconformal index, which we now describe in more detail.

\subsection{The 3d Superconformal Index} \label{subsec: 3dindex}
We now describe the 3d superconformal index of a generic 3d $\mathcal{N}=2$ theory \cite{Imamura:2011su,Kapustin:2011jm} which, in this paper, we introduce as extra d.o.f.s living on the boundary of the $HS^3 \times S^1$ space.
Indeed, as we consider 4d-3d coupled systems, we have to carefully define the 3d index to reflect correctly how the 3d superconformal group on the boundary is embedded into the 4d one. 
Therefore, we make use of slightly unfamiliar definitions, w.r.t.~the commonly used notation in the 3d index literature, which takes into account this subtlety, following the notation of \cite{Dimofte:2011py}.

We define the 3d superconformal index, which is the $S^2 \times S^1$ partition function, as:
\begin{equation}
    \mathcal{I}(\vec{\mu},\vec{m};q) = \Tr_{\mathcal{H}_{\vec{m}}} (-1)^F q^{R/2+j_3} u_i^{T_i}
\end{equation}
where $R$ is the charge under the $U(1)_R$ R-symmetry, suitably embedded into the 4d $SU(2)_R$ R-symmetry in the bulk. $j_3$ is the spin under the $SO(3)$ isometry of $S^2$. The set $u_i$ are fugacities that we turn on w.r.t.~the Cartan subgroup of the global symmetry with generators $T_i$. Notice that the trace is taken over the states of the theory in radial quantization in presence of a disorder operator, that can be freely placed at any point of $S^1$, such that the magnetic flux trough $S^2$ is:
\begin{equation}
    \oint_{S_2} F_i = 2\pi m_i \,,
\end{equation}
where $F_i$ is the field-strength associated to the background gauge field of the $U(1)_{T_i}$ Cartan generator of the global symmetry. The set $m_i \in \mathbb{Z}$ are the magnetic fluxes, or equivalently, they label the holonomy of the gauge fields around $S^1$.
Most of the times we will imply the $q$ dependence of the index and simply write $\mathcal{I}(\vec{\mu},\vec{m})$

The 3d superconformal index can be computed from a collection of free multiplets defining the Lagrangian at zero coupling, projecting onto gauge invariant operators. Since the index is invariant under continous deformations of the action, the index computed in this way is equal to that of the theory at strong coupling, i.e.~at the IR fixed point.
The contribution of the multiplets in interest are the following.
\begin{itemize}
    \item The contribution of a $U(N)$ vector multiplet, together with that of the $U(N)$ Haar measure, it is:
    \begin{equation} \label{eq: 3dind_vec}
        \mathcal{I}_{\text{vec}}^{(N)}(\vec{z},\vec{m}) = \frac{1}{N!} \prod_{i < j = 1}^N (-\sqrt{q})^{-|m_i-m_j|} (1- {\sqrt{q}}^{|m_i-m_j|} z_i/z_j) (1- {\sqrt{q}}^{|m_i-m_j|} z_j/z_i)
    \end{equation}
    Which is considered at generic value of the magnetic flux $(m_1,\ldots,m_N) \in \mathbb{Z}^N$ w.r.t.~the $U(1)^N$ Cartan subgroup of $U(N)$.
    The contribution of a $SU(N)$ vector multiplet is obtained by imposing $\prod_{i=1}^{N} u_i = 1$ and $\sum_{i=1}^{N} m_i = 0$ in the expression above. \\
    As this observation will be useful later on, notice that the contribution of a 3d vector multiplet at zero value of the magnetic fluxes coincides with that of the Haar measure:
    \begin{equation}
        \mathcal{I}_{\text{vec}}^{(N)}(\vec{z},\vec{m}=\vec{0}) = \Delta_N(\vec{z})
    \end{equation}

    \item The contribution of a chiral multiplet with $+1$ charge under a $U(1)$ symmetry, with associated fugacity $\xi$ magnetic flux $m$, and with generic R-charge $R$, it reads:
    \begin{equation} \label{eq: 3dind_chir}
    \begin{split} 
        \mathcal{I}_{\text{chir}}^{(R)}(\xi,m) =&  \left( \xi (-\sqrt{q})^{R-1}\right)^{-\frac{|m|}{2}} \frac{ \left( (-\sqrt{q})^{2-R+|m|} \xi^{-1};q \right)}{ \left( (-\sqrt{q})^{R+|m|} \xi;q \right)}
        \\=&
        \left( \xi  (-\sqrt{q})^{R-1}\right)^{-\frac{ m  }{2}}
        \frac{ \left( (-1)^R \sqrt{q}^{2-R+ m} \xi^{-1} ;q\right)}{\left((-1)^{R } \sqrt{q}^{R+ m  } \xi
         ;q\right)}
    \end{split}
    \end{equation}
\end{itemize}
Notice that in order for the resulting index to only have real coefficients, we have to ensure that the R-charges are always integers. Actually, to be more precise, we have to impose that gauge invariant operators possess only integer R-charges, as the final index will receive contributions only from them. This condition will be automatically satisfied when considering the setup in interest, as the 3d $\mathcal{N}=2$ R-charge is twice the spin under the $SU(2)_R$ 4d $\mathcal{N}=2$ R-symmetry and thus is always a positive integer.

The index also receives classical contributions from FI terms, CS interactions and generic BF couplings. We define a generic contribution as follows:
\begin{equation}
    \mathcal{I}_{\text{class.}} (\vec{u},\vec{m}) =
    \prod_{i}^N u_i^{k_{ii} m_i}
    \prod_{i\neq j}^N u_i^{\frac{k_{ij}m_j}{2}}
    =
    \prod_{i \geq j} 
    \left(u_i^{m_j}
    u_j^{m_i}
    \right)^{\frac{k_{ij}}{2}}
\end{equation}
where we denote by $u_i$ the complete set of fugacities of the theory, that can be associated to gauge or global symmetries. The matrix $k_{ij}$ encodes the level of the interactions so that the diagonal $k_{ii}$ components are the contributions of the (background) CS terms and the non-diagonal $k_{ij}$ are those of BF couplings that represent: mixed CS-levels if the two $U(1)$ symmetries are gauged; FI interactions if only one of the two symmetries is gauged; BF interactions if both the symmetries are not gauged.
We adopt the following convention for the corresponding contribution of a (mixed) CS to the action:
\begin{equation}
    -\frac{ik_{ij}}{4\pi} \int A_i \wedge dA_j \,.
\end{equation}
Notice that in this convention, a standard FI term corresponds to a $k=-2$ level between the gauge symmetry and the topological symmetry. When dealing with non-Abelian gauge groups with non-zero CS levels, we write the contribution of a $(k,k+lN)$ CS-level for a $U(N)$ gauge group as:
\begin{equation}
    \prod_{i=1}^N u_i^{k m_i} \left( \prod_{i=1}^N u_i \right)^{l \sum_{j=1}^N m_j}
\end{equation}
where $k$ satisfies quantization conditions that depend on the matter content of the theory, while $l \in \mathbb{Z}$.

As an example, the superconformal index of a 3d $U(N)$ gauge theory, with CS-level $(k,k+lN)$ and $F$ fundamental chirals is the following.
\begin{equation}
\begin{split}
    \mathcal{I}( \{\vec{v},\vec{m}^{(v)} \} ; \{ \xi,n \}) =& \sum_{\vec{m} \in \mathbb{Z}^N} \oint \prod_{j=1}^N \frac{dz_j ( \xi  z_j^k )^{m_j}}{2 \pi i z_j} 
    \left( \prod_{i=1}^N z_i \right)^{n+l \sum_{j=1}^N m_j}
    \mathcal{I}^{(N)}_{\text{vec}}(\vec{z},\vec{m}) \times \\
    & \times \prod_{j=1}^N \prod_{k=1}^F \mathcal{I}_{\text{chir}}(z_j/v_k,m_j-m^{(v)}_k;1/2)
\end{split}
\end{equation}
where $\xi$ is the fugacity associated to the topological symmetry with associated magnetic flux $n$. $\vec{v}$ is the set of fugacities for the $SU(F)$ flavor symmetry with associated magnetic flux $\vec{m}^{(v)}$. The correct quantization of the Chern-Simons interaction requires $k+F/2$ to be integer, that is $k$ is integer/half-integer for even/odd $F$ and $l \in \mathbb{Z}$.

\subsection{An intermezzo: matching indices at zero background magnetic fluxes}

Before actually describing the computation of the index of the 4d IR effective description, through the Kontsevich-Soibelman operator, to finally discuss in completeness the r.h.s.~of \eqref{eq: rgwall_index_statement}, we present a ``less-refined check" of the boundary duality of Section \ref{sec: bc_proposal}. This computation only involves the Schur half-index and the 3d index and reproduces \eqref{eq: rgwall_index_statement_inverse} at zero value of the magnetic fluxes associated to the $U(1)^{N-1}$ symmetry.

The idea is to translate the expectation of the RG-wall, in \eqref{eq: rgwall_index_statement} or \eqref{eq: rgwall_index_statement_inverse}, as a semi-classical computation where the index of the IR theory is replaced with a Schur half-index of the low-energy effective description. As we will now describe, this matching surprisingly reproduces correctly \eqref{eq: rgwall_index_statement_inverse} at zero value of the magnetic flux $\vec{m}$ and it provides a simpler computation in favor of the proposal in Section \ref{sec: bc_proposal}.
Indeed, this check is less refined that the precise one involving the Kontsevich-Soibelman operator, which is the correct way to compute the index of the $\NN=2$ theory at a generic point of the CB. 
In fact, it is surprising for this matching to work but nonetheless this observation provides an interesting coincidence.

The identity reads:
\begin{equation} \label{eq: rgwall_schurnaive}
    \mathcal{II}^{SW}( \vec{\xi},\vec{m}=\textbf{0} ) = \oint \prod_{i=1}^{N-1} \frac{d u_i}{2\pi i} \Delta_N(\vec{u}) \mathcal{I}^{\Twall[]^{-1}}(\vec{u},\{\vec{\xi},\vec{m}=\textbf{0}\}) \mathcal{II}^{SU(N)}(\vec{u})
\end{equation}
where $\mathcal{II}^{SU(N)}$ is the Schur half-index of $\NN=2$ SYM with $\mathcal{D}$ b.c., which is the contribution of a $SU(N)$ vector multiplet:
\begin{equation}
    \mathcal{II}^{SU(N)}(\vec{u}) = (q;q)^{N-1} \prod_{i<j=1}^N (u_i/u_j q;q) (u_j/u_i q;q) \bigg|_{\prod_{i=1}^N u_i = 1} \,.
\end{equation}
On the other side, $\mathcal{I}^{SW}(\{\vec{\xi},\vec{m}=\vec{0}\})$ is computed as the Schur half-index of the semi-classical description which is given by $N-1$ free vectors and $N-1$ hypermultiplets:
\begin{equation}
    \mathcal{II}^{SW}( \vec{\xi},\vec{m}=\textbf{0} ) = \frac{(q;q)^{N-1}}{\prod_{i=1}^{N-1} (\xi_iq^{1/2} ;q)} \,.
\end{equation}
Finally, $\mathcal{I}^{\Twall[]^{-1}}$ is the 3d index of the $\Twall[]^{-1}$ theory, which is the quiver given in \eqref{eq: twallinv_quiverpic}.
\begin{equation}
\begin{split}
    & \mathcal{I}^{(\Twall[N])^{-1}} (\vec{u}, \{ \vec{\xi},\vec{m}=\textbf{0}\}) = 
    \\ & \qquad \qquad 
    \prod_{r=1}^{N-1} \left[ \sum_{\vec{m}^{(r)} \in \mathbb{Z}^r} \frac{\delta_{\sum_{j=1}^N m^{(r)}_j,0}}{r!} \oint \prod_{j=1}^{r} \frac{d z_j^{(r)}}{2\pi i z_j^{(r)}} \delta \left( \prod_{j=1}^r z_j^{(r)} - 1 \right) \mathcal{I}_{\text{vec}}^{(r)}\left( \vec{z}^{(r)},\vec{m}^{(r)} \right) \right] 
    \\ & \qquad \qquad
    \prod_{r=1}^{N-1} \prod_{j=1}^r \prod_{k=1}^{r+1} \mathcal{I}_{\text{chir}}^{(0)} \left( z^{(r+1)}_k/z^{(r)}_j b_r^{-\frac{1}{r}} b_{r+1}^{\frac{1}{r+1}}, m^{(r+1)}_k - m^{(r)}_j \right) 
\end{split}
\end{equation}
where one has to replace the following convenient definitions:
\begin{equation}
    \vec{z}^{(N)} \equiv \vec{u} \,, \qquad \vec{m}^{(N)} = \textbf{0} \,, \qquad b_{N} = 1 \,.
\end{equation}

We checked the identity in \eqref{eq: rgwall_schurnaive} up to high order in $q$ for fixed low values of the rank, that is for $SU(2),SU(3),SU(4)$.

Although this matching provides a first interesting motivation for the proposal in Section \ref{sec: bc_proposal}, there are obvious limitation. In fact, the identities in \eqref{eq: rgwall_index_statement} and \eqref{eq: rgwall_index_statement_inverse}, provide a finer check as they involve generic non-zero magnetic fluxes.
This can not be done by extending  in some way \eqref{eq: rgwall_schurnaive}, as it is not not known ho to turn on magnetic fluxes for the Schur half-index, while this is known in the more well-defined computation of the IR index with the Kontsevich-Soibelman operator, which we now describe how to compute.

\subsection{Half-index from the IR: Kontsevich-Soibelman operator}

We now describe how to compute the index of a 4d $\mathcal{N}=2$ theory from genuine IR data associated to a given point of the CB, as first prescribed in \cite{Cordova:2015nma,Cordova:2016uwk}.

The Coulomb branch of an $\NN=2$ SQFT is equipped with a charge lattice $\Gamma$ involving electro-magnetic and flavor charges.
The charge lattice is equipped with:
\begin{itemize}
\item the Dirac pairing $\langle \cdot, \cdot \rangle$, namely an integer-valued antisymmetric form:
\begin{equation}
    \langle \cdot, \cdot \rangle:
    \Gamma \times \Gamma \to \mathbb{Z} \,.
\end{equation}
\item The central charge $\mathcal{Z}$, which depends on some CB vacuum coordinate:
\begin{equation}
    \mathcal{Z}: \Gamma \to \mathbb{C} \,.
\end{equation}
In the latter we suppress the dependence on the vacuum coordinate for ease of readability.
\end{itemize}
The sublattice of flavor charges $\Gamma_{\text{flav}}$ is the set of charges with vanishing Dirac pairing with every other charge:
\begin{equation}
    \Gamma_{\text{flav}} = \{ \gamma \in \Gamma \;|\; \langle \gamma, \Gamma \rangle = 0 \}
\end{equation}
or, in other words, $\Gamma_{\text{flav}}$ coincides with the Kernel of the Dirac pairing. Where, the condition $\langle \gamma, \Gamma \rangle = 0$ means that the pairing is zero for all elements in $\Gamma$.
We denote as $\Gamma_{\text{em}}$ the lattice of electromagnetic charges:
\begin{equation}
    \Gamma_{\text{em}} = \Gamma / \Gamma_{\text{flav}}
\end{equation}
To the charge lattice we associate a (non-commutative) quantum torus algebra whose elements $X_{\gamma}$ are labeled by a point of the lattice $\gamma\in \Gamma$ with product given by:
\begin{equation}    \label{eq:QTA_product}
    X_{\gamma_1}X_{\gamma_2} = q^{\tfrac{1}{2} \langle \gamma_1, \gamma_2\rangle}X_{\gamma_1+\gamma_2}
    =
    q^{ \langle \gamma_1, \gamma_2\rangle}X_{\gamma_2}X_{\gamma_1}
\end{equation}
For any charge $\gamma \in \Gamma$ define the quantum Kontsevich-Soibelman factor
\begin{equation}
    K\left(q ; X_\gamma ; \Omega_j(\gamma)\right)=\prod_{n \in \mathbb{Z}} E_q\left((-1)^n q^{n / 2} X_\gamma\right)^{(-1)^n \Omega_n(\gamma)}
\end{equation}
where $\Omega_n (\gamma)$ is the degeneracy of BPS states with spin $n$ and charge $\gamma$. 
The main function of interest is the quantum dilogarithm:
\begin{equation}
    E_q(z)=\left(-q^{\frac{1}{2}} z ; q\right)^{-1}=\prod_{i=0}^{\infty}\left(1+q^{i+\frac{1}{2}} z\right)^{-1}=\sum_{n=0}^{\infty} \frac{\left(-q^{\frac{1}{2}} z\right)^n}{(q;q)_n}
\end{equation}
In this paper we only consider CB chambers where the spectrum is comprised only by hypermultiplets, therefore the KS factors reduce to a single quantum dilogarithm.
The Kontsevich-Soibelman operator is defined as:
\begin{equation}
    \mathcal{O}(q)=\prod_{\gamma \in \Gamma}^{\curvearrowleft} K\left(q ; X_\gamma ; \Omega_j(\gamma)\right)
\end{equation}
where the ordered product is taken in increasing values of the argument of the central charge $\mathcal{Z}$\footnote{In this paper we only consider SQFTs which posses a finite chamber in the Coulomb branch, therefore the product in the Kontsevich-Soibelman operator will involve a finite number of factors.}. 
The KS operator is invariant up to conjugation under wall-crossing as one varies the CB parameter. In \cite{Cordova:2015nma} it was conjectured that the Schur index of an SQFT can be obtained from the KS operator by taking a trace over the quantum torus algebra:
\begin{equation}    \label{eq:KS_Schur_conjecture}
    \mathcal{I}(\vec{\mu};q)=(q;q)^{2 r} \operatorname{Tr}[\mathcal{O}(q)]
\end{equation}
where $r$ is the rank of the theory, $\vec{\mu}$ are fugacities for the Cartan of the global symmetry group and the trace is defined as:
\begin{equation}
\Tr [X_{\gamma}] = 
\begin{cases}
    X_\gamma \qquad &\text{if} \; \gamma \in \Gamma_{\text{flav}}
    \\
    0 \qquad &\text{otherwise}
\end{cases}
\end{equation}
Therefore the trace projects onto the center of the quantum torus algebra, which corresponds to the flavor sublattice $\Gamma_{\text{flav}}$ of the charge lattice $\Gamma$. The elements of the flavor sublattice are identified with the flavor fugacities $\vec{\mu}$. Given a suitable basis $\gamma_{f_i}$ of the flavor sublattice one identifies:
\begin{equation}
    \Tr[X_{\sum_{i} a_i \gamma_{f_i}}] = \prod_{i} \mu_i^{a_i}
\end{equation}

In \cite{Cordova:2016uwk} a similar conjecture was put forward for the case of the half-index on $HS^3 \times S^1$. Here the relevant operator is the quantum spectrum generator defined as:
\begin{equation} \label{eq: half_KSop}
    \mathcal{S}_{\theta}(q)=\prod_{\arg \left(\mathcal{Z}_\gamma\right) \in[\theta, \pi+\theta)}^{\curvearrowleft} K\left(q ; X_\gamma ; \Omega_j(\gamma)\right) .
\end{equation}
which  depends of a choice of angle $\theta$ and is an ordered product of quantum KS factors over the BPS states with $\theta \leq \arg (\mathcal{Z_\gamma}) < \pi + \theta$.
We denote the BPS states involved in the product \eqref{eq: half_KSop} as “particles", in analogy with the BPS quiver program \cite{Alim:2011kw}. We denote as $\{\gamma^{\text{BPS}}_i\}_{i=1}^{r}$ the positive basis for the charges of BPS particles. For each $\gamma^{\text{BPS}}_i$ there is (at least) one BPS hypermultiplet in the BPS spectrum of the theory. All the EM charges of BPS particles are positive integer combinations of the $\gamma^{\text{BPS}}_i$.

Here we consider Dirichlet boundary conditions for the low energy $U(1)^{r}$ gauge fields involved in the SW low energy effective action. 
The gauge symmetry in the bulk becomes a flavor symmetry on the boundary, therefore on top of the flavor fugacities for the global symmetries of the bulk SQFT $\vec{\mu}$ we can also refine the index with fugacities $\vec{\xi}$ for the $U(1)^{r}$ boundary symmetries resulting from the Dirichlet boundary conditions.\footnote{Actually, in this paper we will consider only the case of $SU(N)$ SYM for which there are indeed no global symmetries. Nonetheless, we will include them in the discussion for the moment for the sake of generality.}
Analogously we can define a set of traces of the quantum torus algebra. 
We choose a lattice of ``electric" charges $\Gamma_{\text{ele}} \subset \text{Span}_{\mathbb{R}} (\Gamma_{\text{em}})$ which are mutually local between each other:
\begin{equation}
    \langle \Gamma_{\text{ele}}, \Gamma_{\text{ele}} \rangle = 0 
\end{equation}
and have integer Dirac pairing with local particles:
\begin{equation}
    \langle \Gamma_{\text{ele}}, \Gamma \rangle \in \zz 
\end{equation}
and have rank equal to the number of $U(1)$ gauge groups, namely the rank of the $\NN=2$ theory\footnote{It would interesting to study whether more general choices of $\Ge$ are related to other boundary conditions, we leave this to future work.
}. Equivalently, $\Gamma_{\text{ele}}$ is a choice of maximal Lagrangian sublattice of $\Gamma_{\text{em}}$.
For each choice of ``electric" charges $\Gamma_{\text{ele}}$
we define the following trace operator:
\begin{equation}
\Tr_{\Gamma_{\text{ele}}} [X_{\gamma}] = 
\begin{cases}
    X_\gamma \qquad &\text{if} \; \Dirac[\gamma,\Gamma_{\text{ele}}]=0
    \\
    0 \qquad &\text{otherwise}
\end{cases}
\end{equation}
which project onto elements of the quantum torus algebra which are generated by the chosen basis of $\Gamma_{\text{ele}}$, that are mutually local, and/or by flavor charges.
We then identify:
\begin{equation}
    \Tr[X_{\sum_{i} a_i \gamma_{f_i} + \sum_j b_j \gamma^{\text{ele}}_j}] = \prod_{i} \mu_i^{a_i} \prod_j \xi_j^{b_j}
\end{equation}
We expect that the trace $\Tr_{\Gamma_{\text{ele}}}$ of the quantum spectrum generator $\mathcal{S}_\theta(q)$ is related to the half-index of the theory where we give Dirichlet boundary conditions to the $U(1)^{r}$ gauge fields in an electromagnetic duality frame where the charges in $\Gamma_{\text{ele}}$ are “electric". 

As already proposed in \cite{Cordova:2016uwk} introducing a non-zero flux for the gauge fields on the boundary $S^2$ should instead correspond to inserting $X_{\gamma'}$ in the trace, where $\gamma'$ is a “magnetic" charge w.r.t.~some chosen electric charge.
We choose a lattice of magnetic charges:
\begin{equation}
    \Gamma_{\text{mag}} \subset \text{Span}_{\mathbb{R}} (\Gamma_{\text{em}}/\Gamma_{\text{ele}})
\end{equation}
which contains mutually local charges with integer Dirac pairing with charges of local operators:
\begin{equation}
    \langle \Gamma_{\text{mag}}, \Gamma_{\text{mag}} \rangle = 0,
    \qquad
    \langle \Gamma_{\text{mag}}, \G \rangle \in \zz 
\end{equation}
We also require $\Gamma_{\text{mag}}$ to be a ``minimal" choice of magnetic charges compatible with the Dirac pairing, namely we require that the Dirac pairing (possibly linearly extended) on $\Gamma_{\text{mag}} \oplus \Gamma_{\text{ele}}$ is principal. 
If the Dirac pairing of the $\NN=2$ theory is principal then there is only one consistent choice of $\Gamma_{\text{mag}}$. Otherwise, namely when the $\NN=2$ theory has a non-trivial 1-form symmetry \cite{Gaiotto:2014kfa,Aharony:2013hda}, there are as many choices of $\Gamma_{\text{mag}}$ as there are global forms for the $\NN=2$ theory\cite{DelZotto:2022ras}.

In explicit computations we choose a basis $\{q_i\}_{i=1}^{r}$ of the electric lattice $\Gamma_{\text{ele}}$ such that the charges of BPS particles are positive integer combinations of the $q_i$. Recalling that $r$ is the rank of the theory.
We also choose a “dual" basis $\{p_i\}_{i=1}^{r}$ of the “magnetic" lattice $\Gamma_{\text{mag}}$ satisfying:
\begin{equation}
    \Dirac[q_i,p_i]= -\delta_{i,j},
    \qquad p_i \in \Gamma_{\text{mag}}
\end{equation}
The traces of $X_{q_i}$ are identified with fugacities $\xi_i$. The insertion of $X_{p_i}$ corresponds to turning on a unit of background flux for the same boundary symmetry.

All in all, we then define the following index:
\begin{equation}    \label{eq:SW_index_Dirichlet}
\begin{split}
    & \mathcal{II}^{\text{SW}} (\xi_i, m_i;q) = (q;q)^{r}
    \Tr_{\Gamma_{\text{ele}}} \left[ :\mathcal{S}(q) \prod_{i=1}^{r} (X_{p_i})^{m_i} :\right] \\
    & \text{with} \; \Tr_{\Gamma_{\text{ele}}}[X_{q_i}] = \xi_i
\end{split}
\end{equation}
where the ordered product is defined as $:X_a X_b:\equiv X_{a+b}$, and $m_i$ are integers. Notice that in \eqref{eq:SW_index_Dirichlet} we suppressed the $\theta$ dependence, as is specified by the choice of a basis for $\Gamma_{\text{ele}}$. Most of the times we will also drop the $q$ dependence.

The conjecture is that $\mathcal{II}^{\text{SW}}$ is the half-index of the low energy SW theory with boundary condition defined as follows.
\begin{itemize}
    \item We consider the EM duality frame for the low energy $U(1)^{\text{rank}}$ gauge group where the EM charges in $\Gamma_{\text{ele}}$ are electric.
    This choice is always possible because the charges in $\Gamma_{\text{ele}}$ are mutually local. 
    We give Dirichlet boundary conditions to the corresponding gauge fields.
    
    \item The Dirichlet boundary condition for the $i$-th gauge group $U(1)_i$ has flux $m_i$ on the boundary $S^2$.
    
    \item The BPS hypermultiplet particles $H_i$ with EM charge $\gamma^{\text{BPS}}_i$ split into two chiral fields $(X_i, Y_i)$ under the decomposition from $4d$ $\NN=2$ to $3d$ $\NN=2$.
    Here $X_i$ is the chiral with R-charge $1$ and the same gauge and flavor charges as $H_i$, while $Y_i$ is the chiral with R-charge $1$ and opposite gauge and flavor charges.
    We give $\mathcal{B}_{\chi}$ boundary condition for $H$, as defined in Section \ref{sec: bc_proposal}:
    \begin{equation}
        B_{\chi} \text{ for } H=(X,Y): \qquad
        \begin{cases}
            \partial_{\bot}X|_{\partial} = 0 
            \\
            Y|_{\partial}= 0
        \end{cases}
    \end{equation}
    Notice that the opposite choice, namely giving $\mathcal{B}_{\overline{\chi}}$ b.c. to the BPS particles, corresponds to replacing $\mathcal{S}$ with $\overline{\mathcal{S}}$ in \eqref{eq:SW_index_Dirichlet}, see \cite{Cordova:2016uwk} for more details.
\end{itemize}
This is a refinement of the conjecture already proposed in \cite{Cordova:2016uwk}.
As already discussed in \cite{Dimofte:2013lba} the last point is subtle because the hypermultiplet charges $\gamma^{\text{BPS}}_i$ are not mutually local. Therefore there is no EM frame where all hypemultiplets can be treated as fundamental fields in an effective Lagrangian. On the other hand such boundary conditions can be thought of as allowing monopole line defects to end at the boundary, which can be discussed even when $\gamma^{\text{BPS}}_i$ is not electric. 
We refer the interested reader to the analogous discussion in Section 2.3 of \cite{Dimofte:2013lba}.

\subsection{Example: $SU(2)$ SYM}
The proposal of the RG-wall theory and the matching of half-indices for $SU(2)$ SYM was already discussed in \cite{Cordova:2016uwk}. Here we review their analysis and we expand it by considering different presentations of the RG wall related by Witten's $SL(2,\zz)$ action.

To perform the check of the proposal of Section \ref{subsec: N=2_bc}, we need to specify all the ingredients necessary to compute the various indices appearing in \eqref{eq: rgwall_index_statement} and \eqref{eq: rgwall_index_statement_inverse}.
The Schur half-index of $SU(2)$ SYM is given by:
\begin{equation} \label{eq: su2_halfind}
    \mathcal{II}^{SU(2)} (u) = (q u^2;q) (q u^{-2};q)(q;q) 
\end{equation}
and the 3d superconformal index of the $\Twall[2]$ theory is:
\begin{equation} \label{eq: su2_twall_ind}
    \mathcal{I}^{\Twall[2]}(u,\{ \xi,m\}) = \xi^{-m} \mathcal{I}_{\text{chir}}^{(0)}(u \xi,m) \mathcal{I}_{\text{chir}}^{(0)}(u^{-1} \xi,m) 
\end{equation}
that are two chirals with 0 R-charge and charge $+1$ under $U(1)_\xi$, with a background CS-level $-1$. $u$ is the fugacity of the $SU(2)$ global symmetry.
We also report the index of the inverse-wall $(\Twall[2])^{-1}$:
\begin{equation}\label{eq: su2_twallinv_ind}
\begin{split}
    \mathcal{I}^{(\Twall[2])^{-1}}(u,\{ \xi,m\})
    &= \xi^{m} \mathcal{I}_{\text{chir}}^{(0)}(u \xi^{-1},-m) \mathcal{I}_{\text{chir}}^{(0)}(u^{-1} \xi^{-1},-m)
\end{split}
\end{equation}

To compute the index of the IR effective theory we need to introduce all the data required to compute the half-KS operator.
Pure $SU(2)$ SYM has a strongly coupled chamber in the CB where the BPS spectrum is composed of the monopole, the dyon and the corresponding antiparticles, as originally discussed in \cite{Seiberg:1994rs}. 
Here we discuss the spectrum from the point of view of the more recent BPS quiver approach, which will be more suitable for the generalization to higher $N$. We do not review the formalism in detail and only report the results relevant to our analysis, the interested reader is referred to the original literature \cite{Alim:2011ae,Alim:2011kw,Cecotti:2012gh} and references therein.

The BPS quiver of pure $SU(2)$ SYM is the following:
\begin{equation}
    \begin{tikzpicture}[baseline=(current bounding box).center]
        \node[BPSquivernode,label={[label distance=2mm]above:$\gamma_1$}] (gamma1) at (0,0) {};
        \node[BPSquivernode,label={[label distance=2mm]above:$\gamma_2$}] (gamma2) at (2,0) {};
        \draw[BPSquiverarrow,double,double distance=3pt] (gamma1) -- (gamma2);
    \end{tikzpicture}
\end{equation}
The nodes are associated to the EM charges $\gamma_1$, $\gamma_2$ of BPS hypermultiplets which form a basis for the particles' charges. With respect to the UV Lagrangian $\gamma_1 = (0,1)$ is the monopole, $\gamma_2 = (2,-1)$ is the dyon and the W-bosons have charge $\pm(\gamma_1 + \gamma_2)$. The Dirac pairing, encoded in the BPS quiver, is $\Dirac[\gamma_1,\gamma_2]=2$.
In the strongly coupled chamber:
\begin{equation}
\arg \mathcal{Z}\left(\gamma_2\right)>\arg \mathcal{Z}\left(\gamma_1\right),
\end{equation}
the BPS spectrum is finite and is composed of only four hypermultiplets with charges:
\begin{equation}
    \gamma_1, \gamma_2, -\gamma_1,-\gamma_2
\end{equation}
listed here in increasing values of central charge $\mathcal{Z}$.
We choose a half plane in the $\mathcal{Z}$ plane such that $\gamma_1$ and $\gamma_2$ are particles while $-\gamma_1$ and $-\gamma_2$ are anti-particles. Therefore we define:
\begin{equation}
    \{\gamma^{\text{BPS}}_i\}_{i=1}^{2} = \{ \gamma_1, \gamma_2 \}
\end{equation}
The quantum spectrum generator is:
\begin{equation}
\begin{split}
    \mathcal{S}(q) =& E_q(X_{\gamma_1}) E_q(X_{\gamma_2})
    \\
    &=
    \sum_{p_1,p_2\geq0}
    \frac{(-q)^{\tfrac{1}{2}(p_1+p_2)}}
        {(q)_{p_1} (q)_{p_2}}
    (X_{\gamma_1})^{p_1}(X_{\gamma_2})^{p_2}
        \\
    &=\sum_{p_1,p_2\geq0}
    \frac{(-q)^{\tfrac{1}{2}(p_1+p_2)}}
        {(q)_{p_1} (q)_{p_2}}
        q^{\tfrac{1}{2} A}
        X_{p_1 \gamma_1 + p_2 \gamma_2}
\end{split}
\end{equation}
with:
\begin{equation}
    A= \sum_{1\leq i <j \leq2} \Dirac[\gamma^{\text{BPS}}_i,\gamma^{\text{BPS}}_j] p_i p_j
    = 2 p_1 p_2
\end{equation}
The first few orders in the $q$-expansion read:
\begin{equation}
\begin{split}
    S(q) =
    & 1 - \frac{X_{\gamma_1} + X_{\gamma_2}}{(q;q)_1} q^{1/2} 
    + \frac{X_{2\gamma_1} + X_{2\gamma_2}}{(q;q)_2} q 
    - \frac{X_{3\gamma_1} + X_{3\gamma_2}}{(q;q)_3} q^{3/2} + \left(\frac{X_{4\gamma_1} + X_{4\gamma_2}}{(q;q)_4} + \frac{X_{\gamma_1+\gamma_2}}{(q;q)^2_1}\right) q^2 
    +\\&
    - \frac{X_{5\gamma_1} + X_{5\gamma_2}}{(q;q)_5} q^{5/2} + \frac{X_{6\gamma_1} + X_{6\gamma_2}}{(q;q)_6} q^3 
    - \left(\frac{X_{7\gamma_1} + X_{7\gamma_2}}{(q;q)_7} + \frac{X_{2\gamma_1+\gamma_2} + X_{\gamma_1+2\gamma_2}}{(q;q)_2(q;q)_1}\right) q^{7/2} 
    +\\&
    + \frac{X_{8\gamma_1} + X_{8\gamma_2}}{(q;q)_8} q^4 
    - \frac{X_{9\gamma_1} + X_{9\gamma_2}}{(q;q)_9} q^{9/2} + \left(\frac{X_{10\gamma_1} + X_{10\gamma_2}}{(q;q)_{10}} + \frac{X_{3\gamma_1+\gamma_2} + X_{\gamma_1+3\gamma_2}}{(q;q)_3(q;q)_1}\right) q^5 
    + \\
    & + \mathcal{O}(q^{11/2})
\end{split}
\end{equation}
The conjugate $\overline{\mathcal{S}}(q)$ is obtained from $\mathcal{S}(q)$ by replacing $X_a$ with $X_{-a}$. 
As a sanity check, we can compute the Schur index directly from the quantum spectrun generator as:
\begin{equation}
\begin{split}
    \mathcal{I}(q) &= (q;q)^{2} \Tr [\mathcal{S}(q)\overline{\mathcal{S}}(q) ]
    \\ &= 1 + q^2 + q^6 + q^{12} + q^{20}+\dots
\end{split}
\end{equation}
matching the Schur index on $S^3 \times S^1$ computed as:
\begin{equation}
    \mathcal{I}^{SU(2)}_{S^3 \times S^1} = \frac{1}{2} \oint \frac{du}{2\pi i u} (1-u^2)(1-u^{-2}) \left[ (qu^2;q)(qu^{-2};q) (q;q) \right]^2
\end{equation}

Now that all the ingredients are specified, we perform the check of our proposals in Section \ref{sec: bc_proposal} in various electromagnetic duality frames.


\subsubsection{Boundary conditions in the ``monopole" frame}

We first consider b.c.~in the ``monopole" frame, namely we consider an EM duality frame where the monopole $\gamma_1$ is chosen to be electric.
Therefore we choose the electric and magnetic lattices spanned by the charges $q_1$ and $p_1$, respectively:
\begin{equation}    \label{eq:charges_magnetic_frame}
    q_1 =  \gamma_1,
    \qquad
    p_1= -\frac{\gamma_1 }{2} - \frac{\gamma_2}{2}
\end{equation}
which satisfy $\langle q_1, p_1 \rangle = -1$. 
Notice that, w.r.t.~the UV Lagrangian, $q_1=(0,1)$ is the monopole while $p_1=(-1,0)$ is (minus) a unit of electric charge.
The trace of the electric charge $q_1$ is identified with the fugacity $\xi$ associated to the boundary $U(1)$ symmetry:
\begin{equation}
    \Tr_{\{q_1\}}[X_{q_1}] =  \xi
\end{equation}

\paragraph{Gluing the RG-wall to the UV}
We start by considering the boundary duality in \eqref{eq: rgwall_index_statement_inverse}, 
where we give $\mathcal{D}$ boundary conditions to the IR Seiberg-Witten theory in the monopole frame and, on the other side, we couple the $(S \cdot \Twall[])^{-1}$ theory to the UV Lagrangian.
This setup can be conveniently summarized in the hybrid quiver notation introduced above as:
\begin{equation}
    \begin{tikzpicture}
    \begin{scope}
        \draw (-1,0) node {$\begin{gathered} \NN=2 \\ \text{$SU(2)$ SYM} \end{gathered}$};
        \draw (1,1) -- node[midway, gauge, double, fill=white] (gUV) {$2$} (1,-1);
        \draw[->-] (gUV) -- ++ (1.5,0) node[gauge,fill=white] (g1) {$1$};
        \draw (g1)++(.1,-0.4) node[right,CScolor]{$\scriptstyle 1$};
        \draw (g1)++(0,.5) node[FIcolor]{$\scriptstyle \xi$};
    \end{scope}
    \draw[<->] (4,0)--(5,0);
    \begin{scope}[xshift = 8cm]
        \draw (-1,0) node {$\begin{gathered}SW \\ \text{monopole frame} \end{gathered}$};
        \draw (1,1) -- node[midway, flavor, fill=white] (gIR) {$1$} (1,-1);
        \draw (gIR)++(0.7,0) node[gray] {$\xi$};
    \end{scope}
    \end{tikzpicture}
\end{equation}
We recall that the node sitting at the boundary corresponds to the 4d gauge field which has either Neumann b.c., represented as a circle possibly connected to the 3d theory, or Dirichlet b.c. represented as a square.

In the IR the half-index with $m$ units of flux is thus given by:
\begin{equation} \label{eq: SU2_indDir}
    \mathcal{II}^{IR}(\xi,m) = 
    \mathcal{II}^{SW} (\xi, m) =
    (q;q)\Tr_{\{ \gamma_1\}} \left[ :\mathcal{S}(q) \left(X_{-\frac{\gamma_1 }{2} - \frac{\gamma_2}{2}}\right)^{m}: \right]
\end{equation}
we find that the resulting half-index is non-vanishing only for $m \geq 0$ and even. Below we report the $q$-expansion of the half-index for various values of the flux $m$.

In the UV the half-index of pure 4d $SU(2)$ SYM is coupled to the inverse wall $(S \cdot \Twall[2])^{-1}$ on the boundary is:
\begin{equation}    \label{eq:SU2_UV_Wall_def}
    \mathcal{II}^{UV}(\xi,m) = 
    \oint \frac{du}{ 2\pi i u } \Delta_2(u) \mathcal{II}^{SU(2)}(u) 
    \mathcal{I}^{(S \cdot \Twall[2])^{-1}}(u,\{\xi,m\})
\end{equation}
where $u$ is the Cartan of the $SU(2)$ gauge symmetry which couples to the $SU(2)$ global symmetry of the $(S \cdot \Twall[2])^{-1}$ theory, while $\{\xi,m\}$ is the fugacity/magnetic flux pair associated to the $U(1)$ global symmetry on the boundary. The bulk contribution from pure $SU(2)$ SYM is given in \eqref{eq: su2_halfind}, and the 3d index of the $(S \cdot \Twall[2])^{-1}$ theory reads:
\begin{equation}
    \mathcal{I}^{(S\cdot\Twall[2])^{-1}}(u,\{\xi,m\}) = \sum_{m'} \oint \frac{d\xi'}{2 \pi i \xi'} 
    \xi^{-m'} {\xi'}^{-m} 
    \mathcal{I}^{(S \cdot \Twall[2])^{-1}}(u,\{\xi,m\})
\end{equation}
where the index of the RG-wall is given in \eqref{eq: su2_twallinv_ind}.

We checked that the identity:
\begin{equation}
    \mathcal{II}^{IR}(\xi,m) = \mathcal{II}^{UV}(\xi,m)
\end{equation}
where the two half-indices are defined in \eqref{eq: SU2_indDir} and \eqref{eq:SU2_UV_Wall_def} respectively, agree for $-5 \leq m_{\xi} \leq 5$ up to $\mathcal{O}(q^{10})$.
For completeness we report the $q$-expansion for some values of the magnetic flux $m$:
\begin{equation}
\begin{split}
    m=0:\quad  &
    1-\xi  \sqrt{q}+\left(\xi ^2-1\right) q-\xi ^3 q^{3/2}+\left(\xi ^4-1\right) q^2+\left(\xi -\xi ^5\right) q^{5/2}+\xi ^6 q^3 + \mathcal{O}(q^{7/2}) \\
    m=2:\quad  &
    -\frac{\sqrt{q}}{\xi }+q^2+\frac{q^{5/2}}{\xi }+q^3+\left(\frac{1}{\xi }-\xi \right) q^{7/2}+\left(\frac{1}{\xi }-\xi \right) q^{9/2} + \mathcal{O}(q^{5}) \\
    m = 4: \quad &
    \frac{q}{\xi ^2}-\frac{q^{7/2}}{\xi }-\frac{q^4}{\xi ^2}-\frac{q^{9/2}}{\xi }-\frac{q^5}{\xi ^2}-\frac{q^{11/2}}{\xi }+\left(1-\frac{1}{\xi ^2}\right)
    q^6+\mathcal{O}(q^{7})
\end{split}
\end{equation}
and we notice that for any negative or odd value of $m$, the $q$-expansion of the half-index yields zero.

\paragraph{Gluing the RG-wall to the IR}
We can also consider the setup where we give $\mathcal{D}$ b.c.~to the UV theory. The half-index of this setup is $\mathcal{II}^{SU(2)}(u)$ as defined in \eqref{eq: su2_halfind},
where $u$ parametrizes the Cartan of the boundary $SU(2)$ symmetry.
In the IR we expect to flow to the SW theory coupled to the $S \cdot \Twall[2]$ theory on the boundary. 
This setup can be conveniently summarized in the hybrid quiver notation introduced above as:

\begin{equation}
    \begin{tikzpicture}
    \begin{scope}
        \draw (-1,0) node {$\begin{gathered} \NN=2 \\ \text{$SU(2)$ SYM} \end{gathered}$};
        \draw (1,1) -- node[midway, flavor, double, fill=white] (gUV) {$2$} (1,-1);
    \end{scope}
    \draw[<->] (2,0) -- (3,0);
    \begin{scope}[xshift = 6cm]
        \draw (-1,0) node {$\begin{gathered}SW \\ \text{monopole frame} \end{gathered}$};
        \draw (1,1) -- node[midway, gauge, fill=white] (gIR) {$1$} (1,-1);
        \node[gauge,black] at (2.5,0) (g1) {$1$};
        \draw[dashed] (gIR) -- node[midway,above,BFcolor] {$\scriptstyle +2$} (g1);
        \draw[->-] (g1) -- ++ (1.5,0) node[flavor,fill=white] (gSU2) {$2$};
        \draw (g1)++(0.4,-0.4) node[CScolor]{$\scriptstyle -1$};
    \end{scope}
    \end{tikzpicture}
\end{equation}

Where the dashed line represents a mixed CS coupling at level 2 between the nodes and is a convenient way to encode the coupling of the 3d theory to the bulk: the topological symmetry of the 3d $U(1)$ gauge node is identified with to the 4d $U(1)$ gauge field at the boundary which has Neumann b.c.
The half-index of this boundary configuration can be computed as:
\begin{equation} \label{eq: SU2_IR+wall_index}
    \mathcal{II}^{IR}(u) = (q;q)
    \sum_{m}
    \oint \frac{d\xi}{2\pi i \xi}
    \Tr_{\{ \gamma_1\}} \left[ :\mathcal{S}(q) \left(X_{-\frac{\gamma_1 }{2} - \frac{\gamma_2}{2}}\right)^{m} :\right]
    \mathcal{I}^{S \cdot \Twall[]} (u, \{\xi,m\})
\end{equation}
where the index of the $S \cdot \Twall[2]$ theory is given by:
\begin{equation}
    \mathcal{I}^{S \cdot \Twall[2]}(u, \{\xi,m\}) = \sum_{m'} \oint \frac{d \xi'}{2\pi i \xi'} \xi^{m'} {\xi'}^{m} \mathcal{I}^{\Twall[2]} (u,\{\xi',m'\})
\end{equation}
and $\mathcal{I}^{\Twall[2]}$ is given in \eqref{eq: su2_twall_ind}. The index in the UV is simply given by the half-index contribution of a 4d $SU(2)$ vector, that we give in \eqref{eq: su2_halfind}
We checked the agreement between \eqref{eq: SU2_IR+wall_index} and \eqref{eq: su2_halfind} up to $\mathcal{O}(q^5)$.

As a final comment, we point out that one can, in principle, turn on non-trivial values of the magnetic flux for the $SU(2)$ symmetry. The half-index contribution for pure $SU(2)$ SYM at non-zero magnetic fluxes is expected to be identically zero. Therefore, we also expect that, if we turn on a non-zero magnetic flux for the $SU(2)$ symmetry in \eqref{eq: SU2_IR+wall_index}, we should get zero, which is a non-trivial prediction. Turning on a non-zero magnetic flux associated to the $SU(2)$ symmetry implies that we generalize the 3d index of the $S \cdot \Twall[2]$ theory as: 
\begin{equation}
\begin{split}
    & \mathcal{I}^{S \cdot \Twall[2]}(\{u,n\}, \{\xi,m\}) = \sum_{m'} \oint \frac{d \xi'}{2\pi i \xi'} \xi^{m'} {\xi'}^{m} \mathcal{I}^{\Twall[2]} (\{u,n\},\{\xi',m'\}) \\
    & \mathcal{I}^{\Twall[2]} (\{u,n\},\{\xi,m\}) = \xi^{-m} \mathcal{I}_{\text{chir}}^{(0)}(u \xi,n+m) \mathcal{I}_{\text{chir}}^{(0)}(u^{-1} \xi,-n+m)
\end{split}
\end{equation}
By replacing $\mathcal{I}^{S \cdot \Twall[2]}(u,\{\xi,m\})$ with $\mathcal{I}^{\Twall[2]} (\{u,n\},\{\xi,m\})$ in \eqref{eq: SU2_IR+wall_index} we observe that for any $n \neq 0$ we obtain exactly zero, as expected.

As already discussed in Section \ref{sec: bc_proposal}, the RG-wall has a monopole operator with R-charge 1 and charge $-1$ under the $U(1)_\xi$ global symmetry, which can couple to the BPS monopole $\gamma_1$ via a boundary superpotential.
A simple check for the existence of this operator in the IR is to compute the superconformal index:
\begin{equation}
 \mathcal{I}^{S \cdot \Twall[2]}(u, \{\xi,0\}) =
    1-\frac{q^{1/2}}{\xi}+\mathcal{O}\left(q\right)
\end{equation}
where the $q^{1/2}$ term counts this monopole operator.

Furthermore there is another operator that only exists in the presence of background flux for the $U(1)$ symmetry which has the correct charges to couple to the BPS dyon $\gamma_2$. 
The presence of this operator can be inferred, for example, by computing the 3d index of $S\cdot \Twall[2]$ with background flux $m=2$:
\begin{equation}
    \mathcal{I}^{S \cdot \Twall[2]}(u, \{\xi,2\}) = - q^{1/2} \xi + \mathcal{O} (q^2)
\end{equation}
where the $q^{1/2}$ term counts an operator with the correct charges to couple to the BPS dyon.

A similar analysis can be performed in the other electromagnetic duality frames considered below. 
In all cases we find that the RG wall theory has operators with the correct charges to couple to the BPS monopole and dyon. We do not report the details here.

\subsubsection{Boundary conditions in the ``electric" frame}
We would like to perform the analogous computation in the ``electric" frame. That is we choose the electromagnetic duality for the low energy Seiberg-Witten theory where the W-bosons are electric. 
In this frame the 3d theory on the boundary is given by $\Twall[2]$.
This boundary setup can be obtained from the previous one by performing an $S^{-1}$ transformation in the $SL(2,\mathbb{Z})$ low energy duality group, which is defined as:
\begin{equation}
    S^{-1} = \left(\begin{array}{cc}
        0 & -1 \\
        1 & 0
    \end{array}\right)
\end{equation}
The new choice of electric and magnetic charges is
thus obtained by applying $S^{-1}$ to the choice in \eqref{eq:charges_magnetic_frame} and is
given by\footnote{Compared to the computations in Section 4 of \cite{Cordova:2016uwk}, our choice of electric and magnetic charges is the $S^2T^{-2}$-transformed of their choice. Both choices are “electric" as $q$ is purely electric in terms of the UV Lagrangian. }:
\begin{equation}    \label{eq:charges_electric_frame}
    q_1 = \frac{\gamma_1}{2} +\frac{\gamma_2}{2}
    ,\qquad
    p_1 = \gamma_1
\end{equation}
From the point of view of the UV Lagrangian, $q_1=(1,0)$ is a unit of electric charge while $p_1=(0,1)$ is a unit of magnetic charge. Thus tracing over powers of the $q_1$ state corresponds to projecting over states that are electric from the UV perspective.

\paragraph{Gluing the RG-wall to the UV}
We start by considering the case of $\mathcal{D}$ b.c.~for the IR effective description in the electric frame:
\begin{equation}
    \begin{tikzpicture}
    \begin{scope}
        \draw (-1,0) node {$\begin{gathered} \NN=2 \\ \text{$SU(2)$ SYM} \end{gathered}$};
        \draw (1,1) -- node[midway, gauge, double, fill=white] (gUV) {$2$} (1,-1);
        \draw[->-] (gUV) -- ++ (1.5,0) node[flavor,fill=white] (g1) {$1$};
        \draw (g1)++(0.5,-0.5) node[CScolor]{$\scriptstyle 1$};
        \draw[gray] (g1) ++ (0.5,0) node {$\xi$};
    \end{scope}
    \draw[<->] (4,0) -- (5,0);
    \begin{scope}[xshift = 8cm]
        \draw (-1,0) node {$\begin{gathered}SW \\ \text{electric frame} \end{gathered}$};
        \draw (1,1) -- node[midway, flavor, fill=white] (gIR) {$1$} (1,-1);
        \draw[gray] (gIR)++(0.5,0) node {$\xi$};
    \end{scope}
    \end{tikzpicture}
\end{equation}

The half-index of the IR theory at generic value of the magnetic flux is then given by:
\begin{equation}
    \mathcal{II}^{IR} (\xi,m) = 
    \mathcal{II}^{SW} (\xi, m) =
    (q;q)
    \Tr_{\{  \frac{\gamma_1}{2} + \frac{\gamma_2}{2}\}} \left[ :\mathcal{S}(q) \left(X_{\gamma_1}\right)^{m} :\right]
\end{equation}
with the identification:
\begin{equation}
    \Tr_{\{  \frac{\gamma_1}{2} + \frac{\gamma_2}{2}\}} 
        \left[X_{  \frac{\gamma_1}{2} + \frac{\gamma_2}{2}} \right] = \xi
\end{equation}

On the other hand the half-index of pure $SU(2)$ coupled to $\Twall[2]$ on the boundary is:
\begin{equation}
\mathcal{II}^{UV}(\xi,m) =
    \oint \frac{du}{ 2\pi i u } \Delta_2(u) \mathcal{II}^{SU(2)}(u)
    \mathcal{I}^{(\Twall[2])^{-1}} (u, \{\xi,m\} )
\end{equation}
where $\mathcal{II}^{SU(2)}$ is given in \eqref{eq: su2_halfind} and $\mathcal{I}^{(\Twall[2])^{-1}}$ in \eqref{eq: su2_twallinv_ind}.
We checked that the UV and IR computations agree for $-5\leq m_\xi \leq 5$ up to $\mathcal{O}(q^{15})$.

\paragraph{Gluing the RG-wall to the IR}
We then consider the case of $\mathcal{D}$ b.c.~for the UV theory:

\begin{equation}
    \begin{tikzpicture}
    \begin{scope}
        \draw (-1,0) node {$\begin{gathered} \NN=2 \\ \text{$SU(2)$ SYM} \end{gathered}$};
        \draw (1,1) -- node[midway, flavor, double, fill=white] (gUV) {$2$} (1,-1);
    \end{scope}
    \draw[<->] (2,0) -- (3,0);
    \begin{scope}[xshift = 6cm]
        \draw (-1,0) node {$\begin{gathered}SW \\ \text{electric frame} \end{gathered}$};
        \draw (1,1) -- node[midway, gauge, fill=white] (gIR) {$1$} (1,-1);
        \draw[->-] (gIR) -- ++ (1.5,0) node[flavor,fill=white,solid] (g1) {$2$};
        \draw (gIR)++(0.4,-0.4) node[CScolor]{$\scriptstyle -1$};
    \end{scope}
    \end{tikzpicture}
\end{equation}

Therefore we match the half-index of pure $SU(2)$ SYM, given in \eqref{eq: su2_halfind}, against the index of the IR theory in the electric frame glued to the $\Twall[2]$ theory, which reads:
\begin{equation} \label{eq: SU2_IR+twall_index2}
    \mathcal{II}^{IR}(u) = (q;q) \sum_{m}
    \oint \frac{d\xi}{2\pi i \xi}
    \Tr_{\{ \frac{\gamma_1 }{2} + \frac{\gamma_2}{2} \}} \left[ :\mathcal{S}(q) \left(X_{\gamma_1} \right)^{m} :\right]
    \mathcal{I}^{\Twall[]} (u, \{\xi,m\})
\end{equation}
where the index of the $\Twall[2]$ theory is given in \eqref{eq: su2_twall_ind}. We checked this agreement up to $\mathcal{O}(q^6)$.

Again, as in the monopole frame, one can consider a more generic setup where we turn on a non-zero magnetic flux associated to the $SU(2)$ symmetry. This can be done following the same prescription as described before, by modifying \eqref{eq: SU2_IR+twall_index2}, allowing for a magnetic flux parameterized by an integer $n$. Also in this case we observe that for any value $n \neq 0$ we get a zero IR index, which agrees with the prediction that the half-index of the UV $SU(2)$ SYM theory vanishes for any non-zero magnetic flux.


\subsubsection{Boundary conditions in the ``dyonic" frame}
Finally we consider $\mathcal{D}$ b.c.~in the ``dyonic" frame, namely in an EM duality frame where $\gamma_2$ is chosen to be electric. The corresponding electric and magnetic charges $q_1$ and $p_1$ are obtained from the electric frame \eqref{eq:charges_electric_frame} by applying $S^{-1} T^{2}$ and are given by:
\begin{equation}    \label{eq:charges_dyonic_frame}
    q_1 =  \gamma_2,
    \qquad
    p_1= \frac{\gamma_1 }{2} + \frac{\gamma_2}{2}
\end{equation}
which satisfy $\langle q_1, p_1 \rangle = -1$. 
Notice that, w.r.t.~the UV Lagrangian, $q_1=(2,-1)$ is the dyon while $p_1=(1,0)$ is a unit of electric charge. 
The trace of the electric charge $q_1$ is identified with the fugacity $\xi$ associated to the boundary $U(1)$ symmetry:
\begin{equation}
    \Tr[X_{q_1}] =  \xi
\end{equation}

\paragraph{Gluing the RG-wall to the UV}
We start by considering the boundary duality in \eqref{eq: rgwall_index_statement_inverse}, 
where we give $\mathcal{D}$ boundary conditions to the IR Seiberg-Witten theory in the monopole frame and, on the other side, we couple the $(S^{-1} T^{2} \cdot \Twall[])^{-1}$ theory to the UV Lagrangian.
This setup can be conveniently summarized in the hybrid quiver notation introduced above as:
\begin{equation}
    \begin{tikzpicture}
    \begin{scope}
        \draw (-1,0) node {$\begin{gathered} \NN=2 \\ \text{$SU(2)$ SYM} \end{gathered}$};
        \draw (1,1) -- node[midway, gauge, double, fill=white] (gUV) {$2$} (1,-1);
        \draw[->-] (gUV) -- ++ (1.5,0) node[gauge,fill=white] (g1) {$1$};
        \draw (g1)++(0.4,-0.4) node[CScolor]{$\scriptstyle -1$};
        \draw (g1)++(0,.5) node[FIcolor]{$\scriptstyle \xi^{-1}$};
    \end{scope}
    \draw[<->] (4,0) -- (5,0);
    \begin{scope}[xshift = 8cm]
        \draw (-1,0) node {$\begin{gathered}SW \\ \text{dyonic frame} \end{gathered}$};
        \draw (1,1) -- node[midway, flavor, fill=white] (gIR) {$1$} (1,-1);
        \draw[gray] (gIR)++(0.5,0) node {$\xi$};
    \end{scope}
    \end{tikzpicture}
\end{equation}

In the IR the half-index with $m$ units of flux is thus given by:
\begin{equation} \label{eq: SU2_indDir_dyon}
    \mathcal{II}^{IR} (\xi,m) = 
    \mathcal{II}^{SW} (\xi, m) =
    (q;q) \Tr_{\{ \gamma_2\}} \left[ :\mathcal{S}(q) \left(X_{\frac{\gamma_1 }{2} + \frac{\gamma_2}{2}}\right)^{m} :\right]
\end{equation}
we find that the resulting half-index is non-vanishing only for $m \leq 0$ and even.

On the other side the half-index of pure 4d $SU(2)$ SYM is coupled to the inverse wall $(S^{-1} T^{2} \cdot \Twall[])^{-1}$ on the boundary is:
\begin{equation}    \label{eq:SU2_UV_Wall_def_dyon}
    \mathcal{II}^{UV} (\xi,m) = 
    \oint \frac{du}{ 2\pi i u } \Delta_2(u) \mathcal{II}^{SU(2)}(u) 
    \mathcal{I}^{(S^{-1} \cdot T^2 \cdot \Twall[2])^{-1}}(u,\{\xi,m\})
\end{equation}
where $u$ is the Cartan of the $SU(2)$ gauge symmetry which couples to the $SU(2)$ global symmetry of the $(S^{-1} T^{2}\cdot \Twall[2])^{-1}$ theory, while $\{\xi,m\}$ is the fugacity/magnetic flux pair associated to the $U(1)$ global symmetry on the boundary. The bulk contribution from pure $SU(2)$ SYM is given in \eqref{eq: su2_halfind}, and the 3d index of the $(S^{-1} T^{2} \cdot \Twall[2])^{-1}$ theory reads:
\begin{equation}
\begin{split}
    \mathcal{I}^{(S^{-1} T^2 \cdot\Twall[2])^{-1}}(u,\{\xi,m\}) = \sum_m \oint \frac{d\xi'}{2 \pi i \xi'} \xi^{m'} {\xi'}^{m}  
    \xi'^{-2m'} 
    \mathcal{I}^{(\Twall[2])^{-1}}(u,\{\xi',m'\}) \,,
\end{split}
\end{equation}
where $\mathcal{I}^{(\Twall[2])^{-1}}$ is given in \eqref{eq: su2_twallinv_ind}. We recall that when the RG-wall is glued to the UV theory the $SL(2,\mathbb{Z})$ group acts with the inverse element on the right.

We checked that the two indices \eqref{eq: SU2_indDir_dyon} and \eqref{eq:SU2_UV_Wall_def_dyon} agree for $-5 \leq m_{\xi} \leq 5$ up to $\mathcal{O}(q^{10})$.

\paragraph{Gluing the RG-wall to the IR}
We can also consider the setup where we give $\mathcal{D}$ b.c.~to the UV theory. The half-index of this setup is $\mathcal{II}^{SU(2)}(u)$ as defined in \eqref{eq: su2_halfind},
where $u$ parametrizes the Cartan of the boundary $SU(2)$ symmetry.
In the IR we expect to flow to the SW theory coupled to the $S^{-1} T^{2} \cdot \Twall[2]$ theory on the boundary. 
This setup can be conveniently summarized in the hybrid quiver notation introduced above as:

\begin{equation}
    \begin{tikzpicture}
    \begin{scope}
        \draw (-1,0) node {$\begin{gathered} \NN=2 \\ \text{$SU(2)$ SYM} \end{gathered}$};
        \draw (1,1) -- node[midway, flavor, double, fill=white] (gUV) {$2$} (1,-1);
    \end{scope}
    \draw[<->] (2,0) -- (3,0);
    \begin{scope}[xshift = 6cm]
        \draw (-1,0) node {$\begin{gathered}SW \\ \text{dyonic frame} \end{gathered}$};
        \draw (1,1) -- node[midway, gauge, fill=white] (gIR) {$1$} (1,-1);
        \node[gauge,black] at (2.5,0) (g1) {$1$};
        \draw[dashed] (gIR) -- node[midway,above,BFcolor] {$\scriptstyle -2$} (g1);
        \draw[->-] (g1) -- ++ (1.5,0) node[flavor,double,fill=white] (gSU2) {$2$};
        \draw (g1)++(0.4,-0.4) node[CScolor]{$\scriptstyle +1$};
    \end{scope}
    \end{tikzpicture}
\end{equation}

Where the dashed line represents a mixed CS coupling at level $-2$ between the nodes and is a convenient way to encode the coupling of the 3d theory to the bulk: the topological symmetry of the 3d $U(1)$ gauge node is identified with to the 4d $U(1)$ gauge field at the boundary which has Neumann b.c.
The half-index of this boundary configuration can be computed as:
\begin{equation} \label{eq: SU2_IR+wall_index_dyon}
    \mathcal{II}^{IR}(u) = (q;q)\sum_{m}
    \oint \frac{d\xi}{2\pi i \xi}
    \Tr_{\{ \gamma_2\}} \left[ :\mathcal{S}(q) \left(X_{\frac{\gamma_1 }{2}  \frac{\gamma_2}{2}}\right)^{m} :\right]
    \mathcal{I}^{S^{-1}T^2 \cdot \Twall[]} (u, \{\xi,m\})
\end{equation}
where the index of the $S^{-1}T^2 \cdot \Twall[2]$ theory is given by:
\begin{equation}
    \mathcal{I}^{S^{-1}T^2 \cdot \Twall[2]}(u, \{\xi,m\}) = \sum_{m'} \oint \frac{d \xi'}{2\pi i \xi'} 
    \xi^{-m'} {\xi'}^{-m}
    \xi'^{2m'}
    \mathcal{I}^{\Twall[2]} (u,\{\xi',m'\})
\end{equation}
and $\mathcal{I}^{\Twall[2]}$ is given in \eqref{eq: su2_twall_ind}.
We checked the agreement between \eqref{eq: su2_halfind} and \eqref{eq: SU2_IR+wall_index_dyon} up to $\mathcal{O}(q^5)$.
Similarly to the electric and magnetic frames considered above, one can alto turn on non-zero fluxes for the $SU(2)$ flavor symmetry and check that the half-index vanishes, both in the UV and in the IR computation.
\\

\subsection{RG-walls for $SU(N)$ pure SYM}

In this section we discuss the index identities related to the boundary dualities proposed in Section \ref{sec: bc_proposal} for higher rank, that are \eqref{eq: rgwall_index_statement} and \eqref{eq: rgwall_index_statement_inverse} for the case of $SU(N)$ with $N > 2$.
We show that, in a perturbative expansion in $q$, the half-indices of dual boundary setups agree up to high order. 
The results presented in this section represent the most refined checks supporting our proposal.

As discussed in Section \ref{sec: bc_proposal}, and similarly in the $SU(2)$ example, the effective theory in the IR has different representations related by the $Sp(2(N-1),\zz)$ electromagnetic duality group. 
In this Section we will consider only the ``electric" duality frame, where the low energy $U(1)$ gauge groups are the Cartan of the UV $SU(N)$ gauge group. More generic frames can be obtained using $Sp(2(N-1),\mathbb{Z})$ transformations as described in subsection \ref{subsec: sl2z_transf}.
\\

Before delving into the explicit computations for fixed-rank cases we would like to summarize the relevant formulae and information needed to perform the half-index computations for generic rank.
The contributions to the half index given by the $SU(N)$ vector multiplet with $\mathcal{D}$ b.c. in the UV is given by:
\begin{equation} \label{eq: suN_halfind}
    \mathcal{II}^{SU(N)}(\vec{u}) = (q;q)^{N-1} \prod_{i<j}^N (u_i/u_j q;q)(u_j/u_i q;q) \Bigg|_{\prod_{i=1}^N u_i=1}
\end{equation}

In order to compute the half-index from the IR we need to know the BPS spectrum in one chamber of the Coulomb branch. 
To this end we exploit the following BPS quiver for $SU(N)$ SYM \cite{Alim:2011kw}:
\begin{equation}
\begin{tikzpicture}
	\node[draw,circle, radius=4pt] (g1) at (0,0) {$b_1$};
	\node[draw,circle, radius=4pt] (g2) at (0,2) {$c_1$};
	\node[draw,circle, radius=4pt] (g3) at (2,0) {$b_2$};
	\node[draw,circle, radius=4pt] (g4) at (2,2) {$c_2$};
    \node[draw,circle, radius=4pt] (g5) at (8,0) {$b_{N\text{-}1}$};
	\node[draw,circle, radius=4pt] (g6) at (8,2) {$c_{N\text{-}1}$};

    \path (g4) -- node[midway] {$\dots$} (g5); 
    
	\draw[->] (g1) edge[double] (g2);
	\draw[->] (g3) edge[double] (g4);
    \draw[->] (g5) edge[double] (g6);
	\draw[->] (g2) edge[] (g3);
	\draw[->] (g4) edge[] (g1);
    \draw[->] (g4) edge[] ++ (2,-2);
	\draw[<-] (g3) edge[] ++ (2,2);
    \draw[<-] (g5) edge[] ++ (-2,2);
	\draw[->] (g6) edge[] ++ (-2,-2);
\end{tikzpicture}
\end{equation}
with a quartic superpotential that we do not report here.

As originally discussed in \cite{Alim:2011kw}, there is a strongly coupled chamber where the spectrum is finite and is composed of $N(N-1)$  hypers. 
Here $b_i$ are $N-1$ monopoles associated to a set of simple co-roots, and $c_i$ are the corresponding dyons. 
We refer the reader to \cite{Alim:2011ae,Alim:2011kw} for more details on the case of generic rank, here we report the relevant information for our computations in fixed-rank cases.

Finally we need the contribution to the index given by the 3d theory $\Twall[N]$. This can be computed from the quiver description of these theories as reviewed in subsection \ref{subsec: twall_gsun_generalities}.
We report the quiver lagrangian for the $\Twall[]$ theory for reference below.
\begin{equation}
    \Twall[] \quad : \qquad\qquad
    \begin{tikzpicture}[baseline=(current bounding box).center]
        \node at (0,0) (n1) [flavor,black] {$1$};
        \node at (2,0) (n2) [gauge,double,black] {$2$};
        \node at (5,0) {$\cdots$};
        \node at (8,0) (nnm1) [gauge,double,black] {$\scriptstyle N-1$};
        \node at (10,0) (nn) [flavor,black] {$N$};
    
        \draw[->-] (n1) -- node[midway,above,gray] {$b_1 b_2^{-\tfrac{1}{2}}$} (n2);
        \draw[-<-] (n2)++(2,0) -- node[midway,above,gray] {$b_2^{\tfrac{1}{2}} b_3^{-\tfrac{1}{3}}$} (n2);
        \draw[->-] (nnm1)++(-2,0) -- node[midway,above,gray] {$b_{N-2}^{\tfrac{1}{N-2}} b_{N-1}^{-\tfrac{1}{N-1}}$} (nnm1);
        \draw[->-] (nnm1) -- node[midway,above,gray] {$b_{N-1}^{\tfrac{1}{N-1}}$} (nn);
    \end{tikzpicture}
\end{equation}
with background CS levels:
\begin{equation}
    k_{b_i} = -1,
    \qquad
    k_{b_i, b_j} = 1,
    \qquad
    k_{SU(N)} = - \tfrac{N-1}{2}
\end{equation}
where $b_i$ are the fugacities associated to the baryonic symmetries of the $i$-th $SU(i)$ gauge node.

The theory $(\Twall[])^{-1}$ is given by $CP$-transformed of $\Twall[]$. As a quiver  the theory $(\Twall[])^{-1}$ can be presented as:
\begin{equation}
\Twall[]^{-1} \quad : \qquad\qquad
\begin{tikzpicture}[baseline=(current bounding box).center]
    \node at (0,0) (n1) [flavor,black] {$1$};
    \node at (2,0) (n2) [gauge,double,black] {$2$};
    \node at (5,0) (n3) {$\cdots$};
    \node at (8,0) (nnm1) [gauge,double,black] {$\scriptstyle N-1$};
    \node at (10,0) (nn) [flavor,black] {$N$};

    \draw[-<-] (n1) -- (n2);
    \draw[gray] ($(n1)!0.5!(n2)+(0,.7)$) node {$b_{1}^{-1} b_{2}^{\tfrac{1}{2}}$};
    \draw[->-] (n2)++(2,0) -- (n2);
    \draw[gray] ($(n2)!0.5!(n3)+(-.3,.7)$) node {$b_{2}^{-\tfrac{1}{2}} b_{3}^{\tfrac{1}{3}}$};
    \draw[-<-] (nnm1)++(-2,0) -- (nnm1);
    \draw[gray] ($(n3)!0.5!(nnm1)+(.3,.7)$) node {$b_{N-2}^{-\tfrac{1}{N-2}} b_{N-1}^{\tfrac{1}{N-1}}$};
    \draw[-<-] (nnm1) -- (nn);
    \draw[gray] ($(nnm1)!0.5!(nn)+(0,.7)$) node {$b_{N-1}^{-\tfrac{1}{N-1}}$};
\end{tikzpicture}
\end{equation}
with background CS levels:
\begin{equation}
    k_{b_i} = 1,
    \qquad
    k_{b_i, b_j} = -1,
    \qquad
    k_{SU(N)} = \tfrac{N-1}{2}
\end{equation}

\subsubsection{$SU(3)$}

To perform the matching for $SU(3)$ SYM we need the half-index contribution of the 4d vector multiplet:
\begin{equation} \label{eq: su3_halfind}
    \mathcal{II}^{SU(3)}(\vec{u}) = (q;q)^2 \prod_{i<j}^3 (u_i/u_j q,q)(u_j/u_i q,q) \Bigg|_{\prod_{i=1}^3 u_i=1}
\end{equation}
and the 3d superconformal index of the $\Twall[3]$ theory, which reads:\footnote{For the practical purpose of expanding the index perturbatively, it is convenient to use the alternative Lagrangian description of $\Twall[3]$ which consists of 7 chiral fields with 3 cubic superpotential terms whose index reads:
\begin{equation}
\begin{split}
    \mathcal{I}^{\Twall[3]} (\vec{u}, \{\vec{b}, \vec{m}\})
    =&
    (b_1^{-1} b_2^{1/2})^{m_1} (b_2^{-1} b_1^{1/2})^{m_2} 
    \mathcal{I}_{\text{chir}}^{(2)} \left( b_1^{-1} b_2^{-1} ,-m_1-m_2 \right) 
    \\&
    \prod_{i=1}^{3} \mathcal{I}^{(0)}_{\text{chir}} \left(b_1 u_i^{-1},m_1 \right)  \mathcal{I}^{(0)}_{\text{chir}} \left(b_2 u_i,m_2\right) \Bigg|_{\prod_{i=1}^3 u_i=1}
\end{split}
\end{equation}
}
\begin{equation} \label{eq: su3_twall_index}
\begin{split}
    & \mathcal{I}^{\Twall[3]}(\vec{u},\{\vec{b},\vec{m}\}) = \\
    & \qquad \qquad =
    (b_1^{-1} b_2^{1/2})^{m_1} (b_2^{-1} b_1^{1/2})^{m_2} 
    \sum_{n \in \Gamma} \oint \frac{dz}{2\pi i z} \mathcal{I}_{\text{vec}}^{SU(2)}(z,n) \times \\
    & \qquad \qquad \quad \times \prod_{s=\pm1} \left[ \mathcal{I}_{\text{chir}}^{(0)} (b_1 b_2^{-1/2} z^s, s n + m_1 - m_2/2) \prod_{j=1}^3  \mathcal{I}^{(0)}_{\text{chir}}(b_2^{1/2}z^s u_j, s n+ m_2/2) \right] \\
    & \text{with} \quad \Gamma = \begin{cases}
        \mathbb{Z} &\quad \text{if $m_2 \in \mathbb{Z}_{\text{even}}$ and $m_1 \in \mathbb{Z}$} \\
        \mathbb{Z}+\frac{1}{2} &\quad \text{if $m_2 \in \mathbb{Z}_{\text{odd}}$ and $m_1 \in \mathbb{Z}$}
    \end{cases}
\end{split}
\end{equation}
which is the index of a $SU(2)$ gauge theory with four fundamental flavors, parameterized so that the actual manifest $U(4)$ global symmetry is split into $SU(3)_u \times U(1)_{b_1} \times U(1)_{b_2}$. Also there are background (mixed) CS levels for the $U(1)_{b_i}$ symmetries contributing as the prefactor in the first line.
Notice that the magnetic flux $n$ takes values over a magnetic lattice $\Gamma$ which depend on the choice of $m_2$. The reason being that if we consider $m_1,m_2 \in \mathbb{Z}$, then for odd $m_2$ we have to consider half-integer values of $n$ to ensure that the net magnetic flux seen by the chiral multiplets is integer, which is required by Dirac quantization. Moreover, $\prod_{i=1}^3 u_j = 1$.

Analogously one can write the index of the $(\Twall[3])^{-1}$ from its quiver representation in \eqref{eq: twallinv_quiverpic}:
\begin{equation} \label{eq: su3_twallinv_index}
\begin{split}
    & \mathcal{I}^{(\Twall[3])^{-1}}(\vec{u},\{\vec{b},\vec{m}\}) = \\
    & \qquad \qquad =
    (b_1^{-1} b_2^{1/2})^{-m_1} (b_2^{-1} b_1^{1/2})^{-m_2} 
    \sum_{n \in \Gamma} \oint \frac{dz}{2\pi i z} \mathcal{I}_{\text{vec}}^{SU(2)}(z,n) \times \\
    & \qquad \qquad \quad \times \prod_{s=\pm1} \left[ \mathcal{I}_{\text{chir}}^{(0)} (b_1^{-1} b_2^{1/2} z^s, s n - m_1 + m_2/2) \prod_{j=1}^3  \mathcal{I}^{(0)}_{\text{chir}}(b_2^{-1/2}z^s/u_j, s n - m_2/2) \right] \\
    & \text{with} \quad \Gamma = \begin{cases}
        \mathbb{Z} &\quad \text{if $m_2 \in \mathbb{Z}_{\text{even}}$ and $m_1 \in \mathbb{Z}$} \\
        \mathbb{Z}+\frac{1}{2} &\quad \text{if $m_2 \in \mathbb{Z}_{\text{odd}}$ and $m_1 \in \mathbb{Z}$}
    \end{cases}
\end{split}
\end{equation}

We also need to specify all the IR data necessary to compute the index from the half-KS operator.
Pure $SU(3)$ $\NN=2$ SYM has various finite chambers. Here we consider the finite chamber discussed in \cite{Alim:2011kw}, where the BPS spectrum is encoded in the BPS quiver:
\begin{equation}
\begin{tikzpicture}
	\node[draw,circle, radius=4pt] (g1) at (0,0) {$\gamma_1$};
	\node[draw,circle, radius=4pt] (g2) at (0,2) {$\gamma_2$};
	\node[draw,circle, radius=4pt] (g3) at (2,0) {$\gamma_3$};
	\node[draw,circle, radius=4pt] (g4) at (2,2) {$\gamma_4$};
	\draw[->] (g1) edge[double] (g2);
	\draw[->] (g3) edge[double] (g4);
	\draw[->] (g2) edge[] (g3);
	\draw[->] (g4) edge[] (g1);
\end{tikzpicture}
\end{equation}
From the UV Lagrangian perspective, the $\gamma_i$ states carry charges $(e_1, e_2;\; m_1,m_2)$ as:
\begin{equation}
\begin{split}
    &\gamma_2 = (2,-1;\;-1,0)  \qquad\qquad \gamma_1 = (0,0;\;1,0) \\
    &\gamma_4 = (-1,2;\;0,-1) \qquad\qquad \gamma_3 = (0,0;\;0,1) 
\end{split}
\end{equation}
In this chamber the central charges satisfy:
\begin{equation}
Arg[\mathcal{Z}(\gamma_2)],
Arg[\mathcal{Z}(\gamma_4)]
>
Arg[\mathcal{Z}(\gamma_1)],
Arg[\mathcal{Z}(\gamma_3)]
\end{equation}
and the BPS spectrum is composed of six particles, reported in increasing value for their central charge:\footnote{To be more precise, the first/second/third pair of particles have the same value of central charge but they commute w.r.t.~the Dirac pairing. Therefore the ordering of their KS factors is not important.}
\begin{equation}
\{\gamma^{\text{BPS}}_i \}_{i=1}^6 =
(\gamma_1, \gamma_3, \gamma_1 + \gamma_4, \gamma_2 + \gamma_3, \gamma_2, \gamma_4)
\end{equation}
and the six antiparticles.
The quantum spectrum generator for this choice of particles is given by:
\begin{equation}
\mathcal{S} (q) = \sum_{p_1, \dots, p_6 \geq 0} 
	\frac{(-q)^{\frac{1}{2} \sum_{i=1}^6 p_i} }
		{(q;q)_{p_1} (q;q)_{p_2} \dots (q;q)_{p_6}}
		q^{\frac{1}{2} A}
		X_{ \sum_{i=1}^6 p_i \gamma^{\text{BPS}}_i }
\end{equation}
where:
\begin{equation}
\begin{split}
    A &= \sum_{1\leq i <j \leq6} \Dirac[\gamma^{\text{BPS}}_i,\gamma^{\text{BPS}}_j] p_i p_j \\
    &= 2 p_2 p_3+2 p_6 p_3+2 p_1 p_4+2 p_2 p_5+ 2 p_4 p_5+2 p_1 p_6
    \\&\qquad
    -p_1 p_3-p_5 p_3-p_2 p_4-p_1 p_5  -p_2 p_6-p_4 p_6
\end{split}
\end{equation}

As already discussed in \cite{Cordova:2015nma} such a series needs a regularization procedure to give sensible results when truncated at finite order in $q$. A possible regularization procedure was proposed in \cite{Cordova:2016uwk}. Here we adopt this regularization which gives sensible results for the theories considered in this paper. 
We define a truncated version $\mathcal{S}_k (q)$ of the quantum spectrum generator where we only keep terms involving the variables $X_{a_1 \gamma_1 + a_2 \gamma_2 + a_3 \gamma_3 + a_4 \gamma_4}$ with $a_1+a_2 + a_3 + a_4 \leq k$ for some integer $k$.
$\mathcal{S}_k (q)$ is expected to agree with the quantum spectrum generator up to $\mathcal{O}(q^n)$ for some $n$ as a function of $k$. The precise relation between $k$ and $n$ is theory-dependent and we are not aware of an argument to determine the relation a priori. Experimentally for pure $SU(3)$ SYM we find that $n = \tfrac{k}{4}$, namely $\mathcal{S}_k (q)$ has a good perturbative expansion up to $\mathcal{O}(q^{\tfrac{k}{4}})$ which agrees with the expansion of $\mathcal{S}_{k'>k} (q)$ for higher $k'$.
For generic $SU(N)$, we find experimentally that $n=\frac{k}{N+1}$, namely $S_{(N+1)k}(q)$ gives a good expansion of the quantum spectrum generator $S(q)$ for pure $SU(N)$ SYM up to order $\mathcal{O}(q^k)$.

As an example the truncated $\mathcal{S}_{8} (q)$ can be used to obtain the perturbative expansion of the quantum spectrum generator of $SU(3)$ up to order $\mathcal{O}(q^2)$:
\begin{equation}
\begin{split}
    \scriptstyle S(q) =&
    \scriptstyle
    1+\sqrt{q} \big(-X_{\gamma _1}-X_{\gamma _2}-X_{\gamma _3}-X_{\gamma _4}\big)+q \big(X_{2 \gamma _1}+X_{2 \gamma _2}+X_{2 \gamma _3}+X_{\gamma _1+\gamma
   _3}+X_{2 \gamma _4}
   +X_{\gamma _2+\gamma _4}+X_{\gamma _1+\gamma _2+\gamma _3+\gamma _4}\big)
   +\\&\scriptstyle
   +q^{3/2} 
   \big(-X_{\gamma _1}-X_{3 \gamma _1}-X_{\gamma _2}-X_{3
   \gamma _2}-X_{\gamma _3}-X_{3 \gamma _3}-X_{2 \gamma _1+\gamma _3}+X_{\gamma _2+\gamma _3}-X_{\gamma _1+2 \gamma _3}-X_{\gamma _4}-X_{3 \gamma _4}+X_{\gamma
   _1+\gamma _4}
    +\\&\scriptstyle
    -X_{2 \gamma _2+\gamma _4}-X_{\gamma _1+2 \gamma _2+2 \gamma _3+\gamma _4}
   -X_{\gamma _2+2 \gamma _4}-X_{2 \gamma _1+\gamma _2+\gamma _3+2 \gamma
   _4}\big)
   +\\&\scriptstyle
   +q^2 \big(X_{2 \gamma _1}+X_{4 \gamma _1}+X_{2 \gamma _2}+X_{4 \gamma _2}+X_{\gamma _1+\gamma _2}+X_{2 \gamma _3}+X_{4 \gamma _3}+2 X_{\gamma _1+\gamma
   _3}+X_{3 \gamma _1+\gamma _3}+X_{2 \gamma _1+2 \gamma _3}+X_{\gamma _1+3 \gamma _3}+X_{2 \gamma _4}
   +\\&\scriptstyle
   +X_{4 \gamma _4}+2 X_{\gamma _2+\gamma _4}+X_{3 \gamma _2+\gamma
   _4}+X_{\gamma _3+\gamma _4}+2 X_{\gamma _1+\gamma _2+\gamma _3+\gamma _4}-X_{2 \gamma _1+\gamma _2+\gamma _3+\gamma _4}-X_{\gamma _1+2 \gamma _2+\gamma _3+\gamma
   _4}-X_{\gamma _1+\gamma _2+2 \gamma _3+\gamma _4}
   +\\&\scriptstyle
   +X_{\gamma _1+3 \gamma _2+2 \gamma _3+\gamma _4}+X_{\gamma _1+2 \gamma _2+3 \gamma _3+\gamma _4}+X_{2 \gamma _2+2
   \gamma _4}-X_{\gamma _1+\gamma _2+\gamma _3+2 \gamma _4}+X_{3 \gamma _1+\gamma _2+\gamma _3+2 \gamma _4}+X_{2 \gamma _1+2 \gamma _2+2 \gamma _3+2 \gamma
   _4}
   +\\&\scriptstyle
   +X_{\gamma _2+3 \gamma _4}+X_{2 \gamma _1+\gamma _2+\gamma _3+3 \gamma _4}\big)
   +\mathcal{O}(q^{5/2})
\end{split}
\end{equation}

We checked that the trace of the KS operator reproduces the Schur index of pure $SU(3)$ up to order $\mathcal{O}(q^6)$:
\begin{equation}
\begin{split}
    \mathcal{I}(q) &= (q;q)^{4} \Tr [\mathcal{S}_{24}(q)\overline{\mathcal{S}_{24}}(q) ]
    = 1 +0q+ q^2 + 0q^3 + 2 q^4 + 0q^5 +0q^6 + \mathcal{O}(q^7)
\end{split}
\end{equation}
which matches the Schur index of $SU(3)$ SYM that reads:
\begin{equation}
    \oint \frac{du_1 du_2}{(2\pi i)^2 u_1 u_2} \Delta_3(\vec{u}) (q;q)^4 \prod_{i<j=1}^3 (u_i/u_jq;q)^2(u_j/u_iq;q)^2 \Bigg|_{\prod_{i=1}^3 u_i = 1}
\end{equation}

Next we want to compute the half-index in the electric frame. To compute the IR index we choose the following “electric" and “magnetic" charges:

\begin{equation}    \label{eq:qp_electricframe_SU3}
\begin{cases}
    q_1 = \frac{2}{3}(\gamma_1 + \gamma_2) +\frac{1}{3}(\gamma_3 + \gamma_4)
    \\
    q_2 = \frac{1}{3}(\gamma_1 + \gamma_2) +\frac{2}{3}(\gamma_3 + \gamma_4)
\end{cases}
\qquad
\begin{cases}
    p_1 = \gamma_1
    \\
    p_2 = \gamma_3
\end{cases}
\end{equation}
which satisfy:
\begin{equation}
    \Dirac[q_i, q_j]=\Dirac[p_i,p_j]=0,
    \qquad
    \Dirac[q_i, p_j] = -\delta_{i,j}
\end{equation}
and have integer Dirac pairing with all BPS particles. 
This choice of $q_1,q_2$ corresponds to the ``electric" frame.  In fact, if we express $q_i,p_i$ as UV charges we find:
\begin{equation}
\begin{split}
    q_1 = (1,0;\;0,0) \qquad \qquad p_1 = (0,0;\;1,0) \\
    q_2 = (0,1;\;0,0) \qquad \qquad p_2 = (0,0;\;0,1)
\end{split}
\end{equation}

\paragraph{Gluing the RG-wall to the UV} 
We first consider the case where we impose $\mathcal{D}$ b.c.~to the IR effective theory in the electric frame, which we match with the setup given by $SU(3)$ SYM with $\mathcal{N}$ b.c.~glued to the $(\Twall[3])^{-1}$ theory.

\begin{equation}
    \begin{tikzpicture}
    \begin{scope}
        \draw (-1,0) node {$\begin{gathered} \NN=2 \\ \text{$SU(3)$ SYM} \end{gathered}$};
        \draw (1,1) -- node[midway, gauge, double, fill=white] (gUV) {$3$} (1,-1);
        \draw[->-] (gUV) -- node[midway,above,gray] {$b_2^{-1/2}$} ++ (2,0) node[gauge, double,fill=white] (g1) {$2$};
        \draw[->-] (g1) -- node[midway,above,gray] {$b_2^{1/2} b_1^{-1}$} ++ (2,0) node[flavor,fill=white] (g2) {$1$};
    \end{scope}
    \draw[<->] (6,0)--(7,0);
    \begin{scope}[xshift = 10cm]
        \draw (-1,0) node {$\begin{gathered}SW \\ \text{electric frame} \end{gathered}$};
        \draw (1,1) -- node[midway, flavor, fill=white] (gIR) {$U(1)^2$} (1,-1);
        \draw[gray,right] (gIR)++(.5,0) node {$\vec{b}$};
    \end{scope}
    \end{tikzpicture}
\end{equation}

The index of the IR theory is then given by:
\begin{equation} \label{eq: su3twallinv_ir}
\mathcal{II}^{IR}(\vec{b},\vec{m}) = \mathcal{II}^{SW}(\vec{b}; \vec{m}) =
    (q;q)^2
    \Tr_{\{ q_1, q_2\}} \left[ :\mathcal{S}(q) \left(X_{p_1}\right)^{m_{1}}
    \left(X_{p_2}\right)^{m_{2}}:\right]
\end{equation}
where $\vec b = (b_1, b_2)$ are the fugacities of the boundary $U(1)^2$ symmetry which are identified as:
\begin{equation}
    \Tr_{\{q_1,q_2\}}[X_{q_i}]=b_i
\end{equation}

On the UV side the half-index of  4d $SU(3)$ SYM coupled to $({\Twall[3]})^{-1}$ at the boundary is given by:
\begin{equation} \label{eq: su3twallinv_uv}
\mathcal{II}^{UV}(\vec{b},\vec{m}_b) =
    \oint 
    \Delta_{3}(\vec{u}) 
    \;\;
    \mathcal{II}^{SU(3)}(\vec{u})
    \;\;
    \mathcal{I}^{(\Twall[3])^{-1}} (\vec{u}, \{ \vec{b}, \vec{m}\} )
\end{equation}
where $\mathcal{II}^{SU(3)}$ is the half-index contribution from the bulk $SU(3)$ SYM theory given in \eqref{eq: su3_halfind} and the index of the $(\Twall[3])^{-1}$ theory is given in \eqref{eq: su3_twallinv_index}.

We checked that the half-index agrees between the UV Lagrangian computation \eqref{eq: su3twallinv_uv} and the IR computation \eqref{eq: su3twallinv_ir} for $-4\leq m_{b_{1}},m_{b_{2}} \leq 4$ up to order $\mathcal{O}(q^{6})$. 
For completeness, we report the $q$-expansion of the half-index for some values of the fluxes $m_1, m_2$:
\begin{equation}
\begin{array}{c|c}
    (m_1,m_2) & \text{Half index in \eqref{eq: su3twallinv_uv} or \eqref{eq: su3twallinv_ir}}
    \\ \hline
    (0,0)
    \quad
    & 1+\left(b _1 b _2-2\right) q+\left(b _2^2 b _1^2+\frac{b _1^2}{b _2}+\frac{b _2^2}{b _1}-1\right) q^2 
    +\left(b _1^3 b _2^3-3 b _1 b _2+2\right) q^3+\mathcal{O}\left(q^4\right)
    \\
    (1,0)
    \quad
    & -\frac{b _1^2 \sqrt{q}}{b _2}+\frac{b _1^2 q^{3/2}}{b _2}-b _1^3 q^2+\frac{2 b _1^2 q^{5/2}}{b _2}+\left(-b _1^3-b _2 b _1\right) q^3+\mathcal{O}\left(q^{7/2}\right)
    \\
    (-1,0)
    \quad
    & -\sqrt{q}+q^{3/2}-b _1 b _2 q^2+2 q^{5/2}+\left(-\frac{b _2^2}{b _1}-b _1 b _2\right) q^3+\mathcal{O}\left(q^{7/2}\right)
    \\
    (1,1)
    \quad
    & b _1 b _2 q+\left(b _1^2 b _2^2-2 b _1 b _2\right) q^3+\mathcal{O}\left(q^4\right)
\end{array}
\end{equation}

\paragraph{Gluing the RG-wall to the IR} We can then consider the case of $\mathcal{D}$ b.c.~for $SU(3)$ SYM. On the other side we have a setup given by the IR theory with $\mathcal{N}$ b.c.~in the electric frame glued to the $\Twall[3]$ theory.

The expected half-index identity is between the half-index of $SU(3)$ SYM, given in \eqref{eq: su3_halfind} and the index of the IR setup given by:
\begin{equation} \label{eq: su3_IR+twall_index}
    \mathcal{II}^{IR}(\vec{u}) = (q;q)^2 \sum_{m_1,m_2 \in \mathbb{Z}} \oint \frac{db_1 db_2}{(2 \pi i)^2 b_1 b_2} \Tr_{\{ q_1, q_2\}} \left[ :\mathcal{S}(q) \left(X_{p_1}\right)^{m_{1}}
    \left(X_{p_2}\right)^{m_{2}}:\right]
    \mathcal{I}^{\Twall[3]}(\vec{u},\{\vec{b},\vec{m}\}) 
\end{equation}
We checked that the two expression match to $\mathcal{O}(q^4)$. 

Similarly as in the $SU(2)$ case, we can also turn on magnetic fluxes for the $SU(3)$ symmetry. The half-index of $SU(3)$ SYM in \eqref{eq: su3_halfind} is expected to vanish for any non-zero magnetic flux. In the IR setup this is non-trivially realized as we can turn on a magnetic flux in the $\Twall[3]$ theory. We checked that the generalization of \eqref{eq: su3_IR+twall_index} to non-zero $SU(3)$ magnetic fluxes is identically zero for any non-zero flux.
\\

We can look for operators of the RG-wall theory $\Twall[3]$ that have the correct charges to couple to the BPS particles at the boundary. 
None of the BPS particles is electric in the electromagnetic frame considered here \eqref{eq:qp_electricframe_SU3}, therefore we should look for operators that only exist in the presence of background fluxes for the $U(1)_{b_i}$ symmetries.
Expressing the charges of BPS particles in the basis of $q_i$ and $p_i$ we can identify the background flux corresponding to each BPS particle and we find good operators that couple to $\gamma_1, \gamma_2, \gamma_3$ and $\gamma_4$. The presence of these operators can be inferred, for example, by computing the superconformal index:
\begin{equation}    \label{eq:bdy_W_operators_SU3}
    \begin{array}{c|c}
        \text{Index} & \text{Couples to} \\
        \hline
        \mathcal{I}^{\Twall[3]}(\vec{u},\{\vec{b},(1,0)\}) =  q^{1/2} \left(
        {\color{red}-\frac{b_2}{b_1^2} }
        +f(\vec{u},\vec{b})\right) +\mathcal{O}(q)  & \gamma_1
        \\
        \mathcal{I}^{\Twall[3]}(\vec{u},\{\vec{b},(0,1)\}) =  q^{1/2} \left(
        {\color{red}-\frac{b_1}{b_2^2} }
        +f(\vec{u},\vec{b})\right) +\mathcal{O}(q)  & \gamma_3
        \\
        \mathcal{I}^{\Twall[3]}(\vec{u},\{\vec{b},(0,-1)\}) =  q^{1/2} \left(
        {\color{red}-1 }
        +f(\vec{u},\vec{b})\right) +\mathcal{O}(q)  & \gamma_2
        \\
        \mathcal{I}^{\Twall[3]}(\vec{u},\{\vec{b},(-1,0)\}) =  q^{1/2} \left(
        {\color{red}-1 }
        +f(\vec{u},\vec{b})\right) +\mathcal{O}(q)  & \gamma_4
    \end{array}
\end{equation}
where we highlighted in red the contributions counting operators with R-charge 1 and correct $U(1)_{b_i}$ charges to couple to the corresponding BPS particles. 
Interestingly, we do not find good operators that couple to the composite BPS particles $\gamma_1+\gamma_4$ and $\gamma_2 + \gamma_3$. These are expected to sit in the sectors with $(m_1,m_2) = (1,-1)$ and $(m_1,m_2) = (-1,1)$, respectively, but the $q$-expansion of the superconformal index in these sectors starts at $q^{3/2}$.
These are composite BPS particles, and should naturally couple to analogous composites of the operators highlighted in \eqref{eq:bdy_W_operators_SU3}, as already discussed in Section 2.3 of \cite{Dimofte:2013lba}. It would be interesting to study this issue in more depth, but we leave this to future work.

\subsubsection{$SU(4)$}
The Schur half-index of $SU(4)$ SYM is:
\begin{equation} \label{eq: su4_halfind}
    \mathcal{II}^{SU(4)}(\vec{u}) = (q;q)^3 \prod_{i<j}^4 (u_j/u_jq;q) (u_i/u_jq;q) \Bigg|_{\prod_{i=1}^4 u_i=1}
\end{equation}
The 3d superconformal index of the $\Twall[3]$ theory reads:\footnote{For the practical purpose of expanding the superconformal index perturbatively, it is convenient to consider different duality frames that are simpler gauge theories. The simplest, in terms of amount of matter and rank of the gauge group, consists of an $SU(2)$ theory with 6 fundamentals and 4 singlets, with interactions that break the $U(6)$ symmetry to $SU(4)\times U(1)^3$, already discussed in Section \ref{subsec: twall_gsun_generalities}.}
\begin{equation} \label{eq: su4_twallind}
\begin{split}
    &\mathcal{I}^{\Twall[3]}(\vec{u},\{ \vec{b},\vec{m} \}) = \\
    &\qquad\qquad = 
    (b_1 b_2^{-1/2})^{m_1} (b_2 b_1^{-1/2} b_3^{-1/2})^{m_2} (b_3 b_2^{-1/2})^{m_3} \\
    &\qquad\qquad\quad 
    \sum_{n^{(1)} \in \Gamma_1} \sum_{\vec{n}^{(2)} \in \Gamma_2} \oint \frac{dz^{(1)}}{2\pi i z^{(1)}} \mathcal{I}_{\text{vec}}^{SU(2)}(z^{(1)},n^{(1)}) \oint \frac{d z^{(2)}_1 dz^{(2)}_2}{(2\pi i )^2 z^{(2)}_1 z^{(2)}_2} \mathcal{I}_{\text{vec}}^{SU(3)}(\vec{z}^{(2)},\vec{n}^{(2)}) \\
    &\qquad\qquad\quad 
    \prod_{s=\pm1} \mathcal{I}_{\text{chir}}^{(0)}(b_1 b_2^{-1/2} {z^{(1)}}^s, sn^{(1)}+m_1-m_2/2) \\
    &\qquad\qquad\quad 
    \prod_{s=\pm 1}\prod_{j=1}^3 \mathcal{I}_{\text{chir}}^{(0)}(b_2^{1/2} b_3^{-1/3} {z^{(1)}}^s z^{(2)}_j, sn^{(1)} + n^{(2)}_j +m_2/2-m_3/3) \\
    &\qquad\qquad\quad 
    \prod_{j=1}^3 \prod_{k=1}^4 \mathcal{I}_{\text{chir}}^{(0)}(b_3^{1/3} u_k/z^{(2)}_j, -n^{(2)}_j + m_3/3) \\
    & \text{with} \qquad \Gamma_1 = \mathbb{Z} + \text{Mod}(m_2,2)\frac{1}{2}
    \qquad \Gamma_2 = \mathbb{Z}^2 + \text{Mod}(m_3,3) \left\{ \frac{1}{3},\frac{1}{3} \right\}
\end{split}
\end{equation}
Notice that the magnetic lattices of the $SU(2)$ and $SU(3)$ gauge groups depend on the value of the background fluxes $m_2$ and $m_3$ respectively, due to Dirac quantization.
Analogously, the index of the inverse wall $(\Twall[4])^{-1}$ is:
\begin{equation} \label{eq: su4_twallinvind}
\begin{split}
    &\mathcal{I}^{\Twall[3]}(\vec{u},\{ \vec{b},\vec{m} \}) = \\
    &\qquad\qquad = 
    (b_1 b_2^{-1/2})^{-m_1} (b_2 b_1^{-1/2} b_3^{-1/2})^{-m_2} (b_3 b_2^{-1/2})^{-m_3} \\
    &\qquad\qquad\quad 
    \sum_{n^{(1)} \in \Gamma_1} \sum_{\vec{n}^{(2)} \in \Gamma_2} \oint \frac{dz^{(1)}}{2\pi i z^{(1)}} \mathcal{I}_{\text{vec}}^{SU(2)}(z^{(1)},n^{(1)}) \oint \frac{d z^{(2)}_1 dz^{(2)}_2}{(2\pi i )^2 z^{(2)}_1 z^{(2)}_2} \mathcal{I}_{\text{vec}}^{SU(3)}(\vec{z}^{(2)},\vec{n}^{(2)}) \\
    &\qquad\qquad\quad 
    \prod_{s=\pm1} \mathcal{I}_{\text{chir}}^{(0)}(b_1^{-1} b_2^{1/2} {z^{(1)}}^s, sn^{(1)}-m_1+m_2/2) \\
    &\qquad\qquad\quad 
    \prod_{s=\pm 1}\prod_{j=1}^3 \mathcal{I}_{\text{chir}}^{(0)}(b_2^{-1/2} b_3^{1/3} {z^{(1)}}^s/z^{(2)}_j, sn^{(1)} - n^{(2)}_j - m_2/2 +m_3/3) \\
    &\qquad\qquad\quad 
    \prod_{j=1}^3 \prod_{k=1}^4 \mathcal{I}_{\text{chir}}^{(0)}(b_3^{-1/3} z^{(2)}_j/u_k, n^{(2)}_j - m_3/3) \\
    & \text{with} \qquad \Gamma_1 = \mathbb{Z} + \text{Mod}(m_2,2)\frac{1}{2}
    \qquad \Gamma_2 = \mathbb{Z}^2 + \text{Mod}(m_3,3) \left\{ \frac{1}{3},\frac{1}{3} \right\}
\end{split}
\end{equation}

To compute the index of the IR index we consider a finite chamber whose BPS spectrum is encoded in the BPS quiver:\footnote{To be more precise, each triplet of particles have the same value of central charge but they commute w.r.t.~the Dirac pairing. Therefore the ordering of their KS factors is not important.}
\begin{equation}
\begin{tikzpicture}
	\node[draw,circle, radius=4pt] (g1) at (0,0) {$\gamma_1$};
	\node[draw,circle, radius=4pt] (g2) at (0,2) {$\gamma_2$};
	\node[draw,circle, radius=4pt] (g3) at (2,0) {$\gamma_3$};
	\node[draw,circle, radius=4pt] (g4) at (2,2) {$\gamma_4$};
    \node[draw,circle, radius=4pt] (g5) at (4,0) {$\gamma_5$};
	\node[draw,circle, radius=4pt] (g6) at (4,2) {$\gamma_6$};
	\draw[->] (g1) edge[double] (g2);
	\draw[->] (g3) edge[double] (g4);
    \draw[->] (g5) edge[double] (g6);
	\draw[->] (g2) edge[] (g3);
	\draw[->] (g4) edge[] (g1);
    \draw[->] (g4) edge[] (g5);
	\draw[->] (g6) edge[] (g3);
\end{tikzpicture}
\end{equation}
From the UV Lagrangian perspective the $\gamma_i$ states carry the following charges:
\begin{equation}
\begin{split}
    \gamma_2 = (2,-1,0;\;-1,0,0) \qquad \gamma_1 = (0,0,0;\;1,0,0) \\
    \gamma_4 = (-1,2,-1;\;0,-1,0) \qquad \gamma_3 = (0,0,0;\;0,1,0) \\
    \gamma_6 = (0,-1,2;\;0,0,-1) \qquad \gamma_5 = (0,0,0;\;0,0,1) \\
\end{split}
\end{equation}
In this chamber the central charges satisfy:
\begin{equation}
    Arg(\gamma_1) , Arg(\gamma_3) , Arg(\gamma_5) < Arg(\gamma_2) , Arg(\gamma_4) , Arg(\gamma_6)
\end{equation}
and the BPS spectrum is composed of 12 particles, reported in increasing value for their central charge:
\begin{equation}
\begin{split}
    \{ \gamma_i^{\text{BPS}} \}_{i=1}^{12} = ( &\gamma_1, \gamma_3, \gamma_5, \gamma_3+\gamma_6, \gamma_1+\gamma_4+\gamma_5, \gamma_2+\gamma_3, \\
    & \gamma_1+\gamma_4, \gamma_2+\gamma_3+\gamma_6, \gamma_4+\gamma_5, \gamma_2, \gamma_4, \gamma_6 ) 
\end{split}
\end{equation}

We checked that the truncated quantum spectrum generator $S_{k}(q)$ gives a sensible expansion up to order $\mathcal{O}(q^{k/5})$.
As an example the truncated $S_{15}(q)$ can be used to compute the Shur index up to order $\mathcal{O}(q^3)$:
\begin{equation}
\begin{split}
    \mathcal{I}(q) &= (q;q)^{6} \Tr [\mathcal{S}_{15}(q)\overline{\mathcal{S}_{15}}(q) ]
    = 1 + 0q +  q^2 + 0 q^3+\mathcal{O}(q^4)
\end{split}
\end{equation}

To compute the half-KS operator we pick the following electric and magnetic charges:
\begin{equation}
    \begin{cases}
        q_1 = \frac{1}{4}(3\gamma_1+3\gamma_2+2\gamma_3+2\gamma_4+\gamma_5+\gamma_6) \\
        q_2=\frac{1}{4}(2\gamma_1+2\gamma_2+4\gamma_3+4\gamma_4+2\gamma_5+2\gamma_6) \\
        q_3 = \frac{1}{4}(\gamma_1+\gamma_2+2\gamma_3+2\gamma_4+3\gamma_5+3\gamma_6)
    \end{cases} \qquad 
    \begin{cases}
        p_1 = \gamma_1 \\
        p_2 = \gamma_3 \\
        p_3 = \gamma_5
    \end{cases}
\end{equation}
which satisfy:
\begin{equation}
    \langle q_i,q_j \rangle = \langle p_i,p_j \rangle = 0, \qquad \langle q_i,p_j \rangle = -\delta_{i,j}
\end{equation}
and have integer Dirac pairing with all the BPS particles. This choice corresponds to the ``electric" frame as their charges from the UV Lagrangian perspective read:
\begin{equation}
    \begin{cases}
        q_1 = (1,0,0;\;0,0,0) \\
        q_2 = (0,1,0;\;0,0,0) \\
        q_3 = (0,0,1;\;0,0,0)
    \end{cases} \qquad 
    \begin{cases}
        p_1 = (0,0,0;\;1,0,0) \\
        p_2 = (0,0,0;\;0,1,0) \\
        p_3 = (0,0,0;\;0,0,1)
    \end{cases}
\end{equation}

\paragraph{Gluing the RG-wall to the UV}
We first consider the case of $\mathcal{D}$ b.c.~for the IR effective theory in the electric frame. 
This setup is given by:

\begin{equation}
    \begin{tikzpicture}[scale=.9]
    \begin{scope}
        \draw (-1,0) node {$\begin{gathered} \NN=2 \\ \text{$SU(4)$ SYM} \end{gathered}$};
        \draw (1,1) -- node[midway, gauge, double, fill=white] (gUV) {$4$} (1,-1);
        \draw[->-] (gUV) -- node[midway,above,gray] {$b_3^{-1/3}$} ++ (2,0) node[gauge, double,fill=white] (g1) {$3$};
        \draw[->-] (g1) -- node[midway,above,gray] {$b_3^{1/3}b_2^{-1/2}$} ++ (2,0) node[gauge, double,fill=white] (g2) {$2$};
        \draw[->-] (g2) -- node[midway,above,gray] {$b_2^{1/2} b_1^{-1}$} ++ (2,0) node[flavor,fill=white] (g3) {$1$};
    \end{scope}
    \draw[<->] (8,0)--(9,0);
    \begin{scope}[xshift = 12cm]
        \draw (-1,0) node {$\begin{gathered}SW \\ \text{electric frame} \end{gathered}$};
        \draw (1,1) -- node[midway, flavor, fill=white] (gIR) {$U(1)^3$} (1,-1);
        \draw[gray,right] (gIR)++(.5,0) node {$\vec{b}$};
    \end{scope}
    \end{tikzpicture}
\end{equation}

The corresponding helf-index reads:
\begin{equation}\label{eq: su4_IRindex}
    \mathcal{II}^{SW}(\vec{b},\vec{m}) = (q;q)^3 \Tr_{\{q_i\}}[:S(q)\prod_{i=1}^3 (X_{p_i})^{m_i}:]
\end{equation}
where $b_i$ are the fugacities of the boundary $U(1)^3$ symmetry identified as:
\begin{equation}
    \Tr_{\{q_i\}}[X_{q_i}] = b_i
\end{equation}

The index in \eqref{eq: su4_IRindex} is expected to match wit that of the UV setup, given by $SU(4)$ SYM glued to the $(\Twall[4])^{-1}$ theory, which reads:
\begin{equation}    \label{eq:HI_Lag_SU4}
    \mathcal{II}^{UV}(\vec{b},\vec{m}) = \oint \frac{du_1 du_2 du_3}{(2\pi i )^3 u_1 u_2 u_3} \Delta_4(\vec{u}) \mathcal{II}^{SU(4)}(\vec{u}) \mathcal{I}^{(\Twall[4])^{-1}}(\vec{u}, \{\vec{b},\vec{m}\})
\end{equation}
where $\mathcal{II}^{SU(4)}$ is given in \eqref{eq: su4_halfind} and $\mathcal{I}^{(\Twall[4])^{-1}}$ in \eqref{eq: su4_twallinvind}.

We checked the agreement of the two indices up to $\mathcal{O}(q^3)$ for $-1 \leq m_1,m_2,m_3 \leq 1$.
For completeness, we report the $q$-expansion of the half-index for some values of the fluxes $m_i$, expanded up to $\mathcal{O}(q^2)$ for readability:
\begin{equation}
\begin{array}{c|c}
    (m_1,m_2,m_3) & \text{Half index in \eqref{eq: su4_IRindex} or \eqref{eq:HI_Lag_SU4}}
    \\ \hline
    (0,0,0)
    &
    1+\left(\frac{b _3 b _2}{b _1}+\frac{b _1 b _2}{b _3}-3\right) q+\bigg(\frac{b _1^2}{b _2}+\frac{b _2^2 b _1^2}{b _3^2}-b _3 b _1 
    
    -\frac{b _2 b _1}{b _3}+\frac{b _3^2}{b _2}-\frac{b _2 b _3}{b _1}
    +\frac{b _2^2}{b _3 b _1}+\frac{b _2^2 b _3^2}{b _1^2}\bigg) q^2+\mathcal{O}\left(q^3\right)
    \\
    (1,0,0)
    &-\frac{b _1^2 \sqrt{q}}{b _2}+\frac{2 b _1^2 q^{3/2}}{b _2}+\left(-\frac{b _1^3}{b _3}-b _3 b _1\right) q^2+\mathcal{O}\left(q^{5/2}\right)
    \\
    (0,1,-1)
    &\left(\frac{b _2^2}{b _1 b _3}-\frac{b _2^3}{b _1^2}\right) q^{3/2}+b _2^2 q^2+\mathcal{O}\left(q^{5/2}\right)
    \\
    (1,0,1)
    &\frac{b _1^2 b _3^2 q}{b _2^2}+\left(b _1^2 b _3^2-\frac{b _1^2 b _3^2}{b _2^2}\right) q^2+\mathcal{O}\left(q^3\right)
\end{array}
\end{equation}

\paragraph{Gluing the RG-wall to the IR} 
We then consider the complementary setup, where we give $\mathcal{D}$ b.c.~to $SU(4)$ SYM, whose half-index is simply given by \eqref{eq: su4_halfind}.

In the IR the index is given by that of the effective theory glued to the $\Twall[4]$ theory:
\begin{equation}
    \mathcal{II}^{IR}(\vec{u}) = (q;q)^3 \sum_{m_1,m_2,m_3 \in \mathbb{Z}} \oint \frac{d{b_1} db_2 db_3}{(2 \pi i)^3 b_1 b_2 b_3} \Tr_{\{q_i\}}[: \mathcal{S}(q) \prod_{i=1}^3 (X_{p_i})^{m_i}:] \mathcal{I}^{\Twall[4]}(\vec{u},\{ \vec{b}, \vec{m} \}) 
\end{equation}
We checked the agreement between the UV and IR indices up to order $\mathcal{O}(q^3)$.
\\

Similarly to the lower rank cases discussed above, we can look for operators of the RG-wall theory $\Twall[4]$ that have the correct charges to couple to the BPS particles at the boundary. 
We find good operators in non-trivial background flux sectors that can couple to the BPS particles $\gamma_1,\gamma_2,\gamma_3,\gamma_4,\gamma_5$ and $\gamma_6$. 
However, we do not find good operators that can couple to the composite BPS particles, similarly to the $SU(3)$ case discussed above. We leave a more in-depth analysis of this behavior to future work.

\subsubsection{$SU(5)$}
The Schur half-index of $SU(5)$ SYM is:
\begin{equation} \label{eq: su5_halfind}
    \mathcal{II}^{SU(5)}(\vec{u}) = (q;q)^4 \prod_{i<j}^5 (u_j/u_jq;q) (u_i/u_jq;q) \Bigg|_{\prod_{i=1}^5 u_i=1}
\end{equation}

To compute the index in the IR we consider a finite chamber whose BPS spectrum is encoded in the BPS quiver:
\begin{equation}
\begin{tikzpicture}
	\node[draw,circle, radius=4pt] (g1) at (0,0) {$\gamma_1$};
	\node[draw,circle, radius=4pt] (g2) at (0,2) {$\gamma_2$};
	\node[draw,circle, radius=4pt] (g3) at (2,0) {$\gamma_3$};
	\node[draw,circle, radius=4pt] (g4) at (2,2) {$\gamma_4$};
    \node[draw,circle, radius=4pt] (g5) at (4,0) {$\gamma_5$};
	\node[draw,circle, radius=4pt] (g6) at (4,2) {$\gamma_6$};
    \node[draw,circle, radius=4pt] (g7) at (6,0) {$\gamma_5$};
	\node[draw,circle, radius=4pt] (g8) at (6,2) {$\gamma_6$};
	\draw[->] (g1) edge[double] (g2);
	\draw[->] (g3) edge[double] (g4);
    \draw[->] (g5) edge[double] (g6);
    \draw[->] (g7) edge[double] (g8);
	\draw[->] (g2) edge[] (g3);
	\draw[->] (g4) edge[] (g1);
    \draw[->] (g4) edge[] (g5);
	\draw[->] (g6) edge[] (g3);
    \draw[->] (g6) edge[] (g7);
	\draw[->] (g8) edge[] (g5);
\end{tikzpicture}
\end{equation}
In this chamber the central charges satisfy:
\begin{equation}
    Arg(\gamma_1) , Arg(\gamma_3) , Arg(\gamma_5) , Arg(\gamma_7) < Arg(\gamma_2) , Arg(\gamma_4),Arg(\gamma_6),Arg(\gamma_8)
\end{equation}
and the BPS spectrum is composed of 20 particles, reported in increasing value for their central charge:
\begin{equation}
\begin{split}
    \{ \gamma_i^{\text{BPS}} \}_{i=1}^{20} = ( &\gamma _7,\gamma _5,\gamma _3,\gamma _1,\gamma _2+\gamma _3,\gamma _1+\gamma _4+\gamma _5,\gamma _3+\gamma _6+\gamma _7,\gamma _5+\gamma _8,\gamma _4+\gamma
   _5,
   \\&
   \gamma _2+\gamma _3+\gamma _6+\gamma _7,\gamma _1+\gamma _4+\gamma _5+\gamma _8,\gamma _3+\gamma _6,\gamma _6+\gamma _7,\gamma _4+\gamma _5+\gamma _8,
   \\&
   \gamma
   _2+\gamma _3+\gamma _6,\gamma _1+\gamma _4,\gamma _8,\gamma _6,\gamma _4,\gamma _2 ) 
\end{split}
\end{equation}
We checked that the truncated quantum spectrum generator $S_{k}(q)$ gives a sensible expansion up to order $\mathcal{O}(q^{k/6})$.
The truncated $S_{12}(q)$ can be used to compute the Shur index up to order $\mathcal{O}(q^2)$:
\begin{equation}
\begin{split}
    \mathcal{I}(q) &= (q;q)^{8} \Tr [\mathcal{S}_{12}(q)\overline{\mathcal{S}_{12}}(q) ]
    = 1 + 0q +  q^2 +\mathcal{O}(q^3)
\end{split}
\end{equation}

We choose the following electric and magnetic charges:
\begin{equation}
    \begin{cases}
        q_1 =\frac{1}{5} \left(4 \gamma _1+4 \gamma _2+3 \gamma _3+3 \gamma _4+2 \gamma _5+2 \gamma _6+\gamma _7+\gamma _8\right) \\
        q_2=\frac{1}{5} \left(3 \gamma _1+3 \gamma _2+6 \gamma _3+6 \gamma _4+4 \gamma _5+4 \gamma _6+2\gamma _7+2\gamma _8\right) \\
        q_3 = \frac{1}{5} \left(2 \gamma _1+2 \gamma _2+4 \gamma _3+4 \gamma _4+6 \gamma _5+6 \gamma _6+3 \gamma _7+3 \gamma _8\right) \\
        q_4 = \frac{1}{5} \left(\gamma _1+\gamma _2+2 \gamma _3+2 \gamma _4+3 \gamma _5+3 \gamma _6+4 \gamma _7+4 \gamma _8\right)
    \end{cases} \qquad 
    \begin{cases}
        p_1 = \gamma_1 \\
        p_2 = \gamma_3 \\
        p_3 = \gamma_5 \\
        p_4 = \gamma_7
    \end{cases}
\end{equation}
which satisfy:
\begin{equation}
    \langle q_i,q_j \rangle = \langle p_i,p_j \rangle = 0, \qquad \langle q_i,p_j \rangle = -\delta_{i,j}
\end{equation}
and have integer Dirac pairing with all the BPS particles. In terms of the UV Lagrangian, the charges $q_i$ are purely electric while the charges $p_i$ are purely magnetic.

\paragraph{Gluing the RG-wall to the UV}
We consider the case of $\mathcal{D}$ b.c.~for the IR effective theory in the electric frame. 
This setup is given by:

\begin{equation}
    \begin{tikzpicture}[scale=.9]
    \begin{scope}[xshift=-2cm]
        \draw (-1,0) node {$\begin{gathered} \NN=2 \\ \text{$SU(5)$ SYM} \end{gathered}$};
        \draw (1,1) -- node[midway, gauge, double, fill=white] (gUV) {$4$} (1,-1);
        \draw[->-] (gUV) -- node[midway,above,gray] {$b_4^{-1/4}$} ++ (2,0) node[gauge, double,fill=white] (g1) {$4$};
        \draw[->-] (g1) -- node[midway,above,gray] {$b_4^{1/4}b_3^{-1/3}$} ++ (2,0) node[gauge, double,fill=white] (g2) {$3$};
        \draw[->-] (g2) -- node[midway,above,gray] {$b_3^{1/3} b_2^{-1/2}$} ++ (2,0) node[gauge,double,fill=white] (g3) {$2$};
        \draw[->-] (g3) -- node[midway,above,gray] {$b_2^{1/2} b_1^{-1}$} ++ (2,0) node[flavor,fill=white] (g4) {$1$};
    \end{scope}
    \draw[<->] (8,0)--(9,0);
    \begin{scope}[xshift = 12cm]
        \draw (-1,0) node {$\begin{gathered}SW \\ \text{electric frame} \end{gathered}$};
        \draw (1,1) -- node[midway, flavor, fill=white] (gIR) {$U(1)^4$} (1,-1);
        \draw[gray,right] (gIR)++(.5,0) node {$\vec{b}$};
    \end{scope}
    \end{tikzpicture}
\end{equation}

The corresponding half-index reads:
\begin{equation}\label{eq: su5_IRindex}
    \mathcal{II}^{SW}(\vec{b},\vec{m}) = (q;q)^4 \Tr_{\{q_i\}}[:S(q)\prod_{i=1}^4 (X_{p_i})^{m_i}:]
\end{equation}
where $b_i$ are the fugacities of the boundary $U(1)^4$ symmetry identified as:
\begin{equation}
    \Tr_{\{q_i\}}[X_{q_i}] = b_i
\end{equation}

The index in \eqref{eq: su5_IRindex} is expected to match with that of the UV setup, given by $SU(5)$ SYM glued to the $(\Twall[5])^{-1}$ theory, which reads:
\begin{equation}    \label{eq:HI_lag_SU5}
    \mathcal{II}^{UV}(\vec{b},\vec{m}) = \oint \frac{du_1 du_2 du_3 du_4}{(2\pi i )^4 u_1 u_2 u_3 u_4} \Delta_5(\vec{u}) \mathcal{II}^{SU(5)}(\vec{u}) \mathcal{I}^{(\Twall[5])^{-1}}(\vec{u}, \{\vec{b},\vec{m}\})
\end{equation}
where $\mathcal{II}^{SU(5)}$ is given in \eqref{eq: su5_halfind} and $\mathcal{I}^{(\Twall[5])^{-1}}$ can be computed from one of the quiver Lagrangians for $\Twall[5]$ \eqref{eq:quiv_Twall4} or \eqref{eq:quiv_Twall4_dualized}.

We checked the agreement of the two indices up to $\mathcal{O}(q^2)$ for $-1 \leq m_1,m_2,m_3,m_4 \leq 1$.
We report the $q$-expansion of the half-index for some values of the fluxes $m_i$:
\begin{equation}
    \begin{array}{c|c}
        (m_1,m_2,m_3,m_4) & \text{Half-index in \eqref{eq:HI_lag_SU5} or \eqref{eq: su5_IRindex}} \\
        \hline
         (0,0,0,0) & 
         \begin{array}{r}
         1+\left(\frac{b _1 b _2}{b _3}+\frac{b _3 b _2}{b _1 b _4}+\frac{b _3 b _4}{b _2}-4\right) q+\bigg(\frac{b _1^2}{b _2}+\frac{b _2^2 b _1^2}{b _3^2}-\frac{2 b _2 b _1}{b _3}-\frac{b _3 b _1}{b _4}
         +\frac{b _3^2 b _4^2}{b _2^2}+\frac{b _4^2}{b _3} +
         \\
         +b _2 b _3-\frac{2 b _3 b _4}{b _2}+\frac{b _3^2}{b _2 b _4}-\frac{b _2 b _4}{b _1}+\frac{b _2^2}{b _3 b _1}-\frac{2 b _2 b _3}{b _4 b _1}+\frac{b _2^2 b _3^2}{b _4^2 b _1^2}+2\bigg) q^2+\mathcal{O}\left(q^3\right)
         \end{array}
         \\
         (1,0,0,0) & -\frac{b _1^2 \sqrt{q}}{b _2}+\frac{b _1^2 \left(3 b _2-b _3 b _4\right) q^{3/2}}{b _2^2}+\left(-\frac{b _1^3}{b _3}-\frac{b _3 b _1}{b _4}\right) q^2+\frac{2 b _1^2 \left(b _2+2 b _3 b _4\right) q^{5/2}}{b _2^2}+\mathcal{O}\left(q^3\right)
         \\
         (0,0,1,-1) & \left(\frac{b _3^2}{b _2 b _4}-\frac{b _3^3}{b _2^2}\right) q^{3/2}+\frac{b _3^2 q^2}{b _1}+\left(\frac{4 b _3^3}{b _2^2}-\frac{4 b _3^2}{b _2 b _4}\right) q^{5/2}+\mathcal{O}\left(q^3\right)
         \\
         (1,0,0,1) &\frac{b _1^2 b _4^2 q}{b _2 b _3}-\frac{2 \left(b _1^2 b _4^2\right) q^2}{b _2 b _3}+\mathcal{O}\left(q^3\right)
    \end{array}
\end{equation}

Similarly to the lower rank cases above, one can perform a similar computation by gluing the $\Twall[5]$ theory to the IR and study the existence of operators involved in the boundary superpotential. We have not performed these computations for the case of $SU(5)$.

\section{Gluing two Half-Indices} \label{sec: half_index_interfaces}

Along the lines of the previous section, we now study the partition function of 4d $\mathcal{N}=2$ SYM on $S^3 \times S^1$ with the introduction of defects on the equator of $S^3$.
The basic recipe to construct such index is to glue together two half-indices on $HS^3 \times S^1$ by their $S^2$ boundary.

We will discuss computations that correspond to the proposal of Section \ref{sec: interfaces}, namely we will study setups that are expected to be ``equivalent" to $\mathcal{N}=2$ SYM, and thus match with its Schur index on $S^3 \times S^1$. 

As a zero-th case we can consider the case where we glue two half-indices of $SU(N)$ SYM with a trivial interface as:
\begin{equation} \label{eq: gluingtwohalfindices_susym}
\begin{split}
    \mathcal{I}^{SU(N)}(q) &= \frac{1}{N!} \oint \prod_{j=1}^{N-1} \frac{d u_j}{2 \pi i u_j} \Delta_N(\vec{u}) \mathcal{II}^{SU(N)}(\vec{u}^{-1};q) \mathcal{II}^{SU(N)}(\vec{u};q) \\
    &=  \frac{(q;q)^{2N-2}}{N!} \oint \prod_{j=1}^{N-1} \frac{d u_j}{2 \pi i u_j} \Delta_N(\vec{u})  \prod_{i < j}^N (u_i/u_jq;q)^2 (u_j/u_iq;q)^2 \bigg|_{\prod_{i=1}^N u_i =1}
\end{split}
\end{equation}
where in the last line we recognize precisely the Schur index of $SU(N)$ SYM.
Notice that in the first line we are anti-identifying the $SU(N)$ fugacities to express the action of gauging of that symmetry. Even though the half-index of $SU(N)$ SYM is invariant under the redefinition $\vec{u} \to \vec{u}^{-1}$, it will be useful to take this convention later on.

We consider in particular two setups and their related index identity.
\begin{itemize}
    \item The gluing of the half-index of $\mathcal{N}=2$ SYM and of that the IR effective description, with the addition of the 3d $\mathcal{N}=2$ $\Twall[]$ theory, which is the UV/IR interface. The corresponding index identity reads:
    \begin{equation}    \label{eq:RG_interface_index_identity}
    \begin{split}
        \mathcal{I}^{SU(N)}(q) = & \sum_{\vec{m} \in \mathbb{Z}^{N-1}} \oint \prod_{j=1}^{N-1} \frac{d \xi_j}{2 \pi i \xi_j} \frac{1}{N!} \oint \prod_{j=1}^{N-1} \frac{d u_j}{2 \pi i u_j} \Delta_N(\vec{u}) 
        \times \\ & \qquad \times 
        \mathcal{II}^{SU(N)}(\vec{u}^{-1};q) \mathcal{I}^{\Twall[N]}(\vec{u}, \{\vec{\xi},\vec{m}\};q) \mathcal{II}^{SW}(  \vec{\xi},\vec{m} ;q)
    \end{split}
    \end{equation}
    Which can be taken also in different EM duality frames by transforming the index of the IR effective description and, accordingly, also that of the $\Twall[]$ theory, following the prescription of Section \ref{sec: bc_proposal}.

    \item The gluing of the half-index of two IR descriptions, possibly in different duality frames $\sigma_L$ and $\sigma_R$, with the addition of the corresponding 3d $\mathcal{N}=2$ IR/IR interface theory ${}_{\sigma_L}\mathbb{I}_{\sigma_R}$. The corresponding index identity reads:
    \begin{equation}
    \begin{split}
        \mathcal{I}^{SU(N)}(q) = & \sum_{\vec{m} \in \mathbb{Z}^{N-1}} \sum_{\vec{n} \in \mathbb{Z}^{N-1}} \oint \prod_{j=1}^{N-1} \frac{d \xi_j}{2 \pi i \xi_j} \frac{1}{N!} \oint \prod_{j=1}^{N-1} \frac{d \eta_j}{2 \pi i \eta_j} 
        \times \\ & \qquad \times 
        \mathcal{II}^{SW_{\sigma_L}}(  \vec{\xi},\vec{m} ;q) \mathcal{I}^{{}_{\sigma_L}\mathbb{I}_{\sigma_R}}(\{\vec{\xi},\vec{m}\}, \{\vec{\eta},\vec{n}\} ;q) \mathcal{II}^{SW_{\sigma_R}}(  \vec{\eta},\vec{n} ;q)
    \end{split}
    \end{equation}
\end{itemize}
To simplify the notation we will drop the $q$-dependence from the identities in what follows, therefore the Schur index of $SU(N)$ SYM will be given symply by $\mathcal{I}^{SU(N)}$.

Both setups are expected to match the $S^3 \times S^1$ Schur index of $\mathcal{N}=2$ SYM. We now proceed to describe in more detail these identities that can be both proved analytically as consequence of Section \ref{sec: half-index}, using exact mathematical identities for the index of the $\Twall[]$ theory.  

\subsection{UV/IR interface}
We now describe the first setup, which relates the Schur index of $\mathcal{N}=2$ SYM and the setup given by the UV/IR interface between $\mathcal{N}=2$ SYM and the low energy effective description.
This setup is the same interface represented in \eqref{eq:RG_wall_flow_electric}, taking euclidean $S^3 \times S^1$ as the spacetime and inserting the interface at an $S^2$ equator of $S^3$. 

This case can be simply seen as a direct consequence of the proposal of Section \ref{sec: bc_proposal}, already tested perturbatively in Section \ref{sec: half-index}.
In fact, we already motivated the following identity between half-indices:
\begin{equation} \label{eq: rgwall_halfindidreference}
    \mathcal{II}^{SU(N)}(\vec{u}) = \sum_{\vec{m} \in \mathbb{Z}^{N-1}} \oint \prod_{j=1}^{N-1} \frac{d \xi_j}{2 \pi i \xi_j} \mathcal{I}^{\sigma \cdot \Twall[N]}(\vec{u}, \{ \vec{\xi},\vec{m} \}) \mathcal{II}^{SW_\sigma}( \vec{\xi}, \vec{m} )
\end{equation}
Once this identity is established, we can start from \eqref{eq: gluingtwohalfindices_susym} and it is a direct consequence that:
\begin{equation} \label{eq: halfind_uvirinterf}
\begin{split}
    \mathcal{I}^{SU(N)} & = \frac{1}{N!} \oint \prod_{j=1}^N \frac{d u_j}{2 \pi i u_j} \Delta_N(\vec{u}) \mathcal{II}^{SU(N)}(\vec{u}^{-1}) \mathcal{II}^{SU(N)}(\vec{u}) = 
    \\ & = 
    \sum_{\vec{m} \in \mathbb{Z}^{N-1}} \oint \prod_{j=1}^{N-1} \frac{d \xi_j}{2 \pi i \xi_j} 
    \frac{1}{N!} \oint \prod_{j=1}^{N-1} \frac{d u_j}{2 \pi i u_j} \Delta_N(\vec{u}) 
    \times \\ & \qquad \times 
    \mathcal{II}^{SU(N)}(\vec{u}^{-1}) \mathcal{I}^{\sigma \cdot \Twall[N]}(\vec{u}, \{\vec{\xi},\vec{m}\}) \mathcal{II}^{SW_\sigma}(  \vec{\xi},\vec{m} )
\end{split}
\end{equation}
Which we also tested perturbatively for $N=2,3$ in various duality frames.

The index identity in \eqref{eq:RG_interface_index_identity} is a strong check of our proposal that the 3d theory $\Twall[]$ is the RG-wall of $\NN=2$ SYM. Namely, $\Twall[]$ is the interface theory obtained by letting the UV theory flow only on half of spacetime.

\subsection{IR/IR interface}

The index identity corresponding to the IR/IR interface is also a direct consequence of already established mathematical identities, as we now proceed to explain.

We can start from the last identity of the previous section in \eqref{eq: halfind_uvirinterf} and replace the remaining half-index of $\mathcal{N}=2$ SYM using the identity in \eqref{eq: rgwall_halfindidreference}:
\begin{equation} \label{eq: halfind_iririnterf_step1}
\begin{split}
    \mathcal{I}^{SU(N)} & = 
    \sum_{\vec{m} \in \mathbb{Z}^{N-1}} \oint \prod_{j=1}^{N-1} \frac{d \xi_j}{2 \pi i \xi_j} 
    \frac{1}{N!}
    \oint \prod_{j=1}^{N-1} \frac{d u_j}{2 \pi i u_j} \Delta_N(\vec{u}) 
    \times \\ & \qquad \times 
    \mathcal{II}^{SU(N)}(\vec{u}^{-1}) \mathcal{I}^{\sigma_R \cdot \Twall[N]}(\vec{u}, \{\vec{\xi},\vec{m}\}) \mathcal{II}^{SW_{\sigma_R}}(  \vec{\xi},\vec{m} ) = 
    \\ & = 
    \sum_{\vec{m} \in \mathbb{Z}^{N-1}} \oint \prod_{j=1}^{N-1} \frac{d \xi_j}{2 \pi i \xi_j} \sum_{\vec{n} \in \mathbb{Z}^{N-1}} \oint \prod_{j=1}^{N-1} \frac{d \eta_j}{2 \pi i \eta_j} 
    \times \\ & \qquad \times
    \mathcal{II}^{SW_{\sigma_L}}(  \vec{\xi},\vec{m} ) 
    \times \\ & \qquad \times  
    \frac{1}{N!} \oint \prod_{i=1}^{N-1} \frac{d u_i}{2 \pi i u_i} \Delta_N(\vec{u})
    \mathcal{I}^{\sigma_L \cdot \Twall[N]}(\vec{u}^{-1}, \{\vec{\xi},\vec{m}\})
    \mathcal{I}^{\sigma_R \cdot \Twall[N]}(\vec{u}, \{\vec{\eta},\vec{n}\}) 
    \times \\ & \qquad \times
    \mathcal{II}^{SW_{\sigma_R}}(  \vec{\eta},\vec{n} )
\end{split}
\end{equation}
In the last expression we can recognize the definition of the IR/IR interface by collecting the contribution of the $SU(N)$ gauging of the two $\Twall[]$ theories.
\begin{equation} \label{eq: idwall_ind_definit}
\begin{split}
    & \mathcal{I}^{{}_{\sigma_L}\mathbb{I}_{\sigma_R}} (\{ \vec{\xi},\vec{m} \}, \{\vec{\eta},\vec{n}\}) = 
    \\ & \qquad =
    \frac{1}{N!} \oint \prod_{j=1}^{N-1} \frac{d u_j}{2 \pi i u_j} \Delta_N(\vec{u})
    \mathcal{I}^{\sigma_L \cdot \Twall[N]}(\vec{u}^{-1}, \{\vec{\xi},\vec{m}\})
    \mathcal{I}^{\sigma_R \cdot \Twall[N]}(\vec{u}, \{\vec{\eta},\vec{n}\}) 
\end{split}
\end{equation}
This identity is almost identical to the definition of the IR/IR interface.
In fact, the index of the interface, as it is in \eqref{eq: idwall_ind_definit}, does not properly define the index of a 3d theory. The reason being that we are lacking of a sum over magnetic fluxes for the $SU(N)$ gauge group. However, one can formally introduce a sum over magnetic fluxes of $SU(N)$ in \eqref{eq: halfind_iririnterf_step1}, as their contribution is suppressed by the fact that the half-index of $SU(N)$ SYM is zero for any non-zero magnetic flux, as we pointed out already in Section \ref{sec: half-index}. This is true also for the setup given by the daul half-index of the low energy theory glued to the $\Twall[]$. 
We can therefore formally rewrite the index of the IR/IR interface as:
\begin{equation} \label{eq: idwall_ind_definit_versfluxes}
\begin{split}
    &\mathcal{I}^{{}_{\sigma_L}\mathbb{I}_{\sigma_R}} (\{ \vec{\xi},\vec{m} \}, \{\vec{\eta},\vec{n}\}) = 
    \\ & \qquad = 
    \frac{1}{N!} \sum_{\vec{m}^{(u)} \in \Gamma} \oint \prod_{j=1}^{N-1} \left( \frac{d u_j}{2 \pi i u_j} u_j^{m^{(u)}_j + \sum_{j=1}^{N-1} m_j^{(u)} } 
    \right) 
    \mathcal{I}_{\text{vec}}^{(N)}(\vec{u},\vec{m}^{(u)})
    \times \\ & \qquad \quad \times 
    \mathcal{I}^{\sigma_L \cdot \Twall[N]}( \{ \vec{u}^{-1}, -\vec{m}^{(u)} \} , \{\vec{\xi},\vec{m}\})
    \mathcal{I}^{\sigma_R \cdot \Twall[N]}( \{ \vec{u}, \vec{m}^{(u)} \}, \{\vec{\eta},\vec{n}\}) 
\end{split}
\end{equation}
where we have done the following:
\begin{itemize}
    \item Introduced a sum over a lattice for $\Gamma$ for magnetic fluxes of $SU(N)$:
    \begin{equation}
        \Gamma = \mathbb{Z}^{N-1} + \frac{k}{N} \qquad \text{for $k = 0,\ldots,N-1$}
    \end{equation}
    Which means that we sum not only over integer fluxes but also over fractional fluxes such that $\text{Mod}[m_i^{(u)},1] = k/N$. The fact that we should sum over fractional magnetic fluxes has to do with a subtlety in the quiver in \eqref{eq: magn_ir_interf}, where the actual gauge group should be regarded as 
    \begin{equation}
        \frac{U(1)^2 \times U(2)^2 \times \ldots U(N-1)^2 \times U(N)}{U(1)} = \frac{U(1)^2 \times U(2)^2 \times \ldots U(N-1)^2 \times SU(N)}{\mathbb{Z}_N}
    \end{equation}
    where we decoupled the $U(1)$ factor from the $U(N)$ gauge group, by a suitable redefinition, but we are still left with a gauged $\mathbb{Z}_N$ factor, that implies the sum over fractional magnetic fluxes. 

    \item The Haar-measure $\Delta_N$ is replaced by the contribution of a 3d vector multiplet, that are identical for zero magnetic flux. Also, we introduced the $\vec{m}^{(u)}$ dependence in the index of the $\Twall[]$ theories.

    \item We added the contribution of a CS-level $1$ for the $SU(N)$ gauge symmetry. This choice might seem somehow arbitrary as it can not be detected for $\vec{m}^{(u)} = \textbf{0}$. However, we pick this choice for the CS level based on physical arguments, that we provided in Section \ref{sec: interfaces}, and on the fact that only with this precise choice the 3d theory behaves as expected, as we now describe.
\end{itemize}
We stress that the index of the 3d theory in \eqref{eq: idwall_ind_definit_versfluxes} does change when we sum over the lattice of magnetic fluxes $\Gamma$, w.r.t.~the situation where we do not sum. However, when this index is considered inside the 4d/3d coupled setup in \eqref{eq: halfind_iririnterf_step1}, the extra contributions are suppressed by a delicate interplay with the bulk indices and the gauging.

Following the logic described in Section \ref{sec: interfaces}, we first pick a choice of the pair $(\sigma_L,\sigma_R)$ which is convenient, that is $(S,S)$. In this duality frame the index of the IR/IR interface in \eqref{eq: idwall_ind_definit_versfluxes} reduces to the so-called identity wall theory, that was studied in \cite{Benvenuti:2025huk}, where it was proved that the partition function is a distribution corresponding to $N-1$ free chirals and $N-1$ pairs of periodic Dirac-delta and Kronecker-delta:\footnote{Notice that in \cite{Benvenuti:2025huk} it is reported the identity between $S^3$ partition functions. However, it is described that the identity can be proved using a sequence of IR-dualities, thus proving that is a statement about the whole SCFT. Here we do not prove explicitly the mathematical identity for the $S^2 \times S^1$ index, but simply translate the SCFT statement as the corresponding expected identity. More details about this identity are given in Appendix \ref{app: gsun}.}
\begin{equation} \label{eq: genN_magnmagninterf}
    \mathcal{I}^{{}_S\mathbb{I}_S} (\{ \vec{\xi},\vec{m} \}, \{\vec{\eta},\vec{n}\}) = \prod_{i=1}^{N-1} \delta^{(p)}(\xi_i - \eta_i) \delta_{m_i,n_i} \xi_i^{m_i/2} \mathcal{I}^{(1)}_{\text{chir}}(\xi_i^{-1},-m_i) 
\end{equation}
Where the periodic Delta-function has support whenever the $\xi_i$ and $\eta_i$ fugacities are equal, implying that their arguments are equal modulo $2 \pi $ factors. Also, $\delta_{m,n}$ is the standard Kronecker-delta.

Now, taking this frame as the basic one, we can perform $Sp(2(N-1),\mathbb{Z})$ transformations to find the index of any generic interface. In particular, we can recover the interface between two electric frames as:
\begin{equation}
\begin{split}
    \mathcal{I}^{{}_1\mathbb{I}_1} (\{ \vec{\xi},\vec{m} \}, \{\vec{\eta},\vec{n}\}) & = 
    \sum_{\vec{m}' \in \mathbb{Z}^{N-1}} \oint \prod_{j=1}^{N-1} \frac{d \xi'_j \xi_j^{-m'_j} (\xi'_j)^{-m_j}}{2 \pi i \xi'_j} 
    \sum_{\vec{n}' \in \mathbb{Z}^{N-1}} \oint \prod_{j=1}^{N-1} \frac{d \eta'_j \eta_j^{-n'_j} 
    (\eta'_j)^{-n_j}}{2 \pi i \eta'_j}
    \times \\ & \qquad \times
    \mathcal{I}^{{}_S\mathbb{I}_S} (\{ \vec{\xi},\vec{m} \}, \{\vec{\eta},\vec{n}\};q) = 
    \\ & = 
    \sum_{\vec{m}' \in \mathbb{Z}^{N-1}} \oint \prod_{j=1}^{N-1} \frac{d \xi'_j (\xi_j \eta_j)^{-m'_j} (\xi'_j)^{m'_j/2 -m_j-n_j}}{2 \pi i \xi'_j} \mathcal{I}^{(1)}_{\text{chir}}({\xi'}^{-1}_j, -m'_j)
    \\ & = 
    \prod_{j=1}^{N-1} (-\sqrt{q})^{-(m_j+n_j)/2 } (\xi_j \eta_j)^{-(m_j+n_j)/2} \mathcal{I}^{(0)}_{\text{chir}}( \xi_j \eta_j , m_j+n_j)
\end{split}
\end{equation}
where in the last step we recognized the index of $N-1$ copies of $\mathcal{N}=2$ SQED with a chiral of charge $-1$ and CS level $1/2$. Each of that copies is IR-dual to a single free-chiral with background CS-level $-1$. Adjusting for the extra background terms and keeping track of all the details, we land on the final line,
which is simply the index of $N-1$ free chirals, that from now on we will consider the basic IR/IR interface.

The match of the index obtained from the IR/IR interface and the Schur index is a strong check that gluing two RG-walls produces the trivial defect in the IR, see the discussion in Section \ref{sec: interfaces}. 
Roughly speaking, this also provides evidence that our RG-wall is “as transparent as it can be", as we are able to glue two such walls into a trivial defect in the IR, which is fully transparent to all operators in the IR. As discussed in \cite{Gaiotto:2012np}, this is a necessary property for an interface between a UV and an IR to be a good RG-wall.

In the rest of the section we will describe a selection of examples to describe in more detail the matching between the Schur index of $\mathcal{N}=2$ SYM and the setup given by the IR/IR interface.

\subsubsection{$SU(2)$ electric/electric interface}

As a first example we consider the interface between the IR effective theory of $SU(2)$ SYM in the electric duality frame and itself.

As described above the interface theory consists of a single free chiral that we consider to be charged under two abelian global symmetries $U(1)_\xi \times U(1)_\eta$, with background CS interactions with level:
\begin{equation}
\begin{split}
    & k_{\xi\xi} = k_{\eta \eta} = k_{\xi \eta} = -1/2 \\
    & k_{\xi R} = k_{\eta R} = -1/2
\end{split}
\end{equation}
The associated index reads:
\begin{equation}
    \mathcal{I}^{{}_1\mathbb{I}_1}( \{ \xi,m \}, \{ \eta,n \}) = (-\sqrt{q})^{-(m+n)/2} (\xi \, \eta)^{-(m+n)/2} \mathcal{I}^{(0)}_{\text{chir}}( \xi \, \eta, m+n)
\end{equation}
In the bulk, we consider on both sides of the interface the IR effective theory in the electric frame. As already discussed in Section \ref{sec: half-index}, the corresponding half-index is computed as in \eqref{eq: SU2_indDir}.  The total index thus reads:
\begin{equation}
    \sum_{m,n \in \mathbb{Z}} \oint \frac{d \xi}{2 \pi i \xi} \frac{d \eta}{2 \pi i \eta} \mathcal{II}^{SW}(\xi,m) \mathcal{I}^{{}_1 \mathbb{I}_1}( \{ \xi,m \}, \{ \eta,n \}) \mathcal{II}^{SW}(\eta,n)
\end{equation}
which upon perturbative expansion can be checked to agree with the Schur index of $SU(2)$ SYM, that up to few relevant orders in $q$ reads:
\begin{equation}
    \mathcal{I}^{SU(2)}(q) = 1 +q^2+q^6+q^{12} + ...
\end{equation}

\subsubsection{$SU(2)$ electric/magnetic interface}

By performing a suitable $SL(2,\mathbb{Z})$ transformation on the IR/IR interface between two electric frames we can obtain the interface between the electric and magnetic frames. More explicitly, we perform an $S \in SL(2,\mathbb{Z})$ transformation as described in Section \ref{subsec: sl2z_transf} to get the following 3d index:
\begin{equation}
\begin{split}
    \mathcal{I}^{_1 \mathbb{I}_S}( \{\xi, m\}, \{\eta, n\}) &= \sum_{n' \in \mathbb{Z}} \oint \frac{d \eta'}{2\pi i \eta'} \eta^{n'} {\eta'}^n \mathcal{I}^{_1 \mathbb{I}_1}( \{\xi, m\}, \{\eta', n'\}) \\
    &= \sum_{n' \in \mathbb{Z}} \oint \frac{d \eta'}{2\pi i \eta'} \eta^{n'} {\eta'}^n (-\sqrt{q})^{-(m+n)/2} (\xi \eta)^{-(m+n)/2} \mathcal{I}^{(0)}_{\text{chir}}(\xi\eta,m+n)
\end{split}
\end{equation}
To simplify the computation, we can observe that the index is that of a SQED with CS level $-1/2$ and one chiral with charge $+1$. We can use a basic mirror duality which relates this theory to a single free chiral and, by keeping track of all the details, we obtain the following index:
\begin{equation}
    \mathcal{I}^{_1 \mathbb{I}_S}( \{\xi, m\}, \{\eta, n\}) = \xi^{-n} \eta^{n/2-m} \mathcal{I}^{(1)}_{\text{chir}}(\eta^{-1},-n)
\end{equation}
which simply consists of a free chiral with background CS level $1/2$ and a BF coupling.

The index of the total setup finally reads:
\begin{equation}
    \sum_{m,n \in \mathbb{Z}} \oint \frac{d\xi}{2\pi i \xi} \frac{d \eta}{2\pi i \eta} \mathcal{II}^{SW_1}(\xi,m) \mathcal{I}^{_1 \mathbb{I}_S}( \{\xi, m\}, \{\eta, n\}) \mathcal{II}^{SW_S}(\eta,n) 
\end{equation}
where the two IR indices are computed as follows:
\begin{equation}
\begin{split}
    &\mathcal{II}^{SW_1}(\xi,m) = (q;q) \Tr_{\gamma_1} \left[: S(q) (X_{-\frac{\gamma_1}{2}-\frac{\gamma_2}{2}})^m :\right] \bigg|_{\Tr[X_{\gamma_1}] = \xi} \\
    &\mathcal{II}^{SW_S}(\eta,n) = (q;q) \Tr_{\frac{1}{2}(\gamma_1+\gamma_2)} \left[: S(q) X^n_{\gamma_1} :\right] \bigg|_{\Tr[X_{\frac{1}{2}(\gamma_1+\gamma_2)}] = \eta}
\end{split}
\end{equation}
Expanding perturbatively as a power series in $q$ we find perfect agreement with the Schur index of $SU(2)$ SYM.

\subsubsection{$SU(2)$ magnetic/dyonic interface}

As a last example we consider the interface between the magnetic and dyonic frames.
This can be obtained starting from the previous interface and performing an $S^{-1}T^2$ transformation on the electric frame. The index of the interface becomes:
\begin{equation}
\begin{split}
    \mathcal{I}^{_{S^{-1}T^2} \mathbb{I}_S}( \{\xi,m\}, \{\eta,n\}) &= 
    \sum_{m' \in \mathbb{Z}} \oint \frac{d \xi'}{2\pi i \xi'} \xi^{-m'} {\xi'}^{-m+2m'} \mathcal{I}^{_1 \mathbb{I}_S}( \{\xi', m' \}, \{\eta, n\}) \\
    &= \eta^{n/2} \mathcal{I}^{(1)}_{\text{chir}}(\eta^{-1},-n) \sum_{m' \in \mathbb{Z}} \oint \frac{d \xi'}{2\pi i \xi'} {\xi'}^{2m'-m-n} (\xi \eta)^{-m'}
\end{split}
\end{equation}
which describes a free chiral times a $U(1)_{2}$ TQFT.

The total setup has index:
\begin{equation}
    \sum_{m,n \in \mathbb{Z}} \oint \frac{d\xi}{2\pi i \xi} \frac{d \eta}{2\pi i \eta} \mathcal{II}^{SW_{S^{-1}T^2}}(\xi,m) 
    \mathcal{I}^{_{S^{-1}T^2} \mathbb{I}_S}( \{\xi, m\}, \{\eta, n\}) \mathcal{II}^{SW_S}(\eta,n) 
\end{equation}
where the two IR indices are computed as follows:
\begin{equation}
\begin{split}
    &\mathcal{II}^{SW_{S^{-1}T^2}}(\xi,m) = (q;q) \Tr_{\gamma_2} \left[: S(q) (X_{\frac{\gamma_1}{2}+\frac{\gamma_2}{2}})^m :\right] \bigg|_{\Tr[X_{\gamma_1}] = \xi} \\
    &\mathcal{II}^{SW_S}(\eta,n) = (q;q) \Tr_{\frac{\gamma_1}{2}+\frac{\gamma_2}{2}} \left[ : S(q) X^n_{\gamma_1} : \right] \bigg|_{\Tr[X_{\frac{\gamma_1}{2}+\frac{\gamma_2}{2}}] = \eta}
\end{split}
\end{equation}
Expanding perturbatively as a power series in $q$ we find perfect agreement with the Schur index of $SU(2)$ SYM.

\subsubsection{$SU(3)$ electric/electric interface}

We now move to an higher rank example of an IR/IR interface between two electric frames of $SU(3)$ SYM.

The interface consists of two copies of the $SU(2)$ electric/electric interface, namely it is a theory of two free chirals with BF coulings whose index reads:
\begin{equation}
    \mathcal{I}^{_1 \mathbb{I}_1}( \{ \vec{\xi},\vec{m} \}, \{ \vec{\eta}, \vec{n}\}) = \prod_{i=1}^2 (-\sqrt{q})^{-(m_i+n_i)/2} (\xi_i \eta_i)^{-(m_i+n_i)/2} \mathcal{I}^{(0)}_{\text{chir}}( \xi_i \eta_i, m_i + n_i)
\end{equation}
We stress that this theory is exactly the product of two copies of the $SU(2)$ electric/electric interface as there are no interactions between the them.

The total index reads:
\begin{equation}
    \sum_{\vec{m},\vec{n} \in \mathbb{Z}^2} \oint \prod_{j=1}^2 \frac{d \xi_j}{2 \pi i \xi_j} \frac{d \eta_j}{2 \pi i \eta_j} \mathcal{II}^{SW_1}( \vec{\xi},\vec{m}) \mathcal{I}^{_1 \mathbb{I}_1}( \{ \vec{\xi},\vec{m} \}, \{ \vec{\eta}, \vec{n}\}) \mathcal{II}^{SW_1}( \vec{\eta},\vec{n}) 
\end{equation}
which upon perturbative expansion agrees with the Schur index of $SU(3)$ SYM, which up to few relevant orders in $q$ reads:
\begin{equation}
    \mathcal{I}^{SU(3)}(q) = 1 + q^2 + 2q^4 + ...
\end{equation}

\subsubsection{$SU(3)$ electric/magnetic interface}

We can now construct an interface between the electric and magnetic frames of $SU(3)$, which can be obtained by performing an $S \in Sp(4,\mathbb{Z})$ transformation. The index becomes:
\begin{equation}
\begin{split}
    \mathcal{I}^{_1 \mathbb{I}_S}( \{\vec{\xi} , \vec{m} \}, \{\vec{\eta} , \vec{n} \}) &= \sum_{\vec{n}' \in \mathbb{Z}^2} \oint \prod_{j=1}^2 \frac{d \eta'_j}{2\pi i \eta'_j} \eta_j^{n'_j} {\eta'_j}^{n_j} \mathcal{I}^{_1 \mathbb{I}_1}( \{ \vec{\xi} , \vec{m} \}, \{\vec{\eta} ', \vec{n} '\}) 
\end{split}
\end{equation}
To each copy of SQED with CS level $-1/2$ and one chiral we can apply a basic mirror duality relating it to a free chiral and obtain the following index:
\begin{equation}
    \mathcal{I}^{_1 \mathbb{I}_S}( \{\vec{\xi} , \vec{m} \}, \{\vec{\eta} , \vec{n} \}) = \prod_{i=1}^2 \xi_i^{-n_i} \eta_i^{n_i/2-m_i} \mathcal{I}^{(1)}_{\text{chir}}( \eta_i^{-1},-n_i)
\end{equation}
which is the index of copies of a free chiral with a BF coupling.

The total index reads:
\begin{equation}
    \sum_{\vec{m},\vec{n} \in \mathbb{Z}^2} \oint \prod_{j=1}^2 \frac{d \xi_j}{2 \pi i \xi_j} \frac{d \eta_j}{2 \pi i \eta_j} \mathcal{II}^{SW_1}( \vec{\xi},\vec{m}) \mathcal{I}^{_1 \mathbb{I}_S}( \{ \vec{\xi},\vec{m} \}, \{ \vec{\eta}, \vec{n}\}) \mathcal{II}^{SW_S}( \vec{\eta},\vec{n}) 
\end{equation}
which can be perturbatively expanded and agrees with the Schur index of $SU(3)$ SYM.

\subsection{A TQFT interface formula}

We now describe how one can recover a mathematical identity expressing the gluing of two IR half-indices with an IR/IR interface which formally reads as a TFT.

Let us consider the IR index in the magnetic frame. 
As discussed in Section \ref{sec: half-index}, the half-index is obtained from the quantum spectrum operator $S(q)$ as:
\begin{equation}
    \mathcal{II}^{SW_S}(\vec{\xi},\vec{m};q) = (q;q)^{N-1}\Tr_{\{ \gamma_1,\gamma_3,\ldots \}} \left[
    S(q)\prod_{i=1}^{N-1} X_{\tilde{\gamma}_i}^{m_i}
    \right] 
\end{equation}
where the trace identifies:
\begin{equation}
    \Tr_{\gamma_i}[X_{\gamma_i}] = \xi_i
\end{equation}
and the set $\{ \tilde{\gamma}_i \}$ are the magnetic particles w.r.t.~the set of electric particles $\gamma_{2i+1}$.

One can consider a different identification, where we instead use the inverse fugacities:
\begin{equation}
    \widetilde{\Tr}_{\gamma_i}[X_{\gamma_i}] = \xi_i^{-1}
\end{equation}
And define the following quantity:
\begin{equation}
    \widetilde{\mathcal{II}}^{SW_S}(\vec{\xi},\vec{m};q) = 
    (q;q)^{N-1}
    \widetilde{\Tr}_{\gamma_1,\gamma_3,\ldots} \left[
    S(q)\prod_{i=1}^{N-1} X_{\tilde{\gamma}_i}^{m_i}
    \right] 
\end{equation}

The two different traces satisfy the following index identity:
\begin{equation}
    \mathcal{II}(\vec{\xi},\vec{m};q) = \widetilde{\mathcal{II}}(\vec{\xi},\vec{m};q)  \prod_{i=1}^{N-1} \xi^{-m_i} \frac{(q^{\frac{1}{2}(1+m_i)} \xi_i^{-1};q)}{(q^{\frac{1}{2}(1+m_i)} \xi_i;q)} = 
    \widetilde{\mathcal{II}}(\vec{\xi},\vec{m};q) \prod_{i=1}^{N-1} \xi_i^{-\frac{m_i}{2}} \mathcal{I}_{\text{chir}}^{(1)}(\xi_i,m_i)
\end{equation}
which we tested as a perturbative $q$-expansion.
We then observe that the operation of changing the trace identification, as $\xi_i \to \xi_i^{-1}$, it amounts to the multiplication by the index of $N-1$ free chiral multiplets with R-charge 1 and background CS level $-1/2$ for each $U(1)_{\xi_i}$ symmetry.

We can now use this identity in the definition of the IR/IR interface between two magnetic frames when considered glued to the two IR half-indices, that reads: 
\begin{equation} 
\begin{split}
    & \mathcal{I}^{SU(N)} = 
    \sum_{\vec{m} \in \mathbb{Z}^{N-1}} \oint \prod_{j=1}^{N-1} \frac{d \xi_j}{2 \pi i \xi_j} \sum_{\vec{n} \in \mathbb{Z}^{N-1}} \oint \prod_{j=1}^{N-1} \frac{d \eta_j}{2 \pi i \eta_j} 
    \times \\ 
    & \qquad \qquad \quad \times
    \mathcal{II}^{SW_S}(  \vec{\xi},\vec{m} ) 
    \mathcal{I}^{_S \mathbb{I}_S}( \{\vec{\xi},\vec{m} \}; \{\vec{\eta}, \vec{n} \})
    \mathcal{II}^{SW_S}(  \vec{\eta},\vec{n} ) 
    \\ &
    \mathcal{I}^{_S \mathbb{I}_S}( \{\vec{\xi},\vec{m} \}; \{\vec{\eta}, \vec{n} \}) = \prod_{i=1}^{N-1} \delta^{(p)}(\xi_i - \eta_i) \delta_{m_i,n_i} \xi_i^{m_i/2} \mathcal{I}^{(1)}( \xi_i^{-1},-m_i) 
\end{split}
\end{equation}
We see that we can simply reabsorb the chirals and CS-levels in the definition of the interface to perform the redifinition $\mathcal{II} \to \widetilde{\mathcal{II}}$
for one of the two half-indices, finding the following identity:
\begin{equation} \label{eq: magneti/magnetic_tftinterface}
\begin{split}
    & \mathcal{I}^{SU(N)} = 
    \sum_{\vec{m} \in \mathbb{Z}^{N-1}} \oint \prod_{j=1}^{N-1} \frac{d \xi_j}{2 \pi i \xi_j} \sum_{\vec{n} \in \mathbb{Z}^{N-1}} \oint \prod_{j=1}^{N-1} \frac{d \eta_j}{2 \pi i \eta_j} 
    \times \\ 
    & \qquad \qquad \quad \times
    \mathcal{II}^{SW_S}(  \vec{\xi},\vec{m} ) 
    \mathcal{I}^{_S \mathbb{I}_S}( \{\vec{\xi},\vec{m} \}; \{\vec{\eta}, \vec{n} \})
    \widetilde{\mathcal{II}}^{SW_S}(  \vec{\eta},\vec{n} ) 
    \\ &
    \mathcal{I}^{_S \mathbb{I}_S}( \{\vec{\xi},\vec{m} \}; \{\vec{\eta}, \vec{n} \}) = \prod_{i=1}^{N-1} \delta^{(p)}(\xi_i - \eta_i) \delta_{m_i,n_i}
\end{split}
\end{equation}
which can be interpreted as the gluing of two half-indices with an interface which is a ``trivial" theory, as it is simply a delta function. Notice that one could re-interpret the superconformal index of the interface as that of $N-1$ copies of a $U(1)_0$ TFT, whose index is just a delta-function \cite{Kapustin:1999ha}\footnote{We could further redefine $\eta_i \to \eta_i^{-1}$ as we are integrating over the unitary circle. This reverts the redefinition $\widetilde{\mathcal{II}} \to \mathcal{II}$ and modify slightly the index of the interface. We will however not perform this further transformation for a simpler reading of the index of the interface.}.

As the interface factorizes in $N-1$ copies of the same theory, we now analyze the consequence of \eqref{eq: magneti/magnetic_tftinterface} in the case $N=2$, the generalization should be straightforward. 
If we now apply $SL(2,\mathbb{Z})$ duality tranformations to the interface, as there is no matter in the first place, the index will always be that of a TFT. Let us now look at some examples.
\begin{itemize}
    \item The new magnetic/magnetic interface is now a completely trivial theory, or a delta-function:
    \begin{equation}
        \mathcal{I}^{_S \mathbb{I}_S}( \{ \xi,m \}, \{ \eta,n \}) = \delta(\xi - \eta) \delta_{m,n}
    \end{equation}

    \item Performing an $S^{-1} \in SL(2,\mathbb{Z})$ transformation we can find the interface between the electric and magnetic frames, which is now simply a BF coupling:
    \begin{equation}
        \mathcal{I}^{_1 \mathbb{I}_S}( \{ \xi,m \}, \{ \eta,n \}) = \xi^{-m} \eta^{-n}
    \end{equation}

    \item The magnetic/dyonic interface can be obtained by performing now a $S^{-1}T^2 \in SL(2,\mathbb{Z})$ transformation, which yields a level $U(1)_{-2}$ TQFT, which couples to the bulk trough the gauging of the topological symmetry.
    \begin{equation}
        \mathcal{I}^{_{S^{-1}T^2} \mathbb{I}_S}( \{ \xi,m \}, \{ \eta,n \}) = \sum_{r \in \mathbb{Z}} \oint \frac{d u}{2 \pi i u} u^{2r-m-m} (\xi \eta)^{-r}
    \end{equation}
    Which is reminiscent of the interface for 4d $\mathcal{N}=1$ SYM of \cite{Acharya:2001dz}. It would be interesting further analyze this mathematical fact in the context of the possible connection with the Acharya-Vafa domain walls, as discussed in Section \ref{sec: interfaces}.
\end{itemize}

\acknowledgments

We are grateful to Sara Pasquetti for early collaboration on the project and for many interesting discussions. We are grateful to Antonio Amariti, Noppadol Mekareeya, Sergio Benvenuti and Tudor Dimofte for discussions on this project and related topics.
RC is supported by the STFC grant ST/X000575/1.
SR is supported by the MUR-PRIN grant
No. 2022NY2MXY and the POC grant No. 41355/GRFVG.

\noindent
To the campfire awaiting us at the end of the Universe.

\appendix

\section{Review of 3d Theories} \label{app: 3dwalls}

\subsection{The S-duality wall: $T[SU(N)]$} \label{app: tsun}

The $T[SU(N)]$ theory is a 3d $\mathcal{N}=4$ SCFT. It is notably possible to realize this theory from Type IIB brane setups \cite{Hanany:1996ie,Gaiotto:2008ak} and, as discussed in the bulk of the paper, it is the S-duality wall of 4d $\mathcal{N}=4$ $SU(N)$ SYM.
In this appendix we quickly review notable properties of this theory.

The $T[SU(N)]$ theory is a 3d $\mathcal{N}=4$ SCFT which admits a UV Lagrangian description that is a linear quiver theory:
\begin{equation} \label{eq: tsun_lagrangian}
    T[SU(N)] \quad : \qquad\qquad
    \begin{tikzpicture}[baseline=(current bounding box).center]
        
        \node at (0,0) (n1) [gauge,black] {$1$};
        \node at (1.5,0) (n2) [gauge,black] {$2$};
        \node at (3,0) (n3) {$\cdots$};
        \node at (4.5,0) (nnm1) [gauge,black] {$\scriptstyle N-1$};
        \node at (6,0) (nn) [flavor,black] {$N$};
        
        \draw[-] (n1.north east)      arc[start angle=-45, end angle=225, radius=3mm];
        \draw[-] (n2.north east)      arc[start angle=-45, end angle=225, radius=3mm];
        \draw[-] (nnm1.north east)    arc[start angle=-45, end angle=225, radius=3.1mm];
        
        \draw[->-] (n1.20) --   (n2.160);
        \draw[-<-] (n1.-20) --  (n2.-160);
        \draw[->-] (n2.20) --  ++(.75,0);
        \draw[-<-] (n2.-20) -- ++(.75,0);
        \draw[->-] (nnm1.160) --  ++(-.75,0);
        \draw[-<-] (nnm1.-160) -- ++(-.75,0);
        \draw[->-] (nnm1.20) --   (nn.160);
        \draw[-<-] (nnm1.-20) --  (nn.-160);

        \draw[FIcolor] (n1)++(0,-.5) node {\scriptsize{$y_1/y_2$}};
        \draw[FIcolor] (n2)++(0,-.5) node {\scriptsize{$y_2/y_3$}};
        \draw[FIcolor] (nnm1)++(0,-.5) node {\scriptsize{$y_{N-1}/y_N$}};

        \draw[gray] (nn)++(0,.5) node {\scriptsize{$\vec{x}$}};

        \draw[] (n1)++(0,1) node {\scriptsize{$\phi_1$}};
        \draw[] (n2)++(0,1) node {\scriptsize{$\phi_2$}};

        \draw ($(n1)!0.5!(n2)+(0,.4)$) node {\scriptsize{$Q_1$}};
        \draw ($(n1)!0.5!(n2)+(0,-.4)$) node {\scriptsize{$\tilde{Q}_1$}};
        \draw ($(n2)!0.5!(n3)+(0,.4)$) node {\scriptsize{$Q_2$}};
        \draw ($(n2)!0.5!(n3)+(0,-.4)$) node {\scriptsize{$\tilde{Q}_2$}};

        \draw (n3)++(0,-1.25) node {$\mathcal{W}= \sum_{i=1}^{N-1} \Tr [Q_i \phi_i \tilde{Q}_i] $};
        
    \end{tikzpicture}
\end{equation}
Notice that the theory is depicted in $\mathcal{N}=2^*$ notation, so that $\mathcal{N}=4$ vector multiplets are split into $\mathcal{N}=2$ vectors and an adjoint chiral $\phi_i$, while hypermultiplets are split into pairs of chiral multiplets $(Q_i,\tilde{Q}_i)$ in conjugated representations. We label by $\vec{x}$ the set of fugacities associated to the $SU(N)$ flavor symmetry, so that $\prod_{i=1}^N x_i = 1$. Also, we use a $N$-dimensional vector $\vec{y}$, satisfying $\prod_{i=1}^N y_i =1$, to parameterize the fugacities associated to the topological symmetries as indicated in the figure. 
The theory possesses a superpotential containing only cubic interactions between adjoint chirals and hypermultiplets.

The 3d $\mathcal{N}=4$ R-symmetry group is $SU(2)_H \times SU(2)_C$. In 3d $\mathcal{N}=2^*$ notation we identify an $U(1)_R$ R-symmetry subgroup and, consequently, its commutant $U(1)_A$ is a global symmetry from the $\mathcal{N}=2^*$ point of view.
For the purpose of this paper, we find it convenient to choose $U(1)_R$ to be the Cartan subgroup of $SU(2)_C$ so that $U(1)_A$ can be taken to be the anti-diagonal combination $U(1)_A \equiv U(1)_{H-C}$. In this parameterization hypermultiplets and twisted hypermultiplets have the following charges.
\begin{equation}
    \begin{tabular}{c|cc}
         & $U(1)_R$ & $U(1)_A$ \\
        \hline
        hyper. & 0 & $+1$ \\
        tw. hyper. & $+1$ & $-1$ 
    \end{tabular}
\end{equation}
Notice that this choice is not the superconformal one and at the IR fixed point the $U(1)_A$ symmetry has mixing $+1/2$ with $U(1)_R$. 

\paragraph{Global symmetry and spectrum:} The global symmetry of the $T[SU(N)]$ SCFT is $SU(N)_x \times SU(N)_y \times U(1)_A$.
Where $SU(N)_x$ is the flavor symmetry rotating the flavor hypermultiplets, while $SU(N)_y$ is an emergent symmetry arising from the enhancement of the $U(1)^{N-1}$ topological symmetry.



The spectrum of the theory is composed of two operators. The first is the meson matrix constructed from the flavor hypermultiplet as $ \Tr[Q_{N-1} \tilde{Q}_{N-1}]$, which transforms in the adjoint representation of $SU(N)_x$ with R-charge $0$ and $U(1)_A$ charge $2$. The second is a matrix obtained collecting together $N^2-1$ operators as follows: $N-1$ are the traces of the adjoint chirals; $N^2-N$ are monopole operators with $+1$ or $-1$ magnetic fluxes under a continuous sequence of gauge nodes. This operator transforms in the adjoint representation of the emergent $SU(N)_y$ symmetry, with R-charge $2$ and $U(1)_A$ charge $-2$.

As the spectrum does not contain any operator charged under the center of the global symmetries, we can refine the statement and claim that the actual global symmetry group is $PSU(N)_x \times PSU(N)_y \times U(1)_A$, where $PSU(N) = SU(N)/\mathbb{Z}_N$.
However, this statement is lifted once we allow for generic background configurations, as there are operators in the twisted Hilbert space that transform non-trivially under the center of the global symmetry. This translates into the presence of a mixed anomaly that can be encoded in the following anomaly theory.
\begin{equation}
    \frac{2\pi}{N} \int_{\mathcal{M}_4} w_2^{(x)} \cup w_2^{(y)}
\end{equation}
where $w_2$ is the second Stiefel-Whitney class for $PSU(N)$ and $\mathcal{M}_4$ is a spin-manifold.

\paragraph{Partition functions:}
We now report the partition function of the $T[SU(N)]$ theory for $S^2 \times S^1$ and $S^3$ for future reference. It will be convenient to define the partition function of the $T[U(N)]$ theory, which is simply the $T[SU(N)]$ theory with an extra BF coupling. 

We start from the superconformal index, which is the partition function on $S^2 \times S^1$, following the index convention used in Section \ref{subsec: 3dindex}. 
We express it as a recursive relation so that the index of $T[U(N)]$ is written in terms of the index of $T[U(N-1)]$.
\begin{equation} \label{eq: tsun_superconfind}
\begin{split}
    &\mathcal{I}^{(N)}_{T}(\vec{x},\vec{n}^{(x)};\vec{y},\vec{n}^{(y)};a) = \\
    &\quad = \prod_{i=1}^N \left(
    x_i^{-n^{(y)}_N} y_N^{-n^{(x)}_i}
    \right)
    \sum_{ \vec{m} \in \Gamma(\vec{n}^{(x)})} \frac{1}{(N\text{-}1)!} \oint \prod_{i=1}^{N-1} \left(
    \frac{dz_i}{2\pi i z_i} y_{N}^{m_i} z_i^{n^{(y)}_N}
    \right)
    \mathcal{I}^{(N-1)}_T(\vec{z},\vec{m};\vec{y},\vec{n}^{(y)};a) 
     \\
    & \qquad \mathcal{I}^{(N-1)}_{\text{vec}}(\vec{z},\vec{m})
    \prod_{i,j=1}^{N-1} \mathcal{I}^{(2)}_{\text{chir}}( a^{-2} z_i/z_j, m_i-m_j)
    \prod_{i=1}^{N-1} \prod_{j=1}^N \mathcal{I}^{(0)}_{\text{chir}}\left( a (z_i/x_j)^{\pm}, \pm(m_i-n^{(x)}_j) \right) 
    \\ &
    \mathcal{I}^{(1)}_{T}(x,n^{(x)};y,n^{(y)};a) = x^{-n^{(y)}} y^{-n^{(x)}}
\end{split}
\end{equation}
which becomes the index of the $T[SU(N)]$ theory after imposing: 
\begin{equation}
    \prod_{i=1}^N x_i = \prod_{i=1}^N y_1 = 1 \qquad,\qquad \sum_{i=1}^N n_i^{(x)} = \sum_{i=1}^N n_i^{(y)} = 0 \,.
\end{equation}
If these constraint are not imposed we can consider the global symmetry to be effectively promoted to $U(N)_x \times U(N)_y$.
Notice that in \eqref{eq: tsun_superconfind} the sum over magnetic fluxes for the $U(N-1)$ gauge group is a lattice $\Gamma(\vec{n}^{(x)})$ which depends on the value of the background magnetic flux for the flavor $PSU(N)_x$ symmetry. In fact, as the global symmetry is $PSU(N)_x$ instead of simply $SU(N)_x$, we are allowed to turn on non-integer fluxes that can be classified as follows:
\begin{equation}
    \vec{n}^{(x)} \in \mathbb{Z}^{N} + \frac{k}{N} \qquad \text{with} \qquad k = 0,\ldots,N-1
\end{equation}
in the sense that all the entries of the vector $\vec{n}^{(x)}$ must have the same value of $\text{Mod}[n^{(x)}_i,1] = k/N$. The lattice $\Gamma(\vec{n}^{(x)})$ then is:
\begin{equation} \label{eq: tsun_magneticlattice}
    \Gamma(\vec{n}^{(x)}) = \mathbb{Z}^{N-1} + \frac{k}{N} \qquad \text{if} \qquad \vec{n}^{(x)} \in \mathbb{Z}^{N} + \frac{k}{N} \qquad \text{with} \qquad k = 0,\ldots,N-1
\end{equation}
Notice that this condition propagates inside the gauge group of the $T[SU(N-1)]$ theory used for the recursion. Also the magnetic flux for the topological symmetries can be taken such that:
\begin{equation}
    \text{Mod}[n_i^{(y)},1] = 0,\frac{1}{N},\ldots,\frac{N-1}{N}
\end{equation} 
however this choice does not modify the magnetic lattices.

On $S^3$ the partition function is instead given by:
\begin{equation}
\begin{split}
    &\mathcal{Z}^{(N)}_T (\vec{X},\vec{Y},m_A) = 
    \\ & \quad = 
    e^{\pi i Y_N \sum_{j=1}^N X_j} \oint \prod_{j=1}^N dZ_j e^{-2 \pi i Y_N \sum_{j=1}^{N-1} Z_j} \mathcal{Z}_T^{(N-1)}(\vec{Z},\vec{Y},m_A)
    \\ & \qquad
    \mathcal{Z}_{\text{vec}}^{(N-1)}(\vec{Z}) \prod_{i,j=1}^{N-1} \mathcal{Z}^{(2)}_{\text{chir}}(Z_i - Z_j -2m_A) \prod_{i=1}^{N-1} \prod_{j=1}^N \mathcal{I}^{(0)}_{\text{chir}}( m_A \pm (Z_i-X_j)) 
    \\ &
    \mathcal{Z}^{(1)}_T (X,Y,m_A) = e^{2 \pi i Y X}
\end{split}
\end{equation}
where the partition function of vector and chiral multiplets can be expressed in terms of the hyperbolic gamma function:
\begin{equation}
\begin{split}
    \mathcal{Z}_{\text{vec}}^{(N)}(\vec{Z}) &= \frac{1}{N!}\left[ \prod_{i \neq j = 1}^N \Gamma_h(Z_i - Z_j) \right]^{-1} \\
    \mathcal{Z}_{\text{chir}}^{(r)}(X) &= \Gamma_h( r \omega + X) \\
    \Gamma_h(z) &= \prod_{m,n} \frac{ (m+1)\omega_1 + (n+1)\omega_2 - z}{m \omega_1 + n \omega_2 + z}
\end{split}
\end{equation}
where $\omega=\omega_1+\omega_2$, with $\omega_{1,2}$ related to the squashing parameter of $S^3$. The classical contribution of a mixed CS interaction at level $k$ is instead a phase $e^{-i \pi k A B}$ and we recall that in this paper we consider FI terms as mixed CS interactions between gauge and topological symmetries at level $-2$.

The fugacities appearing in the index can be seen roughly as the exponentiation of the real parameters appearing in the $S^3$ partition function. For example for the axial $U(1)_A$ symmetry $a \sim e^{2 \pi i m_A}$.

\subsubsection{IR dualities}
The $T[SU(N)]$ SCFT notably possesses two independent IR dualities, that are quasi-self-dualities, one being mirror duality \cite{Intriligator:1996ex,Hanany_1997,Gaiotto:2008ak} and the other is called ``flip-flip" duality \cite{Aprile:2018oau}. We now briefly review them.

\paragraph{Self-mirror duality:} 
A notable property of the $T[SU(N)]$ theory is that it is self-dual under the action of mirror duality exchanging the Higgs and the Coulomb branches. More precisely, the statement is that the Lagrangian theory in \eqref{eq: tsun_lagrangian} is IR dual to a theory described by the same Lagrangian but with the non-trivial swap of the global symmetries $SU(N)_x \leftrightarrow SU(N)_y$ and also of the two R-symmetry subgroups $SU(2)_H \leftrightarrow SU(2)_C$. We name the Higgs and Coulomb branch moment maps respectively as $\mu_H, \mu_C$ for the $T[SU(N)]$ theory, and $\mu'_H, \mu'_C$ for the dual theory. Then the duality maps:
\begin{equation}
    \mu_H \leftrightarrow \mu_C' \qquad,\qquad \mu_C \leftrightarrow \mu'_H
\end{equation}
As a partition function identity this duality reads:
\begin{equation}
\begin{split}
    & \mathcal{I}^{(N)}_T(\vec{x},\vec{n}^{(x)};\vec{y},\vec{n}^{(y)};a) = \mathcal{I}^{(N)}_T(\vec{y},\vec{n}^{(y)};\vec{x},\vec{n}^{(x)}; -\sqrt{q}/a) \\
    & \mathcal{Z}^{(N)}_T(\vec{X},\vec{Y},m_A) = \mathcal{Z}^{(N)}_T(\vec{Y},\vec{X},  \omega-m_A)
\end{split}
\end{equation}


\paragraph{Flip-flip duality}

There is a second self-duality enjoyed by the $T[SU(N)]$ SCFT, which is a self-duality modulo singlet fields called ``flip-flip" duality.
The statement of the duality is that the $T[SU(N)]$ theory is IR dual to the same Lagrangian theory with the R-symmetry twisted as $SU(2)_H \leftrightarrow SU(2)_C$ and with the addition of two matrices of singlets $M_H, M_C$, transforming in the adjoint representation of the $SU(N)_x$ Higgs branch symmetry and of the $SU(N)_y$ Couomb branch symmetry, respectively. These singlets couple to the moment maps of the Higgs and Culomb branch $\mu_H,\mu_C$ trough the superpotential:
\begin{equation}
    \mathcal{W} = M_H \mu_H + M_C\mu_C
\end{equation}
This duality is not mirror-like as it does not swap the Higgs and Coulomb branches, even thought it twists the R-symmetry. Under this duality the moment maps are mapped to the matrices of singlets:
\begin{equation}
    \mu_H \leftrightarrow M_H \qquad,\qquad \mu_C \leftrightarrow M_C \,.
\end{equation}
As a partition function identity this duality reads:
\begin{equation}
\begin{split}
    & \mathcal{I}^{(N)}_T(\vec{x},\vec{n}^{(x)};\vec{y},\vec{n}^{(y)};a) = \mathcal{I}^{(N)}_T(\vec{x},\vec{n}^{(x)};\vec{y},\vec{n}^{(y)}; -\sqrt{q}/a) \times \\
    & \qquad\qquad\qquad\qquad\qquad \times
    \prod_{i,j=1}^N \left[ \mathcal{I}_{\text{chir}}^{(2)} (a^{-2} x_i/x_j, n_i^{(x)}-n_j^{(x)})
    \mathcal{I}_{\text{chir}}^{(0)} (a^2 y_i/y_j, n_i^{(y)}-n_j^{(y)})
    \right]
    \\ & \mathcal{Z}^{(N)}_T(\vec{X},\vec{Y},m_A) = \mathcal{Z}^{(N)}_T(\vec{Y},\vec{X},  \omega-m_A) \times \\
    & \qquad\qquad\qquad\qquad \times
    \prod_{i,j=1}^N \left[
    \mathcal{Z}_{\text{chir}}^{(2)}(X_i-X_j-2m_A)
    \mathcal{Z}_{\text{chir}}^{(0)}(Y_i-Y_j+2m_A)
    \right]
\end{split}
\end{equation}


\subsubsection{Fusion to identity} 
The $T[SU(N)]$ theory enjoys a property which can be summarized as the statement that the gluing of two copies, trough the diagonal gauging of a $U(N)$ global symmetry, leads to a theory with a deformed moduli space so that the naively expected $SU(N) \times SU(N)$ global symmetry is spontaneously broken to the diagonal $SU(N)$ subgroup. We will oftentimes refer to the resulting theory as the identity-wall and to this property as the fusion-to-identity. 
The reason being that 
this property is frequently interpreted as the $S S^{-1} = 1$ relation for the $S$ generator of the $SL(2,\mathbb{Z})$ duality group of Type IIB brane setups, or as the expected behavior for the compactification of the 6d $\mathcal{N}=(2,0)$ $\mathfrak{a}_{N+1}$ theory on a cylinder in class-S language \cite{Benini:2010uu}.

To be more precise, let us first give the explicit Lagrangian resulting from the gluing:
\begin{equation} 
    \begin{tikzpicture}[baseline=(current bounding box).center]
        
        \node at (0,0) (n1) [gauge,black] {$1$};
        \node at (1.5,0) (n2) [gauge,black] {$2$};
        \node at (3,0) (n3) {$\cdots$};
        \node at (4.5,0) (nnm1) [gauge,black] {$\scriptstyle N-1$};
        \node at (6,0) (nn) [gauge,black] {$N$};
        \node at (7.5,0) (g1) [gauge,black] {$\scriptstyle N-1$};
        \node at (9,0) (g2) {$\cdots$};
        \node at (10.5,0) (g3) [gauge,black] {$2$};
        \node at (12,0) (g4) [gauge,black] {$1$};
        
        \draw[-] (n1.north east)      arc[start angle=-45, end angle=225, radius=3mm];
        \draw[-] (n2.north east)      arc[start angle=-45, end angle=225, radius=3mm];
        \draw[-] (nnm1.north east)    arc[start angle=-45, end angle=225, radius=3.1mm];

        \draw[-] (nn.north east)      arc[start angle=-45, end angle=225, radius=3mm];

        \draw[-] (g1.north east)      arc[start angle=-45, end angle=225, radius=3.1mm];
        \draw[-] (g3.north east)      arc[start angle=-45, end angle=225, radius=3mm];
        \draw[-] (g4.north east)    arc[start angle=-45, end angle=225, radius=3mm];
        
        \draw[->-] (n1.20) --   (n2.160);
        \draw[-<-] (n1.-20) --  (n2.-160);
        \draw[->-] (n2.20) --  ++(.75,0);
        \draw[-<-] (n2.-20) -- ++(.75,0);
        \draw[->-] (nnm1.160) --  ++(-.75,0);
        \draw[-<-] (nnm1.-160) -- ++(-.75,0);
        \draw[->-] (nnm1.20) --   (nn.160);
        \draw[-<-] (nnm1.-20) --  (nn.-160);

        \draw[->-] (nn.20) --   (g1.160);
        \draw[-<-] (nn.-20) --  (g1.-160);
        \draw[->-] (g1.20) --  ++(.75,0);
        \draw[-<-] (g1.-20) -- ++(.75,0);
        \draw[->-] (g3.160) --  ++(-.75,0);
        \draw[-<-] (g3.-160) -- ++(-.75,0);
        \draw[->-] (g3.20) --   (g4.160);
        \draw[-<-] (g3.-20) --  (g4.-160);

        \draw[FIcolor] (n1)++(0,-.5) node {\scriptsize{$y_1/y_2$}};
        \draw[FIcolor] (n2)++(0,-.5) node {\scriptsize{$y_2/y_3$}};
        \draw[FIcolor] (nnm1)++(0,-.5) node {\scriptsize{$y_{N-1}/y_N$}};
        \draw[FIcolor] (nn)++(0,-.5) node {\scriptsize{$y_N/x_N$}};
        \draw[FIcolor] (g1)++(0,-.5) node {\scriptsize{$x_N/x_{N-1}$}};
        \draw[FIcolor] (g3)++(0,-.5) node {\scriptsize{$x_3/x_2$}};
        \draw[FIcolor] (g4)++(0,-.5) node {\scriptsize{$x_2/x_1$}};
        
    \end{tikzpicture}
\end{equation}
where the gluing is performed through the diagonal $U(N)$ gauging of the two flavor symmetries\footnote{Actually, this property still hold when we consider the gluing trough the gauging of the Coulomb branch symmetries or a mixed situation where we glue an Higgs branch to a Coulomb branch. However, we will not need to consider these situations.}. The resulting theory is expected to have a $SU(N)_y \times SU(N)_x$ global symmetry on the Coulomb branch. However, this theory is bad, in the terminology of \cite{Gaiotto:2008ak}, as the scaling dimension of certain monopole operators violate the unitarity bound if computed w.r.t.~the UV R-symmetry. As a non-trivial consequence, the theory, as first noticed in \cite{Benini:2010uu}, has a deformed moduli space and the global symmetry $SU(N)_y \times SU(N)_x$ is spontaneously broken to the diagonal $SU(N)$ subgroup and, in the deep IR, the theory behaves as a free $\mathcal{N}=4$ vector multiplet. Therefore, if one attempts to gauge one of the two $SU(N)$ global symmetries, this gauge symmetry is Higgsed and the resulting theory is empty.

It was shown in \cite{Bottini:2021vms,Comi:2022aqo} that this property can be proved, in a sense, only assuming Aharony duality \cite{Aharony:1997gp} and the fact that the path integral of the 3d $\mathcal{N}=4$ pure $U(1)$ theory is a functional delta \cite{Kapustin:1999ha}. 

As a superconformal index identity this property reads:
\begin{equation} \label{eq: tsun_idwallindex}
\begin{split}
    & \frac{1}{N!} \sum_{\vec{m}\in \mathbb{Z}^N} \oint \frac{dz_i}{2\pi i z_i} \mathcal{I}^{(N)}_{\text{vec}}(\vec{z},\vec{m})
    \prod_{i,j=1}^{N} \mathcal{I}^{(2)}_{\text{chir}}( a^{-2} z_i/z_j, m_i-m_j) \times \\
    & \quad \times \mathcal{I}^{(N)}_T( \vec{z}; \vec{m};\vec{y},\vec{n}^{(y)};a) \mathcal{I}^{(N)}_T( \vec{z}; \vec{m}; \vec{x}^{-1},-\vec{n}^{(x)}; a) = \\
    & \qquad \qquad \qquad  =  \frac{\sum_{\sigma \in S_N} \prod_{i=1}^N \delta^{(p)}(x_i/ y_{\sigma(i)}) \delta_{n^{(x)}_i, n^{(y)}_{\sigma(i)}}}{\mathcal{I}^{(N)}_{\text{vec}}(\vec{x},\vec{n}^{(x)})
    \prod_{i,j=1}^{N} \mathcal{I}^{(0)}_{\text{chir}}( a^2 x_i/x_j, n^{(x)}_i-n^{(x)}_j)}
\end{split}
\end{equation}
where $\delta^{(p)}(x)$ is the periodic delta, so that it has support over $x = e^{2 \pi i k}$, for $k \in \mathbb{Z}$, while $\delta_{a,b}$ is the standard Kronecker-delta. Also, the sum is taken over the elements $\sigma$ of the $S_N$ permutation group. Therefore this identity clearly shows that the theory has the distributional effect of identifying the fugacities $\vec{x}$ and $\vec{y}$, and similarly their magnetic fluxes, modulo permutations. 
In the spirit of \cite{Spiridonov:2014cxa}, one can interpret the resulting singular functional partition function as that of a theory with spontaneous symmetry breaking \cite{Comi:2025zwu}.
The factor at the denominator can be interpreted in two ways. It is there to ensure the normalization of the distribution so that upon gauging with a $\mathcal{N}=4$ measure it yields:
\begin{equation}
\begin{split}
    & \frac{1}{N!}\sum_{\vec{n}^{(x)}\in \mathbb{Z}^N} \oint \frac{dz_i}{2\pi i x_i} \mathcal{I}^{(N)}_{\text{vec}}(\vec{x},\vec{n}^{(x)})
    \prod_{i,j=1}^{N} \mathcal{I}^{(0)}_{\text{chir}}( a^2 x_i/x_j, n^{(x)}_i-n^{(x)}_j) \times \\
    & \qquad \times \frac{\sum_{\sigma \in S_N} \prod_{i=1}^N \delta^{(p)}(x_i y_{\sigma(i)}) \delta_{n^{(x)}_i, n^{(y)}_{\sigma(i)}}}{\mathcal{I}^{(N)}_{\text{vec}}(\vec{x},\vec{n}^{(x)})
    \prod_{i,j=1}^{N} \mathcal{I}^{(0)}_{\text{chir}}( a^2 x_i/x_j, n^{(x)}_i-n^{(x)}_j)} = 1
\end{split}
\end{equation}
or, on a more physical side, it can be seen as the contribution of the supersymmetric version of the Goldstone bosons generated by the spontanous breaking of the global symmetry.

Similarly, for the $S^3$ partition function the fusion-to-identity property reads:
\begin{equation} \label{eq: tsun_idwallparfun}
\begin{split}
    & \int \prod_{i=1}^N dZ_i \mathcal{Z}_{\text{vec}}^{(N)}(\vec{Z}) \prod_{i,j=1}^N \mathcal{Z}_{\text{chir}}^{(2)}(Z_i-Z_j-2m_A) 
    \\ & \qquad
    \mathcal{Z}^{(N)}_T(\vec{Z},\vec{X},m_A) \mathcal{Z}^{(N)}_T(\vec{Z},-\vec{Y},m_A) = 
    \\ & \qquad \qquad \qquad = 
    \frac{\sum_{\sigma \in S_N} \prod_{i=1}^N \delta(X_i - Y_{\sigma(i)})}{ \mathcal{Z}_\text{vec}^{(N)}(\vec{X}) \prod_{i,j=1}^N \mathcal{Z}_{\text{chir}}^{(0)}(X_i - X_j +2m_A)}
\end{split}
\end{equation}
which can be interpreted analogously as the superconformal index identity.

\subsection{The $G[SU(N)]$ theory} \label{app: gsun}

In this appendix we review in more detail properties of the $G[SU(N)]$ theory, starting from its definition as a massive deformation of the $T[SU(N)]$ theory.

\paragraph{Real mass deformation of the $T[SU(N)]$ theory}

The $G[SU(N)]$ theory can be defined as the result of a real mass deformation of the $T[SU(N)]$ theory associated to the axial subgroup of the R-symmetry $U(1)_A$, which has the effect of breaking the R-symmetry as $SU(2)_H \times SU(2)_C \to U(1)_R$ and therefore supersymmetry from $\mathcal{N}=4$ to $\mathcal{N}=2$.

We observe that for non-zero value of the axial real mass parameter $m_A$, there is an interacting vacuum where the flavor $SU(N)$ symmetry is not broken, only if we turn on non-zero FI parameters satisfying:
\begin{equation} \label{eq: tsun_fimasses}
    Y_{i+1}-Y_i = 2m_A 
\end{equation}
and non-zero CB parameters. That is we turn on a VEV for the $\sigma^{(n)}$ real scalar component in the dynamical vector multiplets, that transform in the adjoint representation of the $U(n)$ gauge node, as:
\begin{equation}
    \sigma^{(n)} = (N-n)m_A  \mathbb{I}^{(n)}
\end{equation}
where $\mathbb{I}^{(n)}$ is the $n\times n$ identity matrix. Indeed, this deformation does not Higgs the gauge groups as the VEV is diagonal.
From the data of the deformation it is possible to compute effective masses for the chiral fields and thus the Lagrangian of the resulting theory, as well as Chern-Simons interactions produced by integrating out massive chiral multiplets. 
For example, all the bifundamental hypermultiplets acquire an effective mass:
\begin{equation}
    m_{\text{eff}} = m_A \pm m_A \,,
\end{equation}
where the plus sign refers to the chiral which is in the fundamental representation of $U(n)$ and antifundamental of $U(n+1)$. Oppositely, the anti-chiral is represented by the minus sign and is in the antifundamental of $U(n)$ and fundamental of $U(n+1)$. Consequently, we conclude that all the chirals acquire a positive mass while anti-chirals remain massless. A chiral in the bifundamental representation which acquires a positive mass generates a positive $1/2$ unit of CS level for both groups and $-1/2$ unit of mixed CS interaction. Summing all these contributions together and repeating the analysis for adjoint chirals, which completely acquire a mass, we are led to the Lagrangian theory describing the theory in this interacting vacuum.

\paragraph{The Lagrangian description}
The resulting theory is named $G[SU(N)]$ and is represented by the following quiver theory:
\begin{equation}    \label{eq: gsun_chiralquiver}
\begin{array}{l}
    \begin{tikzpicture}[baseline=(current bounding box).center]
        \begin{scope}[yshift=-3cm]
        \node at (0,0) (n1) [gauge,black] {$1$};
        \node at (2,0) (n2) [gauge,black] {$2$};
        \node at (8,0) (nnm1) [gauge,black] {$\scriptstyle N-1$};
        \node at (10,0) (nn) [flavor,black] {$N$};

        \draw[CScolor] (n1.south east) node[anchor=west] {$\scriptstyle 1$};
        \draw[CScolor] (n2.south east) node[anchor=west] {$\scriptstyle (0,2)$};
        \draw[CScolor] (nnm1.south east) node[anchor=west] {$\scriptstyle (0,N-1)$};
        \draw[CScolor] (nn.south east) node[anchor=west] {$\scriptstyle \tfrac{N-1}{2}$};

        \draw[FIcolor] (n1.north) node[anchor=south] {$\scriptstyle \xi_1$};
        \draw[FIcolor] (n2.north) node[anchor=south] {$\scriptstyle \xi_2$};
        \draw[FIcolor] (nnm1.north) node[anchor=south] {$\scriptstyle \xi_{N-1}$};

        \draw[black] (nn.north) node[anchor=south] {$\scriptstyle \vec{X}$};
        
        \draw[-<-] (n1) -- node[midway,above,BFcolor] {$\scriptstyle -1$} (n2);
        \draw[->-] (n2)++(2,0) -- node[midway,above,BFcolor] {$\scriptstyle -1$} (n2);
        \draw[-<-] (nnm1)++(-2,0) -- node[midway,above,BFcolor] {$\scriptstyle -1$} (nnm1);
        \draw[-<-] (nnm1) --  (nn);
        \end{scope}
    \end{tikzpicture}
\end{array}
\end{equation}
This theory is characterized by the sequence of gauge groups with increasing ranks and matter given by chiral multiplets in the bifundamental $\bar{\Box} \times \Box$ of the $U(n) \times U(n+1)$ groups. The theory is also characterized by a matrix of Chern-Simons couplings, such that the level associated to each $SU(n) \in U(n)$ subgroup of the gauge symmetries is $0$, while the levels of the $U(1) \in U(n)$ subgroups are specified by the following (symmetric) matrix:
\begin{equation}
    {\bf k} = \begin{pmatrix}
        1 & -1 & 0 & 0 &\cdots & 0 \\
         & 1 & -1 & 0 & \cdots & 0 \\
         &  & 1 & -1 & \cdots & 0 \\
        & & & \cdots & &
    \end{pmatrix}
\end{equation}

The global symmetry of the theory is $U(1)^{N-1} \times SU(N)$, where the former is the topological symmetry and the latter is the flavor symmetry. We stress that the enhancement $U(1)^{N-1} \to SU(N)$ no longer occurs in the $G[SU(N)]$ theory, like in the parent $T[SU(N)]$, since the real mass deformation explicitly breaks it. We conventionally parameterize the topological symmetry so that $\xi_i$ is the fugacity associated to the topological symmetry of the $i$-th node, which is related to the fugacities for the topological symmetries in the parent $T[SU(N)]$ theory as $y_1/y_{i+1} = \xi_i$.

The R-charge of all the chiral multiplets is taken to be 0. This doesn't necessarily coincide with the superconformal R-symmetry at the SCFT point and, possibly, Abelian global symmetries mix with the trial R-symmetry along the R-flow. The mixing coefficient can be computed, in principle, using F-extremization \cite{Jafferis:2010un}, we will however not be interested in fixing them.

The spectrum of the theory is composed only of the monopole operators $M_i$ with GNO fluxes $-1$ for the $i$-th gauge group and zero for the rest. These operators, with the choice of parameterization explained before, carry $+1$ R-charge and $+1$ charge under $U(1)_{\xi_i}$. We recall that we are treating FI terms as mixed CS interactions between the gauge symmetry and the topological symmetry, therefore we have the slightly counterintuitive situation that a monopole with negative magnetic flux carries positive charge under the topological symmetry.
The theory then possesses a conserved current for the $SU(N)$ flavor symmetry, but no chiral operators charged under it.

\paragraph{Partition function}
We now provide the explicit partition function of the $G[SU(N)]$ theory on $S^2 \times S^1$ and on $S^3$.

Let us start from the partition function on $S^2 \times S^1$, which is the superconformal index. We follow the notation introduced in Section \ref{subsec: 3dindex}. The superconformal index of the $G[SU(N)]$ theory reads:
\begin{equation}
\begin{split}
    & \mathcal{I}^{(N)}_G ( \{ \vec{x},\vec{n}^{(x)} \}, \{ \vec{\xi},\vec{n}^{(\xi)} \}) = 
    \\ & \quad =
    \prod_{r=1}^{N-1} \left[ \sum_{\vec{m}^{(r)} \in \Gamma^{(r)}} \frac{1}{r!} \oint \prod_{i=1}^{r} \frac{d z^{(r)}_i \xi_i^{-m^{(r)}_i} (z_i^{(r)})^{m_i^{(r)}-n_i^{(\xi)}}}{2 \pi i z^{(n)}_i} \mathcal{I}_{\text{vec}}^{(r)}(\vec{z}^{(r)},\vec{m}^{(r)})
    \right]
    \\ & \qquad
    \prod_{r=1}^{N-1} \left[
    \prod_{i=1}^r \prod_{j=1}^{r+1} (z_i^{(r)})^{-\frac{m^{(r+1)}_j}{2}} (z_j^{(r+1)})^{-\frac{m^{(r)}_i}{2}} \mathcal{I}_{\text{chir}}^{(0)} ( z_i^{(r)}/z_j^{(r)}, m_i^{(r)}-m_j^{(r+1)}) 
    \right]
    \\ & 
    \mathcal{I}^{(1)}_G ( \{ x,n^{(x)} \}, \{ \xi,n^{(\xi)} \}) = 1
\end{split}
\end{equation}
setting $\vec{z}^{(N)} \equiv \vec{x}$ and $\vec{m}^{(N)} \equiv \vec{n}^{(x)}$, with $\prod_{i=1}^N x_i =1$ and $\sum_{i=1}^N n_i^{(x)}=0$.
Where the second line simply encodes the contribution of vector multiplets, FI terms and CS levels. The third line encodes the contribution from mixed CS-levels and chiral multiplets. 
The lattice of magnetic charges follows the same rule as for the $T[SU(N)]$ case, reported in \eqref{eq: tsun_magneticlattice}, namely:
\begin{equation}
    \Gamma_n = \mathbb{Z}^n + \frac{k}{N} \qquad \text{if} \qquad \vec{n}^{(x)} \in \mathbb{Z}^N + \frac{k}{N} \qquad \text{for} \qquad k=0,\ldots,N-1
\end{equation}

Similarly, we can also write the $S^3$ partition function.
\begin{equation}
\begin{split}
    & \mathcal{Z}^{(N)}_G(\vec{X},\vec{\Xi}) = 
    \\ & \quad =
    \prod_{r=1}^{N-1} \left[ \frac{1}{r!}
    \int \prod_{j=1}^r dZ^{(r)}_j e^{2\pi i \Xi_r \sum_{j=1}^r Z_j - i\pi \left( \sum_{j=1}^r Z_j \right)^2} \mathcal{Z}_{\text{vec}}^{(r)}(\vec{Z}^{(n)})
    \right]
    \\ & \qquad
    \prod_{r=1}^{N-1} \left[ e^{\pi i \sum_{j=1}^r \sum_{k=1}^{r+1} Z_j^{(r)} Z_k^{(r+1)}} 
    \prod_{j=1}^r \prod_{k=1}^{r+1} \mathcal{Z}_{\text{chir}}^{(0)} (Z_j^{(r)} - Z_k^{(r+1)})
    \right]
    \\ & 
    \mathcal{Z}^{(1)}_G(X, \Xi ) = 1
\end{split}
\end{equation}
with $\vec{Z}^{(N)} \equiv \vec{X}$ and $\sum_{i=1}^N X_i = 0$.

\subsubsection{IR dualities}
The $G[SU(N)]$ theory enjoys three type of IR dualities, all originating from dualities and properties of the parent $T[SU(N)]$ theory. We now proceed to describe them.

\paragraph{The Abelian-planar IR dual description} 

It was found in \cite{Benvenuti:2025huk} that the $G[SU(N)]$ theory admits a second Lagrangian UV completion.
This second, IR dual, description can be obtained by combining the massive deformation of the $T[SU(N)]$ theory with mirror duality. In fact, as it exists an interacting vacuum on the Coulomb branch as a result of the deformation, in virtue of the self-mirror duality for the $T[SU(N)]$, there must exists also a mirror-like dual vacuum on the Higgs branch. We can find the Lagrangian description of this dual vacuum by first performing mirror duality on the theory in \eqref{eq: tsun_lagrangian} and then using the same data of the deformation as described above, where now the FI parameters become real-masses for the flavor $SU(N)$ symmetry. We discover that in this case the dual interacting vacuum exists only at a point of the moduli space such that the real scalar components of the vector multiplet $\sigma^{(n)}$ satisfy:
\begin{equation}
    \sigma_i^{(n)} = \left(i-\frac{N+n}{2} \right)2m_A
\end{equation}
where $\sigma_i^{(n)}$ is the $i-th$ diagonal component of the real scalar associated to the $U(n)$ gauge group. We then observe that this VEV breaks the gauge symmetries as $U(n) \to U(1)^{n}$, so that the final theory is Abelian. Also, due to the Higgsing the final theory possesses a monopole superpotential that ensures the non-proliferation of global symmetries, as we will describe in a moment.

The resulting theory is described by the following Abelian planar quiver theory:
\begin{equation} \label{eq: gsun_planarquiver}
    \includegraphics[]{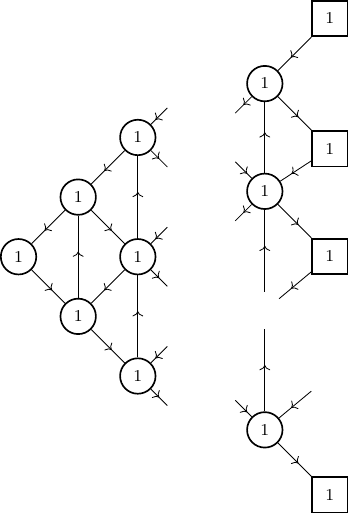}
\end{equation}
where we have not explicitly reported superpotential and Chern-Simons interactions. These can be reconstructed with the following rules.
The superpotential can be written as follows.
\begin{itemize}
    \item To each triangular face of the quiver we write a cubic interaction between the three chirals forming the loop.
    \item To each pair of gauge nodes connected by a vertical arrow we write a linear superpotential for the monopole carrying GNO flux $+1$ for the upper gauge node and $-1$ for the lower one.
\end{itemize}
Similarly, the level of CS interactions can be written starting from the quiver as follows.
\begin{itemize}
    \item Two nodes connected by a diagonal chiral have a mixed CS interaction with level $-1$. Similarly, two nodes connected by a vertical arrow have a mixed CS interaction with level $+1$.
    \item The CS level of each gauge node can be computed from the mixed CS interaction with the following formula:
    \begin{equation}
        k_i = -\frac{1}{2} \sum_{j \neq i} k_{ij}
    \end{equation}
    where $k_i$ is the CS level of a chosen $i$-th node and $k_{ij}$ is the level of the mixed CS interaction between the $i$-th and $j$-th nodes. The sum runs over all nodes except $i$.
\end{itemize}

The planar Abelian dual has a flavor symmetry which is $U(1)^{N-1}$, after taking into account mesonic superpotentials and gauge redefinitions, that map to the topological symmetry of the $G[SU(N)]$ theory in \eqref{eq: gsun_chiralquiver}. There are gauge invariant mesonic operators charged under this flavor symmetry that are constructed by taking the product of two chirals forming the shortest path between two square nodes. Each of these chirals carries $+1$ charge under one $U(1)_{\xi_i}$ factor, thus mapping to the gauge invariant monopoles of the $G[SU(N)]$ theory in \eqref{eq: gsun_chiralquiver}. Longer path connecting two square nodes do not give rise to independent gauge invariant operators but they are either composite operators or zero due to F-term relation set by the mesonic superpotential.

The topological symmetry of the planar Abelian quiver in \eqref{eq: gsun_planarquiver} is $U(1)^{N-1}$ which is one factor for each $U(1)$ gauge group minus one for each monopole superpotential term. This topological symmetry enhances in the IR to $SU(N)$ and maps to the flavor symmetry of the $G[SU(N)]$ theory in \eqref{eq: gsun_chiralquiver}.
The only gauge invariant chiral monopole operators in the planar Abelian theory in \eqref{eq: gsun_planarquiver} are those that appear linearly in the superpotential, thus the theory does not possess any chiral monopole operator in its spectrum.
However it is possible to construct a matrix of $1/4$-BPS monopole operators that are expected to have exact R-charge 2 and compose the conserved current of an emergent $SU(N)$ global symmetry.

\paragraph{$N!$ duality frames:} 
There is a second type of IR duality that is enjoyed by the $G[SU(N)]$ theory described by the quiver in \eqref{eq: gsun_chiralquiver}. This second IR duality is qualitatively different than the one presented above. Namely, the Abelian-planar description in \eqref{eq: gsun_planarquiver} should be regarded as an $\mathcal{N}=2$ mirror-like duality, while the one we now present is an ``electric" duality, in the sense that it does not exchange flavor and topological symmetries.

The origin of this duality is also very different in spirit as it does not originate from an IR duality of the $T[SU(N)]$ theory. The idea is that since in the $T[SU(N)]$ theory it is expected an enhancement of topological symmetries to $SU(N)$, then there must be an emergent Weyl group that permutes the FI parameters as:
\begin{equation}
    Y_i \to Y_{\sigma(i)} \quad \text{where} \quad \sigma \in S_N \,.
\end{equation}
Therefore applying this transformation on the input of the massive deformation in \eqref{eq: tsun_fimasses}, we expect that all the deformation defined by the limit:
\begin{equation}
    Y_{\sigma(i+1)} - Y_{\sigma(i)} = 2m_a
\end{equation}
should lead to an equivalent, i.e.~IR dual, vacuum.
We thus expect an orbit of $N! = |S_N|$ theories that are all IR dual to the $G[SU(N)]$ theory\footnote{As we will see in a moment, we actually find a lower number of IR dual Lagrangians as some element of $S_N$ lead to IR dualities that simply consist in the swapping of two symmetries and thus relate two identical Lagrangians with non-trivial map of global symmetries. If one considers also these frames as inequivalent then we find precisely $N!$ theories.}.

One can derive all these dual frames following the strategies discussed in \cite{Benvenuti:2025huk}, leveraging on the ideas of \cite{Aharony:2013dha,Benini:2011mf}, we will however simply state the result here.
We claim that all the duality frames of the theory in \eqref{eq: gsun_chiralquiver} can be produced with a simple algorithmic strategy. We find that there are two mutations that we are allowed to perform starting from the quiver of the $G[SU(N)]$ theory in \eqref{eq: gsun_chiralquiver}, with the result being a Lagrangian that is IR dual to the original.\footnote{Notice that actually we do not produce exactly $N!$ different Lagrangians as swapping the $A \leftrightarrow B$ parameters produces the same result up to arrow flippings. However, we still get a number of Lagrangian asymptotically scaling as $N!$.}
\begin{itemize}
    \item \textbf{Tail splitting:} We pick a gauge group $U(M)$ in the theory \eqref{eq: gsun_chiralquiver} and two positive integers $A,B$ such that $A+B=M$. We replace the $U(1), \ldots, U(M)$ tail with two tails as follows:
    \begin{equation} \label{eq: tailsplittingmove}
        \includegraphics[]{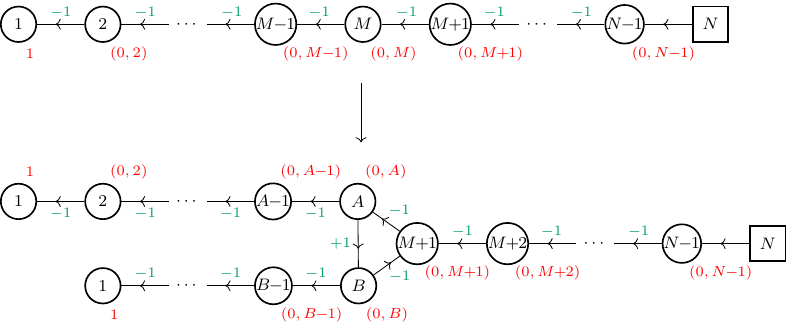}
    \end{equation}
    This move generates a cubic superpotential interaction between the chirals forming the triangular loop. The R-charge of all chirals is still 0 apart from that of the vertical chiral closing the loop, which is $2$, which is compatible with the new superpotential term.
    After the move the CS level of each $U(n)$ gauge node is still $(0,n)$ and all the mixed CS levels are still $-1$, besides that connecting vertically the $U(A)$ and $U(B)$ nodes which is $+1$.
    Indeed, this procedure can be iterated on the sub-tails starting from $U(1)$ to $U(A)$ or $U(B)$, to create more bifurcations in the quiver. 

    \item \textbf{Arrow inversion:} We pick a bifundamental in the quiver and we invert the orientation of the arrow, that is, for example, that we swap the representation from $\bar{\square} \times \square$ with $\square \times \bar{\square}$. 
    \begin{equation}
        \includegraphics[]{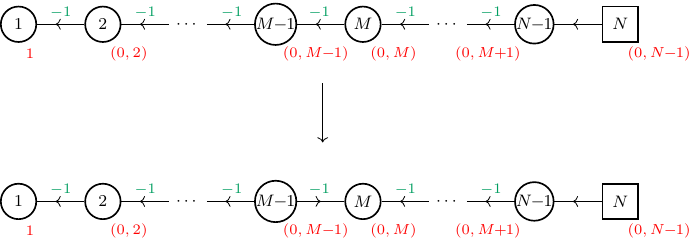}
    \end{equation}
    We can also apply this move to the triangular loop in the quiver in \eqref{eq: tailsplittingmove}, by inverting the orientation of all the arrows in the loop. Notice that this move does not change CS levels and also the R-charge of the chirals.
\end{itemize} 
\begin{figure}[h!]
    \centering
    \includegraphics[width=\textwidth]{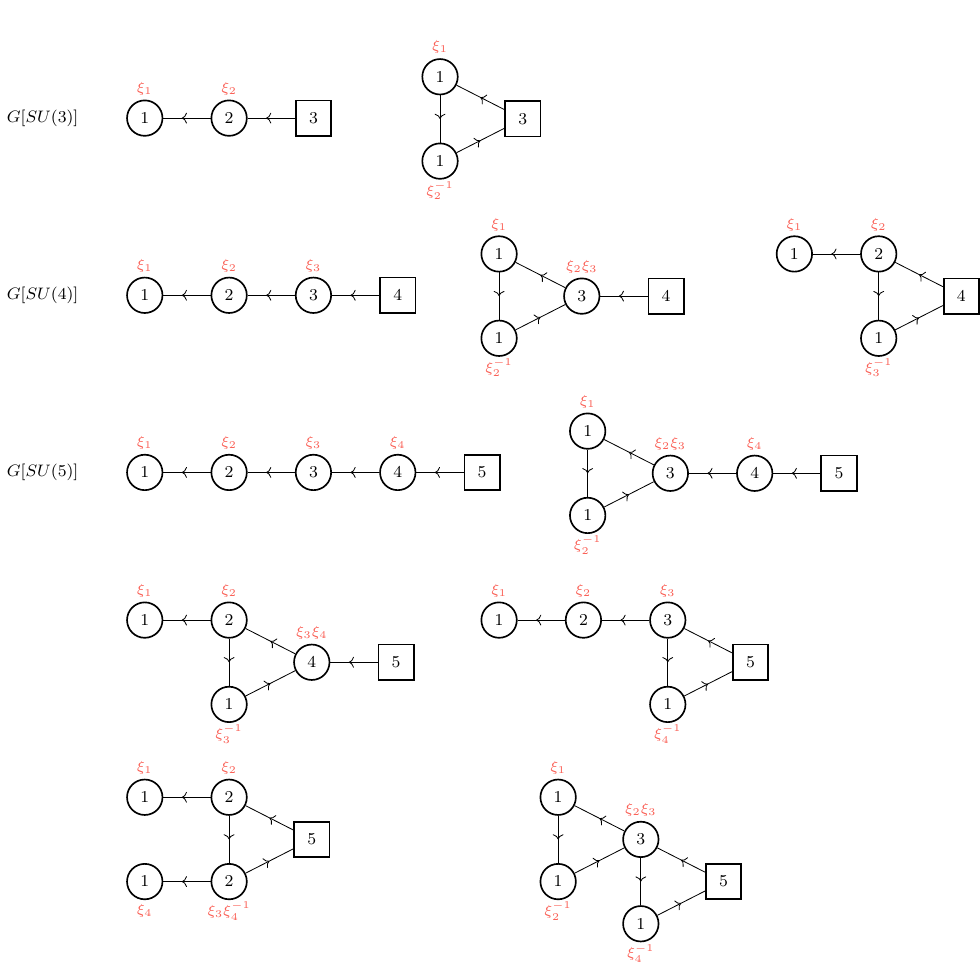}
    \caption{Duality frames of the $G[SU(N)]$ theory generated by the tail-splitting move, for $N=3,4,5$. Each $U(n)$ gauge ndoe carries lavel $(0,n)$ and each pair of gauge nodes connected by a chiral have a mixed CS level of $-1$, apart from those connected vertically that have mixed CS level $+1$. To each triangular loop we also assign a cubic superpotential term. Each chiral has R-charge 0 apart from the vertical arrows that have R-charge 2, compatible with the superpotential. In orange we provide the FI parmaters to keep track of the mapping of topologicla symmetries across the various dualities.}
    \label{fig: N!dualities_examples}
\end{figure}
Notice that each time we repeat the tail splitting, we reduce the total sum of the ranks of gauge groups by 1, but the total number of gauge groups is constant. Therefore the theory always possess $N-1$ gauge nodes providing a $U(1)^{N-1}$ topological symmetry and $N-1$ gauge invariant monopole operators, one for each gauge node. 

Each of the two duality moves has a complicated map of the topological symmetries that we do not attempt to give in full generality. We however present the case where we start from the $G[SU(N)]$ theory and perform the arrow-flip move to all the chiral multiplets:
\begin{equation}
    \includegraphics[]{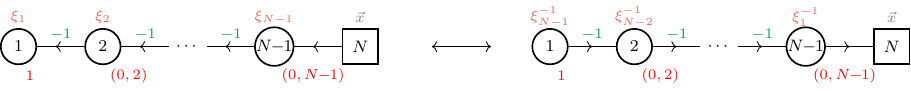}
\end{equation}
which maps the topological symmetries as indicated by the FI paramters reported in orange. Namely it changes the fugacity of the topological symmetry of a $U(n)$ gauge node as $\xi_n \to \xi_{N-n}^{-1}$. This duality is used in Section \ref{sec: interfaces} to read the identity-wall formed by gluing a $S \cdot \Twall[]$ and a $(S \cdot \Twall[])^{-1}$.

The duality frames obtained using the tail-splitting move are instead quite useful to consider when performing computations. For example computational time to perform the perturbative expansion of the superconformal index decreases notably.
Also, this duality shows that, for example, in the case $N=3$ the theory can be described by a pure Abelian model with a manifest $SU(3)$ symmetry and might be useful for similar insight for higher rank cases. In Figure \ref{fig: N!dualities_examples} we report all possible tail-splitting dualities, but not arrow-flipped ones, for $N=3,4,5$ along with the map for the topological fugacities.

One can indeed chain together the Abelian-planar duality with this orbit of dualities to find also $N!$ ``electric" dualities for the theory in \eqref{eq: gsun_planarquiver}. However, we discover that this does not give rise to new Lagrangian theories. The reason is that when we perform the massive deformation of the $T[SU(N)]$ theory on the real parameters associated to the flavor symmetry, the Weyl group is manifest. With this computation we simply discover that on the planar-Abelian side it is an IR duality to simply any number $U(1)$ symmetries, which indeed leads to $N!$ possible parameterizations that are all IR dual to each other.

\paragraph{Flip-Flip duality}

There is another IR duality enjoyed by the $G[SU(N)]$ theory, which is inherited from the flip-flip duality of the parent $T[SU(N)]$ theory.

The statement of this duality is that the $G[SU(N)]$ theory is dual to an almost identical Lagrangian which is the following:
\begin{equation}
    \includegraphics[]{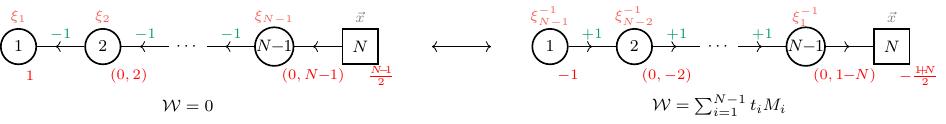}
\end{equation}
Notice that in the dual frame not only we change the sign of CS interactions and the orientation of the arrows, but we also introduce $N-1$ singlets coupling to the gauge invariant monopole operators with positive magnetic flux. These singlets map to the gauge invariant monopoles on the l.h.s.~, that carry negative magnetic flux. Also the CS levels change as follows:
\begin{equation}
    k_{SU(N)} = \frac{N-1}{2} \to -\frac{N+1}{2} ,\quad k_{\xi_i} = 0 \to -1/2, \quad k_{\xi_i \xi_j} = 0 \to 0
\end{equation}

This is another ``electric" dual frame as it does not exchange mesonic and topological symmetries and it can be indeed combined with the two other dualities to form an intricated web. In particular, there is a combination of flip-flip and arrow-flip dualities that shows that the $S \cdot \Twall[]$ and $(S \cdot \Twall[])^{-1}$ are dual modulo flipping fields and background interactions.

\subsubsection{Fusion to identity}

The $G[SU(N)]$ theory inherits a fusion-to-identity property from the parent $T[SU(N)]$ theory. In other words, a suitable gluing of two $G[SU(N)]$ theories leads to a theory with a deformed moduli space whose partition function behaves as a delta function.

There are however two different ways to glue together two $G[SU(N)]$ theories, namely we can perform the diagonal gauging of their $SU(N)$ or $U(1)^{N-1}$ global symmetries. These two possibilities lead to similar, although very different, behavior.

As it will be useful in the bulk of the paper, we consider instead of the $G[SU(N)]$ the $CP$-transformed, which we recall is related to the $\Twall[]$ theory by and $S$-transformation. 
Let us first consider the case where we perform the diagonal gauging of the $SU(N)$ global symmetries with an additional CS-level 1. Namely we consider the following quiver:
\begin{equation} 
    \includegraphics[]{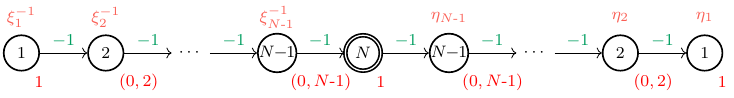}
\end{equation}
Actually, this theory should be regarded as having gauge group:
\begin{equation}
\begin{split}
    &\left[ U(1)^2 \times U(2)^2 \times... \times U(N-1)^2 \times U(N) \right] /U(1) = \\
    & \qquad\qquad 
    = S\left[ U(1)^2 \times U(2)^2 \times... \times U(N-1)^2 \times U(N) \right] / \mathbb{Z}_N
\end{split}
\end{equation}
Where we can solve the $U(1)$ decoupling constraint by considering central gauge node $SU(N)$, but there is still a gauged $\mathbb{Z}_N$ 1-form symmetry to include. Most of the times this detail will not be relevant as when this theory is coupled to a 4d bulk the contribution of any non-trivial $\mathbb{Z}_N$ background is suppressed, but it is fundamental to consider in the purely 3d setup and in particular in the superconformal index identity, that we will explain in a moment.

This theory was studied in \cite{Benvenuti:2025huk} and it was argued that it exibhit a deformed moduli space where the topological symmetry is spontaneously broken from $U(1)^{N-1} \times U(1)^{N-1}$ to $U(1)^{N-1}$ such that, intuitively, the topological symmetries of the gauge nodes with the same ranks are broken to their diagonal subgroup. Also, the theory in the deep IR flows to a model of $N-1$ free chirals.

As a partition function identity this property translates in the fact that the partition function behaves as a delta-function.
\begin{equation} \label{eq: gsun_idwallindex}
\begin{split}
    & \frac{1}{N!} \sum_{\vec{m}\in \Gamma} \oint \frac{dz_i z_i^{m_i}}{2\pi i z_i} \mathcal{I}^{SU(N)}_{\text{vec}}(\vec{z},\vec{m})
    \times \\
    & \quad \times \mathcal{I}^{(N)}_G( \{ \vec{z}; \vec{m} \}; \{\vec{y},\vec{n}^{(\xi)} \} ) \mathcal{I}^{(N)}_G( \{ \vec{z}^{-1}; -\vec{m} \}; \{ \vec{\eta},\vec{n}^{(\eta)} \} ) = \\
    & \qquad \qquad \qquad  =  \prod_{i=1}^{N-1} 
    \left[
    \delta^{(p)}(\xi_i/ \eta_i) \delta_{n^{(\xi)}_i, n^{(\eta)}_i} \xi_i^{m_i/2} \mathcal{I}^{(1)}_{\text{chir}}( \xi_i^{-1}, -n^{(\xi)}_i)
    \right]
\end{split}
\end{equation}
Where the sum over magnetic fluxes is done over the lattice:
\begin{equation}
    \Gamma = \mathbb{Z}^{N-1} + \frac{k}{N} \qquad \text{with} \quad k=0,\ldots,N-1
\end{equation}
which includes magnetic fluxes $\vec{m}$ such that $\text{Mod}[m_i,N]=k/N$ for all $i$. A choice of an integer lattice does not give rise to the expected mathematical identity.

As an $S^3$ partition function identity it reads:
\begin{equation} \label{eq: gsun_idwallparfun}
\begin{split}
    & \int \prod_{i=1}^{N-1} dZ_i \mathcal{Z}_{\text{vec}}^{SU(N)}(\vec{Z}) 
    \mathcal{Z}^{(N)}_G(\vec{Z},\vec{\Xi}) \mathcal{Z}^{(N)}_G(-\vec{Z},\vec{\Theta}) =  
     \prod_{i=1}^{N-1} 
     \left[
     \delta(\Xi_i - \Theta_i ) e^{-\frac{i\pi}{2} \Xi_i^2} \mathcal{Z}_{\text{chir}}^{(1)}(-\Xi_i) 
     \right]
\end{split}
\end{equation}

\newpage

\bibliographystyle{JHEP}
\bibliography{References}

\providecommand{\href}[2]{#2}\begingroup\raggedright\begin{thebibliography}{10}

\bibitem{Gaiotto:2008sa}
D.~Gaiotto and E.~Witten, \emph{{Supersymmetric Boundary Conditions in N=4 Super Yang-Mills Theory}}, \href{https://doi.org/10.1007/s10955-009-9687-3}{\emph{J. Statist. Phys.} {\bfseries 135} (2009) 789} [\href{https://arxiv.org/abs/0804.2902}{{\ttfamily 0804.2902}}].

\bibitem{Gaiotto:2008ak}
D.~Gaiotto and E.~Witten, \emph{{S-Duality of Boundary Conditions In N=4 Super Yang-Mills Theory}}, \href{https://doi.org/10.4310/ATMP.2009.v13.n3.a5}{\emph{Adv. Theor. Math. Phys.} {\bfseries 13} (2009) 721} [\href{https://arxiv.org/abs/0807.3720}{{\ttfamily 0807.3720}}].

\bibitem{Hosomichi:2010vh}
K.~Hosomichi, S.~Lee and J.~Park, \emph{{AGT on the S-duality Wall}}, \href{https://doi.org/10.1007/JHEP12(2010)079}{\emph{JHEP} {\bfseries 12} (2010) 079} [\href{https://arxiv.org/abs/1009.0340}{{\ttfamily 1009.0340}}].

\bibitem{Gaiotto:2012np}
D.~Gaiotto, \emph{{Domain Walls for Two-Dimensional Renormalization Group Flows}}, \href{https://doi.org/10.1007/JHEP12(2012)103}{\emph{JHEP} {\bfseries 12} (2012) 103} [\href{https://arxiv.org/abs/1201.0767}{{\ttfamily 1201.0767}}].

\bibitem{Dimofte:2013lba}
T.~Dimofte, D.~Gaiotto and R.~van~der Veen, \emph{{RG Domain Walls and Hybrid Triangulations}}, \href{https://doi.org/10.4310/ATMP.2015.v19.n1.a2}{\emph{Adv. Theor. Math. Phys.} {\bfseries 19} (2015) 137} [\href{https://arxiv.org/abs/1304.6721}{{\ttfamily 1304.6721}}].

\bibitem{Seiberg:1994rs}
N.~Seiberg and E.~Witten, \emph{{Electric - magnetic duality, monopole condensation, and confinement in N=2 supersymmetric Yang-Mills theory}}, \href{https://doi.org/10.1016/0550-3213(94)90124-4}{\emph{Nucl. Phys. B} {\bfseries 426} (1994) 19} [\href{https://arxiv.org/abs/hep-th/9407087}{{\ttfamily hep-th/9407087}}].

\bibitem{Seiberg:1994aj}
N.~Seiberg and E.~Witten, \emph{{Monopoles, duality and chiral symmetry breaking in N=2 supersymmetric QCD}}, \href{https://doi.org/10.1016/0550-3213(94)90214-3}{\emph{Nucl. Phys. B} {\bfseries 431} (1994) 484} [\href{https://arxiv.org/abs/hep-th/9408099}{{\ttfamily hep-th/9408099}}].

\bibitem{Benvenuti:2024seb}
S.~Benvenuti, R.~Comi, S.~Pasquetti, G.~Pedde~Ungureanu, S.~Rota and A.~Shri, \emph{{Planar Abelian Mirror Duals of $\mathcal{N}=2$ SQCD$_3$}},  \href{https://arxiv.org/abs/2411.05620}{{\ttfamily 2411.05620}}.

\bibitem{Benvenuti:2025huk}
S.~Benvenuti, R.~Comi, S.~Pasquetti, G.~Pedde~Ungureanu, S.~Rota and A.~Shri, \emph{{A chiral-planar dualization algorithm for 3d $ \mathcal{N} $ = 2 Chern-Simons-matter theories}}, \href{https://doi.org/10.1007/JHEP10(2025)211}{\emph{JHEP} {\bfseries 10} (2025) 211} [\href{https://arxiv.org/abs/2505.02913}{{\ttfamily 2505.02913}}].

\bibitem{Witten:2003ya}
E.~Witten, \emph{{SL(2,Z) action on three-dimensional conformal field theories with Abelian symmetry}},  in \emph{{From Fields to Strings: Circumnavigating Theoretical Physics: A Conference in Tribute to Ian Kogan}}, pp.~1173--1200, 7, 2003 [\href{https://arxiv.org/abs/hep-th/0307041}{{\ttfamily hep-th/0307041}}].

\bibitem{Gang:2012ff}
D.~Gang, E.~Koh and K.~Lee, \emph{{Superconformal Index with Duality Domain Wall}}, \href{https://doi.org/10.1007/JHEP10(2012)187}{\emph{JHEP} {\bfseries 10} (2012) 187} [\href{https://arxiv.org/abs/1205.0069}{{\ttfamily 1205.0069}}].

\bibitem{Dimofte:2011py}
T.~Dimofte, D.~Gaiotto and S.~Gukov, \emph{{3-Manifolds and 3d Indices}}, \href{https://doi.org/10.4310/ATMP.2013.v17.n5.a3}{\emph{Adv. Theor. Math. Phys.} {\bfseries 17} (2013) 975} [\href{https://arxiv.org/abs/1112.5179}{{\ttfamily 1112.5179}}].

\bibitem{Cordova:2016uwk}
C.~Cordova, D.~Gaiotto and S.-H.~Shao, \emph{{Infrared Computations of Defect Schur Indices}}, \href{https://doi.org/10.1007/JHEP11(2016)106}{\emph{JHEP} {\bfseries 11} (2016) 106} [\href{https://arxiv.org/abs/1606.08429}{{\ttfamily 1606.08429}}].

\bibitem{Gaiotto:2019jvo}
D.~Gaiotto and T.~Okazaki, \emph{{Dualities of Corner Configurations and Supersymmetric Indices}}, \href{https://doi.org/10.1007/JHEP11(2019)056}{\emph{JHEP} {\bfseries 11} (2019) 056} [\href{https://arxiv.org/abs/1902.05175}{{\ttfamily 1902.05175}}].

\bibitem{Pasquetti:2011fj}
S.~Pasquetti, \emph{{Factorisation of N = 2 Theories on the Squashed 3-Sphere}}, \href{https://doi.org/10.1007/JHEP04(2012)120}{\emph{JHEP} {\bfseries 04} (2012) 120} [\href{https://arxiv.org/abs/1111.6905}{{\ttfamily 1111.6905}}].

\bibitem{Benini:2012ui}
F.~Benini and S.~Cremonesi, \emph{{Partition Functions of ${\mathcal{N}=(2,2)}$ Gauge Theories on S$^{2}$ and Vortices}}, \href{https://doi.org/10.1007/s00220-014-2112-z}{\emph{Commun. Math. Phys.} {\bfseries 334} (2015) 1483} [\href{https://arxiv.org/abs/1206.2356}{{\ttfamily 1206.2356}}].

\bibitem{Nieri:2015yia}
F.~Nieri and S.~Pasquetti, \emph{{Factorisation and holomorphic blocks in 4d}}, \href{https://doi.org/10.1007/JHEP11(2015)155}{\emph{JHEP} {\bfseries 11} (2015) 155} [\href{https://arxiv.org/abs/1507.00261}{{\ttfamily 1507.00261}}].

\bibitem{Doroud:2012xw}
N.~Doroud, J.~Gomis, B.~Le~Floch and S.~Lee, \emph{{Exact Results in D=2 Supersymmetric Gauge Theories}}, \href{https://doi.org/10.1007/JHEP05(2013)093}{\emph{JHEP} {\bfseries 05} (2013) 093} [\href{https://arxiv.org/abs/1206.2606}{{\ttfamily 1206.2606}}].

\bibitem{Beem:2012mb}
C.~Beem, T.~Dimofte and S.~Pasquetti, \emph{{Holomorphic Blocks in Three Dimensions}}, \href{https://doi.org/10.1007/JHEP12(2014)177}{\emph{JHEP} {\bfseries 12} (2014) 177} [\href{https://arxiv.org/abs/1211.1986}{{\ttfamily 1211.1986}}].

\bibitem{Cordova:2015nma}
C.~Cordova and S.-H.~Shao, \emph{{Schur Indices, BPS Particles, and Argyres-Douglas Theories}}, \href{https://doi.org/10.1007/JHEP01(2016)040}{\emph{JHEP} {\bfseries 01} (2016) 040} [\href{https://arxiv.org/abs/1506.00265}{{\ttfamily 1506.00265}}].

\bibitem{Cecotti:2015lab}
S.~Cecotti, J.~Song, C.~Vafa and W.~Yan, \emph{{Superconformal Index, BPS Monodromy and Chiral Algebras}}, \href{https://doi.org/10.1007/JHEP11(2017)013}{\emph{JHEP} {\bfseries 11} (2017) 013} [\href{https://arxiv.org/abs/1511.01516}{{\ttfamily 1511.01516}}].

\bibitem{Kontsevich:2008fj}
M.~Kontsevich and Y.~Soibelman, \emph{{Stability structures, motivic Donaldson-Thomas invariants and cluster transformations}},  \href{https://arxiv.org/abs/0811.2435}{{\ttfamily 0811.2435}}.

\bibitem{Gaiotto:2010okc}
D.~Gaiotto, G.W.~Moore and A.~Neitzke, \emph{{Four-dimensional wall-crossing via three-dimensional field theory}}, \href{https://doi.org/10.1007/s00220-010-1071-2}{\emph{Commun. Math. Phys.} {\bfseries 299} (2010) 163} [\href{https://arxiv.org/abs/0807.4723}{{\ttfamily 0807.4723}}].

\bibitem{Dimofte:2009bv}
T.~Dimofte and S.~Gukov, \emph{{Refined, Motivic, and Quantum}}, \href{https://doi.org/10.1007/s11005-009-0357-9}{\emph{Lett. Math. Phys.} {\bfseries 91} (2010) 1} [\href{https://arxiv.org/abs/0904.1420}{{\ttfamily 0904.1420}}].

\bibitem{Dimofte:2009tm}
T.~Dimofte, S.~Gukov and Y.~Soibelman, \emph{{Quantum Wall Crossing in N=2 Gauge Theories}}, \href{https://doi.org/10.1007/s11005-010-0437-x}{\emph{Lett. Math. Phys.} {\bfseries 95} (2011) 1} [\href{https://arxiv.org/abs/0912.1346}{{\ttfamily 0912.1346}}].

\bibitem{Pestun:2007rz}
V.~Pestun, \emph{{Localization of gauge theory on a four-sphere and supersymmetric Wilson loops}}, \href{https://doi.org/10.1007/s00220-012-1485-0}{\emph{Commun. Math. Phys.} {\bfseries 313} (2012) 71} [\href{https://arxiv.org/abs/0712.2824}{{\ttfamily 0712.2824}}].

\bibitem{Gava:2016oep}
E.~Gava, K.S.~Narain, M.N.~Muteeb and V.I.~Giraldo-Rivera, \emph{{$N = 2$ gauge theories on the hemisphere $HS^4$}}, \href{https://doi.org/10.1016/j.nuclphysb.2017.04.007}{\emph{Nucl. Phys. B} {\bfseries 920} (2017) 256} [\href{https://arxiv.org/abs/1611.04804}{{\ttfamily 1611.04804}}].

\bibitem{Gaiotto:2010be}
D.~Gaiotto, G.W.~Moore and A.~Neitzke, \emph{{Framed BPS States}}, \href{https://doi.org/10.4310/ATMP.2013.v17.n2.a1}{\emph{Adv. Theor. Math. Phys.} {\bfseries 17} (2013) 241} [\href{https://arxiv.org/abs/1006.0146}{{\ttfamily 1006.0146}}].

\bibitem{Gaiotto:2014kfa}
D.~Gaiotto, A.~Kapustin, N.~Seiberg and B.~Willett, \emph{{Generalized Global Symmetries}}, \href{https://doi.org/10.1007/JHEP02(2015)172}{\emph{JHEP} {\bfseries 02} (2015) 172} [\href{https://arxiv.org/abs/1412.5148}{{\ttfamily 1412.5148}}].

\bibitem{Aharony:2013hda}
O.~Aharony, N.~Seiberg and Y.~Tachikawa, \emph{{Reading between the lines of four-dimensional gauge theories}}, \href{https://doi.org/10.1007/JHEP08(2013)115}{\emph{JHEP} {\bfseries 08} (2013) 115} [\href{https://arxiv.org/abs/1305.0318}{{\ttfamily 1305.0318}}].

\bibitem{DelZotto:2022ras}
M.~Del~Zotto and I.~Garc{\'\i}a~Etxebarria, \emph{{Global structures from the infrared}}, \href{https://doi.org/10.1007/JHEP11(2023)058}{\emph{JHEP} {\bfseries 11} (2023) 058} [\href{https://arxiv.org/abs/2204.06495}{{\ttfamily 2204.06495}}].

\bibitem{Argyres:2022kon}
P.C.~Argyres, M.~Martone and M.~Ray, \emph{{Dirac pairings, one-form symmetries and Seiberg-Witten geometries}}, \href{https://doi.org/10.1007/JHEP09(2022)020}{\emph{JHEP} {\bfseries 09} (2022) 020} [\href{https://arxiv.org/abs/2204.09682}{{\ttfamily 2204.09682}}].

\bibitem{Closset:2023pmc}
C.~Closset and H.~Magureanu, \emph{{Reading between the rational sections: Global structures of 4d $\mathcal{N}=2$ KK theories}}, \href{https://doi.org/10.21468/SciPostPhys.16.5.137}{\emph{SciPost Phys.} {\bfseries 16} (2024) 137} [\href{https://arxiv.org/abs/2308.10225}{{\ttfamily 2308.10225}}].

\bibitem{Garding:2023unh}
E.R.~G{\r{a}}rding, \emph{{Defect groups of class $ \mathcal{S} $ theories from the Coulomb branch}}, \href{https://doi.org/10.1007/JHEP01(2025)148}{\emph{JHEP} {\bfseries 01} (2025) 148} [\href{https://arxiv.org/abs/2311.16224}{{\ttfamily 2311.16224}}].

\bibitem{Arias-Tamargo:2023duo}
G.~Arias-Tamargo and M.~De~Marco, \emph{{Disconnected gauge groups in the infrared}}, \href{https://doi.org/10.1007/JHEP06(2024)050}{\emph{JHEP} {\bfseries 06} (2024) 050} [\href{https://arxiv.org/abs/2312.13360}{{\ttfamily 2312.13360}}].

\bibitem{LeFloch:2015bto}
B.~Le~Floch, \emph{{S-duality wall of SQCD from Toda braiding}}, \href{https://doi.org/10.1007/JHEP10(2020)152}{\emph{JHEP} {\bfseries 10} (2020) 152} [\href{https://arxiv.org/abs/1512.09128}{{\ttfamily 1512.09128}}].

\bibitem{Garozzo:2019xzi}
I.~Garozzo, N.~Mekareeya and M.~Sacchi, \emph{{Duality walls in the 4d $ \mathcal{N} $ = 2 SU(N) gauge theory with $2N$ flavours}}, \href{https://doi.org/10.1007/JHEP11(2019)053}{\emph{JHEP} {\bfseries 11} (2019) 053} [\href{https://arxiv.org/abs/1909.02832}{{\ttfamily 1909.02832}}].

\bibitem{Bason:2026qbc}
D.~Bason and R.~Valandro, \emph{{A class of half-BPS boundary conditions for $A_{K-1}$ circular quivers}},  \href{https://arxiv.org/abs/2606.03339}{{\ttfamily 2606.03339}}.

\bibitem{Cecotti:2011iy}
S.~Cecotti, C.~Cordova and C.~Vafa, \emph{{Braids, Walls, and Mirrors}},  \href{https://arxiv.org/abs/1110.2115}{{\ttfamily 1110.2115}}.

\bibitem{Acharya:2001dz}
B.S.~Acharya and C.~Vafa, \emph{{On domain walls of N=1 supersymmetric Yang-Mills in four-dimensions}},  \href{https://arxiv.org/abs/hep-th/0103011}{{\ttfamily hep-th/0103011}}.

\bibitem{Dimofte:2011ju}
T.~Dimofte, D.~Gaiotto and S.~Gukov, \emph{{Gauge Theories Labelled by Three-Manifolds}}, \href{https://doi.org/10.1007/s00220-013-1863-2}{\emph{Commun. Math. Phys.} {\bfseries 325} (2014) 367} [\href{https://arxiv.org/abs/1108.4389}{{\ttfamily 1108.4389}}].

\bibitem{Dimofte:2013iv}
T.~Dimofte, M.~Gabella and A.B.~Goncharov, \emph{{K-Decompositions and 3d Gauge Theories}}, \href{https://doi.org/10.1007/JHEP11(2016)151}{\emph{JHEP} {\bfseries 11} (2016) 151} [\href{https://arxiv.org/abs/1301.0192}{{\ttfamily 1301.0192}}].

\bibitem{Benvenuti:2026usm}
S.~Benvenuti, R.~Comi, G.~Pedde~Ungureanu, S.~Rota and A.~Shri, \emph{{Universal Planar Abelian Duals for 3d $\mathcal{N}=2$ Unitary CS-SQCD}},  \href{https://arxiv.org/abs/2603.08842}{{\ttfamily 2603.08842}}.

\bibitem{Benvenuti:2026xcv}
S.~Benvenuti, V.~Cagioni, S.~Rota and A.~Shri, \emph{{Universal Planar Abelian Duals for 3d $\mathcal{N}=2$ Symplectic CS-SQCD}},  \href{https://arxiv.org/abs/2605.06776}{{\ttfamily 2605.06776}}.

\bibitem{Hashimoto:2014nwa}
A.~Hashimoto, P.~Ouyang and M.~Yamazaki, \emph{{Boundaries and defects of $ \mathcal{N}=4 $ SYM with 4 supercharges. Part II: Brane constructions and 3d $ \mathcal{N}=2 $ field theories}}, \href{https://doi.org/10.1007/JHEP10(2014)108}{\emph{JHEP} {\bfseries 10} (2014) 108} [\href{https://arxiv.org/abs/1406.5501}{{\ttfamily 1406.5501}}].

\bibitem{Bason:2025qsw}
D.~Bason, \emph{{Boundary conditions of four-dimensional N=2 gauge theories}}, Ph.D. thesis, Trieste U, 2025.

\bibitem{Bason:2025zpy}
D.~Bason, C.~Copetti, L.~Di~Pietro and Z.~Ji, \emph{{$\mathcal{N}=2$ Super Yang-Mills in AdS$_4$ and $F_{\text{AdS}}$-maximization}},  \href{https://arxiv.org/abs/2506.05162}{{\ttfamily 2506.05162}}.

\bibitem{Bason:2025sxb}
D.~Bason, C.~Copetti, L.~Di~Pietro, Z.~Ji and S.~Komatsu, \emph{{F-theorem for Quantum Field Theories in Anti-de Sitter Space}},  \href{https://arxiv.org/abs/2512.18392}{{\ttfamily 2512.18392}}.

\bibitem{Erdmenger:2002ex}
J.~Erdmenger, Z.~Guralnik and I.~Kirsch, \emph{{Four-dimensional superconformal theories with interacting boundaries or defects}}, \href{https://doi.org/10.1103/PhysRevD.66.025020}{\emph{Phys. Rev. D} {\bfseries 66} (2002) 025020} [\href{https://arxiv.org/abs/hep-th/0203020}{{\ttfamily hep-th/0203020}}].

\bibitem{Cordova:2016emh}
C.~Cordova, T.T.~Dumitrescu and K.~Intriligator, \emph{{Multiplets of Superconformal Symmetry in Diverse Dimensions}}, \href{https://doi.org/10.1007/JHEP03(2019)163}{\emph{JHEP} {\bfseries 03} (2019) 163} [\href{https://arxiv.org/abs/1612.00809}{{\ttfamily 1612.00809}}].

\bibitem{Argyres:1995jj}
P.C.~Argyres and M.R.~Douglas, \emph{{New phenomena in SU(3) supersymmetric gauge theory}}, \href{https://doi.org/10.1016/0550-3213(95)00281-V}{\emph{Nucl. Phys. B} {\bfseries 448} (1995) 93} [\href{https://arxiv.org/abs/hep-th/9505062}{{\ttfamily hep-th/9505062}}].

\bibitem{Hanany:1996ie}
A.~Hanany and E.~Witten, \emph{{Type IIB superstrings, BPS monopoles, and three-dimensional gauge dynamics}}, \href{https://doi.org/10.1016/S0550-3213(97)00157-0}{\emph{Nucl. Phys. B} {\bfseries 492} (1997) 152} [\href{https://arxiv.org/abs/hep-th/9611230}{{\ttfamily hep-th/9611230}}].

\bibitem{Closset:2025akk}
C.~Closset, W.~Gu, O.~Khlaif, E.~Sharpe, H.~Zhang and H.~Zou, \emph{{Schubert line defects in 3d GLSMs. Part I. Complete flag manifolds and quantum Grothendieck polynomials}}, \href{https://doi.org/10.1007/JHEP04(2026)074}{\emph{JHEP} {\bfseries 04} (2026) 074} [\href{https://arxiv.org/abs/2512.19802}{{\ttfamily 2512.19802}}].

\bibitem{Closset:2026bnk}
C.~Closset, W.~Gu, O.~Khlaif, E.~Sharpe, H.~Zhang and H.~Zou, \emph{{Schubert line defects in 3d GLSMs. Part II. Partial flag manifolds and parabolic quantum polynomials}}, \href{https://doi.org/10.1007/JHEP04(2026)075}{\emph{JHEP} {\bfseries 04} (2026) 075} [\href{https://arxiv.org/abs/2601.18881}{{\ttfamily 2601.18881}}].

\bibitem{Argyres:1994xh}
P.C.~Argyres and A.E.~Faraggi, \emph{{The vacuum structure and spectrum of N=2 supersymmetric SU(n) gauge theory}}, \href{https://doi.org/10.1103/PhysRevLett.74.3931}{\emph{Phys. Rev. Lett.} {\bfseries 74} (1995) 3931} [\href{https://arxiv.org/abs/hep-th/9411057}{{\ttfamily hep-th/9411057}}].

\bibitem{Douglas:1995nw}
M.R.~Douglas and S.H.~Shenker, \emph{{Dynamics of SU(N) supersymmetric gauge theory}}, \href{https://doi.org/10.1016/0550-3213(95)00258-T}{\emph{Nucl. Phys. B} {\bfseries 447} (1995) 271} [\href{https://arxiv.org/abs/hep-th/9503163}{{\ttfamily hep-th/9503163}}].

\bibitem{Alim:2011kw}
M.~Alim, S.~Cecotti, C.~Cordova, S.~Espahbodi, A.~Rastogi and C.~Vafa, \emph{{$\mathcal{N} = 2$ quantum field theories and their BPS quivers}}, \href{https://doi.org/10.4310/ATMP.2014.v18.n1.a2}{\emph{Adv. Theor. Math. Phys.} {\bfseries 18} (2014) 27} [\href{https://arxiv.org/abs/1112.3984}{{\ttfamily 1112.3984}}].

\bibitem{Aharony:1997gp}
O.~Aharony, \emph{{IR duality in d = 3 N=2 supersymmetric USp(2N(c)) and U(N(c)) gauge theories}}, \href{https://doi.org/10.1016/S0370-2693(97)00530-3}{\emph{Phys. Lett. B} {\bfseries 404} (1997) 71} [\href{https://arxiv.org/abs/hep-th/9703215}{{\ttfamily hep-th/9703215}}].

\bibitem{Aharony:2014uya}
O.~Aharony and D.~Fleischer, \emph{{IR Dualities in General 3d Supersymmetric SU(N) QCD Theories}}, \href{https://doi.org/10.1007/JHEP02(2015)162}{\emph{JHEP} {\bfseries 02} (2015) 162} [\href{https://arxiv.org/abs/1411.5475}{{\ttfamily 1411.5475}}].

\bibitem{Benini:2011mf}
F.~Benini, C.~Closset and S.~Cremonesi, \emph{{Comments on 3d Seiberg-like dualities}}, \href{https://doi.org/10.1007/JHEP10(2011)075}{\emph{JHEP} {\bfseries 10} (2011) 075} [\href{https://arxiv.org/abs/1108.5373}{{\ttfamily 1108.5373}}].

\bibitem{Closset:2023vos}
C.~Closset and O.~Khlaif, \emph{{Twisted indices, Bethe ideals and 3d $ \mathcal{N} $ = 2 infrared dualities}}, \href{https://doi.org/10.1007/JHEP05(2023)148}{\emph{JHEP} {\bfseries 05} (2023) 148} [\href{https://arxiv.org/abs/2301.10753}{{\ttfamily 2301.10753}}].

\bibitem{Spiridonov:2014cxa}
V.P.~Spiridonov and G.S.~Vartanov, \emph{{Vanishing superconformal indices and the chiral symmetry breaking}}, \href{https://doi.org/10.1007/JHEP06(2014)062}{\emph{JHEP} {\bfseries 06} (2014) 062} [\href{https://arxiv.org/abs/1402.2312}{{\ttfamily 1402.2312}}].

\bibitem{Giacomelli:2023zkk}
S.~Giacomelli, C.~Hwang, F.~Marino, S.~Pasquetti and M.~Sacchi, \emph{{Probing bad theories with the dualization algorithm. Part I}}, \href{https://doi.org/10.1007/JHEP04(2024)008}{\emph{JHEP} {\bfseries 04} (2024) 008} [\href{https://arxiv.org/abs/2309.05326}{{\ttfamily 2309.05326}}].

\bibitem{Comi:2025zwu}
R.~Comi, S.~Garavaglia, S.~Giacomelli, S.~Pasquetti and P.~Singh, \emph{{Breaking bad theories of class $ \mathcal{S} $}}, \href{https://doi.org/10.1007/JHEP05(2026)075}{\emph{JHEP} {\bfseries 05} (2026) 075} [\href{https://arxiv.org/abs/2508.21071}{{\ttfamily 2508.21071}}].

\bibitem{Kapustin:1999ha}
A.~Kapustin and M.J.~Strassler, \emph{{On mirror symmetry in three-dimensional Abelian gauge theories}}, \href{https://doi.org/10.1088/1126-6708/1999/04/021}{\emph{JHEP} {\bfseries 04} (1999) } [\href{https://arxiv.org/abs/hep-th/9902033}{{\ttfamily hep-th/9902033}}].

\bibitem{Benini:2010uu}
F.~Benini, Y.~Tachikawa and D.~Xie, \emph{{Mirrors of 3d Sicilian theories}}, \href{https://doi.org/10.1007/JHEP09(2010)063}{\emph{JHEP} {\bfseries 09} (2010) 063} [\href{https://arxiv.org/abs/1007.0992}{{\ttfamily 1007.0992}}].

\bibitem{Bottini:2021vms}
L.E.~Bottini, C.~Hwang, S.~Pasquetti and M.~Sacchi, \emph{{4d S-duality wall and SL(2, \ensuremath{\mathbb{Z}}) relations}}, \href{https://doi.org/10.1007/JHEP03(2022)035}{\emph{JHEP} {\bfseries 03} (2022) 035} [\href{https://arxiv.org/abs/2110.08001}{{\ttfamily 2110.08001}}].

\bibitem{Hwang:2021ulb}
C.~Hwang, S.~Pasquetti and M.~Sacchi, \emph{{Rethinking mirror symmetry as a local duality on fields}}, \href{https://doi.org/10.1103/PhysRevD.106.105014}{\emph{Phys. Rev. D} {\bfseries 106} (2022) } [\href{https://arxiv.org/abs/2110.11362}{{\ttfamily 2110.11362}}].

\bibitem{Comi:2022aqo}
R.~Comi, C.~Hwang, F.~Marino, S.~Pasquetti and M.~Sacchi, \emph{{The SL(2, \ensuremath{\mathbb{Z}}) dualization algorithm at work}}, \href{https://doi.org/10.1007/JHEP06(2023)119}{\emph{JHEP} {\bfseries 06} (2023) 119} [\href{https://arxiv.org/abs/2212.10571}{{\ttfamily 2212.10571}}].

\bibitem{Giveon:2008zn}
A.~Giveon and D.~Kutasov, \emph{{Seiberg Duality in Chern-Simons Theory}}, \href{https://doi.org/10.1016/j.nuclphysb.2008.09.045}{\emph{Nucl. Phys. B} {\bfseries 812} (2009) 1} [\href{https://arxiv.org/abs/0808.0360}{{\ttfamily 0808.0360}}].

\bibitem{Okazaki:2019ony}
T.~Okazaki, \emph{{Mirror symmetry of 3D $\mathcal{N}=4$ gauge theories and supersymmetric indices}}, \href{https://doi.org/10.1103/PhysRevD.100.066031}{\emph{Phys. Rev. D} {\bfseries 100} (2019) 066031} [\href{https://arxiv.org/abs/1905.04608}{{\ttfamily 1905.04608}}].

\bibitem{Hatsuda:2025yzp}
Y.~Hatsuda and T.~Okazaki, \emph{{S-duality of boundary lines in $ \mathcal{N} $ = 4 SYM theories and supersymmetric indices}}, \href{https://doi.org/10.1007/JHEP08(2025)127}{\emph{JHEP} {\bfseries 08} (2025) 127} [\href{https://arxiv.org/abs/2505.14962}{{\ttfamily 2505.14962}}].

\bibitem{Hatsuda:2025zvi}
Y.~Hatsuda and T.~Okazaki, \emph{{Interface line operators in $ \mathcal{N}=4 $ SYM theories and supersymmetric indices}}, \href{https://doi.org/10.1007/JHEP02(2026)104}{\emph{JHEP} {\bfseries 02} (2026) 104} [\href{https://arxiv.org/abs/2510.25168}{{\ttfamily 2510.25168}}].

\bibitem{Hatsuda:2026ysv}
Y.~Hatsuda and T.~Okazaki, \emph{{Quarter-indices for basic ortho-symplectic corners}},  \href{https://arxiv.org/abs/2604.26418}{{\ttfamily 2604.26418}}.

\bibitem{Imamura:2011su}
Y.~Imamura and S.~Yokoyama, \emph{{Index for three dimensional superconformal field theories with general R-charge assignments}}, \href{https://doi.org/10.1007/JHEP04(2011)007}{\emph{JHEP} {\bfseries 04} (2011) 007} [\href{https://arxiv.org/abs/1101.0557}{{\ttfamily 1101.0557}}].

\bibitem{Kapustin:2011jm}
A.~Kapustin and B.~Willett, \emph{{Generalized Superconformal Index for Three Dimensional Field Theories}},  \href{https://arxiv.org/abs/1106.2484}{{\ttfamily 1106.2484}}.

\bibitem{Gadde:2011uv}
A.~Gadde, L.~Rastelli, S.S.~Razamat and W.~Yan, \emph{{Gauge Theories and Macdonald Polynomials}}, \href{https://doi.org/10.1007/s00220-012-1607-8}{\emph{Commun. Math. Phys.} {\bfseries 319} (2013) 147} [\href{https://arxiv.org/abs/1110.3740}{{\ttfamily 1110.3740}}].

\bibitem{Kinney:2005ej}
J.~Kinney, J.M.~Maldacena, S.~Minwalla and S.~Raju, \emph{{An Index for 4 dimensional super conformal theories}}, \href{https://doi.org/10.1007/s00220-007-0258-7}{\emph{Commun. Math. Phys.} {\bfseries 275} (2007) 209} [\href{https://arxiv.org/abs/hep-th/0510251}{{\ttfamily hep-th/0510251}}].

\bibitem{Romelsberger:2007ec}
C.~Romelsberger, \emph{{Calculating the Superconformal Index and Seiberg Duality}},  \href{https://arxiv.org/abs/0707.3702}{{\ttfamily 0707.3702}}.

\bibitem{Alim:2011ae}
M.~Alim, S.~Cecotti, C.~Cordova, S.~Espahbodi, A.~Rastogi and C.~Vafa, \emph{{BPS Quivers and Spectra of Complete N=2 Quantum Field Theories}}, \href{https://doi.org/10.1007/s00220-013-1789-8}{\emph{Commun. Math. Phys.} {\bfseries 323} (2013) 1185} [\href{https://arxiv.org/abs/1109.4941}{{\ttfamily 1109.4941}}].

\bibitem{Cecotti:2012gh}
S.~Cecotti and M.~Del~Zotto, \emph{{4d N=2 Gauge Theories and Quivers: the Non-Simply Laced Case}}, \href{https://doi.org/10.1007/JHEP10(2012)190}{\emph{JHEP} {\bfseries 10} (2012) 190} [\href{https://arxiv.org/abs/1207.7205}{{\ttfamily 1207.7205}}].

\bibitem{Intriligator:1996ex}
K.A.~Intriligator and N.~Seiberg, \emph{{Mirror symmetry in three-dimensional gauge theories}}, \href{https://doi.org/10.1016/0370-2693(96)01088-X}{\emph{Phys. Lett. B} {\bfseries 387} (1996) 513} [\href{https://arxiv.org/abs/hep-th/9607207}{{\ttfamily hep-th/9607207}}].

\bibitem{Hanany_1997}
A.~Hanany and E.~Witten, \emph{{Type IIB superstrings, BPS monopoles, and three-dimensional gauge dynamics}}, \href{https://doi.org/10.1016/s0550-3213(97)80030-2}{\emph{Nuclear Physics B} {\bfseries 492} (1997) 152}.

\bibitem{Aprile:2018oau}
F.~Aprile, S.~Pasquetti and Y.~Zenkevich, \emph{{Flipping the head of $T[SU(N)]$: mirror symmetry, spectral duality and monopoles}}, \href{https://doi.org/10.1007/JHEP04(2019)138}{\emph{JHEP} {\bfseries 04} (2019) 138} [\href{https://arxiv.org/abs/1812.08142}{{\ttfamily 1812.08142}}].

\bibitem{Jafferis:2010un}
D.L.~Jafferis, \emph{{The Exact Superconformal R-Symmetry Extremizes Z}}, \href{https://doi.org/10.1007/JHEP05(2012)159}{\emph{JHEP} {\bfseries 05} (2012) 159} [\href{https://arxiv.org/abs/1012.3210}{{\ttfamily 1012.3210}}].

\bibitem{Aharony:2013dha}
O.~Aharony, S.S.~Razamat, N.~Seiberg and B.~Willett, \emph{{3d dualities from 4d dualities}}, \href{https://doi.org/10.1007/JHEP07(2013)149}{\emph{JHEP} {\bfseries 07} (2013) 149} [\href{https://arxiv.org/abs/1305.3924}{{\ttfamily 1305.3924}}].

\end{thebibliography}\endgroup
\end{document}